*Antiferroelectric Fluids with Hierarchical Polar Organization*

Hiroya Nishikawa*, Hina Saito, Tetsu Katayama, Atsuko Nihonyanagi, Dennis Kwaria, Hiroyuki Inuzuka, Junxi Liu, Go Watanabe and Fumito Araoka*

H. Nishikawa, A. Nihonyanagi, D. Kwaria, H. Inuzuka, J. Liu, F. Araoka

RIKEN Center for Emergent Matter Science, 2-1 Hirosawa, Wako, Saitama 351-0198, Japan

E-mail: hiroya.nishikawa@riken.jp; fumito.araoka@riken.jp

T. Katayama, H. Saito, G. Watanabe

School of Frontier Engineering, Kitasato University, 1-15-1 Minami-ku, Kitazato, Sagamihara, Kanagawa 252-0373, Japan

**Abstract**

The discovery of the ferroelectric nematic ($N_F$) phase has accelerated research on fluids exhibiting polar order along the molecular long axis. However, their antiferroelectric (AFE) counterparts remain less explored. Here, we investigate how terminal alkyl-chain elongation competes with polar core–core interactions in a homologous series. This series exhibits modulated AFE order accompanied by SmC-type cybotactic correlations, with the AFE phase persisting in homologs bearing terminal alkyl chains of up to 17 carbon atoms. Three distinct phase sequences on cooling lead to two types of two-dimensional (2D) AFE lattices. Multiple sharp X-ray diffraction (XRD) peaks reveal rectangular and oblique lattice structures, supporting their classification as AFE columnar phases. XRD analysis and molecular dynamics (MD) simulations support a hierarchical structural model in which locally aligned polar molecules occupy thin segments of finite width within polar columns. The columns occupy the sites of a 2D lattice, with neighboring columns polarized in opposite directions. This model provides a structural framework for understanding the distinct precursor phase sequences and links packing between columns to the resulting AFE superstructures.

## 1. Introduction

Local antiparallel order is a fundamental mode of organization in condensed matter. Representative examples include antiparallel spin order in collinear antiferromagnets and antipolar dipole order in molecular antiferroelectric (AFE) crystals.[1,2] Ferroelectrics (FEs) develop macroscopic spontaneous polarization through long-range ordering of local dipoles. In AFEs, antipolar ordering of local dipoles cancels the macroscopic polarization. In molecular AFE crystals, dipolar arrangements can range from collinear to noncollinear, but their spatial organization remains constrained by the underlying crystal lattice.[3,4] By contrast, soft matter can reorganize both dipole orientations and the positional order without such constraints. This flexibility allows AFE structures to emerge through coupling among orientational, positional, and polar order.

Liquid crystals (LCs) provide a soft-matter platform for AFE order, with field-induced transition from antipolar to polar states observed in several systems. Most of the conventional AF and AFE LCs are based on lateral polar order, in which the local polarization orients perpendicular to the molecular long axis. In chiral rod-like smectics, FE, ferrielectric, and AFE states arise from interlayer correlations of in-layer polarization generated by molecular tilt and chirality.[5,6] In bent-core LCs, lateral polar order develops within layers through molecular shape and directed packing. The resulting assemblies can evolve into modulated, ribbon, or columnar structures.[3,7–10] Terminal alkyl-chain length alters core–chain segregation and the balance of packing interactions. It thereby influences smectic, modulated, and two-dimensional (2D) columnar structures, as well as FE/AFE correlations (Fig. 1a).

Unlike these conventional systems with lateral polar order, the ferroelectric nematic ($N_F$) phase exhibits longitudinal polar order along the molecular long axis.[11–14] Subsequently, other hierarchical phases based on such local longitudinal polar order have been reported, including the $N_X$[15] (also termed $N_S$[16,17] or $SmZ_A$[18]) phase and the $SmA_{AF}$ phase[19] with antiparallel polar modulation (Fig. 1b). In rod-like mesogens with longitudinal polarity, terminal-chain elongation promotes lamellar organization and shifts the balance between polar-core interactions and terminal-chain packing. These changes can lead to competition between polar and positional order. Recent studies on related longitudinally polar mesogens with terminal alkyl chains extending up to $n$ = 10 have reported modulated AFE smectic phases characterized by satellite reflections, with the resulting structures described using 2D or even three-dimensional (3D) lattices.[20] However, how local polar motifs relate to the resulting long-range structures remains unclear.

Here, a longitudinal polar system, ***n*-012-CN** homologous series, which shares a common highly polar core, was studied by systematically varying the terminal chain over a wide range from $n$ = 1 to 20 (Fig. 1d). Local polar order persisted up to $n$ = 17, and notably, two distinct 2D AFE columnar phases were observed for $n$ = 9–17 across three cooling phase sequences, given as Pathway I–III (Fig. 1c,e). Their X-ray diffraction (XRD) patterns differed from those of previously reported modulated smectic phases, with multiple sharp reflections consistently indexed to rectangular or oblique 2D lattices. In Pathway I, a rectangular AFE columnar phase with local SmC-type packing ($Col_{r_AF}$) emerged from the $N_{X_cybC}$ phase with SmC-type cybotactic correlations. Pathway II also reached the same $Col_{r_AF}$ phase through a reentrant N (reN) sequence with SmC-type cybotactic correlations. In Pathway III, observed in the longest polar homologs, an oblique AFE columnar phase ($Col_{ob_AF}$) developed directly from the antiferroelectric smectic A ($SmA_{AF}$) phase. However, the $SmA_{AF}$ precursor alone does not determine the local packing within the $Col_{ob_AF}$ structure. Our MD model suggests tilted, SmC-type local packing within the polar columns of this phase. To track structural and polar evolution across these phase sequences, we combined X-ray diffraction (XRD) analysis and molecular dynamics (MD) simulations with optical, polarization, dielectric, and nonlinear-optical measurements. These results support a hierarchical model in which thin, finite-width regions of locally polar packing constitute segments of columns arranged on distinct 2D AFE lattices.

## 2. Results and Discussion

### 2.1. Molecular Design

In longitudinal polar order of LCs, terminal alkyl-chain elongation generally destabilizes the $N_F$ phase and related polar phases. Terminal chain elongation alters the balance in the electrostatic and dispersive interactions among polar mesogens, and this alteration may destabilize the parallel alignment preferred for macroscopic FE order. For example, in the well-known $N_F$ LC, the DIO system, its $N_F$ phase almost vanishes when the terminal chain exceeds four carbon atoms.[21] Similarly, another well-known $N_F$ LC, the RM734 system, also exhibits such destabilization of the $N_F$ phase for axial terminal chains longer than three carbon atoms.[13]

However, this trend is in fact not universal. In the **MUT_JK10n** series, which has a relatively long mesogenic core and a large molecular dipole moment, terminal chain elongation stabilizes lamellar polar phases. These phases include $SmA_F$ and $SmC_F$, and their formation is accompanied by nanosegregation between the rigid polar core and flexible alkyl chains.[22] Similarly, the more fluorinated **SR-n-Re** series retains the $N_F$ phase even up to $n = 6$.[23] Such behaviors suggest that a highly polar mesogenic core can offset and partially reduce the destabilizing effect of terminal chain elongation. Thus, the role of terminal chain elongation might not be just diluting polarity and changes with the balance among polar-core interactions, dispersion interactions, alkyl-chain packing, and the nanosegregation effect.

Molecular dipole moment alone is also insufficient to explain the formation of longitudinal polar LC phases. Recent studies have highlighted the importance of molecular charge distribution and intermolecular association.[24–26] In particular, Mandle and co-workers proposed a pairing-based model linking both ferroelectric nematic and ferroelectric smectic order to a preference for syn-parallel molecular pairing.[26] A preference for a specific lateral pairing geometry imposes geometrical constraints that promote lamellar order. By contrast, the availability of multiple syn-parallel pairing modes favors ferroelectric nematic order without long-range positional order. Terminal chain elongation may therefore modify the balance among these local pairing motifs and alter molecular packing, rather than simply reducing longitudinal polar interactions.

In the previously reported ***n*-012** series, the $N_F$ phase persists up to $n = 3$[27] but disappears at $n = 4$ (not shown here). In contrast, in the slightly modified ***n*-012-CN** series with cyano substitution at the opposite end to the terminal chain, the $N_F$ phase remains stable beyond $n = 4$. This result may reflect the increased longitudinal polarity introduced by the cyano group, together with changes in molecular charge topology and local intermolecular association. DFT calculations yielded molecular dipole moments of ca. 13.5–14.0 D across the ***n*-012-CN** series, oriented at ca. 16–19° to the molecular long axis (Supplementary Table 1). Therefore, we decided to thoroughly examine the ***n*-012-CN** series over a wide range of terminal chain length, $n$ = 1–20, to elucidate how local polar order evolves beyond the $N_F$ stability range.

We determined the phase transition temperatures and sequences of each synthesized homolog using polarized optical microscopy (POM), differential scanning calorimetry (DSC), and XRD under magnetic-field application. The corresponding experimental data are shown in Supplementary Figs. 1–6, 7–11, and 12–52, respectively. Broadband dielectric spectroscopy (BDS) was also performed across the homologous series to track the dielectric response accompanying phase evolution (Supplementary Figs. 53–56). Variation of terminal chain length produced several distinct phase regimes, revealing a complex evolution rather than the simple destabilization of the $N_F$ phase.

**2.2. Development of Cybotactic Nematic Order**

Homologs of ***n*-012-CN** with short terminal alkyl chains ($n$ = 1–3) exhibited the $N_F$ phase over a wide temperature range (e.g., $\Delta T \approx 120$ °C for $n$ = 1). This phase showed $\pi$-twist walls[28] (Supplementary Fig. 1) and banded textures in antiparallel-rubbed (AP) and syn-parallel-rubbed (SP) cells, respectively. With increasing the terminal chain length, the $N_F$ phase was immediately destabilized. Instead, the $N_X$ phase was progressively stabilized. The $N_X$ phase exhibited a typical zigzag texture[16] on POM (Fig. 2a), equatorial XRD spots with $q_M$ perpendicular to the magnetically aligned director **n**, and an AFE response in polarization reversal current (PRC) measurements. These observations indicate that chain elongation promotes a transformation from the synparallel polar order of the $N_F$ phase to the antiparallel polar order of the $N_X$ phase.

Small-angle scattering spots in the **n** direction were observed in the $N_X$ phase of the short terminal-chain homologs $n$ = 1–2 (Supplementary Fig. 12b for $n$ = 2). For $n$ = 3–8, however, the small-angle scattering spots were diffused, and then, further cooling yielded a split pattern with four symmetric spots throughout the $N_X$–$N_F$ region upon cooling (Fig. 2c–e). Such split patterns observed in nematic phases are characteristic of SmC-type cybotactic correlations, indicating short-range, tilted layer-like packing.[29,30] Hereafter, the $N_X$ and $N_F$ phases with these correlations are denoted as $N_{X_cybC}$ and $N_{F_cybC}$ phases, respectively. To the best of our knowledge, the $N_{X_cybC}$ phase has not been reported previously. The POM textures of the $N_{F_cybC}$ and $N_{X_cybC}$ phases are not so different from those of the original $N_F$ and $N_X$ phases. Hence, these cybotactic phases are hardly distinguished by POM observation without XRD analysis. For $n$ = 6–8 upon cooling in the $N_{X_cybC}$ phase, defect lines sluggishly appeared, followed by a transition to the lower temperature $N_{F_cybC}$ phase (Fig. 2b). This $N_{F_cybC}$ phase differs slightly from those observed in the shorter-chain homologs, as evidenced by the sluggish defect growth described above and the slightly larger transition enthalpy compared with that of the $N_{X_cybC}$–$N_{F_cybC}$ transition for $n$ = 3–5 (Fig. 3e). Thus, the $N_{F_cybC}$ phase in $n$ = 6–8 is further classified and denoted as the $N_F{}'_{_cybC}$ phase.

### 2.3. Evolution of Local SmC-Type Polar Packing

In this section, to characterize the cybotactic nematic phases, $N_{X_cybC}$ and $N_F{}'_{cybC}$, we focus on the representative homolog, **7-012-CN**. Figs. 2c,d show the 2D XRD patterns of the $N_{X_cybC}$ phase at higher (100 °C) and lower (76 °C) temperatures, respectively (see also Supplementary Fig. 17). On cooling, the small-angle equatorial spots at $q_M$ progressively sharpened (Fig. 2f), with a corresponding periodicity $w$ of approximately 4.7 nm (Fig. 2g), which was considerably smaller than the similar spot at ca. 8.8 nm, corresponding to the electron-density modulation period $w_M$ of the conventional $N_X$ phase of DIO (Fig. 2h).[18] We propose a model for the $N_{X_cybC}$ phase as shown in Fig. 2i, in which localized regions of tilted, polar packing (cybotactic clusters) are embedded in an antipolar $N_X$-like matrix.

From the 2D XRD patterns, the molecular tilt angle was estimated as $\beta = \Delta\chi/2$, where $\Delta\chi$ is the angular separation between the splitting spots (Fig. 2e). In the cybotactic model, this angle describes molecular tilt relative to the local layer normal. Fig. 3a shows temperature dependences of the layer spacing $d$ (= 2π/$q$) and the molecular tilt angle $\beta$, both of which were estimated from the same 2D XRD pattern. On cooling through the $N_{cybC}$– $N_{X_cybC}$ phase transition, $\beta$ increased while $d$ decreased. However, $\beta$ reached a plateau of ca. 40°, while $d$ kept decreasing, inconsistently with the generally used relation for SmC phases, $d = L_{mol}\cos\beta$, where $L_{mol}$ is the molecular length. As one possible explanation, we speculated that this inconsistency arises from offset molecular packing, which increases the layer thickness. To discuss this point, we introduced an apparent molecular length $L_{app}$, defined as $L_{app} = d/\cos\beta$ (Fig. 3b), and an offset parameter $\delta$, defined as $\delta = [\{L_{app}/L_{mol}\} - 1] \times 100$. Positive and negative $\delta$ values indicate that $L_{app}$ exceeds or falls below $L_{mol}$, respectively, corresponding to layer thickening or thinning relative to the simple tilt projection. For $n$ = 7, $L_{app}$ exceeded $L_{mol}$ by approximately 20–25% in the high-temperature $N_{cybC}$ phase, consistent with substantial longitudinal offsets within the locally tilted layers. On cooling, $L_{app}$ decreased, suggesting a progressive reduction in these offsets. At the $N_{X_cybC}$–$N_{F'_cybC}$ transition, $L_{app}$ dropped abruptly below $L_{mol}$, so $\delta$ became negative. This change suggests that local molecular packing within the tilted layers reorganizes during the AFE-to-FE transition.

The same analysis was applicable to $n$ = 5, 6, and 8. For $n$ = 5, $\delta$ remained positive throughout the $N_{X_cybC}$ to $N_{F_cybC}$ phases (Figs. 3c,d), while $n$ = 6 and 8 presented the same trend as $n$ = 7 (Supplementary Figs. 47 and 48). The crossover point from molecular offset to inset therefore is located between $n$ = 5 and 6. The larger transition enthalpy for n = 6–8 (Fig. 3e) is consistent with a more substantial rearrangement of local packing at the $N_{X_cybC}$–$N_{F'_cybC}$ transition.

Fig. 3f summarizes the evolution of molecular offset/inset in the two distinct regions of terminal-chain length, $n$=3–5 and 6–8. For $n$ = 3–5, the apparent molecular length in the $N_{F_cybC}$ phase approaches the molecular length ($\delta \approx 0$). This suggests that enthalpic contributions associated with the core-to-core attraction and the terminal chain organization can overcome the energetic penalty of reducing the offset. By contrast, $n$ = 6–8 undergoes a discontinuous transition to the molecular inset state ($\delta < 0$), possibly meaning molecular interpenetration between the layers or folding of the terminal chains in the $N_{F}'{}_{_cybC}$ phase.

The $N_{F}'{}_{_cybC}$ phase was no longer observed at $n = 9$, whereas the $N_{X_cybC}$ phase persisted up to $n = 12$ (Supplementary Figs. 13–19, 21, 23, and 26). This persistence indicates enhanced stability of the antiparallel polar order at longer terminal chain lengths, and homologs with $n \geq 9$ developed more complex ordered phases at lower temperatures as will be mentioned below.

## 2.4. Local Packing Evolution into 2D Order

### 2.4.1. 12-012-CN: The $N_{cybC}$–$N_{X_cybC}$–$^{2D}X$ Sequence

On cooling, homologs with $n = 9$–12 underwent a transition from the $N_{X_cybC}$ phase to a phase with greater positional order. Before the transition to this complex phase, a zigzag texture appeared in the $N_{X_cybC}$ phase and became blockier toward the transition point (Fig. 4a,b). For detailed analysis, ***12*-012-CN** was chosen because of its sufficiently wide temperature range and better alignment characteristics of this complex phase. Fig. 4c,d shows the 2D XRD patterns for the $N_{X_cybC}$ phase and the complex phase, and the latter exhibits multiple discrete spots at specific azimuthal angles, indicating a long-range 2D lattice structure, temporarily denoted as the $^{2D}X$ phase. Interestingly, the behavior of $w_M$ along $q_M$ differed from those of the same phase in the shorter homolog ***7*-012-CN**, whose $w_M$ remained nearly constant within the $N_{X_cybC}$ phase and increased as the transition to the lower temperature $N_{F}'{}_{_cybC}$ phase was approached (Fig. 2f). On the other hand, $w_M$ of ***12*-012-CN** decreased monotonically on cooling and dropped at the transition to the $^{2D}X$ phase (Fig. 4e). Although the $^{2D}X$ phase has a complex lattice structure as will be described in Section 2.5, the split scattering characteristic of SmC-type cybotactic correlations persisted in the $^{2D}X$ phase (Fig. 4d), meaning that the local SmC-type motif is still retained. The $\beta$ and $\delta$ analysis across the $N_{cybC}$–$N_{X_cybC}$–$^{2D}X$ sequence (Figs. 4f; Supplementary Fig. 49) revealed that molecular offset existed at lower temperatures, as positive $\delta$ was retained throughout the sequence. $\delta$ reached ca. 25% at maximum in the $N_{X_cybC}$ phase and then decreased discontinuously to approximately 12–17% on entering the $^{2D}X$ phase. The formation of the long-range 2D lattice order is therefore accompanied by a discontinuous reduction in the molecular offset, while local SmC-type packing persists.

### 2.4.2. ***13*-012-CN: The N–SmA$_d$–reN–reN$_{cybC}$–$^{2D}$X Sequence**

***13*-012-CN** reached a similar $^{2D}$X phase, but through a different phase sequence, N–SmA$_d$–reN–reN$_{cybC}$–$^{2D}$X (Fig. 5g), where the SmA$_d$ phase showed a parabola-like texture on POM and higher-order XRD peaks arising from its lamellar order (Fig. 5a,d; Supplementary Fig. 28). The SmA$_d$ phase exhibits a layer spacing greater than the molecular length, commonly attributed to longitudinally offset antiparallel molecular associations. Both the reN and reN$_{cybC}$ (Fig. 5b) phases are reentrant nematic phases, but the latter shows the split scattering characteristic of SmC-type cybotactic correlations (Fig. 5e). The $^{2D}$X phase of ***13*-012-CN** displayed a broken fan-like texture (Fig. 5c) on POM. Although this texture differed from the blocky zigzag texture of ***12*-012-CN** (Fig. 4b), their 2D XRD patterns were closely similar to each other (Fig. 5f). Although the $\beta$ analysis is not applicable to the N and reN phases, their $d$-spacings provide further insight into local positional correlations. In the N, SmA$_d$, and reN phases, the characteristic spacing d exceeded the molecular length ($L_{mol}$ = 4.12 nm; Supplementary Fig. 50). Although $d$ might not represent a true head-to-tail spacing in the N and reN phases, the persistence of this characteristic spacing suggests that local positional correlations are maintained not only in the SmA$_d$ phase but also in these nematic phases. These correlations may involve antiparallel molecular association.

The polar response also evolved across the phase sequence. No distinct PRC peaks were observed in the N, SmA$_d$, or reN phases (Supplementary Fig. 57b). However, field-induced second-harmonic-generation (SHG) was detectable even in this high-temperature N phase at 8 V $\mu m^{-1}$, and its intensity increased toward reN$_{cybC}$ upon cooling (Supplementary Fig. 58). The dielectric response showed a similar temperature dependence, increasing from the high-temperature N phase to a relative permittivity of ca. 200 in the reN$_{cybC}$ phase (Supplementary Fig. 55).

Reentrant N–SmA–N sequences have long been observed in polar LCs[31] and are generally attributed to temperature-dependent changes in molecular association.[32] In the model proposed by Madhusudana and co-workers, longitudinally offset antiparallel molecular associations stabilize the SmA ($SmA_d$) phase. On cooling, parallel associations become more favorable, destabilizing smectic order and giving rise to a reentrant nematic phase. Further cooling can stabilize the monolayer SmA phase, in which the layer spacing is close to the molecular length.[33,34] ***13*-012-CN** also undergoes a $SmA_d$–reN transition but follows a different path on further cooling, passing through $reN_{cybC}$ before entering the $^{2D}X$ phase. The dielectric and field-induced SHG responses suggest that local polar correlations are already present in reN and persist throughout the reentrant sequence.

**2.5.Structure and Antiferroelectric Order of the Rectangular and Oblique 2D Columnar Phases**

Indexed XRD reflections and AFE switching responses identify two low-temperature AFE columnar phases with distinct lattice geometries: rectangular and oblique. In our model, a polar column segment is a thin slice across a column, with finite lateral extent and a common local direction of polarization. SmC-type local packing refers to molecular tilt relative to a local layer normal, which is distinct from the geometric obliquity of the 2D lattice.

We first analyzed the long-range 2D lattice structure and its polar organization of the low-temperature ($^{2D}X$) phase in ***12*-012-CN**. Cooling from the $N_{X_cybC}$ phase produced several sharp diffraction spots that showed strong alignment under an applied magnetic field. These spots were indexed as (*hk*) = (11), (20), (02), (21), and (12) (Fig. 6a,b; Supplementary Table 5). This indexing defines a rectangular 2D lattice with constants *a* and *b*, which showed similar temperature dependences (Fig. 6c), consistent with a common ordered structure. The indexed reflections had widths comparable to that of (11), supporting their assignment to a common long-range 2D lattice rather than to diffuse scattering from short-range correlations. Diffuse scattering from local SmC-type packing also persisted and partially overlapped the (11) reflection (Fig. 6a). Together with the continuous evolution of $\beta$ described in Section 2.4.1, this persistence indicates that local SmC-type correlations coexist with long-range 2D order. The PRC measurements revealed a distinct AFE response, with switching peaks that evolved continuously across the $N_{X_cybC}$–$^{2D}X$ phase transition (Fig. 6g; Supplementary Fig. 57a).

Therefore, we classify this phase as a rectangular AFE columnar phase, $\mathrm{Col_{r_AF}}$, while proposing $^{\mathrm{2D}r}\mathrm{SmC_{AF}}$ as a complementary designation of its local SmC-type packing. In the proposed model (Fig. 6i), polar columns with local SmC-type packing occupy a rectangular lattice, with neighboring columns polarized in opposite directions. This arrangement yields zero net macroscopic polarization in the absence of an applied field. A similar rectangular lattice and AFE response were observed in ***13*-012-CN** after the N–$\mathrm{SmA_d}$–reN–$\mathrm{reN_{cybC}}$ phase sequence (Fig. 6d,e,h; Supplementary Fig. 57b; Supplementary Table 6). Miscibility with the $^{\mathrm{2D}}$X phase of ***12*-012-CN** further supported their assignment to the same phase (Supplementary Fig. 59).

In the longest polar homologs, $n = 16$ and 17, the $\mathrm{N_{cybC}}$ phase was no longer stabilized. Instead, the $\mathrm{SmA_{AF}}$ phase transformed into a low-temperature phase, as observed by POM and XRD (Figs. 7a–f). The reflections from this phase were indexed using an oblique 2D setting defined by $a$, $b$, and $\gamma \neq 90°$ (Figs. 7g,h,k; Supplementary Tables 9 and 10). Sharp reflections from each indexed orientation domain supported long-range 2D positional order. Mirror-related orientation domains were analyzed separately in reciprocal space (Supplementary Fig. 60). The PRC measurements also confirmed a clear AFE switching response (Fig. 7i,j; Supplementary Fig. 57c). Together, these observations support the assignment of this phase as an oblique AFE columnar phase, $\mathrm{Col_{ob_AF}}$ (Fig. 7l). A similar $\mathrm{Col_{ob_AF}}$ phase occurred in ***17*-012-CN**, and miscibility experiments supported the assignment of both homologs to the same phase family (Supplementary Figs. 6, 41, 51, and 61). Although ***16*-012-CN** and ***17*-012-CN** belong to the same $\mathrm{Col_{ob_AF}}$ phase family, their oblique lattice angles differed substantially: $\gamma = 114.0°$ at 64 °C for ***16*-012-CN** and 104.8° at 59 °C for ***17*-012-CN**. To examine this difference, we applied basis transformations to the original oblique lattices and explored nearly orthogonal unit-cell representations (Supplementary Note 2). The diffraction patterns were consistent with pseudo-simple rectangular and pseudo-centered rectangular representations for ***16*-012-CN** and ***17*-012-CN**, respectively (Supplementary Fig. 62). These representations suggest differences in the relative positions of the columns within the same AFE columnar phase family. In ***12*-012-CN**, diffuse scattering associated with local SmC-type correlations could be distinguished from the sharp lattice reflection.

By contrast, ***16*-012-CN** showed no independently resolvable diffuse component from which the local layer spacing, molecular tilt *β*, or longitudinal offset could be reliably determined by XRD. Consequently, although $Col_{ob_AF}$ develops directly from $SmA_{AF}$, the measurements do not establish whether local SmA-like packing is retained or reorganizes into a tilted SmC-type arrangement during the transition.

The longer-chain homologs followed three pathways to the two AFE columnar lattices. Homologs with $n$ = 9–12 followed Pathway I (Supplementary Figs. 20, 22, and 25; Supplementary Tables 2–5). Homologs with $n$ = 13–15 followed the reentrant Pathway II (Supplementary Figs. 35 and 39; Supplementary Tables 6–8). Homologs with $n$ = 16 and 17 followed Pathway III, in which $SmA_{AF}$ directly preceded $Col_{ob_AF}$. We next examine local molecular packing, particularly in the $Col_{r_AF}$ and $Col_{ob_AF}$ phase, using all-atom molecular dynamics (MD) simulations.

### 2.6. Molecular Basis of Offset Packing in the Rectangular and Oblique AFE Columnar Phases

We performed MD simulations for ***12*-012-CN** and ***16*-012-CN** to examine molecular packing within the proposed polar columns on rectangular and oblique AFE lattices, respectively. Particular attention was given to ***16*-012-CN**, for which neither the local molecular tilt nor the longitudinal offset could be independently determined from XRD analysis. The initial configurations contained polar columns with locally layered molecular packing (Supplementary Fig. 64 and Supplementary Note 4). For ***16*-012-CN**, simulations with 14 molecules per layer yielded similar tilted molecular arrangements from both SmA-like and SmC-like starting configurations (Supplementary Fig. 65). We used the 16-molecule-per-layer model for the analysis in Fig. 8d–f because its equilibrated lattice parameters agreed more closely with experiment. In both compounds, molecules within the equilibrated columns were tilted and longitudinally offset (Fig. 8a,d). Details of the model-selection procedure and the rationale for the chosen molecular populations are provided in Supplementary Notes 4 and 5.

Based on the equilibrated structures, we calculated electron-density profiles along the directions indicated by arrows 1 and 2 in Fig. 8a,d (parallel to the c-axis). Both ***12*-012-CN** and ***16*-**

**012-CN** exhibited periodic density modulations (Fig. 8b,e), consistent with local layering of the polar mesogenic cores. This modulation is also compatible with the longitudinal charge modulation revealed by the 3D electrostatic potential (ESP) analysis (Supplementary Note 3 and Supplementary Fig. 63). The electron-density modulation is weaker in simulations of ***16*-012-CN** than in those of ***12*-012-CN** (Fig. 8b,e), possibly reflecting greater molecular disorder within the layers. Lattice distortion may also contribute to this difference: in ***16*-012-CN**, the observed (31) spacing consistently exceeds the value calculated for the oblique lattice (Supplementary Table 9). Cylindrical distribution function (CDF) analysis (Supplementary Note 4) was carried out to discuss the arrangement of neighboring molecules. The three red spots in Fig. 8c,f suggested that neighboring molecules occupy both side-by-side and longitudinally offset positions within the locally layered packing. Similar offset arrangements have been reported in some $N_F$ phases.[35] For molecules with longitudinally modulated charge distributions, such offsets may lower the electrostatic potential energy, thereby contributing to the stabilization of polar ordering. Overall, these simulations provide further support for the proposed local molecular packing models in the $Col_{r_AF}$ and $Col_{ob_AF}$ phases.

## 2.7. Competition between Polar-Core Interactions and Terminal-Chain Packing

Pathways I and II lead to the Colr_AF phase, whereas Pathway III produces the Colob_AF phase. The XRD and electrical results support a hierarchical model in which locally polar molecules occupy finite-width segments of columns arranged on a 2D lattice. Within this model, the antiparallel polarization of neighboring columns accounts for the AFE response. MD simulations provide supporting molecular detail for the proposed local packing. The experimental data do not resolve the relative timing of column formation and 2D lattice ordering.

Local molecular packing may reflect a balance between favorable intermolecular interactions and the loss of positional and conformational entropy. Variation of the electrostatic potential along the polar core provides a possible basis for favorable contacts between longitudinally displaced molecules. SmC-type cybotactic correlations in both the $N_F$ and $N_X$ phases indicate that local layer-like packing can coexist with different forms of polar order. Such organization may preserve favorable contacts without requiring long-range layer registry. On further cooling, intermolecular interactions may outweigh the entropic cost of ordering polar columns on a 2D lattice. Deviations of the observed spacing from the simple projection $d = L_{mol}\cos\beta$ also indicate that molecular tilt alone is insufficient to describe the local packing. In ***12*-012-CN**, the apparent offset parameter δ decreases discontinuously at the onset of 2D order while the tilted local motif persists, suggesting additional rearrangement of longitudinal molecular positions and/or molecular conformations.

Related competition between polar interactions and lamellar organization has been discussed for modulated smectic phases. Strachan et al. proposed that molecular tilt alleviates unfavorable dipolar interactions and that competition between polar-core interactions and the layering tendency of terminal chains promotes modulated structures.[20] They further reported lateral density modulation in the $SmA_{AF_Mod}$ phase and proposed a three-dimensional arrangement of distorted, finite smectic-layer blocks for the $SmC_{AF_Mod}$ phase. Hobbs et al. interpreted satellite reflections in the $SmA_{AF}$ and $SmC_{AF}$ phases as evidence of one-dimensional lateral density modulation associated with periodic polarization splay and, in the tilted phase, alternating molecular tilt.[36] Similar satellite reflections can also arise in undulated smectic structures.[7] These studies thus describe periodic modulations superimposed on an underlying smectic organization. In contrast, the combined XRD observations, AFE switching response, and MD-supported local packing in the present study support the assignment of the observed phases to rectangular and oblique lattices of polar columns. In ***12*-012-CN**, the coexistence of diffuse SmC-type scattering and sharp lattice reflections reveals distinct levels of local and long-range positional organization. The model assigns diffuse scattering to packing within columns and sharp reflections to positional order between columns.

In ***12*-012-CN**, the inferred rearrangement of local packing at the onset of 2D order may reflect frustration between the packing requirements of the polar cores and terminal chains. A polar column segment could relieve this frustration by limiting the lateral extent of the tilted packing. Extending locally favored tilt and longitudinal displacement throughout a continuous smectic layer could hinder simultaneous optimization of core contacts and chain packing. The lateral edges of a segment may accommodate local adjustments of molecular positions, orientations, and chain conformations. Favorable contacts between columns could stabilize their relative positions on the lattice despite the associated loss of positional entropy.

Terminal-chain length may influence lattice geometry by altering molecular conformation, local packing, and contacts between columns. The MD models of ***12*-012-CN** and ***16*-012-CN** illustrate local arrangements compatible with the rectangular and oblique lattices, respectively. Molecular populations are model inputs. Column segment dimensions have not been measured independently. The free-energy balance governing column width and lattice selection remains unresolved.

## Conclusion

By systematically varying the terminal alkyl-chain length in the ***n*-012-CN** homologous series ($n$ = 1–20), we traced the evolution of longitudinal polar order. We identified three cooling pathways leading to two types of AFE columnar lattices. In Pathways I and II, SmC-type cybotactic correlations developed before the onset of the rectangular AFE columnar phase $Col_{r_AF}$, through direct and reentrant phase sequences, respectively. In Pathway III, $SmA_{AF}$ transformed directly into the oblique AFE columnar phase $Col_{ob_AF}$.

Combined XRD analysis and MD simulations provided a molecular interpretation of these phase sequences and lattice structures. In $Col_{r_AF}$, SmC-type local correlations persisted across the transition. In ***16*-012-CN**, MD simulations initialized with SmA-like and SmC-like arrangements relaxed toward similar oblique arrays of polar columns. These results support tilted, SmC-type local packing in the simulated oblique phase.

Based on these results, we propose a hierarchical structural model in which locally polar molecules occupy thin, finite-width segments of polar columns. The columns occupy rectangular or oblique 2D lattices, with neighboring columns polarized in opposite directions. The resulting framework links terminal-chain length and precursor phase sequences to local molecular packing and the long-range arrangement of polar columns.

## Method

**Polarized optical microscopy (POM).** Polarized optical microscopy was performed on a polarizing microscope (Eclipse LV100 POL, Nikon) with a hot stage (HSC402, INSTEC) on the rotation stage. Unless otherwise noted, the sample temperature was controlled using the INSTEC temperature controller (mk2000, INSTEC).

**Differential scanning calorimetry (DSC).** Differential scanning calorimetry was performed on a calorimeter (DSC3+, Mettler-Toledo). Rate: 10 K $min^{-1}$. Cooling/heating profiles were recorded and analyzed using the Mettler-Toledo STARe software system.

**Broadband Dielectric spectroscopy (BDS).** Dielectric relaxation spectroscopy was performed ranging between 1 Hz and 1 MHz using an impedance/gain-phase analyzer (SI 1260, Solartron Metrology) and a dielectric interface (SI 1296, Solartron Metrology). Prior to starting the measurement of the LC sample, the capacitance of the empty cell was determined as a reference. The temperature was controlled using a homemade heater block and temperature controller. The measurements were automatically performed by the LabVIEW software programming.

**Polarization reversal current (PRC) measurement.** PRC measurements were performed in the temperature range of the polar phase under a triangular-wave *E*-field using a ferroelectricity evaluation system (FCE 10, TOYO Corporation), which is composed of an arbitrary waveform generator (2411B), an IV/QV amplifier (model 6252) and a simultaneous A/D USB device (DT9832).

**Second harmonic generation (SHG) measurements.** The SHG investigation was carried out using a Q-switched DPSS Nd: YAG laser (NL201-SH-ATTN, EKSPLA) at $\lambda$ = 1064 nm with a 5 ns pulse width (pulse energy: ≥ 900 μJ) at 2.5 kHz repetition rate. The fundamental beam was incident on the ITO-coated glass LC cell (22 μm) at an oblique angle of 30° relative to the substrate normal, followed by the detection of the SHG signal using a photomultiplier tube (Hamamatsu Photonics).

Wide- and small-angle X-ray scattering (WAXD, SAXS) analysis. 2D SAXS and WAXD measurements were carried out at BL38B1 in the SPring-8 synchrotron radiation facility (Hyogo, Japan). The samples held in a homemade holder with magnets[27] were measured at a constant temperature using a temperature controller and a hot stage (mk2000, INSTEC) with high-temperature-resistance samarium cobalt magnets (~1 T, Shimonishi Seisakusho Co., Ltd.). The scattering vector q ($q = 4\pi \sin\theta\, \lambda^{-1}$; $2\theta$ and $\lambda$ = scattering angle and wavelength of an incident X-ray beam (1.0 Å)) and position of an incident X-ray beam on the detector were calibrated using several orders of layer diffractions from silver behenate ($d$ = 58.380 Å). The sample-to-detector distances were ca. 240 mm for WAXS and ca. 2600 mm for SAXS. 1D scattering profiles were obtained by integrating the 2D patterns using Igor Pro with the Nika plugin. Integration was performed over a single azimuthal range of $\Delta\varphi$ = 180°, centered on the meridional direction ($\varphi$ = 90°). The rectangular and oblique 2D lattices were indexed using the d-spacings of independent SAXS reflections. The resulting assignments were checked against the spacings and relative azimuthal positions of the remaining reflections, as described in Supplementary Note 1.

**Data availability**

The authors declare that the data supporting the findings of this study are available within the paper and its supplementary information files. All other information is available from the corresponding authors upon reasonable request.

**Acknowledgements**

We are grateful to Dr. H. Masunaga and Dr. N. Shimizu (JASRI/SPring-8) for the use of BL38B1 at SPring-8, and to Dr. H. Koshino (RIKEN, CSRS) for allowing us to use the JNM-ECZ500 (500 MHz, JEOL). We wish to thank Dr. Nogawa (RIKEN, CSRS) for the HRESI-MS measurement. We would like to acknowledge the Hokusai GreatWave Supercomputing Facility (project no. RB240057) at the RIKEN Advanced Center for assistance in computing and communication. This work was partially supported by JSPS KAKENHI (JP22K14594, JP26K01483, JP26K22864: H.N.; JP26K01327: G.W; JP23K17341: F.A.), JST CREST (JPMJCR23O1: F.A.) and JST SICORP EIG CONCERT-Japan (JPMJSC22C3: F.A.). The X-ray scattering measurements were performed at BL38B1 in SPring-8 with the approval of the RIKEN SPring-8 Center (proposal 20250081, 20260055).

**Contributions**

H.N. conceived the project and designed the experiments. H.N. designed the molecules and performed DFT calculations. H.S., T.K. and G.W. performed DFT calculations, MD simulations and CDF analyses. A.N. synthesized and characterized all compounds. H.N. performed DSC, POM, XRD and PRC studies. F.A., J.L. and D.K. assisted with the XRD measurements. H.N. analyzed the XRD data. H.I. and H.N. performed the BDS studies. D.K. and H.N. performed SHG measurements. H.N. and F.A. wrote the manuscript, and all authors approved the final manuscript.

**Corresponding authors**

Correspondence to Hiroya Nishikawa or Fumito Araoka

**Competing interests**

The authors declare no competing interests.

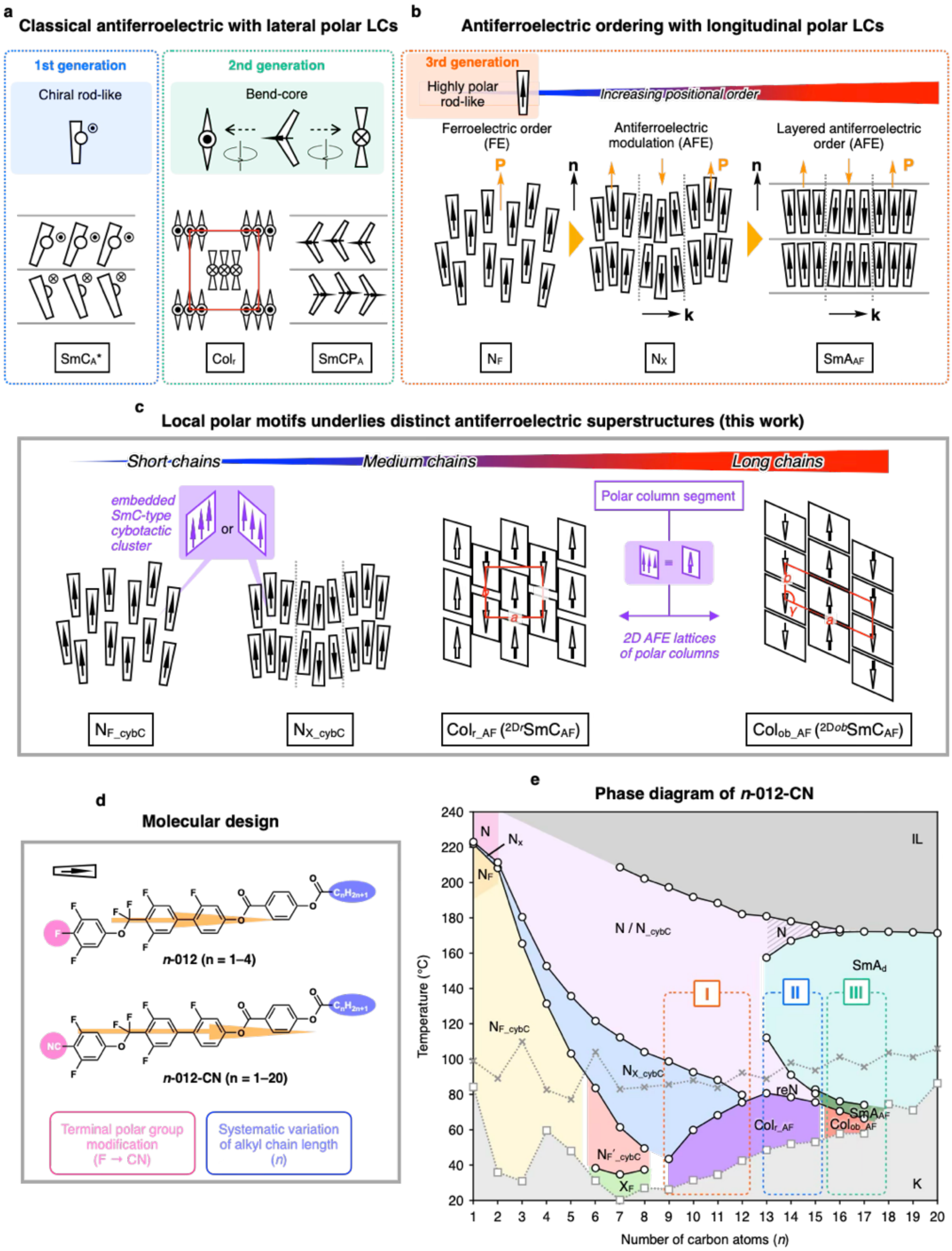

**Fig. 1 | Molecular design and hierarchical pathways to antiferroelectric superstructures.**

(a) Schematic representations of classical AFE LCs with lateral polarization perpendicular to the molecular long axis, including chiral rod-like and bent-core systems. (b) Representative milestones in the development of rod-like LCs with longitudinal polarization along the molecular long axis: ferroelectric nematic ($N_F$) phase, antiferroelectric modulated mesophase ($N_X$), and layered antiferroelectric smectic A ($SmA_{AF}$) phase. (c) Proposed hierarchical structure. Thin, finite-width segments represent local molecular packing within polar columns arranged on 2D AFE lattices. Experimental observations indicate that SmC-type cybotactic correlations persist in the rectangular $Col_{r_AF}$ phase. $Col_{ob_AF}$ develops from $SmA_{AF}$, and MD simulations suggest that the polar columns can accommodate tilted, SmC-type local packing. (d) Molecular structures of the ***n*-012** and ***n*-012-CN** homologous series, highlighting modification of the terminal polar group and systematic variation of the alkyl-chain length n. (e) Phase diagram of the ***n*-012-CN** series as a function of temperature and *n*. The regions corresponding to Pathways I–III are indicated.

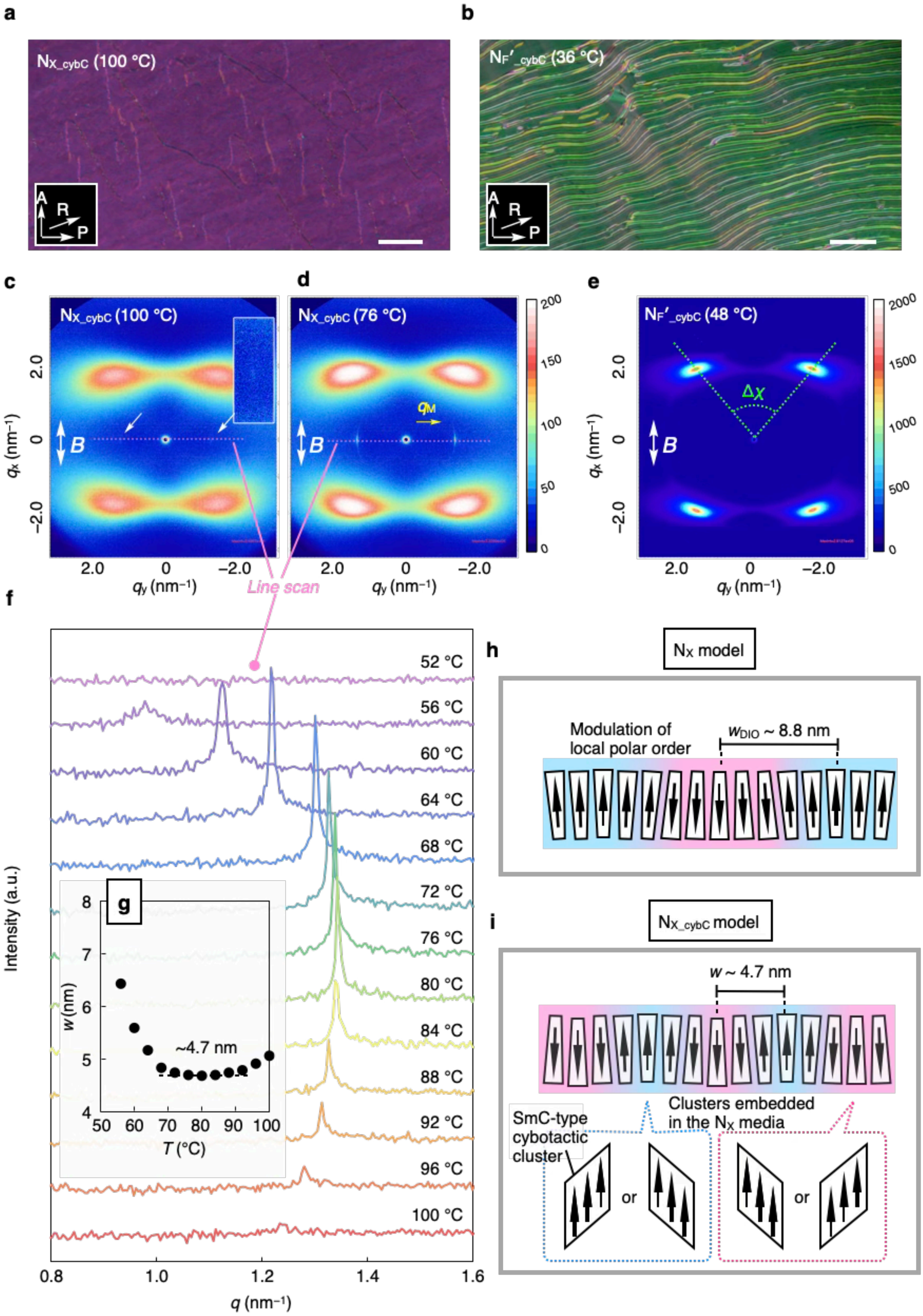

**Fig. 2 | Development of SmC-type cybotactic order in *7*-012-CN.**

(a,b) Polarized optical microscopy images of the $N_{X_cybC}$ phase at 100 °C and the $N_{F}'_{_cybC}$ phase at 36 °C, respectively. (c–e) 2D XRD patterns obtained under magnetic alignment in $N_{X_cybC}$ at 100 and 76 °C and $N_{F}'_{_cybC}$ at 48 °C. The angular separation $\Delta\chi$ of the split small-angle reflections is used to determine the local tilt angle $\beta = \Delta\chi/2$. (f) Temperature-dependent small-angle XRD profiles on cooling. (g) Modulation periodicity w determined from the position of the small-angle reflection. (h) Schematic representation of the conventional $N_X$ structure of DIO, with an electron-density modulation period $w_M$ of ca. 8.8 nm and a full antiferroelectric repeat of ca. 17.5 nm. (i) Proposed $N_{X_cybC}$ structure comprising polar cybotactic clusters with local SmC-type packing embedded within an $N_X$-like antipolar medium, giving a shorter characteristic periodicity of ca. 4.7 nm.

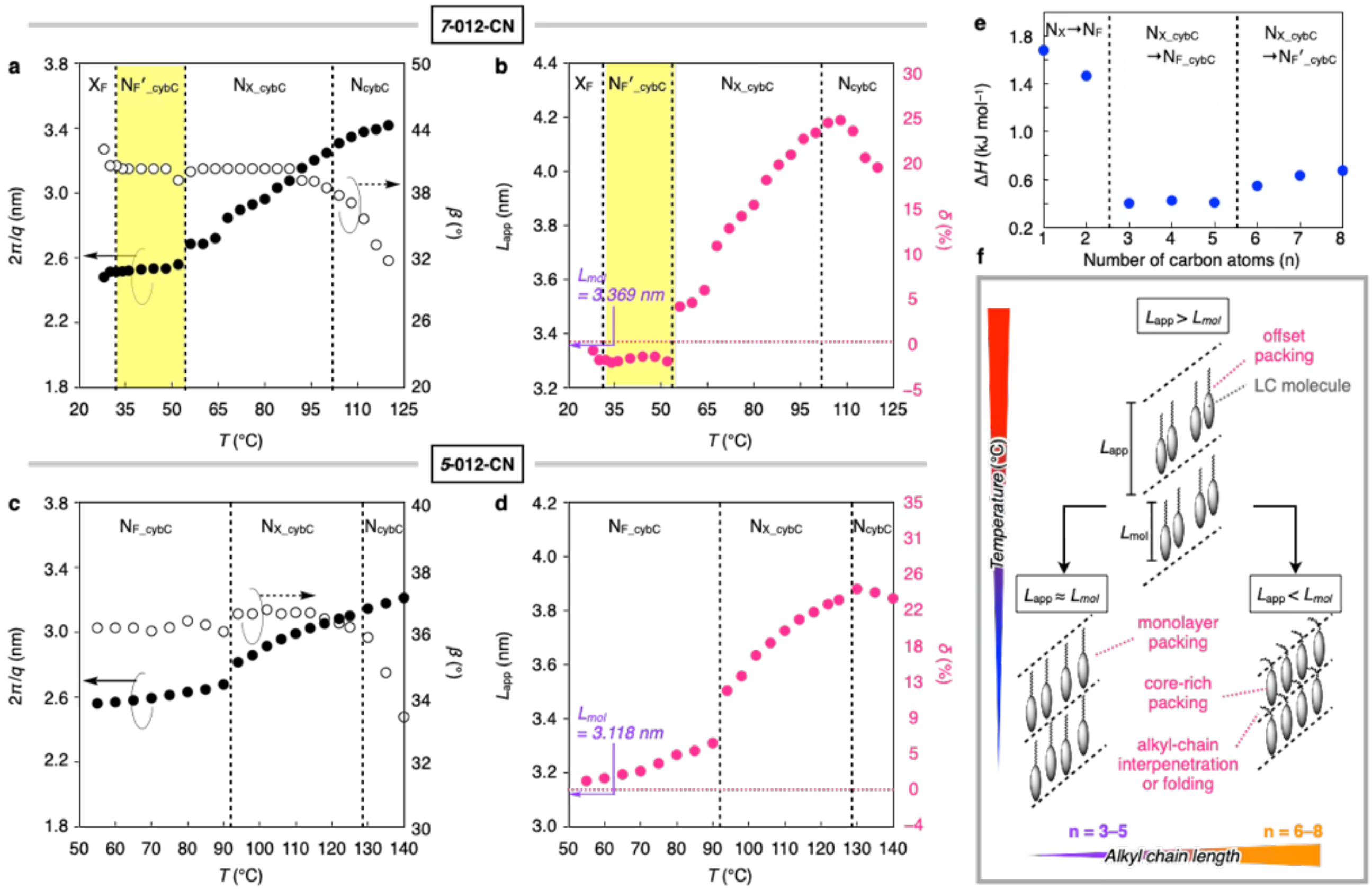

**Fig. 3 | Chain-length-dependent reconstruction of local SmC-type packing.**

(a,b) Temperature dependences of the characteristic spacing d, local tilt angle β, apparent molecular length $L_{app}$, and offset parameter δ for **7-012-CN**. (c,d) Corresponding quantities for **5-012-CN**. The calculated molecular lengths $L_{mol}$ are indicated. (e) Transition enthalpy $\Delta H$ as a function of terminal alkyl-chain length n, showing the larger enthalpy change associated with the $N_{X_cybC}$–$N_{F}'_{_cybC}$ transition for n = 6–8. (f) Schematic representation of the chain-length-dependent evolution of local packing. Cooling reduces the longitudinal offset; shorter homologs approach a monolayer-like arrangement, whereas n = 6–8 enter a contracted packing state with $L_{app} < L_{mol}$. The latter may involve terminal-chain interpenetration and/or conformational rearrangement.

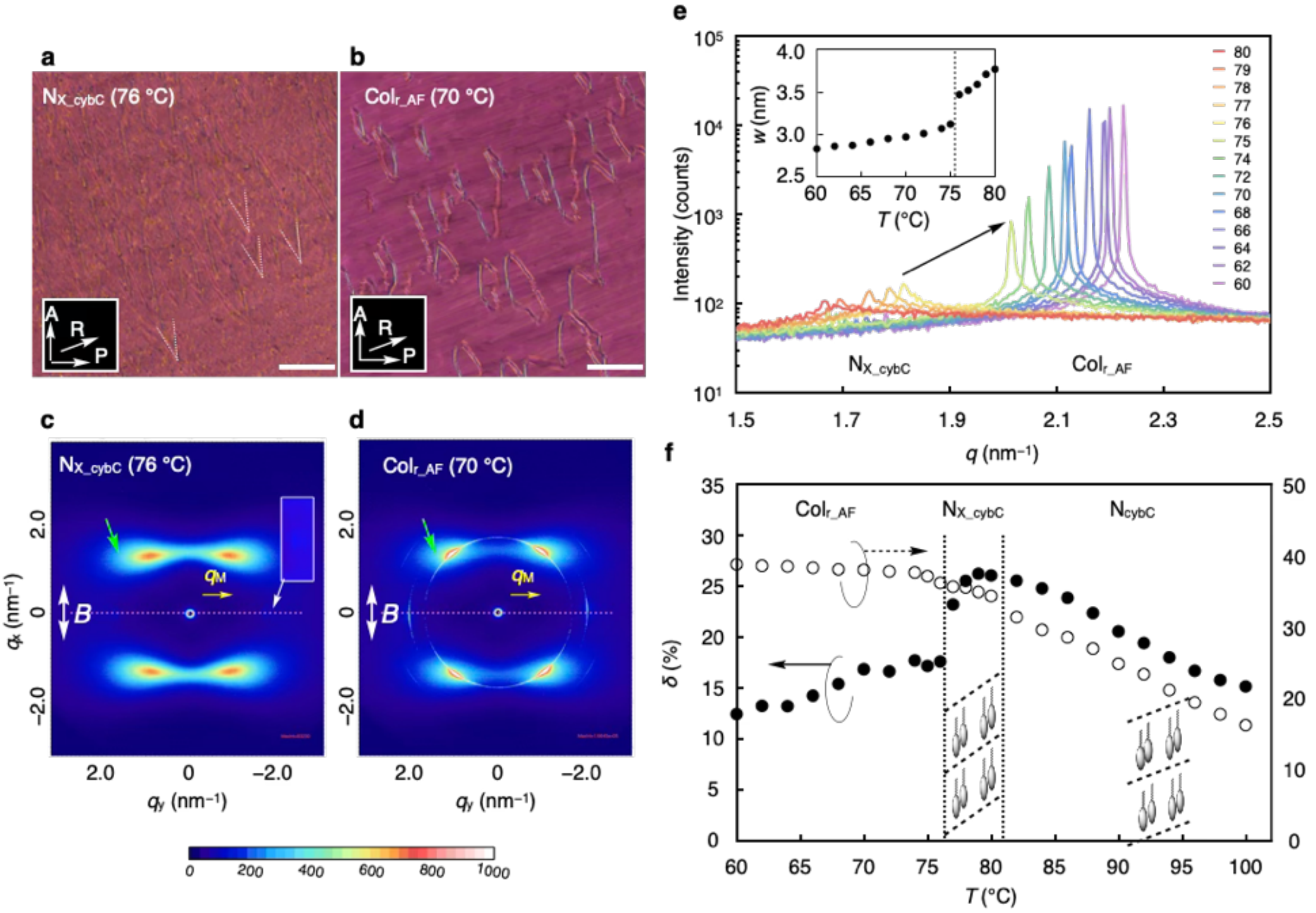

**Fig. 4 | Evolution of SmC-type cybotactic packing into the Col$_{r_AF}$ phase in *12*-012-CN.**

(a,b) Polarized optical microscopy images of N$_{X_cybC}$ at 76 °C and Col$_{r_AF}$ ($^{2Dr}$SmC$_{AF}$) at 70 °C. (c,d) Corresponding 2D XRD patterns under magnetic alignment. Split small-angle scattering characteristic of SmC-type cybotactic packing persists after the development of long-range 2D order. (e) Temperature-dependent small-angle XRD profiles across the N$_{X_cybC}$–Col$_{r_AF}$ transition; the inset shows the modulation periodicity $w$. (f) Temperature dependences of the local tilt angle $\beta$ and offset parameter δ through the N$_{cybC}$–N$_{X_cybC}$–Col$_{r_AF}$ sequence. The discontinuous decrease in $\delta$ at the transition indicates reorganization of the longitudinal offset while the tilted SmC-type local motif is retained.

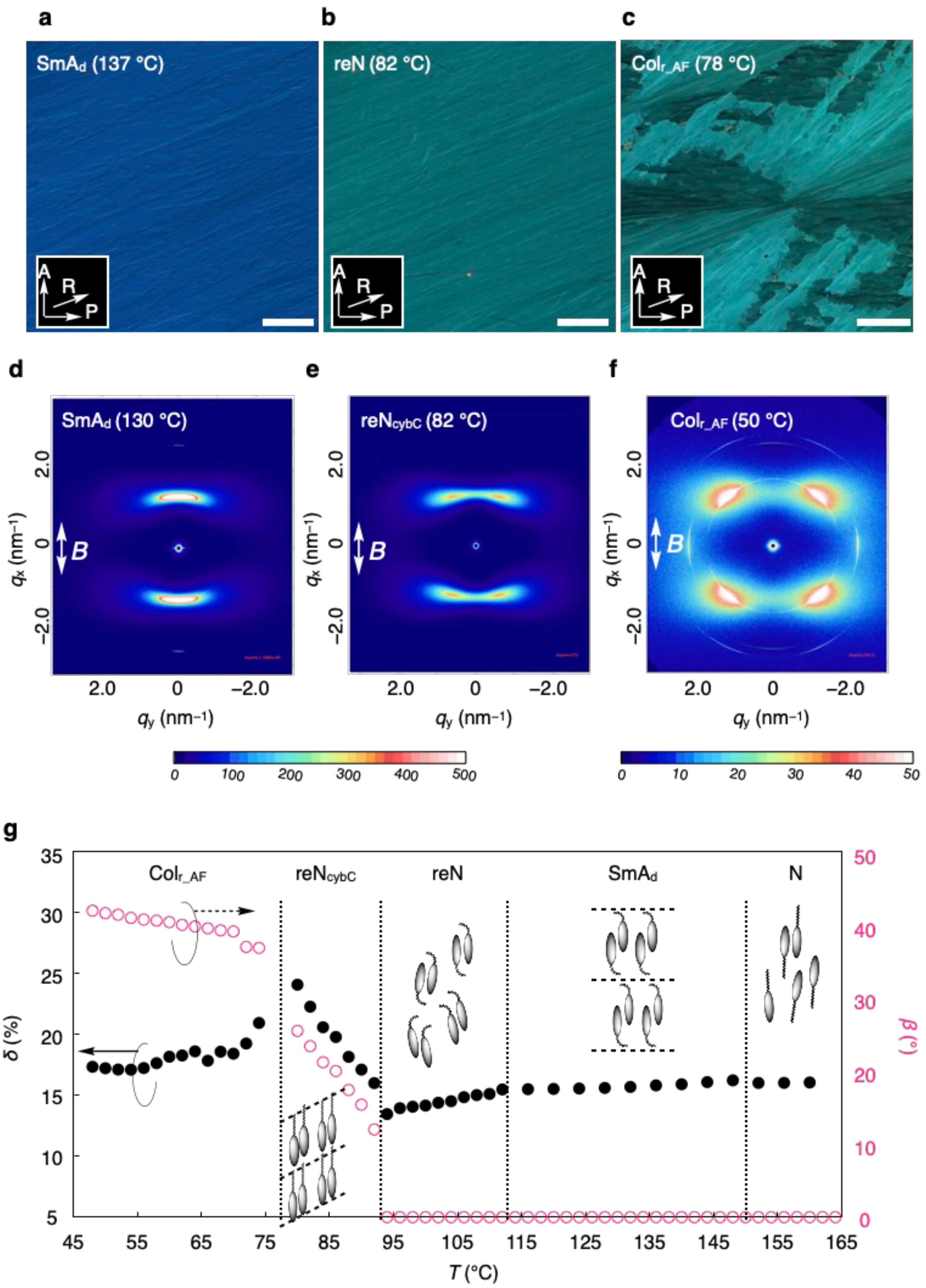

**Fig. 5 | Reentrant pathway to the Col$_{r_AF}$ phase in *13*-012-CN.**
(a–c) POM images of the SmA$_d$, reN$_{cybC}$, and Col$_{r_AF}$ ($^{2Dr}$SmC$_{AF}$) phases at the indicated temperatures. (d–f) 2D XRD patterns at 130, 82, and 50 °C, respectively. Split small-angle reflections in reN$_{cybC}$ indicate SmC-type cybotactic correlations, and sharp lattice reflections appear in Col$_{r_AF}$. (g) Temperature dependence of the offset parameter $\delta$ and local tilt angle $\beta$ through the N–SmA$_d$–reN–reN$_{cybC}$–Col$_{r_AF}$ sequence. Schematics show extended antiparallel association in SmA$_d$, tilted SmC-type packing in reN$_{cybC}$ and rectangular 2D order at low temperature. Without resolvable SmC-type splitting, $L_{app}$ was approximated by $d$, corresponding to $\beta = 0°$.

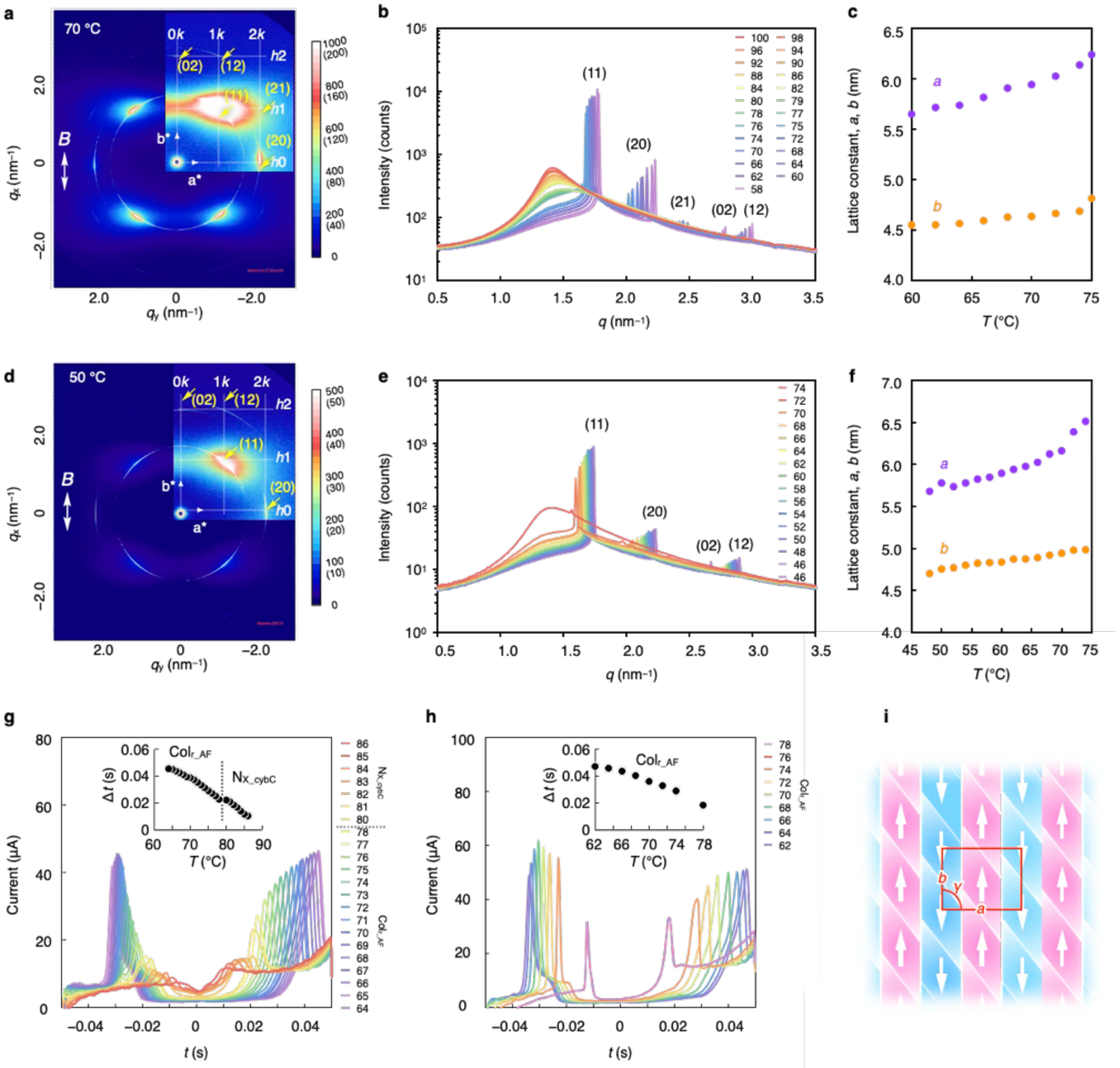

**Fig. 6 | Rectangular 2D antiferroelectric columnar order in *12*-012-CN and *13*-012-CN.**
(a,d) Magnetically aligned 2D XRD patterns of the $Col_{r_AF}$ phases of ***12*-012-CN** and ***13*-012-CN**, respectively, with indexed reciprocal lattices. (b,e) Temperature-dependent radial XRD profiles showing the development of the indexed reflections. Their comparable widths support assignment to a common long-range 2D lattice. (c,f) Temperature dependences of the lattice constants *a* and *b*. (g,h) Polarization-reversal-current profiles across the transitions into $Col_{r_AF}$ for ***12*-012-CN** and ***13*-012-CN**, respectively; the insets show the corresponding switching times. (i) Proposed structural model of $Col_{r_AF}$ ($^{2Dr}SmC_{AF}$), in which polar columns with local SmC-type packing occupy a rectangular lattice, with neighboring columns polarized in opposite directions. Note: for clarity, {*hk*} is used in the diffraction patterns to collectively label symmetry-equivalent (*hk*)-type reflections.

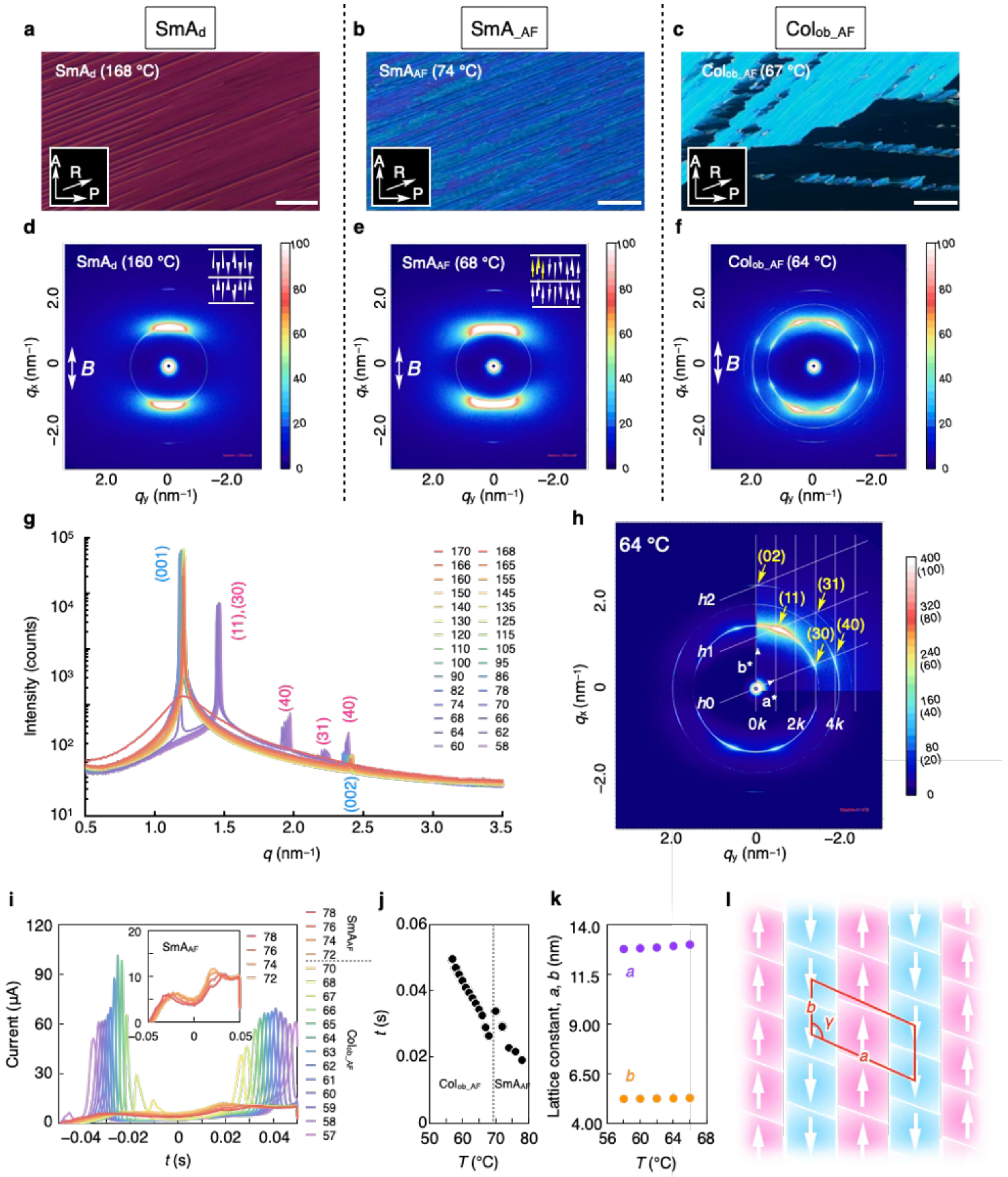

**Fig. 7 | Formation and structure of the Col$_{ob}$_AF phase in the longest polar homologs.**

(a–c) POM images of the SmA$_d$, SmA$_{AF}$, and Col$_{ob}$_AF phases at the indicated temperatures. (d–f) Corresponding magnetically aligned 2D XRD patterns. (g) Radial XRD profiles showing additional reflections in the Col$_{ob}$_AF phase. (h) Oblique reciprocal lattice and indexed reflections. (i) PRC across the SmA$_{AF}$–Col$_{ob}$_AF transition, with the SmA$_{AF}$ response shown in the inset. (j) Switching time. (k) Lattice constants *a* and *b*. (l) Proposed antiparallel arrangement of polar columns on an oblique 2D lattice.

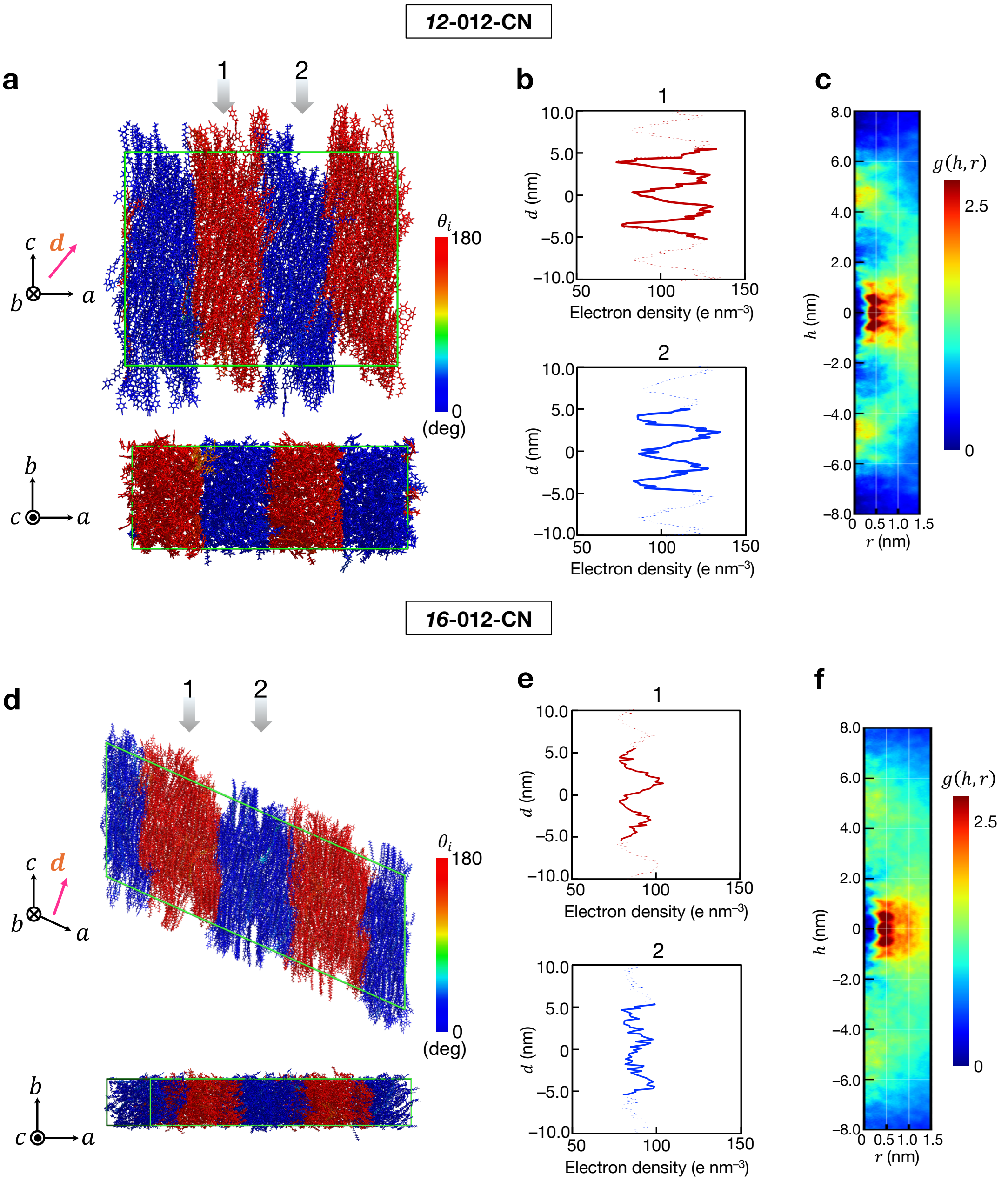

**Fig. 8 | Molecular packing in the $Col_{r_AF}$ phase of *12*-012-CN and the $Col_{ob_AF}$ phase of *16*-012-CN examined by all-atom MD simulations.** (a,d) MD snapshots showing the relaxed molecular packing in the modeled polar columns. Red and blue distinguish columns with opposite polar orientations. (b,e) Electron-density profiles along selected directions of local layer-like modulation in two oppositely polarized columns, shown in red and blue. (c,f) CDFs evaluated for molecules in the simulated columns. Note: Panel *d* shows the MD snapshots obtained from a simulation initialized with locally tilted, SmC-type packing.

# Supplementary Information

*Antiferroelectric Fluids with Hierarchical Polar Organization*

*Hiroya Nishikawa*, Hina Saito, Tetsu Katayama, Atsuko Nihonyanagi, Dennis Kwaria, Hiroyuki Inuzuka, Junxi Liu, Go Watanabe and Fumito Araoka**

*To whom correspondence should be addressed.

E-mail: hiroya.nishikawa@riken.jp (H.N.), fumito.araoka@riken.jp (F.A.)

**Table of contents**

## Methods

### 1. General and materials

**Nuclear magnetic resonance (NMR) spectroscopy**: $^{1}$H, $^{13}$C, and $^{19}$F NMR spectra were recorded on JNM-ECZ500 (JEOL) operating at 500 MHz, 126 MHz, and 471 MHz for $^{1}$H [$^{1}$H{$^{19}$F}], $^{13}$C{$^{1}$H} [$^{13}$C{$^{1}$H,$^{19}$F}] and $^{19}$F [$^{19}$F{$^{1}$H}] NMR, respectively, using the TMS (trimethylsilane) as an internal standard for $^{1}$H NMR and the deuterated solvent for $^{13}$C NMR. The absolute values of the coupling constants are given in Hz, regardless of their signs. Signal multiplicities were abbreviated by s (singlet), d (doublet), t (triplet), q (quartet), quint (quintet), sext (sextet), and dd (double–doublet), respectively.

**High-resolution mass (HRMS) spectroscopy**: HR-ESI-TOF-MS spectra were recorded on a Waters Cyclic IMS system using flow injection analysis with a Waters UPC$^{2}$ system. The mobile phase was liquid $CO_2$/MeOH (6:4, v/v).

**Density Functional Theory (DFT) Calculation**: DFT calculations were performed using the Gaussian 09 revision E01 package at the B3LYP/6-31G(d,p) level of theory.[S1] The calculation method is as follows: opt b3lyp /6-31g(d,p) SCF=tight pop=MK iop(6/33=2,6/42=6).

**Information of used liquid crystalline (LC) cells**:

*ITO glass cell (EHC)*:

- ITO-coated type, electrode area: 5 × 10 mm
- Experiments: POM (thickness: 5.0 μm) and BDS (thickness: 10.0 μm)

*ITO glass cell (homemade)*:

- Experiments: PRC studies (thickness: 4.6 μm)

*ITO glass cell (homemade)*:

- Experiments: SHG studies (thickness: 22 μm)
- The ITO glass plates were baked at 400 °C for 1 h before fabrication.

*Antiparallel-rubbed (AP)and Synparallel-rubbed (SP) cell (EHC)*:

- PI-coated type, thickness: 5.0 μm
- Alignment layer: LX-1400
- Experiments: POM studies

## 2. Synthesis of Compounds.

### 2.1. Synthetic route

All compounds were synthesized according to the protocols described below (Scheme S1). NMR and MS spectra for final target ***n*-012-CN** (n = 1–20) are shown in Supplementary Figs. 66–105.

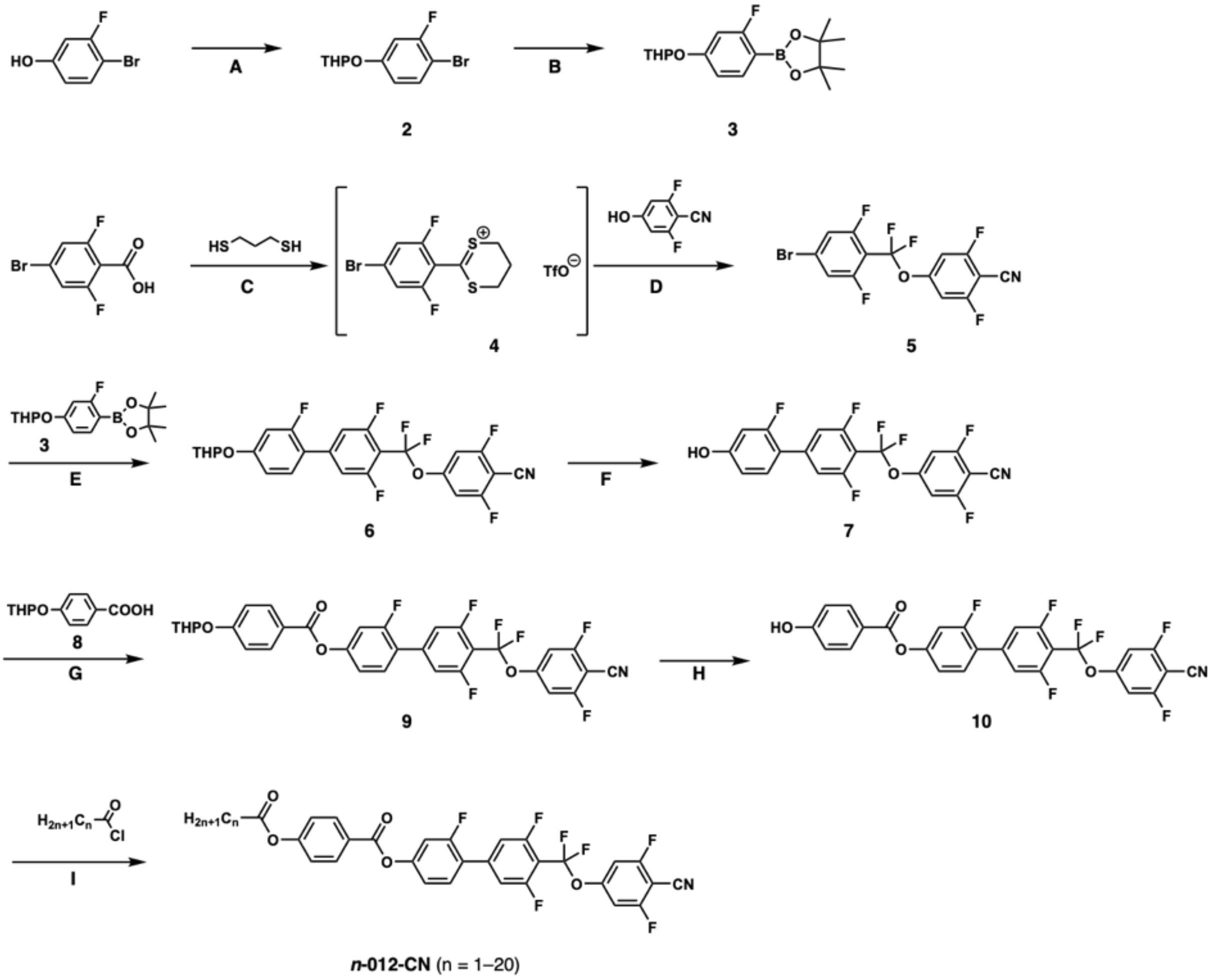

**Supplementary Scheme 1** Synthetic pathway of ***n*-012-CN** (*n* = 1–20). (A) 3,4-dihydro-2H-pyran, PPTS, $CH_2Cl_2$, r.t., 4.5 h, (B) $B_2pin_2$, Pd(dppf)$Cl_2$-$CH_2Cl_2$, KOAc, 1,4-dioxane, 80 °C 2 h; 100 °C 8 h, (C) 1,3-propanedithiol, trifluoromethanesulfonic acid, isooctane, *tert*-butyl methyl ether, toluene, 60 °C 0.5 h; 110 °C 2 h; 0 °C 0.5 h, (D) 2,6-Difluoro-4-hydroxybenzonitrile, $Et_3N$, $Et_3N$-3HF, $Br_2$, −78 °C 0.5 h; 0 °C 2 h, (E) 3, Pd(dppf)$Cl_2$-$CH_2Cl_2$, $K_2CO_3$, 1,4-dioxane/$H_2O$, 80 °C 7 h, (F) *p*-TsOH, $CH_2Cl_2$/MeOH, 40 °C, 2.5 h, (G) WSCI, DMAP, $CH_2Cl_2$, 0 °C, 1 h; r.t., 14 h, (H) *p*-TsOH, $CH_2Cl_2$/MeOH, 40 °C, 2 h, (I) $RCO_2Cl$, $Et_3N$, $CH_2Cl_2$, 0 °C, or $RCO_2H$, WSCI, DMAP, $CH_2Cl_2$, 0 °C; r.t., overnight.

**2.2.1.** ***Synthesis of 2-(4-bromo-3-fluorophenoxy)tetrahydro-2H-pyran*** **(2)**

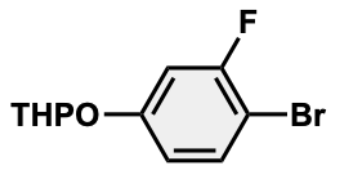

This compound was synthesized according to the procedure described in the literature.[S2]

$^{1}H\{^{19}F\}$-NMR (500 MHz, $CDCl_3$): d 7.40 (d, $J$ = 8.9 Hz, 1H), 6.88 (d, $J$ = 2.7 Hz, 1H), 6.75 (dd, $J$ = 8.9, 2.7 Hz, 1H), 5.37 (t, $J$ = 3.0 Hz, 1H), 3.86–3.82 (m, 1H), 3.63–3.59 (m, 1H), 2.00–1.93 (m, 1H), 1.87–1.84 (m, 2H), 1.74–1.65 (m, 2H), 1.62–1.58 (m, 1H);

$^{19}F\{^{1}H\}$-NMR (471 MHz, $CDCl_3$): d −105.4 (s, 1F)

**2.2.2.** ***Synthesis of 2-(2-fluoro-4-((tetrahydro-2H-pyran-2-yl)oxy)phenyl)-4,4,5,5-tetramethyl-1,3,2-dioxaborolane*** **(3)**

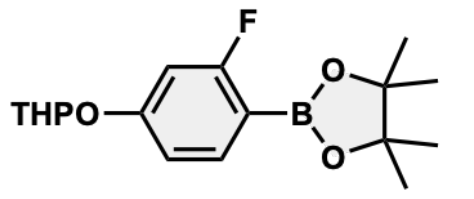

This compound was synthesized according to the procedure described in the literature.[S2]

$^{1}H\{^{19}F\}$-NMR (500 MHz, $CDCl_3$): d 7.64 (d, $J$ = 8.3 Hz, 1H), 6.82 (dd, $J$ = 8.3, 2.3 Hz, 1H), 6.75 (d, $J$ = 2.1 Hz, 1H), 5.45 (t, $J$ = 3.0 Hz, 1H), 3.87-3.82 (m, 1H), 3.63-3.59 (m, 1H), 2.02–1.94 (m, 1H), 1.87–1.84 (m, 2H), 1.73–1.63 (m, 2H), 1.62–1.58 (m, 1H), 1.34 (s, 12H);

$^{19}F\{^{1}H\}$-NMR (471 MHz, $CDCl_3$): d −100.5 (s, 1F)

**2.2.3.** ***Synthesis of 2-(4-bromo-2,6-difluorophenyl)-5,6-dihydro-4H-1,3-dithiyn-1-ylium triflate*** **(4)**

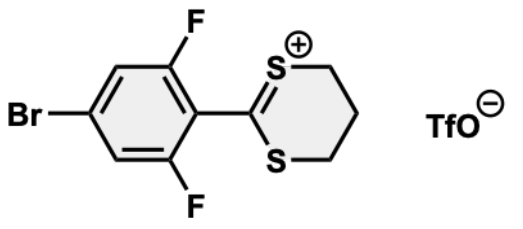

This compound was synthesized according to the procedure described in the literature.[S3] This compound was used in next step without purification.

**2.2.4.** ***Synthesis of 4-((4-bromo-2,6-difluorophenyl)difluoromethoxy)-2,6-difluorobenzonitrile*** **(5)**

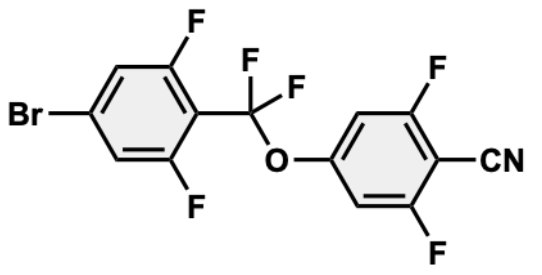

This compound was synthesized according to the procedure described in the literature. [S3]

$^{1}H\{^{19}F\}$-NMR (500 MHz, $CDCl_3$): d 7.24 (s, 2H), 7.00 (s, 2H);

$^{19}F\{^{1}H\}$-NMR (471 MHz, $CDCl_3$): d −62.0 (t, $J$ = 25.7 Hz, 2F), −101.1 (s, 2F), −108.6 (t, $J$ = 27.5 Hz, 2F)

**2.2.5.** ***Synthesis of 4-(difluoro(2',3,5-trifluoro-4'-((tetrahydro-2H-pyran-2-yl)oxy)-[1,1'-biphenyl]-4-yl)methoxy)-2,6-difluorobenzonitrile*** **(6)**

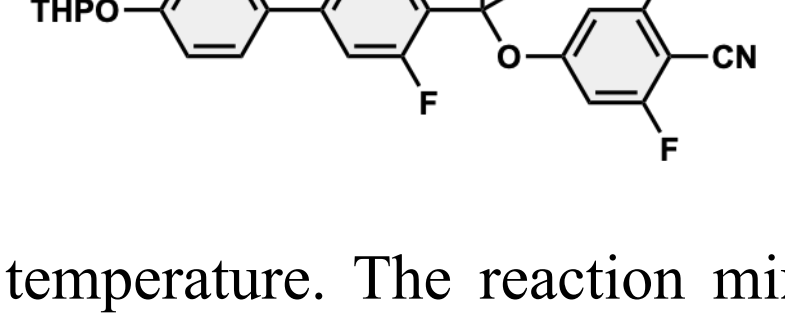

To a solution of **3** (1.22 g, 3.77 mmol) and 5 (1.36 g, 3.43 mmol) in 1,4-dioxane/$H_2O$ (15 mL/1.5 mL) were added Pd(dppf)$Cl_2$-$CH_2Cl_2$ (154 mg, 0.189 mmol) and $K_2CO_3$ (1.42 g, 10.3 mmol) at room temperature. The reaction mixture was stirred at 80 °C under Ar for 2 h. After cooling to room temperature, the mixture was filtered through a Celite pad and washed with AcOEt. The filtrate was concentrated under reduced pressure, and the residue was purified by flash column chromatography on silica gel ($CH_2Cl_2$/hexane, 25:70–50:50) to afford **6** (1.36 g, 2.65 mmol, 77.3 %) as a white solid.
$^1H\{^{19}F\}$-NMR (500 MHz, $CDCl_3$): d 7.34 (d, *J* = 8.6 Hz, 1H), 7.20 (s, 2H), 7.02 (s, 2H), 6.96–6.93 (m, 2H), 5.47 (t, *J* = 3.0 Hz, 1H), 3.87 (td, *J* = 10.7, 2.9 Hz, 1H), 3.67-3.63 (m, 1H), 2.05–1.96 (m, 1H), 1.91–1.88 (m, 2H), 1.76–1.68 (m, 2H), 1.64-1.61 (m, 1H);
$^{19}F\{^1H\}$-NMR (471 MHz, $CDCl_3$): d −61.7 (t, *J* = 25.7 Hz, 2F), -101.3 (s, 2F), −110.8 (t, *J* = 25.7 Hz, 2F), −114.4 (s, 1F).

### 2.2.6. *Synthesis of 4-(difluoro(2',3,5-trifluoro-4'-hydroxy-[1,1'-biphenyl]-4-yl)methoxy)-2,6-difluorobenzonitrile* (7)

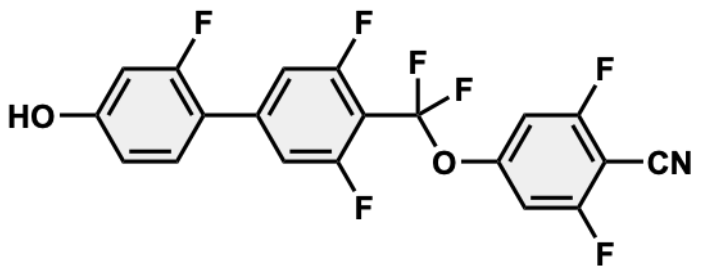

To a solution of **6** (1.24 g, 2.42 mmol) in $CH_2Cl_2$/MeOH (5 mL/5 mL) was added *p*-TsOH-$H_2O$ (23.0 mg, 0.121 mmol) at room temperature and the reaction mixture was stirred at 40 °C under Ar for 2 h. After cooling to room temperature, the solvent was removed under reduced pressure, and the resulting residue was purified by flash column chromatography on silica gel (AcOEt/hexane, 40:60–80:20) to afford **7** (1.04 g, 2.42 mmol) as a white solid.
$^1H\{^{19}F\}$-NMR (500 MHz, $CDCl_3$): d 7.31 (d, *J* = 8.6 Hz, 1H), 7.19 (s, 2H), 7.02 (s, 2H), 6.75–6.71 (m, 2H), 5.47 (s, 1H);
$^{19}F\{^1H\}$-NMR (471 MHz, $CDCl_3$): d −61.7 (t, *J* = 27.5 Hz, 2F), -101.3 (s, 2F), −110.8 (t, *J* = 25.7 Hz, 2F), −114.3 (s, 1F).

### 2.2.7. *Synthesis of 4-((tetrahydro-2H-pyran-2-yl)oxy)benzoic acid* (8)

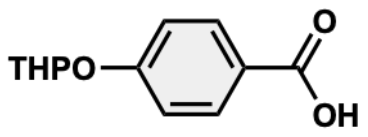

This compound was previously synthesized in our laboratory.[S2]

### 2.2.8. *Synthesis of 4'-((4-cyano-3,5-difluorophenoxy)difluoromethyl)-2,3',5'-trifluoro-[1,1'-biphenyl]-4-yl 4-((tetrahydro-2H-pyran-2-yl)oxy)benzoate* (9)

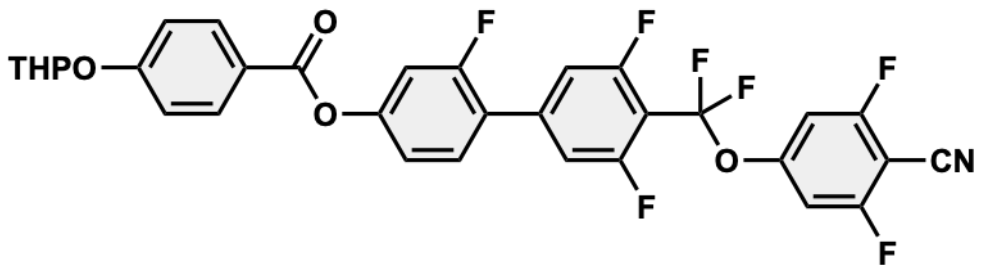

A solution of **7** (1.28 g, 3.00 mmol) and 8 (0.80 g, 3.60 mmol) in $CH_2Cl_2$ (20 mL) was treated with WSCI (690 mg, 3.60 mmol) and DMAP (36.7 mg, 0.300 mmol) at 0 °C. After stirring for 1 h at 0 °C under Ar atmosphere, the reaction mixture was warmed to room temperature and stirred for 16 h. The solvent was removed under reduced pressure, and the residue was purified by flash column chromatography on silica gel ($CH_2Cl_2$/hexane, 40:60–70:30) to give **9** (1.71 g, 2.71 mmol, 90.4%) as a white solid.
$^1H\{^{19}F\}$-NMR (500 MHz, $CDCl_3$): d 8.14 (dd, *J* = 6.9, 2.1 Hz, 2H), 7.48 (dd, *J* = 8.0, 0.9 Hz, 1H), 7.25 (m, 2H), 7.18–7.15 (m, 4H), 7.04 (s, 2H), 5.57 (t, *J* = 3.0 Hz, 1H), 3.89–3.85 (m, 1H), 3.67–3.63 (m, 1H), 2.06–1.99 (m, 1H), 1.93-1.90 (m, 2H), 1.77–1.68 (m, 2H), 1.66-1.62 (m, 1H);
$^{19}F\{^1H\}$-NMR (471 MHz, $CDCl_3$): d −61.8 (t, *J* = 27.5 Hz, 2F), −101.2 (s, 2F), −110.2 (t, *J* = 25.7 Hz, 2F), −113.9 (s, 1F).

**2.2.9.** ***Synthesis of 4'-((4-cyano-3,5-difluorophenoxy)difluoromethyl)-2,3',5'-trifluoro-[1,1'-biphenyl]-4-yl 4-hydroxybenzoate*** **(10)**

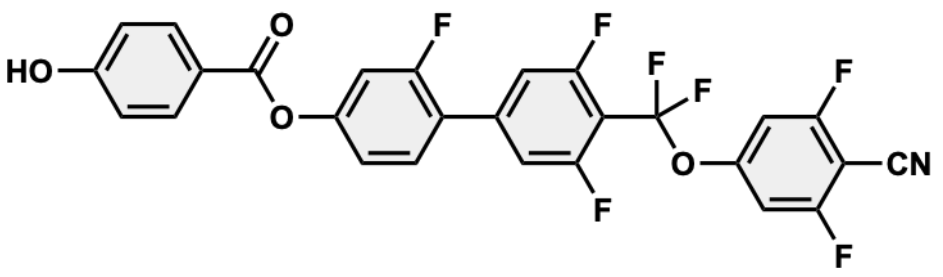

To a solution of **9** (2.70 g, 4.28 mmol) in $CH_2Cl_2$/MeOH (10 mL/10 mL) was added *p*-TsOH-$H_2O$ (40.7 mg, 0.214 mmol) at room temperature. The resulting mixture was stirred at 40 °C under Ar atmosphere for 3 h. After cooling to room temperature, the solvent was removed under reduced pressure, and the resulting residue was purified by flash column chromatography on silica gel (EtOAc/hexane, 40:60–90:10) to afford **10** (1.04 g, 2.42 mmol) as a white solid.

$^{1}H\{^{19}F\}$-NMR (500 MHz, $CDCl_3$): d 8.12 (d, *J* = 8.7 Hz, 2H), 7.49 (d, *J* = 8.2 Hz, 1H), 7.24 (s, 2H), 7.18–7.16 (m, 2H), 7.04 (s, 2H), 6.95 (d, *J* = 8.6 Hz, 2H), 5.78 (s, 1H);

$^{19}F\{^{1}H\}$-NMR (471 MHz, $CDCl_3$): d -61.8 (t, *J* = 27.5 Hz, 2F), −101.2 (s, 2F), −110.2 (t, *J* = 25.7 Hz, 2F), −113.9 (s, 1F).

**2.2.10.** ***Synthesis of 4'-((4-cyano-3,5-difluorophenoxy)difluoromethyl)-2,3',5'-trifluoro-[1,1'-biphenyl]-4-yl 4-acetoxybenzoate*** **(*1-012-CN*)**

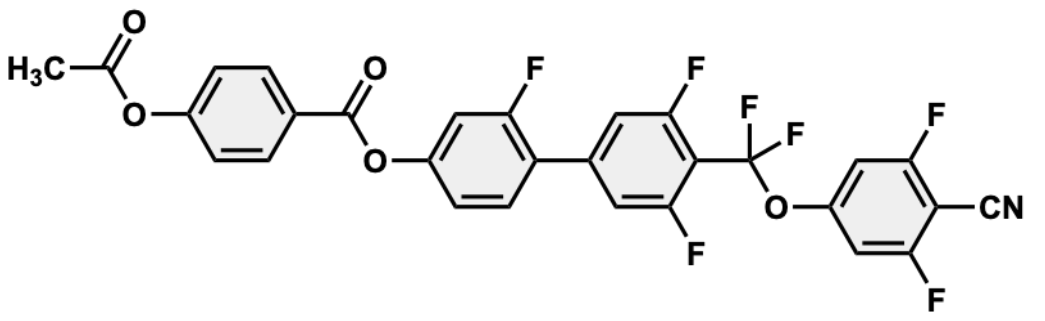

$^{1}H\{^{19}F\}$-NMR (500 MHz, $CDCl_3$): d 8.24 (d, *J* = 8.6 Hz, 2H), 7.51–7.49 (d, *J* = 7.9 Hz, 1H), 7.29–7.25 (m, 4H), 7.19–7.17 (m, 2H), 7.04 (s, 2H), 2.36 (s, 3H);

$^{19}F\{^{1}H\}$-NMR (471 MHz, $CDCl_3$): d −61.8 (t, *J* = 27.5 Hz, 2F), −101.2 (s, 2F), −110.1 (t, *J* = 25.7 Hz, 2F), −113.6 (s, 1F);

$^{13}C\{^{1}H,^{19}F\}$-NMR (126 MHz, $CDCl_3$): d 168.7, 163.8, 163.6, 159.9, 159.5, 155.2, 152.4, 141.3, 131.9, 130.5, 126.2, 123.1, 122.1, 120.3 (t, *J* = 265.6 Hz), 118.5, 113.2, 110.9, 108.6, 108.2 (t, *J* = 31.4 Hz), 105.7, 89.9, 21.1;

ESI-HRMS (m/z): $[M-H]^-$ calc for 588.0682; found, 588.0680.

**2.2.11.** ***Synthesis of 4'-((4-cyano-3,5-difluorophenoxy)difluoromethyl)-2,3',5'-trifluoro-[1,1'-biphenyl]-4-yl 4-(propionyloxy)benzoate*** **(*2-012-CN*)**

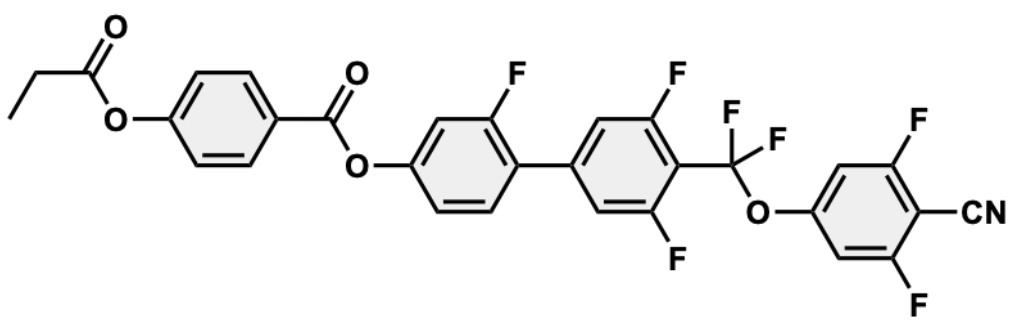

$^{1}H\{^{19}F\}$-NMR (500 MHz, $CDCl_3$): d 8.24 (d, *J* = 8.5 Hz, 2H), 7.50 (d, *J* = 7.9 Hz, 1H), 7.29–7.25 (m, 4H), 7.19–7.17 (m, 2H), 7.04 (s, 2H), 2.65 (q, J = 7.5 Hz, 2H), 1.30 (t, *J* = 7.52 Hz, 3H);

$^{19}F\{^{1}H\}$-NMR (471 MHz, $CDCl_3$): d −61.8 (t, *J* = 25.7 Hz, 2F), −101.2 (s, 2F), −110.1 (t, *J* = 25.7 Hz, 2F), −113.6 (s, 1F);

$^{13}C\{^{1}H,^{19}F\}$-NMR (126 MHz, $CDCl_3$): d 172.3, 163.8, 163.6, 159.9, 159.5, 155.4, 155.2, 152.4, 141.3, 131.9, 130.5, 126.0, 123.1, 122.0, 120.3 (t, *J* = 265.6 Hz), 118.5, 113.2, 110.9, 108.6, 108.2 (t, *J* = 31.4 Hz), 105.7, 89.9, 27.8, 8.90;

ESI-HRMS (m/z): $[M-H]^-$ calc for 602.0839; found, 602.0836.

**2.2.12.** ***Synthesis of 4'-((4-cyano-3,5-difluorophenoxy)difluoromethyl)-2,3',5'-trifluoro-[1,1'-***

***biphenyl]-4-yl 4-(butyryloxy)benzoate* (*3-012-CN*)**

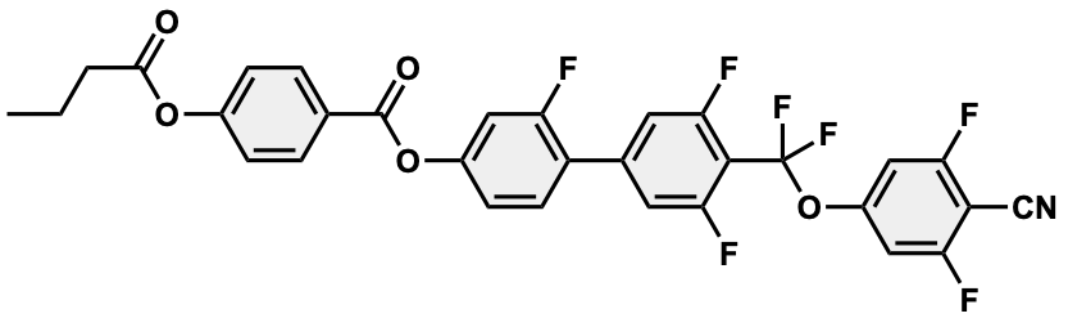

$^1H\{^{19}F\}$-NMR (500 MHz, $CDCl_3$): d 8.24 (d, *J* = 8.6 Hz, 2H), 7.50 (d, *J* = 7.9 Hz, 1H), 7.29–7.25 (m, 4H), 7.19–7.17 (m, 2H), 7.04 (s, 2H), 2.60 (t, *J* = 7.4 Hz, 2H), 1.86-1.78 (m, 2H), 1.07 (t, *J* = 7.3 Hz, 3H);

$^{19}F\{^1H\}$-NMR (471 MHz, $CDCl_3$): d −61.8 (t, *J* = 25.7 Hz, 2F), −101.2 (s, 2F), −110.1 (t, *J* = 25.7 Hz, 2F), −113.7 (s, 1F);

$^{13}C\{^1H,^{19}F\}$-NMR (126 MHz, $CDCl_3$): d 171.5, 163.8, 163.6, 159.9, 159.5, 155.3, 155.2, 152.4, 141.3, 131.9, 130.5, 126.0, 123.1, 122.1, 120.3 (t, *J* = 265.6 Hz) , 118.5, 113.2, 111.0, 108.6, 108.2 (t, *J* = 31.4 Hz), 105.7, 89.9, 36.2, 18.3, 13.6;

ESI-HRMS (m/z): $[M-H]^-$ calc for 616.0995; found, 616.0999.

**2.2.13. *Synthesis of 4'-((4-cyano-3,5-difluorophenoxy)difluoromethyl)-2,3',5'-trifluoro-[1,1'-biphenyl]-4-yl 4-(pentanoyloxy)benzoate* (*4-012-CN*)**

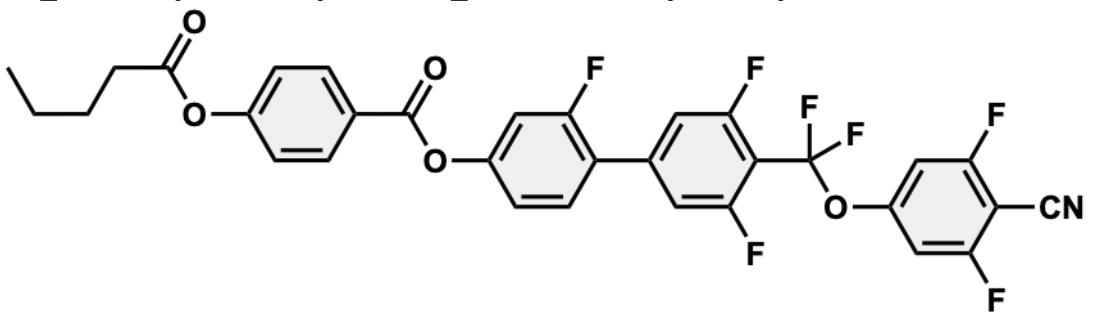

$^1H\{^{19}F\}$-NMR (500 MHz, $CDCl_3$): d 8.24 (d, *J* = 8.6 Hz, 2H), 7.50 (d, *J* = 7.9 Hz, 1H), 7.28–7.25 (m, 4H), 7.19–7.17 (m, 2H), 7.04 (s, 2H), 2.62 (t, *J* = 7.5 Hz, 2H), 1.80–1.74 (m, 2H), 1.47 (td, *J* = 14.9, 7.4 Hz, 2H), 0.99 (t, *J* = 7.3 Hz, 3H);

$^{19}F\{^1H\}$-NMR (471 MHz, $CDCl_3$): d −61.8 (t, *J* = 25.7 Hz, 2F), −101.2 (s, 2F), −110.1 (t, *J* = 25.7 Hz, 2F), −113.7 (s, 1F);

$^{13}C\{^1H,^{19}F\}$-NMR (126 MHz, $CDCl_3$): d 171.6, 163.8, 163.6, 159.9, 159.5, 155.4, 155.2, 152.4, 141.3, 131.9, 130.5, 126.0, 123.1, 122.1, 120.3 (t, *J* = 265.6 Hz), 118.5, 113.2, 110.9, 108.6, 108.2 (t, *J* = 31.4 Hz), 105.7, 89.9, 34.1, 26.8, 22.2, 13.7;

ESI-HRMS (m/z): $[M-H]^-$ calc for 630.1152; found, 630.1150.

**2.2.14. *Synthesis of 4'-((4-cyano-3,5-difluorophenoxy)difluoromethyl)-2,3',5'-trifluoro-[1,1'-biphenyl]-4-yl 4-(hexanoyloxy)benzoate* (*5-012-CN*)**

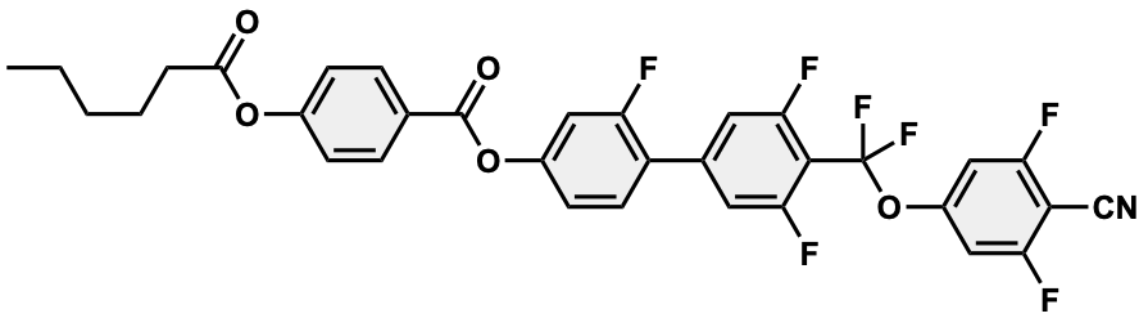

$^1H\{^{19}F\}$-NMR (500 MHz, $CDCl_3$): d 8.24 (m, 2H), 7.50 (d, *J* = 7.9 Hz, 1H), 7.28–7.25 (m, 4H), 7.19–7.17 (m, 2H), 7.04 (s, 2H), 2.61 (t, *J* = 7.5 Hz, 2H), 1.82–1.76 (m, 2H), 1.46–1.36 (m, 4H), 0.95 (t, *J* = 7.0 Hz, 3H);

$^{19}F\{^1H\}$-NMR (471 MHz, $CDCl_3$): d -61.8 (t, *J* = 25.7 Hz, 2F), −101.2 (s, 2F), −110.1 (t, *J* = 25.7 Hz, 2F), −113.7 (s, 1F);

$^{13}C\{^1H,^{19}F\}$-NMR (126 MHz, $CDCl_3$): d 171.6, 163.8, 163.6, 159.9, 159.5, 155.4, 155.2, 152.4, 141.3, 131.9, 130.5, 126.0, 123.1, 122.4, 122.1, 120.3 (t, *J* = 265.6 Hz), 118.5, 118.2, 113.2, 111.0, 108.6, 108.5, 108.2 (t, *J* = 31.4 Hz), 108.0, 105.7, 89.9, 34.3, 31.2, 24.5, 22.3, 13.9;

ESI-HRMS (m/z): $[M-H]^-$ calc for 644.1308; found, 644.1306.

**2.2.15. *Synthesis of 4'-((4-cyano-3,5-difluorophenoxy)difluoromethyl)-2,3',5'-trifluoro-[1,1'-***

***biphenyl]-4-yl 4-(heptanoyloxy)benzoate* (*6-012-CN*)**

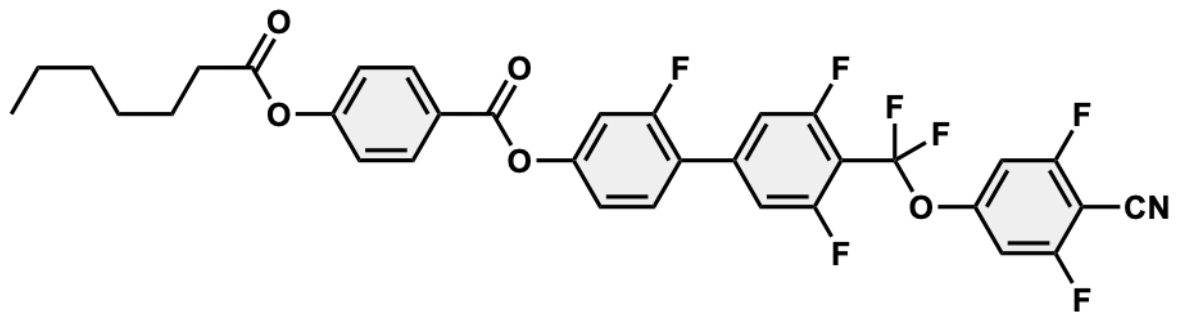

$^1$H{$^{19}$F}-NMR (500 MHz, $CDCl_3$): d 8.25–8.22 (m, 2H), 7.50 (d, *J* = 8.7 Hz, 1H), 7.28–7.25 (m, 4H), 7.19–7.17 (m, 2H), 7.04 (s, 2H), 2.61 (t, *J* = 7.5 Hz, 2H), 1.81-1.75 (m, 2H), 1.45–1.41 (m, 2H), 1.38–1.34 (m, 4H), 0.92 (t, *J* = 6.9 Hz, 3H);

$^{19}$F{$^1$H}-NMR (471 MHz, $CDCl_3$): d −61.8 (t, *J* = 25.7 Hz, 2F), −101.2 (s, 2F), −110.1 (t, *J* = 25.7 Hz, 2F), −113.7 (s, 1F);

$^{13}$C{$^1$H,$^{19}$F}-NMR (126 MHz, $CDCl_3$): d 171.6, 163.8, 163.6, 159.9, 159.5, 155.4, 155.2, 152.4, 141.3, 131.9, 130.5, 126.0, 123.1, 122.1, 120.3 (t, *J* = 265.6 Hz), 118.5, 113.2, 110.9, 108.6, 108.2 (t, *J* = 31.4 Hz), 105.7, 89.9, 34.4, 31.4, 28.7, 24.8, 22.5, 14.0;

ESI-HRMS (m/z): [M-H]$^-$ calc for 658.1464; found, 658.1463.

**2.2.16.** ***Synthesis of 4'-((4-cyano-3,5-difluorophenoxy)difluoromethyl)-2,3',5'-trifluoro-[1,1'-biphenyl]-4-yl 4-(octanoyloxy)benzoate* (*7-012-CN*)**

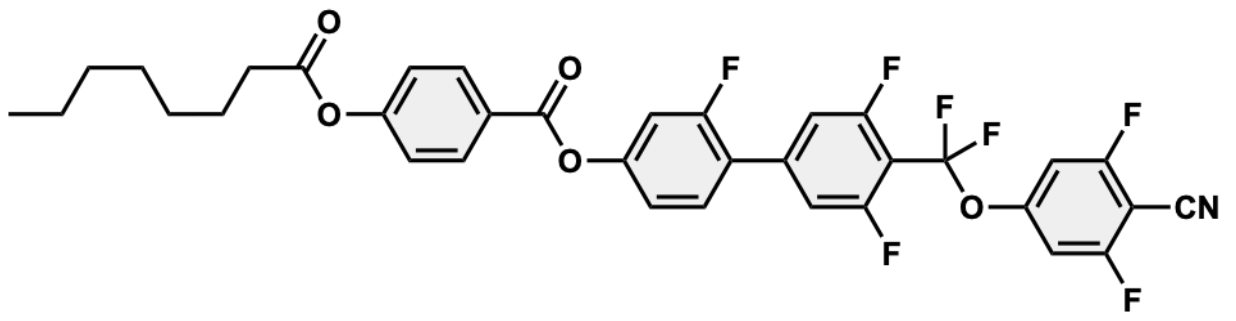

$^1$H{$^{19}$F}-NMR (500 MHz, $CDCl_3$): d 8.24 (d, *J* = 8.6 Hz, 2H), 7.50 (d, *J* = 8.3 Hz, 1H), 7.28–7.25 (m, 4H), 7.19–7.17 (m, 2H), 7.04 (s, 2H), 2.61 (t, J = 7.5 Hz, 2H), 1.81–1.75 (m, 2H), 1.46–1.31 (m, 8H), 0.91 (t, *J* = 6.8 Hz, 3H);

$^{19}$F{$^1$H}-NMR (471 MHz, $CDCl_3$): d −61.8 (t, *J* = 25.7 Hz, 2F), −101.2 (s, 2F), −110.1 (t, *J* = 25.7 Hz, 2F), −113.7 (s, 1F);

$^{13}$C{$^1$H,$^{19}$F}-NMR (126 MHz, $CDCl_3$): d 171.6, 163.8, 163.6, 159.9, 159.5, 155.4, 155.2, 152.4, 141.3, 131.9, 130.5, 126.0, 123.1, 122.1, 120.3 (t, *J* = 265.6 Hz), 118.5, 113.2, 110.9, 108.6, 108.2 (t, *J* = 31.4 Hz), 105.7, 89.9, 34.4, 31.6, 29.0, 28.9, 24.8, 22.6, 14.0;

ESI-HRMS (m/z): [M-H]$^-$ calc for 672.1621; found, 672.1619.

**2.2.17.** ***Synthesis of 4'-((4-cyano-3,5-difluorophenoxy)difluoromethyl)-2,3',5'-trifluoro-[1,1'-biphenyl]-4-yl 4-(nonanoyloxy)benzoate* (*8-012-CN*)**

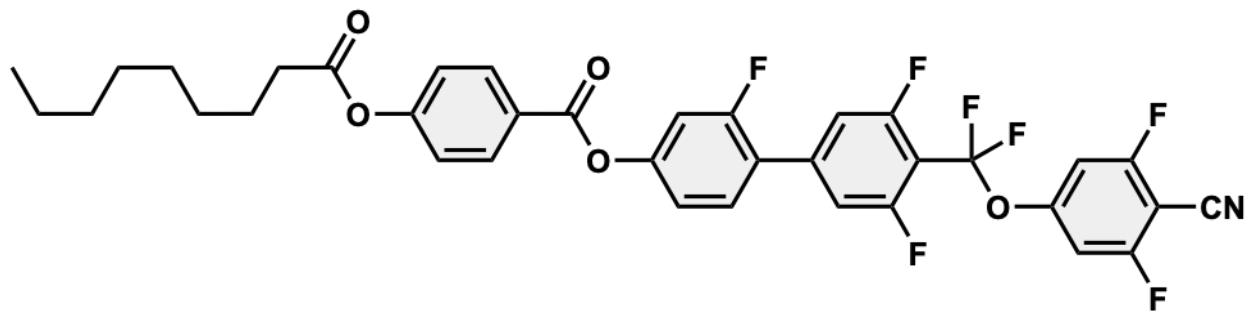

$^1$H{$^{19}$F}-NMR (500 MHz, $CDCl_3$): d 8.24 (d, *J* = 8.9 Hz, 2H), 7.50 (d, *J* = 8.0 Hz, 1H), 7.28–7.25 (m, 4H), 7.19–7.17 (m, 2H), 7.04 (s, 2H), 2.61 (t, *J* = 7.6 Hz, 2H), 1.81–1.75 (m, 2H), 1.46–1.30 (m, 10H), 0.90 (t, *J* = 6.7 Hz, 3H);

$^{19}$F{$^1$H}-NMR (471 MHz, $CDCl_3$): d −61.8 (t, *J* = 25.7 Hz, 2F), -101.2 (s, 2F), −110.1 (t, *J* = 25.7 Hz, 2F), −113.7 (s, 1F);

$^{13}$C{$^1$H,$^{19}$F}-NMR (126 MHz, $CDCl_3$): d 171.6, 163.8, 163.6, 159.9, 159.5, 155.4, 155.2, 152.4, 141.3, 131.9, 130.5, 126.0, 123.1, 122.1, 120.3 (t, *J* = 264.4 Hz), 118.5, 113.2, 110.9, 108.6, 108.2 (t, *J* = 31.4 Hz), 105.7, 89.9, 34.4, 31.8, 29.2, 29.1, 29.1, 24.8, 22.6, 14.1;

ESI-HRMS (m/z): [M-H]$^{-}$ calc for 686.1777; found, 686.1773.

### **2.2.18.** ***Synthesis of 4'-((4-cyano-3,5-difluorophenoxy)difluoromethyl)-2,3',5'-trifluoro-[1,1'-biphenyl]-4-yl 4-(decanoyloxy)benzoate*** **(*9-012-CN*)**

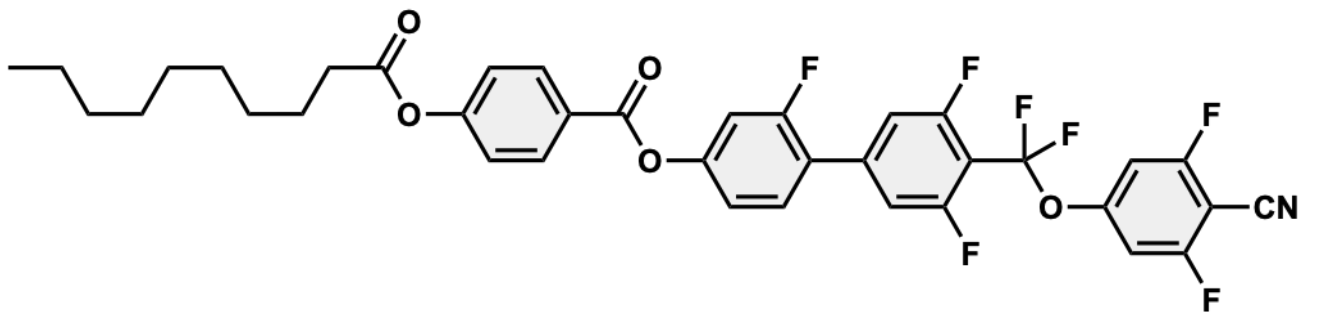

$^{1}$H{$^{19}$F}-NMR (500 MHz, $CDCl_3$): d 8.24 (m, 2H), 7.50 (d, *J* = 8.0 Hz, 1H), 7.28–7.25 (m, 4H), 7.19–7.17 (m, 2H), 7.04 (s, 2H), 2.61 (t, *J* = 7.5 Hz, 2H), 1.81–1.75 (m, 2H), 1.46–1.30 (m, 12H), 0.89 (t, *J* = 6.3 Hz, 3H);

$^{19}$F{$^{1}$H}-NMR (471 MHz, $CDCl_3$): d −61.8 (t, *J* = 25.7 Hz, 2F), −101.2 (s, 2F), −110.1 (t, *J* = 27.5 Hz, 2F), −113.7 (s, 1F);

$^{13}$C{$^{1}$H,$^{19}$F}-NMR (126 MHz, $CDCl_3$): d 171.6, 163.8, 163.6, 159.9, 159.5, 155.4, 155.2, 152.4, 141.3, 131.9, 130.5, 126.0, 123.1, 122.1, 120.3 (t, *J* = 264.4 Hz), 118.5, 113.2, 111.0, 108.6, 108.5, 108.2 (t, *J* = 31.4 Hz), 108.0, 105.8, 89.9, 34.4, 31.8, 29.4, 29.2, 29.1, 24.8, 22.7, 14.1;

ESI-HRMS (m/z): [M-H]$^{-}$ calc for 700.1934; found, 700.1931.

### **2.2.19.** ***Synthesis of 4'-((4-cyano-3,5-difluorophenoxy)difluoromethyl)-2,3',5'-trifluoro-[1,1'-biphenyl]-4-yl 4-(undecanoyloxy)benzoate*** **(*10-012-CN*)**

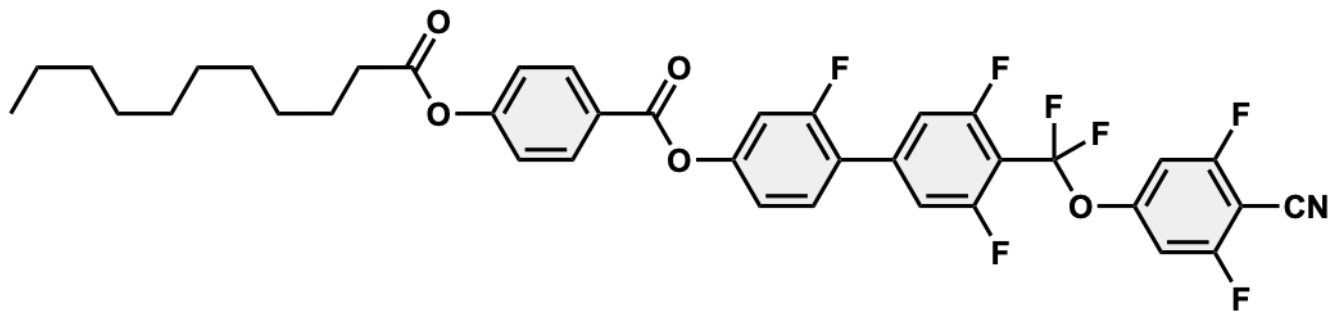

$^{1}$H{$^{19}$F}-NMR (500 MHz, $CDCl_3$): d 8.24 (d, *J* = 8.7 Hz, 2H), 7.50 (d, *J* = 8.6 Hz, 1H), 7.28–7.25 (m, 4H), 7.19–7.17 (m, 2H), 7.04 (s, 2H), 2.61 (t, *J* = 7.6 Hz, 2H), 1.81-1.75 (m, 2H), 1.46–1.28 (m, 14H), 0.89 (t, *J* = 6.9 Hz, 3H);

$^{19}$F{$^{1}$H}-NMR (471 MHz, $CDCl_3$): d −61.8 (t, *J* = 25.7 Hz, 2F), −101.2 (s, 2F), −110.1 (t, *J* = 25.7 Hz, 2F), −113.7 (s, 1F);

$^{13}$C{$^{1}$H,$^{19}$F}-NMR (126 MHz, $CDCl_3$): d 171.6, 163.8, 163.6, 159.9, 159.5, 155.4, 155.2, 152.4, 141.3, 131.9, 130.5, 126.0, 123.1, 122.1, 120.3 (t, *J* = 264.4 Hz), 118.5, 113.2, 111.0, 108.6, 108.5, 108.2 (t, *J* = 30.8 Hz), 108.0, 105.8, 89.9, 34.4, 31.9, 29.5, 29.4, 29.3, 29.2, 29.1, 24.8, 22.7, 14.1;

ESI-HRMS (m/z): [M-H]$^{-}$ calc for 714.2090; found, 714.2086.

### **2.2.20.** ***Synthesis of 4'-((4-cyano-3,5-difluorophenoxy)difluoromethyl)-2,3',5'-trifluoro-[1,1'-biphenyl]-4-yl 4-(dodecanoyloxy)benzoate*** **(*11-012-CN*)**

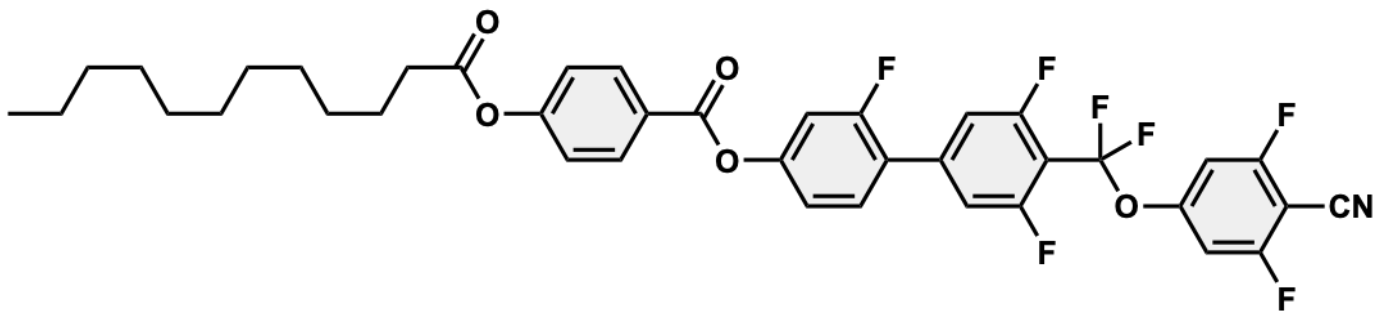

$^{1}$H{$^{19}$F}-NMR (500 MHz, $CDCl_3$): d 8.25–8.22 (m, 2H), 7.50 (d, *J* = 8.2 Hz, 1H), 7.28–7.25 (m, 6H), 7.19–7.17 (m, 2H), 7.04 (s, 2H), 2.61 (t, *J* = 7.5 Hz, 2H), 1.81–1.75 (m, 2H), 1.46-1.41 (m, 2H), 1.37–1.27 (m, 16H), 0.89 (t, *J* = 6.9 Hz, 3H);

$^{19}$F{$^{1}$H}-NMR (471 MHz, $CDCl_3$): d −61.8 (t, *J* = 27.5 Hz, 2F), −101.2 (s, 2F), −110.1 (t, *J* = 25.7 Hz, 2F), −113.7 (s, 1F);

$^{13}$C{$^{1}$H,$^{19}$F}-NMR (126 MHz, $CDCl_3$): d 171.6, 163.8, 163.6, 159.9, 159.6, 155.4, 155.2, 152.4, 141.3,

131.9, 130.5, 126.0, 123.1, 122.1, 120.3 (t, *J* = 268.0 Hz), 118.5, 113.2, 111.0, 108.6, 108.5, 108.2 (t, *J* = 31.4 Hz), 108.0, 105.7, 89.9, 34.4, 31.9, 29.6, 29.4, 29.3, 29.2, 29.1, 24.8, 22.7, 14.1;
ESI-HRMS (m/z): $[M-H]^-$ calc for 728.2247; found, 728.2249.

**2.2.21.** ***Synthesis of 4'-((4-cyano-3,5-difluorophenoxy)difluoromethyl)-2,3',5'-trifluoro-[1,1'-biphenyl]-4-yl 4-(tridecanoyloxy)benzoate*** **(*12-012-CN*)**

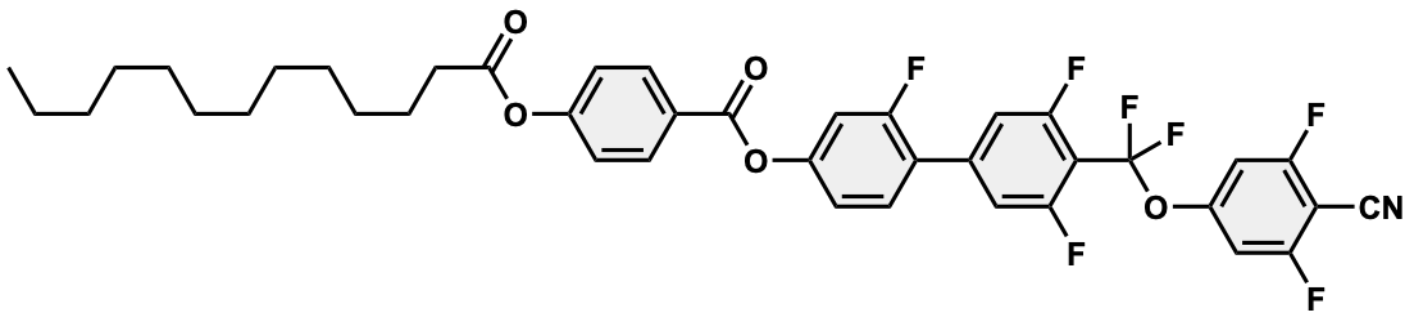

$^1H\{^{19}F\}$-NMR (500 MHz, $CDCl_3$): d 8.24 (d, J = 8.4 Hz, 2H), 7.50 (d, J = 7.9 Hz, 1H), 7.28–7.25 (m, 4H), 7.19–7.17 (m, 2H), 7.04 (s, 2H), 2.60 (t, J = 7.5 Hz, 2H), 1.81–1.75 (m, 2H), 1.46–1.27 (m, 18H), 0.88 (t, *J* = 6.9 Hz, 3H);
$^{19}F\{^1H\}$-NMR (471 MHz, $CDCl_3$): d −61.8 (t, J = 27.5 Hz, 2F), −101.2 (s, 2F), −110.1 (t, *J* = 25.7 Hz, 2F), −113.7 (s, 1F);
$^{13}C\{^1H,^{19}F\}$-NMR (126 MHz, $CDCl_3$): d 171.6, 163.8, 163.6, 159.9, 159.5, 155.4, 155.2, 152.4, 141.3, 131.9, 130.5, 126.0, 123.1, 122.1, 120.3 (t, *J* = 265 Hz), 118.5, 113.2, 111.0, 108.6, 108.5, 108.2 (t, *J* = 31.4 Hz), 108.0, 105.8, 89.9, 34.4, 31.9, 29.6, 29.6, 29.4, 29.3, 29.2, 29.1, 24.8, 22.7, 14.1;
ESI-HRMS (m/z): $[M-H]^-$ calc for 742.2404; found, 742.2408.

**2.2.22.** ***Synthesis of 4'-((4-cyano-3,5-difluorophenoxy)difluoromethyl)-2,3',5'-trifluoro-[1,1'-biphenyl]-4-yl 4-(tetradecanoyloxy)benzoate*** **(*13-012-CN*)**

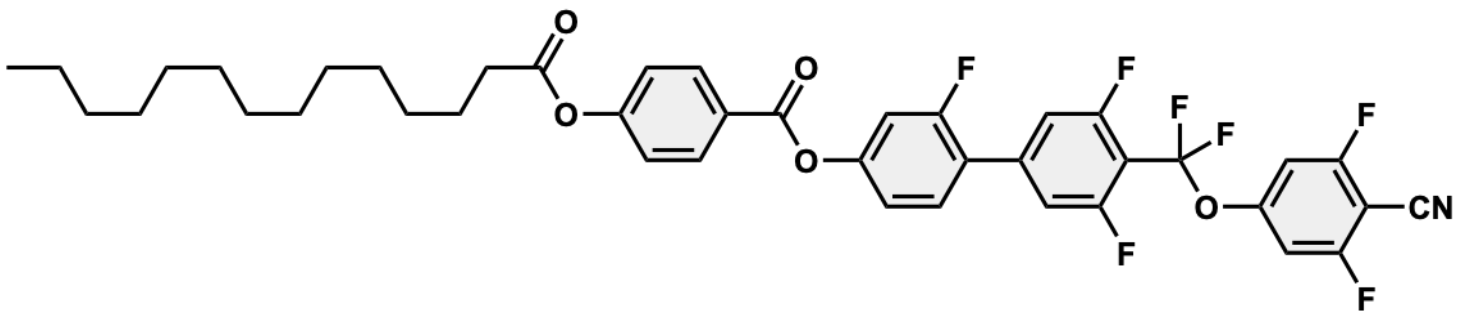

$^1H\{^{19}F\}$-NMR (500 MHz, $CDCl_3$): d 8.24 (d, *J* = 8.9 Hz, 2H), 7.50 (d, *J* = 7.9 Hz, 1H), 7.28–7.25 (m, 4H), 7.19–7.17 (m, 2H), 7.04 (s, 2H), 2.61 (t, *J* = 7.5 Hz, 2H), 1.81-1.75 (m, 2H), 1.45–1.27 (m, 20H), 0.88 (t, *J* = 6.9 Hz, 3H);
$^{19}F\{^1H\}$-NMR (471 MHz, $CDCl_3$): d −61.8 (t, *J* = 27.5 Hz, 2F), −101.2 (s, 2F), −110.1 (t, *J* = 25.7 Hz, 2F), −113.7 (s, 1F);
$^{13}C\{^1H,^{19}F\}$-NMR (126 MHz, $CDCl_3$): d 171.7, 163.8, 163.6, 159.9, 159.5, 155.4, 155.2, 152.4, 141.2, 131.9, 130.5, 126.0, 123.1, 122.1, 120.3 (t, *J* = 264.4 Hz), 118.5, 113.2, 111.0, 108.6, 108.2 (t, *J* = 30.8 Hz), 105.7, 89.9, 34.4, 31.9, 29.7, 29.6, 29.6, 29.4, 29.4, 29.2, 29.1, 24.8, 22.7, 14.1;
ESI-HRMS (m/z): $[M-H]^-$ calc for 756.2560; found, 756.2552.

**2.2.23.** ***Synthesis of 4'-((4-cyano-3,5-difluorophenoxy)difluoromethyl)-2,3',5'-trifluoro-[1,1'-biphenyl]-4-yl 4-(pentadecanoyloxy)benzoate*** **(*14-012-CN*)**

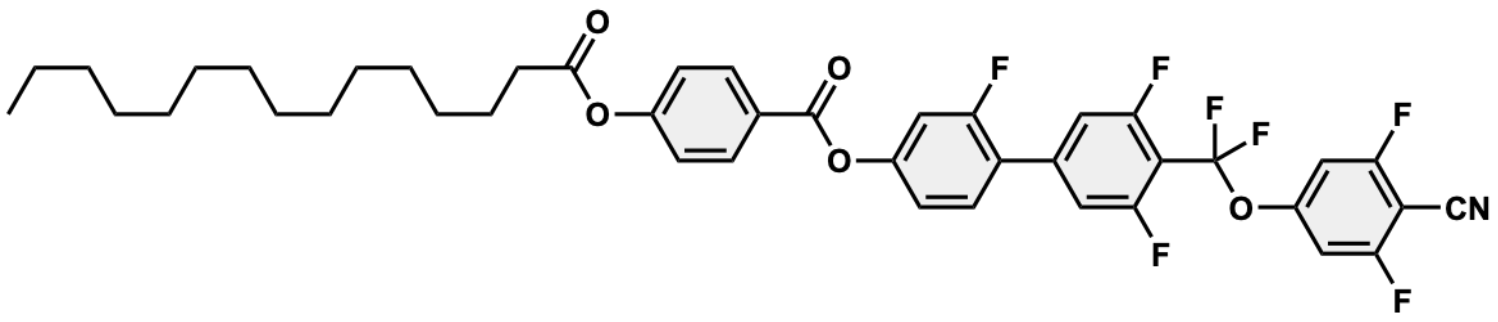

$^1H\{^{19}F\}$-NMR (500 MHz, $CDCl_3$): d 8.24 (d, *J* = 8.4 Hz, 2H), 7.50 (d, *J* = 8.0 Hz, 1H), 7.28–7.25 (m, 4H), 7.19–7.17 (m, 2H), 7.04 (s, 2H), 2.61 (t, *J* = 7.4 Hz, 2H), 1.81-1.75 (m, 2H), 1.46–1.27 (m, 22H), 0.88 (t, J = 6.8 Hz, 3H);
$^{19}F\{^1H\}$-NMR (471 MHz, $CDCl_3$): d −61.8 (t, *J* = 25.7 Hz, 2F), −101.2 (s, 2F), −110.1 (t, *J* = 25.7 Hz,

2F), −113.7 (s, 1F);
$^{13}C\{^1H,^{19}F\}$-NMR (126 MHz, $CDCl_3$): d 171.6, 163.8, 163.6, 159.9, 159.5, 155.4, 155.2, 152.4, 141.3, 131.9, 130.5, 126.0, 123.1, 122.1, 120.3 (t, *J* = 264.4 Hz), 118.5, 113.2, 111.0, 108.6, 108.2 (t, *J* = 31.4 Hz), 105.8, 89.9, 34.4, 31.9, 29.7, 29.7, 29.6, 29.6, 29.4, 29.3, 29.2, 29.1, 24.8, 22.7, 14.1;
ESI-HRMS (m/z): $[M-H]^-$ calc for 770.2717; found, 770.2715.

### 2.2.24. *Synthesis of 4'-((4-cyano-3,5-difluorophenoxy)difluoromethyl)-2,3',5'-trifluoro-[1,1'-biphenyl]-4-yl 4-(palmitoyloxy)benzoate* (*15-012-CN*)

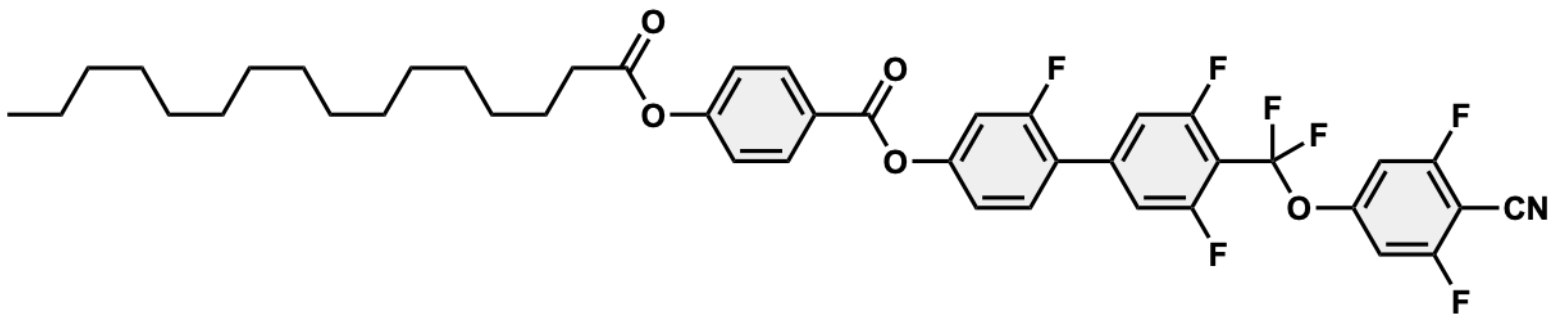

$^1H\{^{19}F\}$-NMR (500 MHz, $CDCl_3$): d 8.24 (d, *J* = 8.6 Hz, 2H), 7.50 (d, *J* = 8.0 Hz, 1H), 7.29–7.25 (m, 4H), 7.19–7.17 (m, 2H), 7.04 (s, 2H), 2.60 (t, *J* = 7.5 Hz, 2H), 1.81–1.75 (m, 2H), 1.46–1.26 (m, 24H), 0.88 (t, *J* = 6.8 Hz, 3H);
$^{19}F\{^1H\}$-NMR (471 MHz, $CDCl_3$): d −61.8 (t, *J* = 25.7 Hz, 2F), −101.2 (s, 2F), −110.1 (t, *J* = 25.7 Hz, 2F), −113.7 (s, 1F);
$^{13}C\{^1H,^{19}F\}$-NMR (126 MHz, $CDCl_3$): d 171.7, 163.8, 163.6, 159.9, 159.5, 155.4, 155.2, 152.4, 141.2, 131.9, 130.5, 126.0, 123.1, 122.1, 120.3 (t, J = 265 Hz), 118.5, 113.2, 111.0, 108.6, 108.2 (t, *J* = 30.8 Hz), 105.7, 89.9, 34.4, 31.9, 29.7, 29.6, 29.6, 29.4, 29.4, 29.2, 29.1, 24.8, 22.7, 14.1;
ESI-HRMS (m/z): $[M-H]^-$ calc for 784.2873; found, 784.2877.

### 2.2.25. *Synthesis of 4'-((4-cyano-3,5-difluorophenoxy)difluoromethyl)-2,3',5'-trifluoro-[1,1'-biphenyl]-4-yl 4-(heptadecanoyloxy)benzoate* (*16-012-CN*)

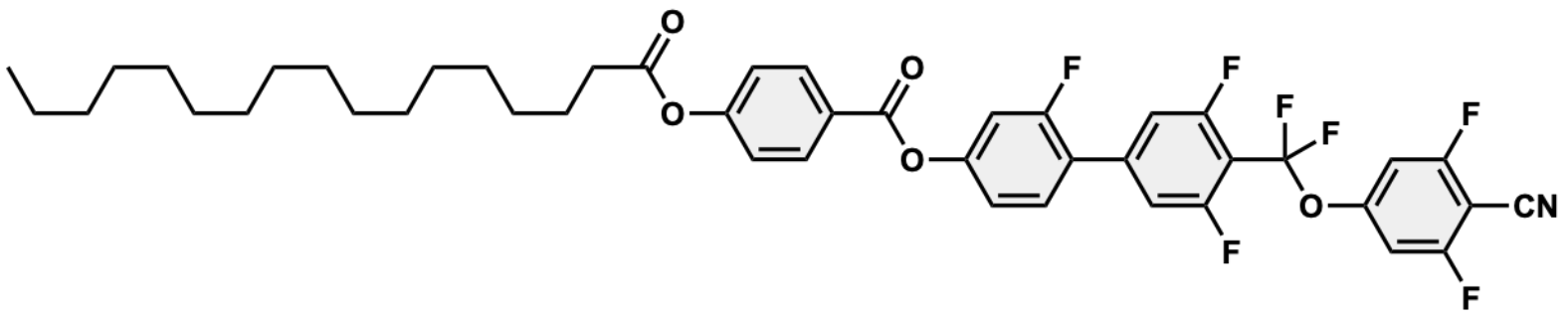

$^1H\{^{19}F\}$-NMR (500 MHz, $CDCl_3$): d 8.24 (d, *J* = 8.6 Hz, 2H), 7.50 (d, *J* = 7.9 Hz, 1H), 7.28–7.25 (m, 4H), 7.18 (d, *J* = 8.0 Hz, 2H), 7.04 (s, 2H), 2.61 (t, *J* = 7.5 Hz, 2H), 1.81–1.75 (m, 2H), 1.44–1.26 (m, 26H), 0.88 (t, *J* = 6.7 Hz, 3H);
$^{19}F\{^1H\}$-NMR (471 MHz, $CDCl_3$): d −61.8 (t, *J* = 25.7 Hz, 2F), −101.2 (s, 2F), −110.1 (t, *J* = 25.7 Hz, 2F), −113.7 (s, 1F);
$^{13}C\{^1H,^{19}F\}$-NMR (126 MHz, $CDCl_3$): d 171.6, 163.8, 163.6, 159.9, 159.5, 155.4, 155.2, 152.4, 141.3, 131.9, 130.5, 126.0, 123.1, 122.1, 120.3 (t, *J* = 268 Hz), 118.5, 113.2, 111.0, 108.6, 108.2 (t, *J* = 31.4 Hz), 105.8, 89.9, 34.4, 31.9, 29.7, 29.6, 29.6, 29.4, 29.4, 29.2, 29.1, 24.8, 22.7, 14.1;
ESI-HRMS (m/z): $[M-H]^-$ calc for 798.3029; found, 798.3032.

### 2.2.26. *Synthesis of 4'-((4-cyano-3,5-difluorophenoxy)difluoromethyl)-2,3',5'-trifluoro-[1,1'-biphenyl]-4-yl 4-(stearoyloxy)benzoate* (*17-012-CN*)

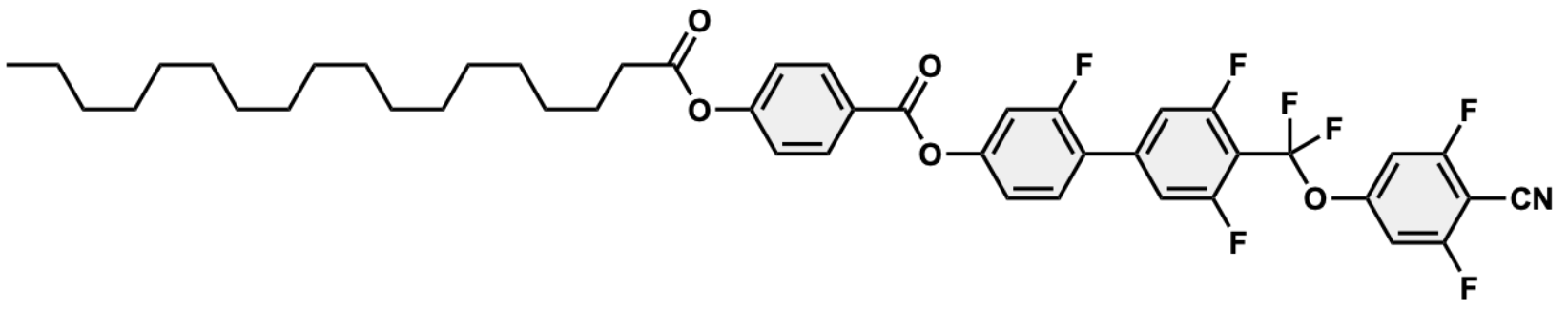

$^1H\{^{19}F\}$-NMR (500 MHz, $CDCl_3$): d 8.24 (d, *J* = 8.7 Hz, 2H), 7.50 (d, *J* = 8.0 Hz, 1H), 7.28–7.25 (m, 4H), 7.19–7.17 (m, 2H), 7.04 (s, 2H),

2.61 (t, $J$ = 7.5 Hz, 2H), 1.81-1.75 (m, 2H), 1.46–1.26 (m, 28H), 0.88 (t, $J$ = 6.9 Hz, 3H);

$^{19}$F{$^{1}$H}-NMR (471 MHz, $CDCl_3$): d −61.8 (t, $J$ = 25.7 Hz, 2F), −101.2 (s, 2F), −110.1 (t, $J$ = 25.7 Hz, 2F), −113.7 (s, 1F);

$^{13}$C{$^{1}$H,$^{19}$F}-NMR (126 MHz, $CDCl_3$): d 171.7, 163.8, 163.6, 159.9, 159.5, 155.3, 155.2, 152.4, 141.2, 131.9, 130.5, 126.0, 123.1, 122.1, 120.3 (t, $J$ = 265 Hz), 118.5, 113.2, 111.0, 108.6, 108.2 (t, $J$ = 30.8 Hz), 105.7, 89.9, 34.4, 31.9, 29.7, 29.6, 29.6, 29.4, 29.4, 29.2, 29.1, 24.8, 22.7, 14.1;

ESI-HRMS (m/z): [M-H]$^{-}$ calc for 812.3186; found, 812.3186.

**2.2.27.** ***Synthesis of 4'-((4-cyano-3,5-difluorophenoxy)difluoromethyl)-2,3',5'-trifluoro-[1,1'-biphenyl]-4-yl 4-(nonadecanoyloxy)benzoate*** **(*18-012-CN*)**

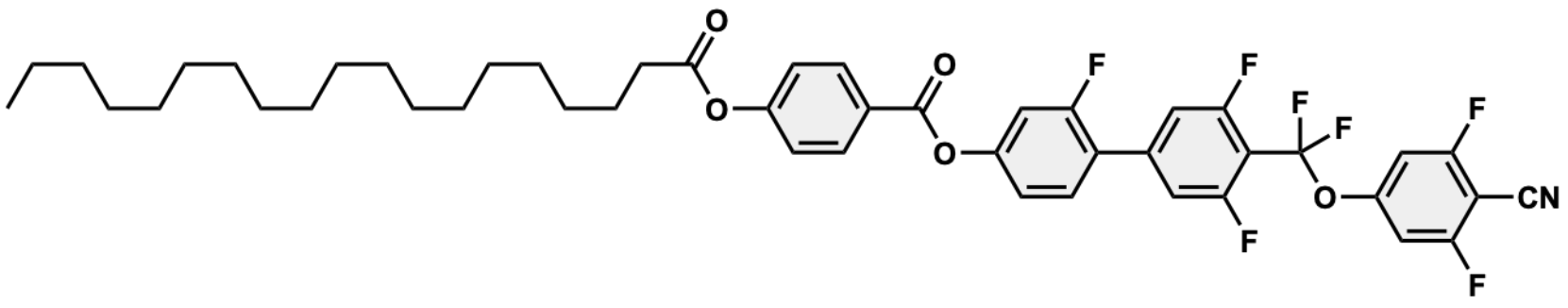

$^{1}$H{$^{19}$F}-NMR (500 MHz, $CDCl_3$): d 8.24 (d, $J$ = 8.6 Hz, 2H), 7.50 (d, $J$ = 8.0 Hz, 1H), 7.28–7.25 (m, 4H), 7.19–7.17 (m, 2H), 7.04 (s, 2H), 2.61 (t, $J$ = 7.5 Hz, 2H), 1.81–1.75 (m, 2H), 1.46–1.26 (m, 30H), 0.88 (t, $J$ = 6.7 Hz, 3H);

$^{19}$F{$^{1}$H}-NMR (471 MHz, $CDCl_3$): d −61.8 (t, $J$ = 27.5 Hz, 2F), −101.2 (s, 2F), −110.1 (t, $J$ = 25.7 Hz, 2F), −113.7 (s, 1F);

$^{13}$C{$^{1}$H,$^{19}$F}-NMR (126 MHz, $CDCl_3$): d 171.6, 163.8, 163.6, 159.9, 159.5, 155.4, 155.2, 152.4, 141.3, 131.9, 130.5, 126.0, 123.1, 122.1, 120.3 (t, $J$ = 264.4 Hz), 118.5, 113.2, 111.0, 108.6, 108.2 (t, $J$ = 30.8 Hz), 105.8, 89.9, 34.4, 31.9, 29.7, 29.6, 29.4, 29.4, 29.2, 29.1, 24.8, 22.7, 14.1;

ESI-HRMS (m/z): [M-H]$^{-}$ calc for 826.3342; found, 826.3338.

**2.2.28.** ***Synthesis of 4'-((4-cyano-3,5-difluorophenoxy)difluoromethyl)-2,3',5'-trifluoro-[1,1'-biphenyl]-4-yl 4-(icosanoyloxy)benzoate*** **(*19-012-CN*)**

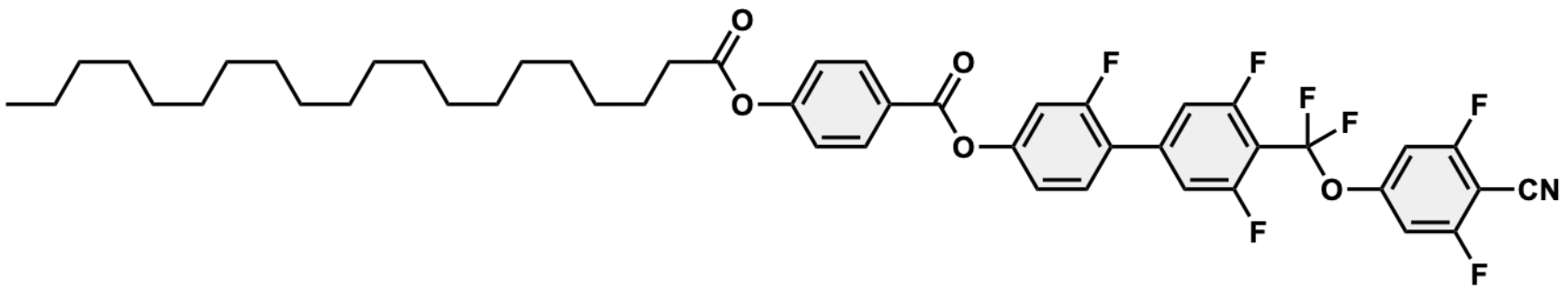

$^{1}$H{$^{19}$F}-NMR (500 MHz, $CDCl_3$): d 8.23 (d, $J$ = 8.7 Hz, 2H), 7.50 (d, $J$ = 8.4 Hz, 1H), 7.28–7.25 (m, 4H), 7.19–7.17 (m, 2H), 7.04 (s, 2H), 2.60 (t, $J$ = 7.5 Hz, 2H), 1.81-1.75 (m, 2H), 1.45–1.26 (m, 32H), 0.88 (t, $J$ = 6.9 Hz, 3H);

$^{19}$F{$^{1}$H}-NMR (471 MHz, $CDCl_3$): d −61.8 (t, J = 27.5 Hz, 2F), −101.2 (s, 2F), −110.1 (t, $J$ = 25.7 Hz, 2F), −113.6 (s, 1F);

$^{13}$C{$^{1}$H,$^{19}$F}-NMR (126 MHz, $CDCl_3$): d 171.6, 163.8, 163.6, 159.9, 159.5, 155.4, 155.2, 152.4, 147.8, 141.3, 131.9, 130.5, 126.0, 123.1, 122.1, 120.3 (t, $J$ = 263.2 Hz), 118.5, 113.2, 111.0, 108.6, 108.2 (t, $J$ = 31.4 Hz), 105.8, 89.9, 34.4, 31.9, 29.7, 29.6, 29.6, 29.4, 29.3, 29.2, 29.1, 24.8, 22.7, 14.1;

ESI-HRMS (m/z): [M-H]$^{-}$ calc for 840.3499; found, 840.3502.

**2.2.29.** ***Synthesis of 4'-((4-cyano-3,5-difluorophenoxy)difluoromethyl)-2,3',5'-trifluoro-[1,1'-biphenyl]-4-yl 4-(henicosanoyloxy)benzoate* (*20-012-CN*)**

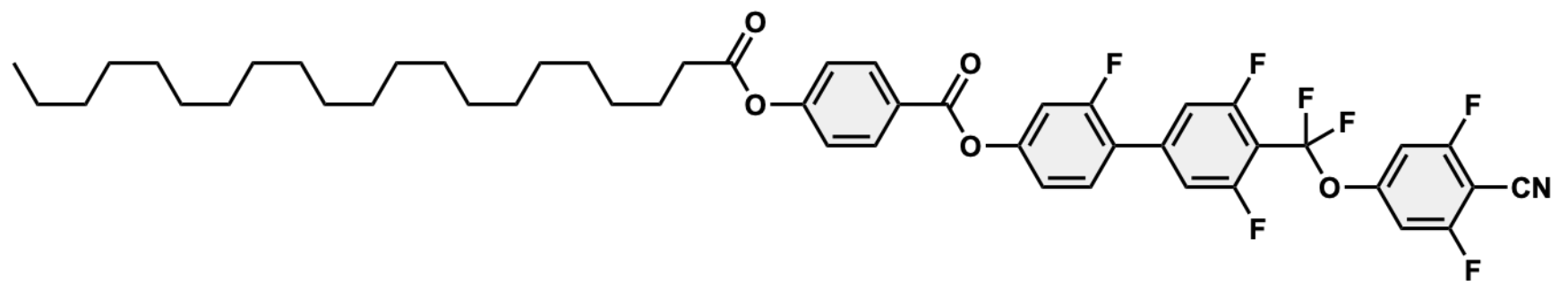

$^{1}H\{^{19}F\}$-NMR (500 MHz, $CDCl_3$): d 8.24 (d, $J$ = 8.6 Hz, 2H), 7.50 (d, $J$ = 8.1 Hz, 1H), 7.28–7.25 (m, 4H), 7.19–7.17 (m, 2H), 7.04 (s, 2H), 2.60 (t, $J$ = 7.6 Hz, 2H), 1.81–1.75 (m, 2H), 1.44–1.22 (m, 34H), 0.88 (t, J = 6.58 Hz, 3H);

$^{19}F\{^{1}H\}$-NMR (471 MHz, $CDCl_3$): d −61.8 (t, $J$ = 25.7 Hz, 2F), −101.2 (s, 2F), −110.1 (t, $J$ = 25.7 Hz, 2F), −113.6 (s, 1F);

$^{13}C\{^{1}H,^{19}F\}$-NMR (126 MHz, $CDCl_3$): d 171.6, 163.8, 163.6, 159.9, 159.5, 155.4, 155.2, 152.4, 141.3, 131.9, 130.5, 126.0, 123.1, 122.4, 122.1, 120.3 (t, $J$ = 264.4 Hz), 118.5, 118.2, 113.2, 111.0, 108.6, 108.2 (t, $J$ = 30.8 Hz), 105.8, 89.9, 34.4, 31.9, 29.7, 29.6, 29.4, 29.3, 29.2, 29.1, 24.8, 22.7, 14.1;

ESI-HRMS (m/z): $[M-H]^-$ calc for 854.3655; found, 854.3657.

**Supplementary Note (Notes 1–4)**

**Supplementary Note 1 | XRD Analysis: Indexing of 2D lattices and determination of lattice parameters.**

For each diffraction spot in the 2D X-ray scattering pattern, the magnitude of the scattering vector, $q$, and the azimuthal position were determined. The corresponding $d$-spacing was calculated as

$$d = \frac{2\pi}{q}.$$

The 2D lattices were indexed from the relationships between the experimental $d$-spacings, Miller indices, and lattice parameters, following standard procedures used for LC 2D lattices [S5]. Candidate indices were assigned to independent reflections, the lattice parameters were obtained from their experimental $d$-spacings, and the remaining reflections were used to test the assignment.

*Rectangular lattice*

For a rectangular lattice, $\gamma = 90^\circ$ and

$$\frac{1}{d_{hk}^2} = \frac{h^2}{a^2} + \frac{k^2}{b^2}.$$

Independent $h0$ and $0k$ reflections were used to obtain $a$ and $b$. For the representative indexing shown in Fig. 6a,d (main text), the $(20)$ and $(02)$ reflections give

$$\frac{1}{d_{20}^2} = \frac{4}{a^2}, \qquad \frac{1}{d_{02}^2} = \frac{4}{b^2},$$

and therefore

$$a = 2d_{20}, \qquad b = 2d_{02}.$$

The calculated spacing of any assigned $(hk)$ reflection was then obtained from

$$d_{hk,\mathrm{cal}} = \left(\frac{h^2}{a^2} + \frac{k^2}{b^2}\right)^{-1/2}.$$

The resulting values were compared with the experimental spacings of the remaining reflections, such as $(11)$ and $(12)$. A single pair of $a$ and $b$ was required to reproduce the full set of indexed reflections.

*Oblique lattice*

For an oblique lattice, the unit cell is defined by $a$, $b$, and $\gamma$, and

$$\frac{1}{d_{hk}^2} = \frac{1}{\sin^2\gamma}\left(\frac{h^2}{a^2} + \frac{k^2}{b^2} - \frac{2hk\cos\gamma}{ab}\right).$$

For convenience, we define

$$A = \frac{1}{a\sin\gamma}, \qquad B = \frac{1}{b\sin\gamma},$$

so that

$$\frac{1}{d_{hk}^2} = h^2A^2 + k^2B^2 - 2hkAB\cos\gamma.$$

Independent $h0$, $0k$, and mixed-index reflections were used to determine the three unknown lattice parameters. For the representative oblique pattern shown in Fig. 7h (main text), the $(30)$, $(02)$, and $(11)$ reflections were used. From $(30)$ and $(02)$,

$$A = \frac{1}{3d_{30}}, \qquad B = \frac{1}{2d_{02}}.$$

The lattice angle was then obtained from the $(11)$ reflection as

$$\cos\gamma = \frac{A^2 + B^2 - 1/d_{11}^2}{2AB}.$$

The real-space lattice constants follow from

$$a = \frac{1}{A\sin\gamma}, \qquad b = \frac{1}{B\sin\gamma}.$$

The derived $a$, $b$, and $\gamma$ were used to calculate the remaining reflections, including $(31)$ and $(40)$. For reflections containing negative indices, the signs of $h$ and $k$ were retained because the cross-term changes sign when one index is negative.

The $n$ = 9–15 homologues were analyzed as rectangular lattices, whereas $n$ = 16 and 17 were analyzed as oblique lattices.

**Supplementary Note 2 | Pseudo-rectangular representations of the oblique 2D lattices**

The two-dimensional (2D) oblique lattices of ***16*-012-CN** and ***17*-012-CN** were re-examined using alternative unit-cell settings to compare their relative positions of the columns on the lattice (Supplementary Fig. 62). The original lattice parameters were taken from Supplementary Tables 9 (***16*-012-CN**, 64 °C) and Supplementary Tables 10 (***17*-012-CN**, 59 °C). These transformations change the lattice basis without introducing additional reflections or new periodicities.

For the original real-space lattice vectors **a** and **b**, separated by an angle *γ*, the transformed vectors **A** and **B** were defined in terms of an integer *m*:

$$\mathbf{A} = m\mathbf{a} + \mathbf{b},$$
$$\mathbf{B} = \mathbf{b}$$

The transformed lattice parameters were calculated from the scalar products:

$$A^2 = m^2a^2 + b^2 + 2mabcos\gamma,$$
$$B = b,$$
$$cos\Gamma = \frac{macos\gamma + b}{A}$$

For ***16*-012-CN** ($a$ = 14.17 nm, $b$ = 5.78 nm, $\gamma$ = 114.0°), $m$ = 1 gives $A$ = 12.949 nm, $B$ = 5.780 nm, and $\Gamma \approx$ 90°. The cell area is unchanged, yielding a nearly orthogonal primitive setting, here termed pseudo-simple (primitive) rectangular. For ***17*-012-CN** ($a$ = 11.27 nm, $b$ = 5.77 nm, $\gamma$ = 104.8°), $m$ = 2 gives $A$ = 21.789 nm, $B$ = 5.770 nm, and $\Gamma \approx$ 90°. The transformed cell has twice the area of the original primitive cell. Because $a$ = ($A$−$B$)/2, it contains an additional lattice point equivalent to fractional coordinates (1/2, 1/2), giving a pseudo-centered rectangular representation.

The original reflection indices ($h$, $k$) and transformed indices ($H$, $K$) are related by

$$H = mh + k, K = k$$

For $m$ = 2, $H + K = 2(h + k)$, so only reflections with even $H + K$ occur in the doubled-cell representation.

**Supplementary Note 3 | 3D ESP mapping.**

The ESP mapped molecular surfaces were created using DFT calculations with the GAUSSIAN 09 program. The conditions for DFT calculation are the same as those for determining the RESP charge. The 3D ESP surface is displayed at an electron density iso-surface of 0.0004. The ESP was averaged as a function of the long molecular axis according to the previous study. [S6] The 3D electron density and ESP cube data were converted to 1D by slicing along the z-axis, aligned with the molecule's long axis. For each *z*-plane, an electron-density iso-contour (isovalue 0.0004) was traced to represent the 3D surface, and the corresponding ESP values on that contour were extracted. The ESP values were averaged over the contour and subsequently multiplied by the contour length (loop length) to compensate for the varying molecular cross section along the molecular axis

**Supplementary Note 4 | All-atom molecular dynamics (MD) simulations.**

All-atom MD simulations of ***12*-012-CN** and ***16*-012-CN** used the generalized Amber force field (GAFF)[S7], following our previous studies[S8]. Initial molecular geometries were generated using Winmostar. Atomic charges were obtained using the restrained electrostatic potential (RESP) method[S9] following geometry optimization at the B3LYP/6-31G(d,p) level in Gaussian 09[S1].

Initial assemblies were constructed using structural models informed by XRD analysis (Figs. 6a,i and 7h,l, main text). Each system comprised eight polar columns, with molecular arrays in the *xz* plane repeated ten times along the y axis. For ***12*-012-CN**, each array contained seven molecules, giving 560 molecules in total. The initial SmC-like configuration incorporated a molecular tilt of 38° and a longitudinal offset estimated from XRD. For ***16*-012-CN**, neither the local molecular tilt nor the longitudinal offset could be independently determined by XRD; both were therefore treated as model parameters. Two initial configurations with fourteen molecules per layer (1,120 molecules in total) were first examined, using SmA-like ($\beta$ = 0°) and SmC-like ($\beta$ = 19°) packing. Both relaxed toward similar tilted arrangements of polar columns at 335 K. Supplementary Fig. 65 shows the initial and relaxed configurations of both models.

The relaxed fourteen-molecule model initialized with SmC-like packing yielded lattice parameters of $a$ = 12.30 nm and $b$ = 5.40 nm. These were smaller than the experimental values of $a$ = 14.09 nm and $b$ = 5.77 nm at 62 °C. A model with sixteen molecules per layer (1,280 molecules in total) was therefore constructed using modified SmC-like initial packing. Its relaxed lattice parameters, $a$ = 13.86 nm and $b$ = 5.67 nm, were closer to experiment. This model was used for the structural analyses in Fig. 8d–f (main text). Supplementary Fig. 64a,b shows the initial configurations of the seven-molecule ***12*-012-CN** and sixteen-molecule ***16*-012-CN** models, respectively. The rationale for the molecular populations is provided in Supplementary Note 5, and the total molecule counts and simulation-box dimensions are listed in Supplementary Table 11.

Following steepest-descent energy minimization, each system was pre-equilibrated for 5 ns at 250 K, followed by 5 ns of pre-equilibration and 100 ns of equilibration at 300 K. Further equilibration was performed at 340 K for ***12*-012-CN** and 335 K for ***16*-012-CN**. All runs were conducted at 1 bar. During pre-equilibration, a Berendsen barostat[S10] and stochastic velocity-rescaling thermostat[S11] were used with coupling times of 2.0 and 0.2 ps, respectively. During equilibration, a Parrinello–Rahman barostat[S12] and Nosé–Hoover thermostat[S13–15] were used with coupling times of 5.0 and 1.0 ps, respectively. Long-range electrostatic interactions were calculated using the smooth particle-mesh Ewald (PME) method[S16] with a grid spacing of 0.30 nm. Real-space cutoffs for Coulomb and van der Waals interactions were both 1.4 nm. The time step was 2 fs, and covalent bonds involving hydrogen atoms were constrained using LINCS[S4].

To examine layer-like positional organization, electron-density profiles were calculated along the local density-modulation direction, denoted by $d$ and treated as the putative local layer normal[S17,S18]. For ***16*-012-CN**, these profiles describe the simulated local structure rather than an independently established experimental layer periodicity. Intermolecular positional correlations were characterized using cylindrical distribution functions (CDFs), following the previously reported method[S19].

**Supplementary Note 5 | Estimation of the number of molecules per polar column segment.**
Conventional columnar LC phases typically comprise stacked disc-like molecules.[S20] Alternatively, rod-like molecules can assemble into multimolecular units that form columns, as previously described for phasmidic and polycatenar mesogens.[S21] The $\mathrm{Col_{r_AF}}$ and $\mathrm{Col_{ob_AF}}$ phases of the ***n*-012-CN** series are interpreted using the latter structural model. The number of molecules, $N$ in a columnar unit-cell slice is related to its density and volume by

$$N = \frac{\rho V N_{\mathrm{A}}}{M}, \tag{1}$$

where $\rho$ is the density, $V$ is the slice volume, $N_{\mathrm{A}}$ is the Avogadro constant, and $M$ is the molar mass.[S5] For a two-dimensional lattice with parameters $a$, $b$, and $\gamma$, the cross-sectional area is $ab\sin\gamma$. The slice volume is therefore $V = abc\sin\gamma$, where $c$ is the effective thickness along the column axis. Here, $c$ was assumed to be 0.4 nm, corresponding to the assumed nearest-neighbor intermolecular separation within a polar bundle. For a unit-cell slice containing $Z_b$ bundles, each comprising $n_b$ molecules, $N = Z_b n_b$, giving

$$n_b = \frac{\rho abc\sin\gamma\, N_{\mathrm{A}}}{Z_b M}. \tag{2}$$

Candidate molecular populations were evaluated for ***12*-012-CN** in the $\mathrm{Col_{r_AF}}$ phase at 68 °C and ***16*-012-CN** in the $\mathrm{Col_{ob_AF}}$ phase at 62 °C. Assuming two columns per unit cell ($Z_b = 2$), the density corresponding to each candidate population was calculated by rearranging Eq. (2).

For ***12*-012-CN**, the rectangular lattice parameters $a = 5.91$ nm and $b = 4.63$ nm give a cross-sectional area of 27.36 nm$^2$. 7 molecules per column segment correspond to a calculated density of 1.58 g cm$^{-3}$. For ***16*-012-CN**, the oblique lattice parameters $a = 14.09$ nm, $b = 5.77$ nm, and $\gamma = 114.02°$ give a cross-sectional area of 74.26 nm$^2$. Populations of 16–17 molecules per column segment correspond to calculated densities of 1.43–1.52 g cm$^{-3}$.

For comparison, highly densities of ca. 1.3–1.54 g cm$^{-3}$ have been reported for materials exhibiting the $\mathrm{N_F}$ phase incorporating the offset alignment.[S22, S23] On this basis, initial MD models containing 7 and 16 molecules per polar bundle were adopted for ***12*-012-CN** and ***16*-012-CN**, respectively. These populations are model-building choices consistent with the assumed geometry and density range, rather than independently determined structural parameters. They depend on the assumed density, effective axial thickness, and number of columns per unit cell.

**Supplementary Figures (Figs. 1–105)**

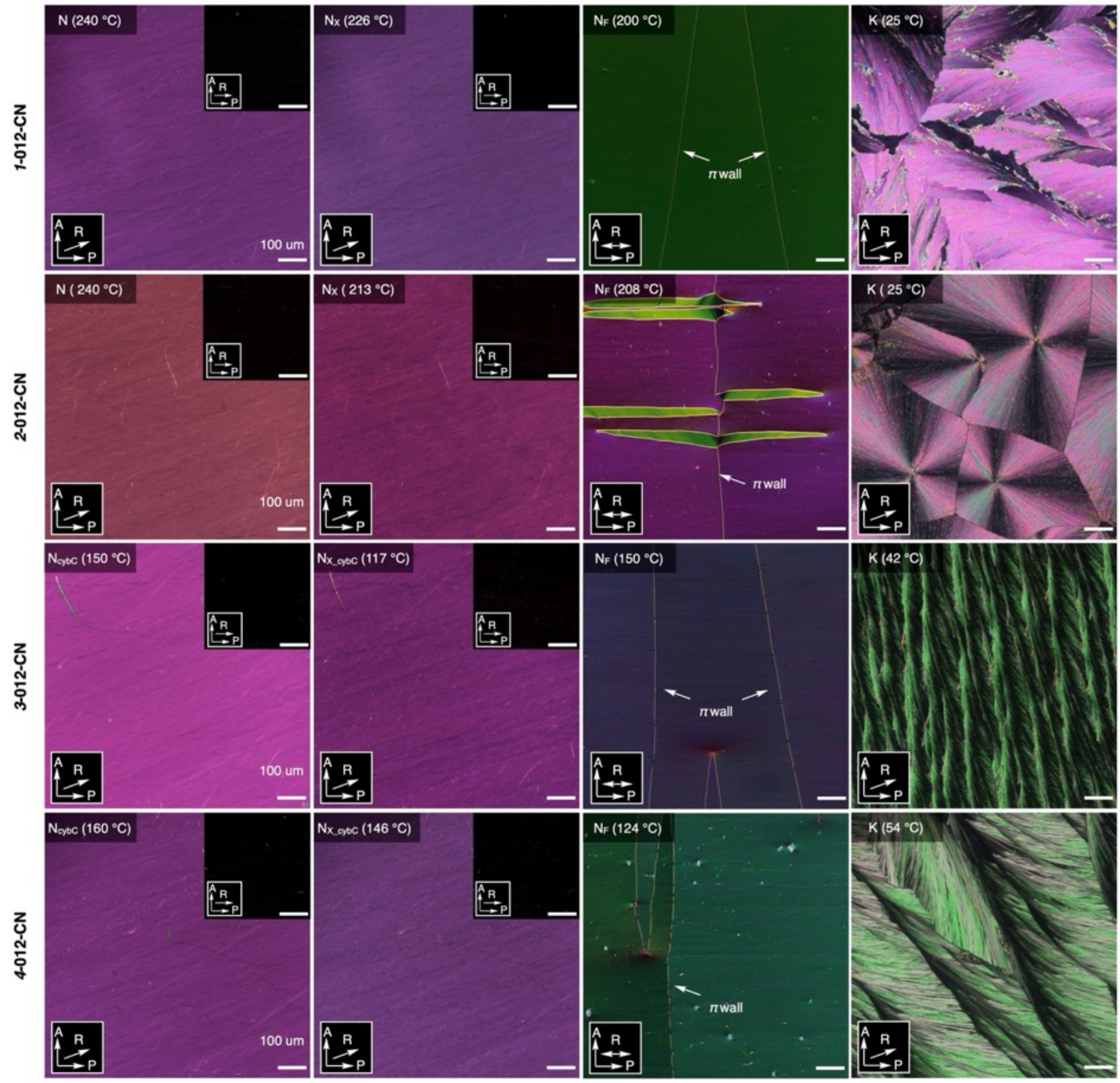

**Supplementary Fig. 1 |** POM images taken under cross polarizers in SP and AP cells for ***n*-012-CN** (n = 1–4). Abbrev.: A, P and R denote analyzer, polarizer and rubbing direction, respectively. For R, single and double arrows indicate unidirectional and antiparallel rubbing, respectively.

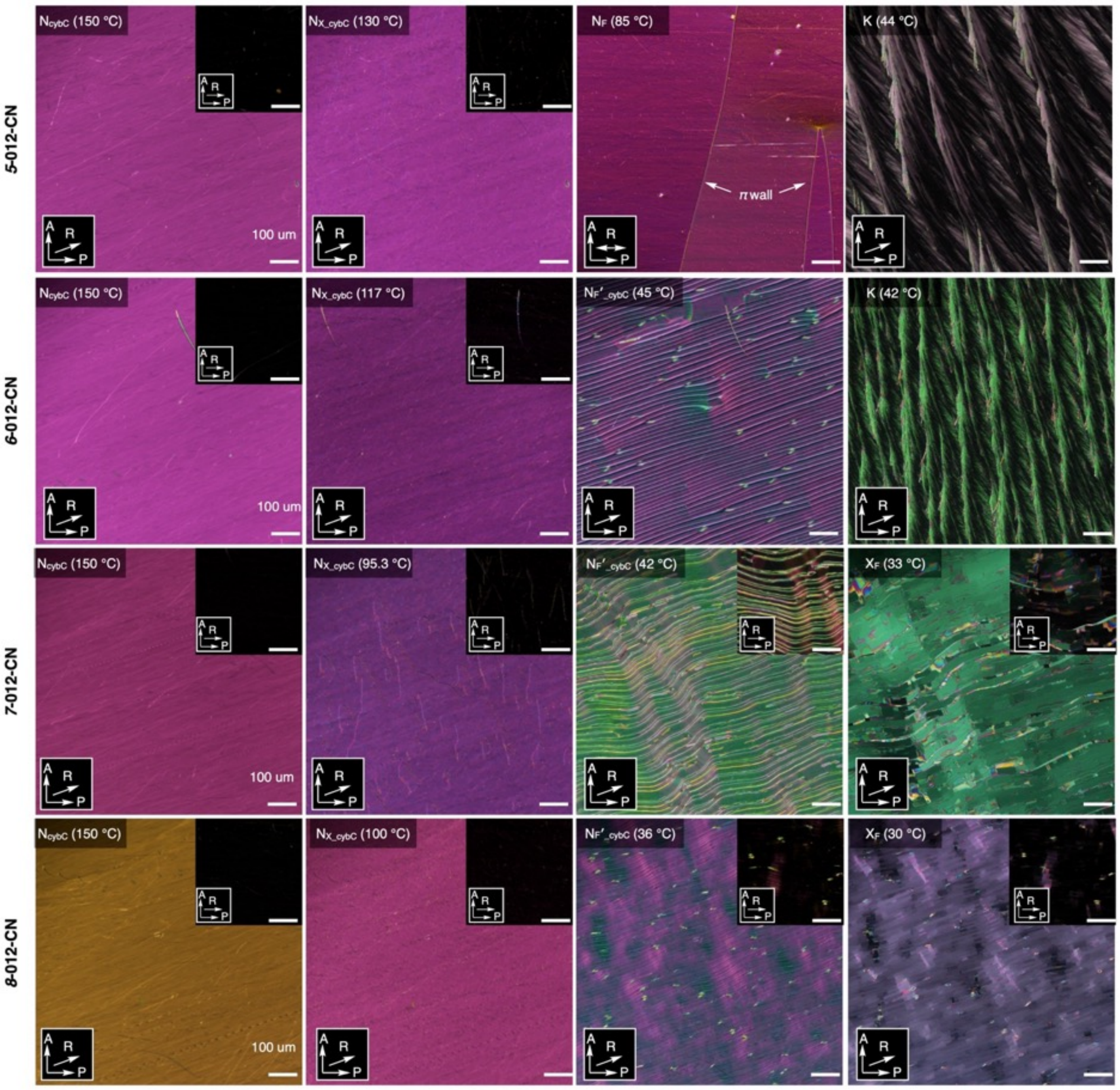

**Supplementary Fig. 2 |** POM images taken under cross polarizers in SP and AP cells for ***n*-012-CN** (n = 5–8). Abbrev.: A, P and R denote analyzer, polarizer and rubbing direction, respectively. For R, single and double arrows indicate unidirectional and antiparallel rubbing, respectively.

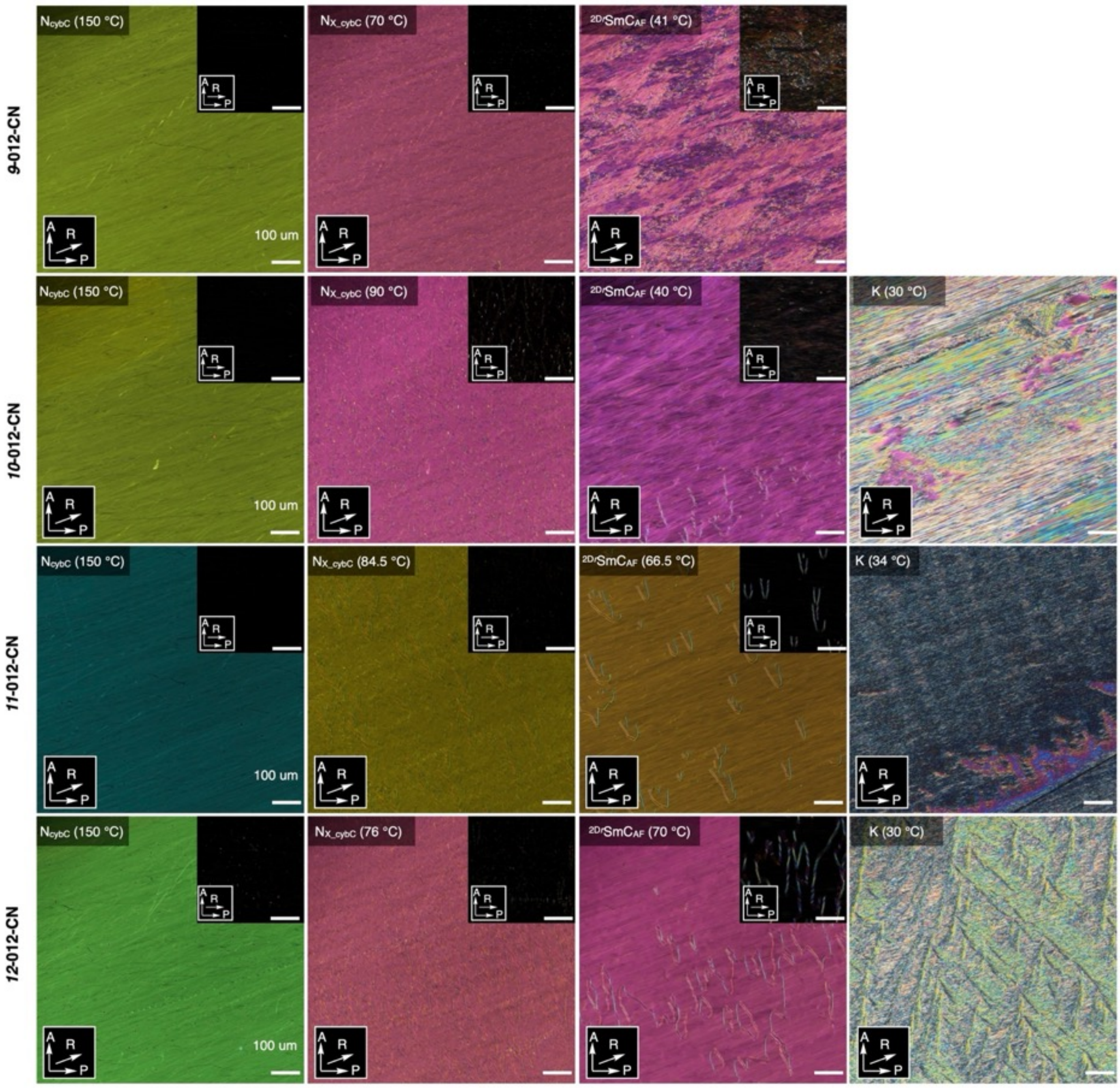

**Supplementary Fig. 3 |** POM images taken under cross polarizers in SP and AP cells for ***n*-012-CN** (n = 9–12). Abbrev.: A, P and R denote analyzer, polarizer and rubbing direction, respectively. For R, single and double arrows indicate unidirectional and antiparallel rubbing, respectively.

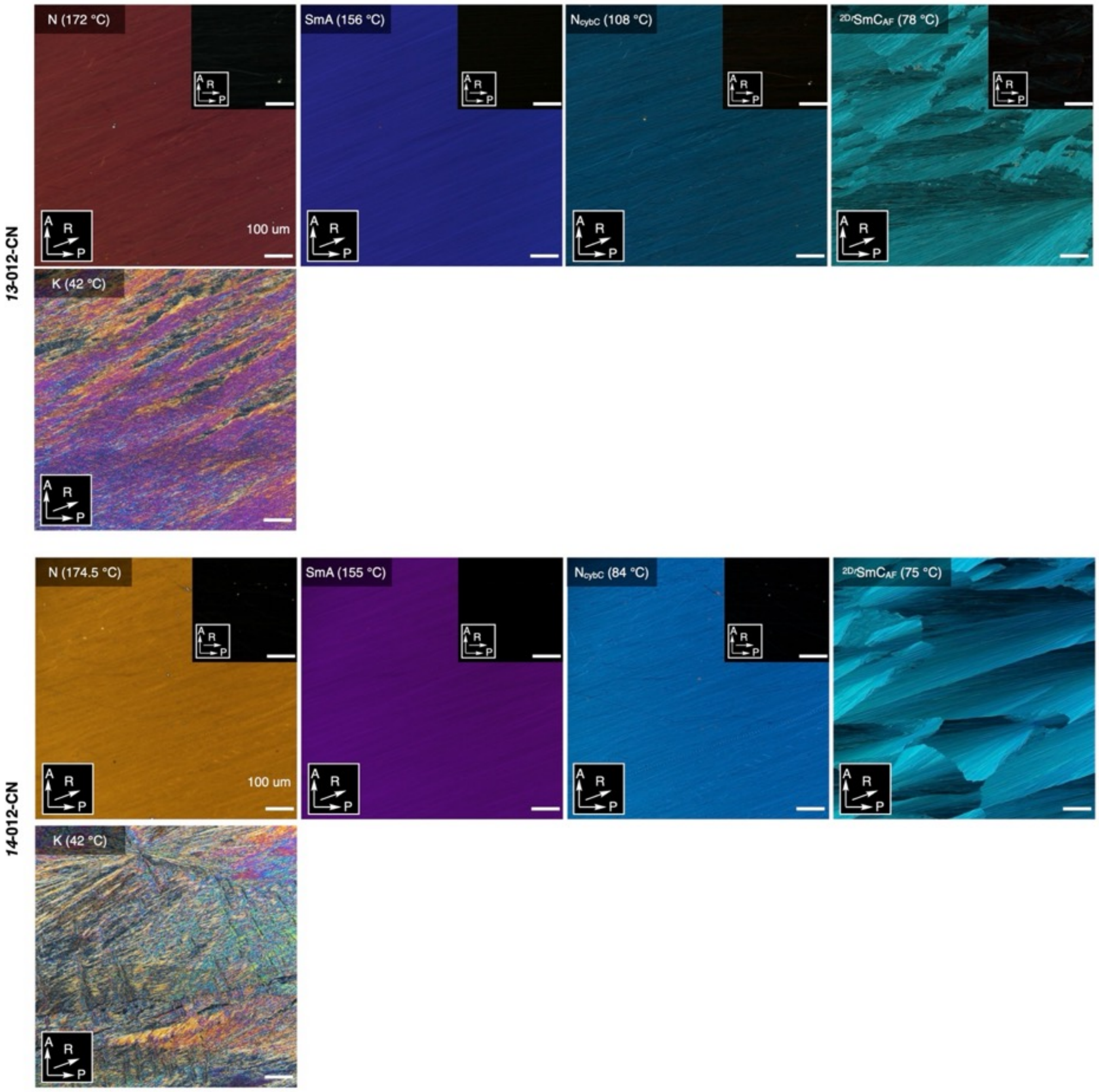

**Supplementary Fig. 4 |** POM images taken under cross polarizers in SP and AP cells for ***n*-012-CN** (n = 13–14). Abbrev.: A, P and R denote analyzer, polarizer and rubbing direction, respectively. For R, single and double arrows indicate unidirectional and antiparallel rubbing, respectively.

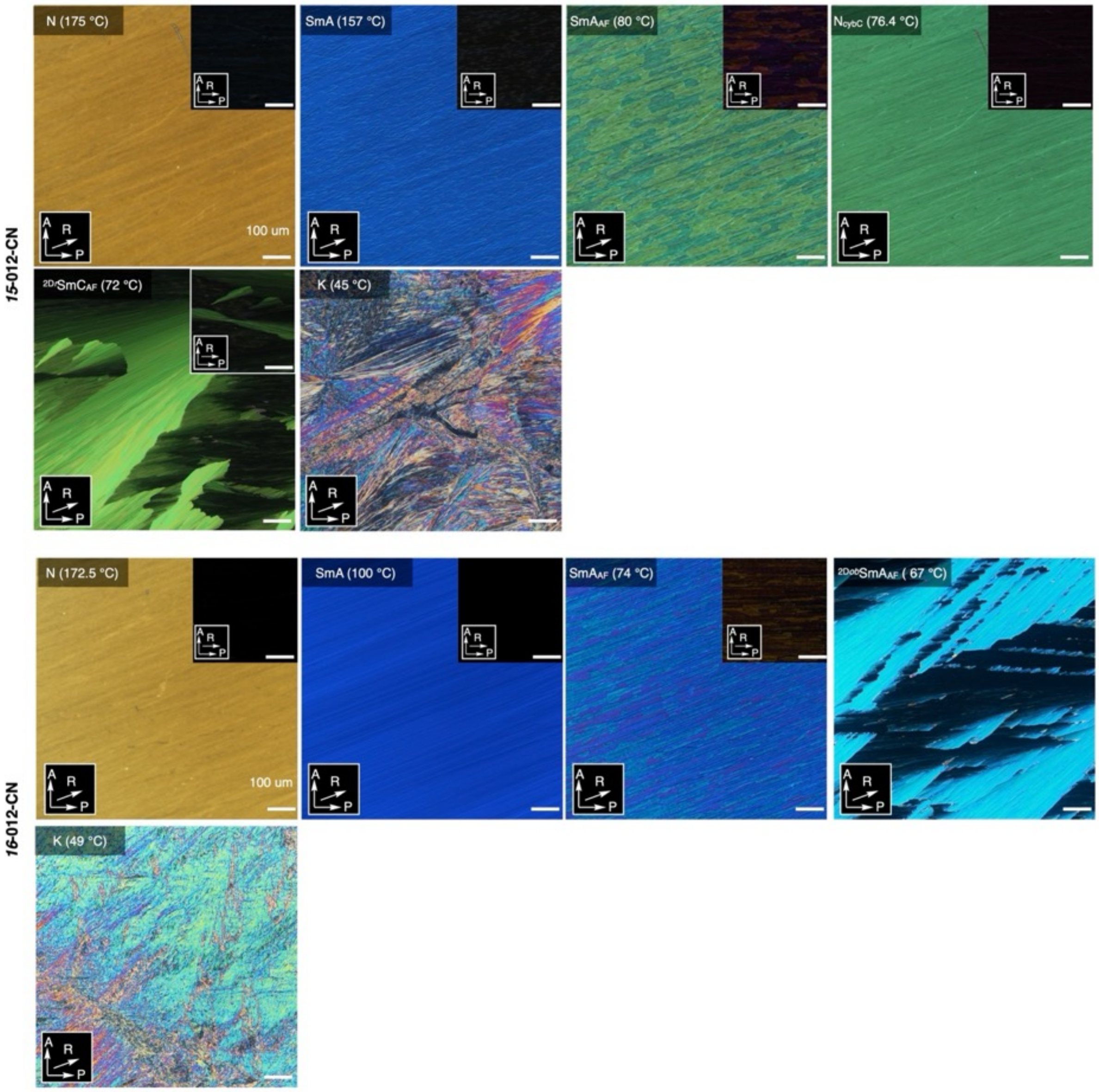

**Supplementary Fig. 5 |** POM images taken under cross polarizers in SP and AP cells for ***n*-012-CN** (n = 15–16). Abbrev.: A, P and R denote analyzer, polarizer and rubbing direction, respectively. For R, single and double arrows indicate unidirectional and antiparallel rubbing, respectively.

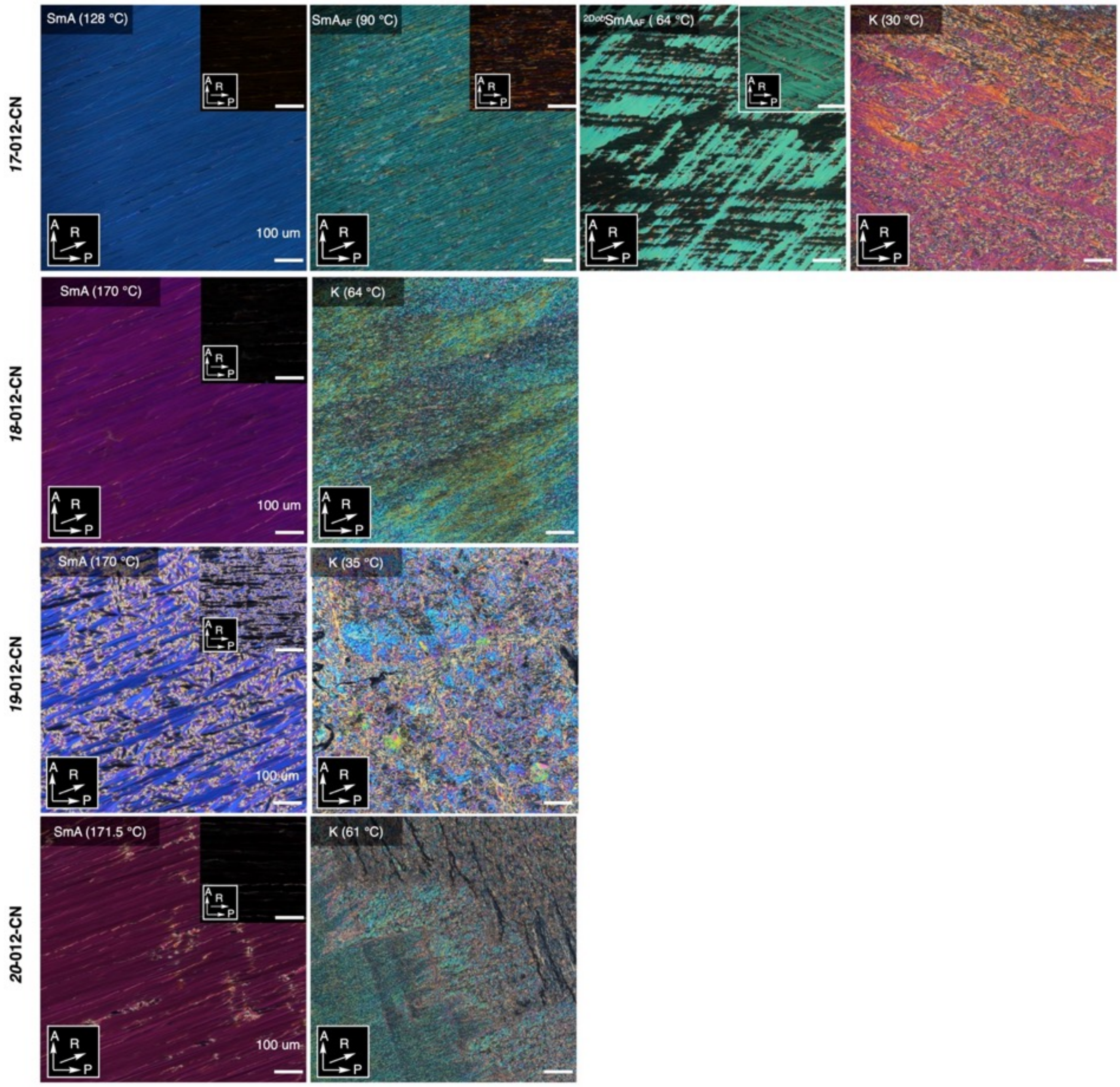

**Supplementary Fig. 6 |** POM images taken under cross polarizers in SP and AP cells for ***n*-012-CN** (n = 17–20). Abbrev.: A, P and R denote analyzer, polarizer and rubbing direction, respectively. For R, single and double arrows indicate unidirectional and antiparallel rubbing, respectively.

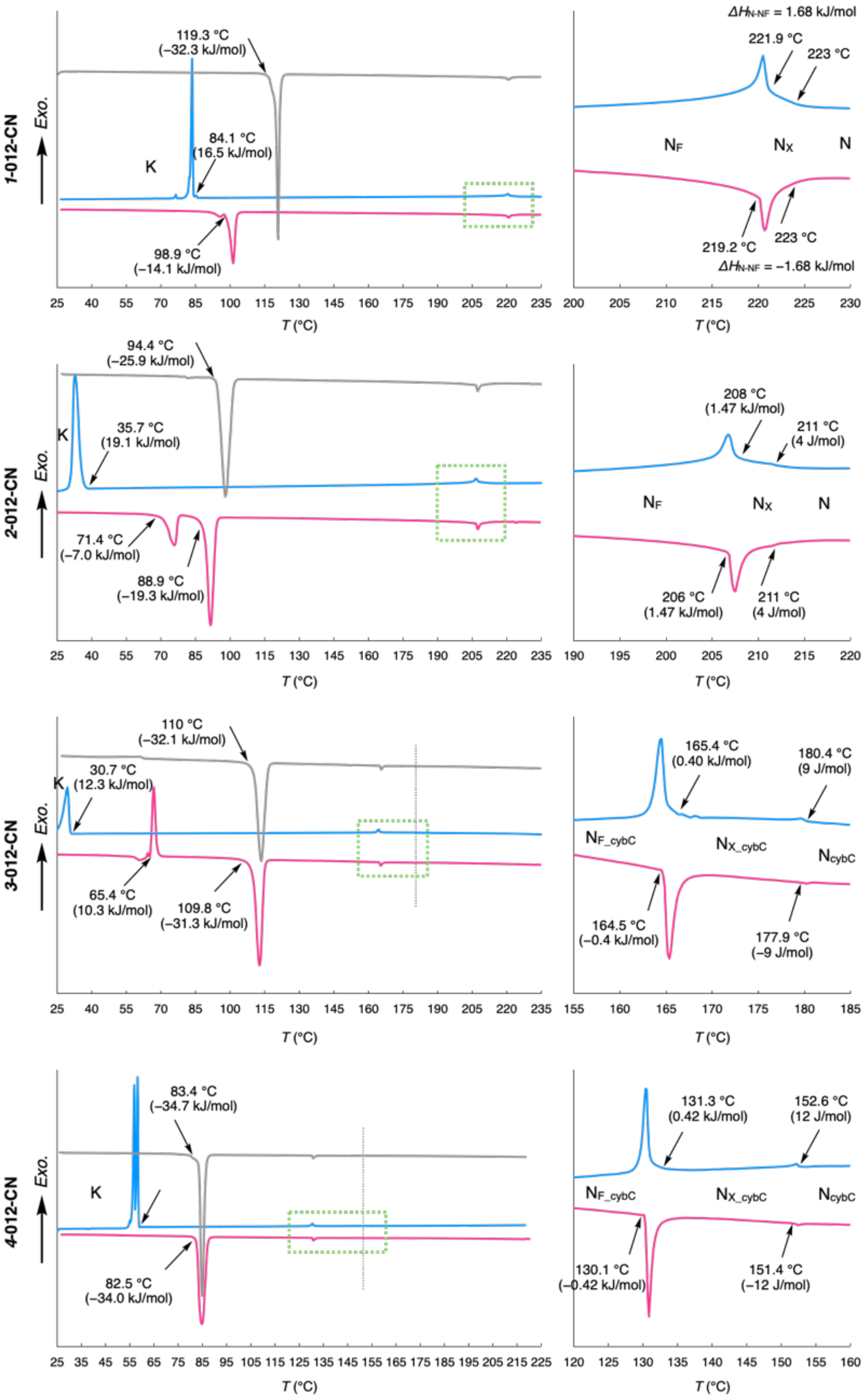

**Supplementary Fig. 7 |** DSC curves for ***n*-012-CN** (n = 1–4). The values in parentheses indicate the phase-transition enthalpies. The gray, blue, and magenta curves denote the first heating, first cooling, and second heating scans, respectively. Scan rate: 10 K min$^{-1}$.

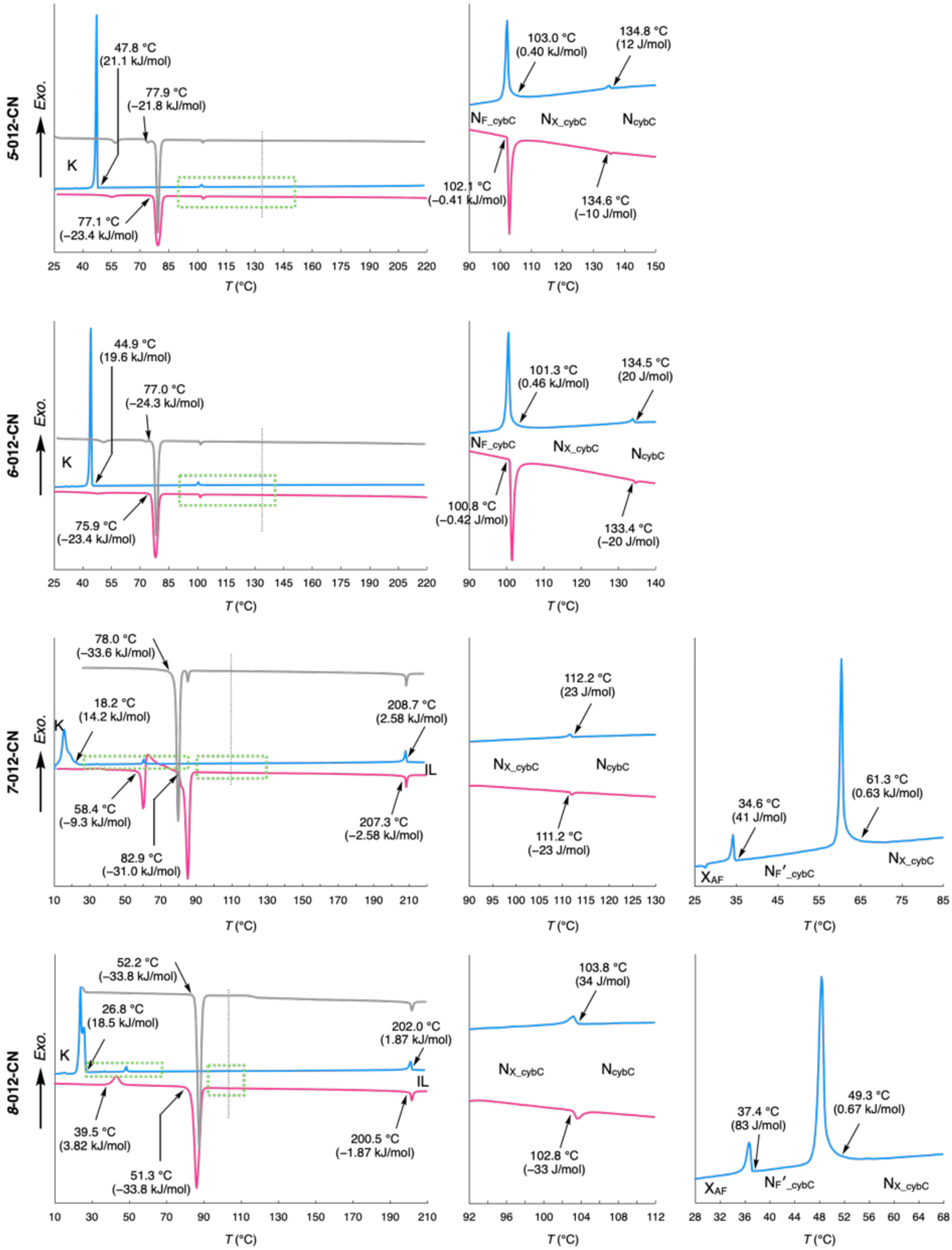

**Supplementary Fig. 8 |** DSC curves for ***n*-012-CN** (n = 5–8). The values in parentheses indicate the phase-transition enthalpies. The gray, blue, and magenta curves denote the first heating, first cooling, and second heating scans, respectively. Scan rate: 10 K min$^{-1}$.

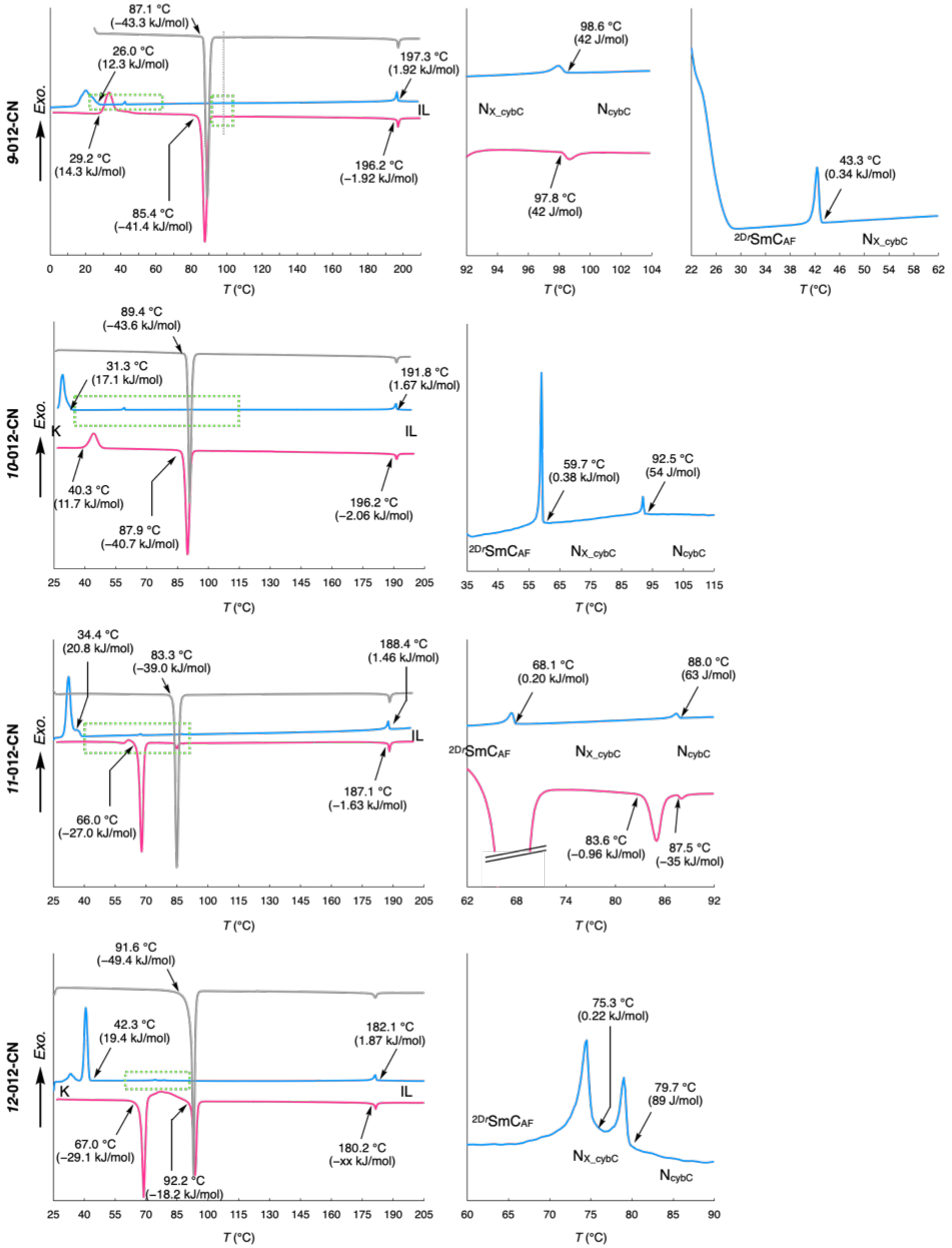

**Supplementary Fig. 9 |** DSC curves for ***n*-012-CN** (n = 9–12). The values in parentheses indicate the phase-transition enthalpies. The gray, blue, and magenta curves denote the first heating, first cooling, and second heating scans, respectively. Scan rate: 10 K min$^{-1}$.

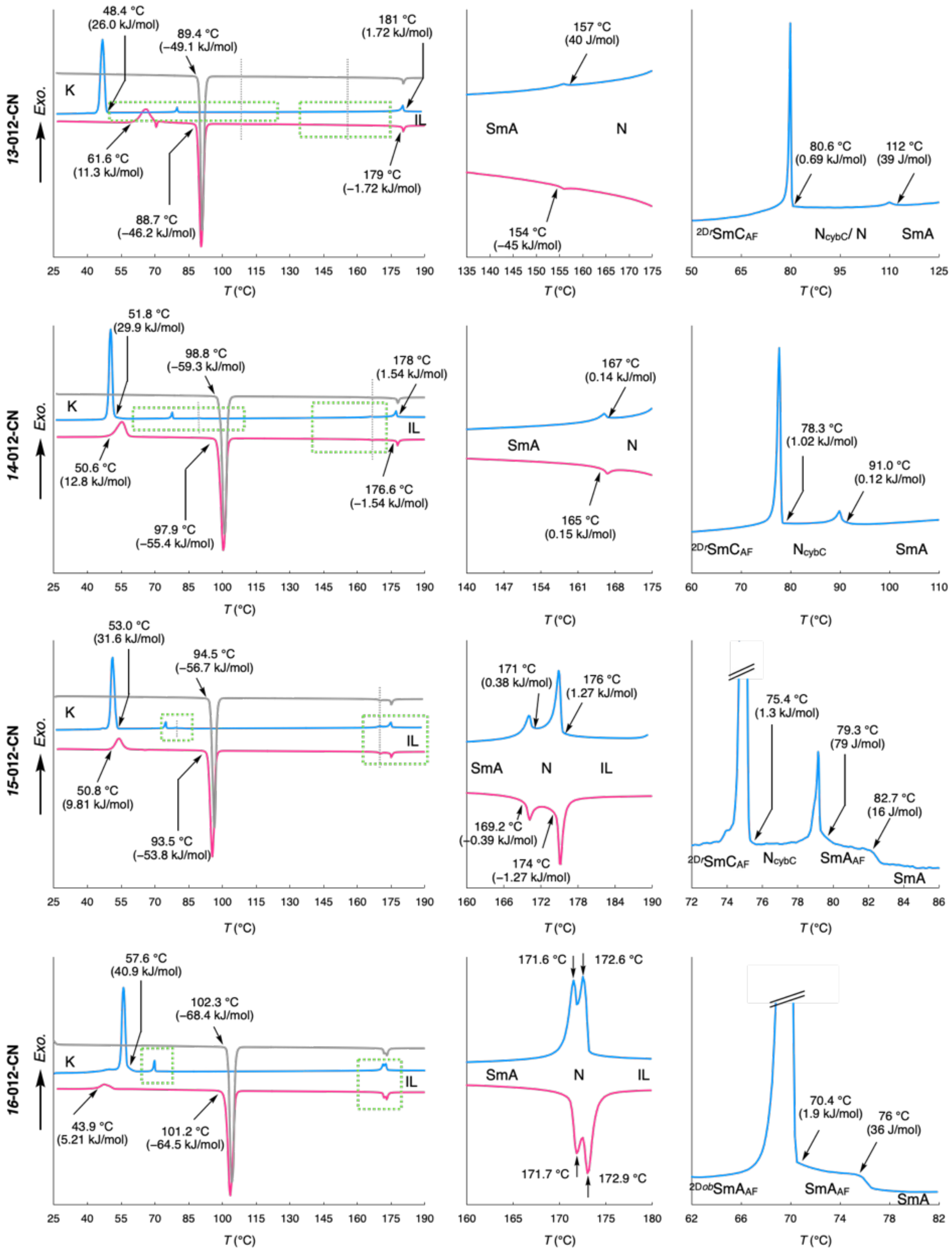

**Supplementary Fig. 10 |** DSC curves for ***n*-012-CN** (n = 13–16). The values in parentheses indicate the phase-transition enthalpies. The gray, blue, and magenta curves denote the first heating, first cooling, and second heating scans, respectively. Scan rate: 10 K min$^{-1}$.

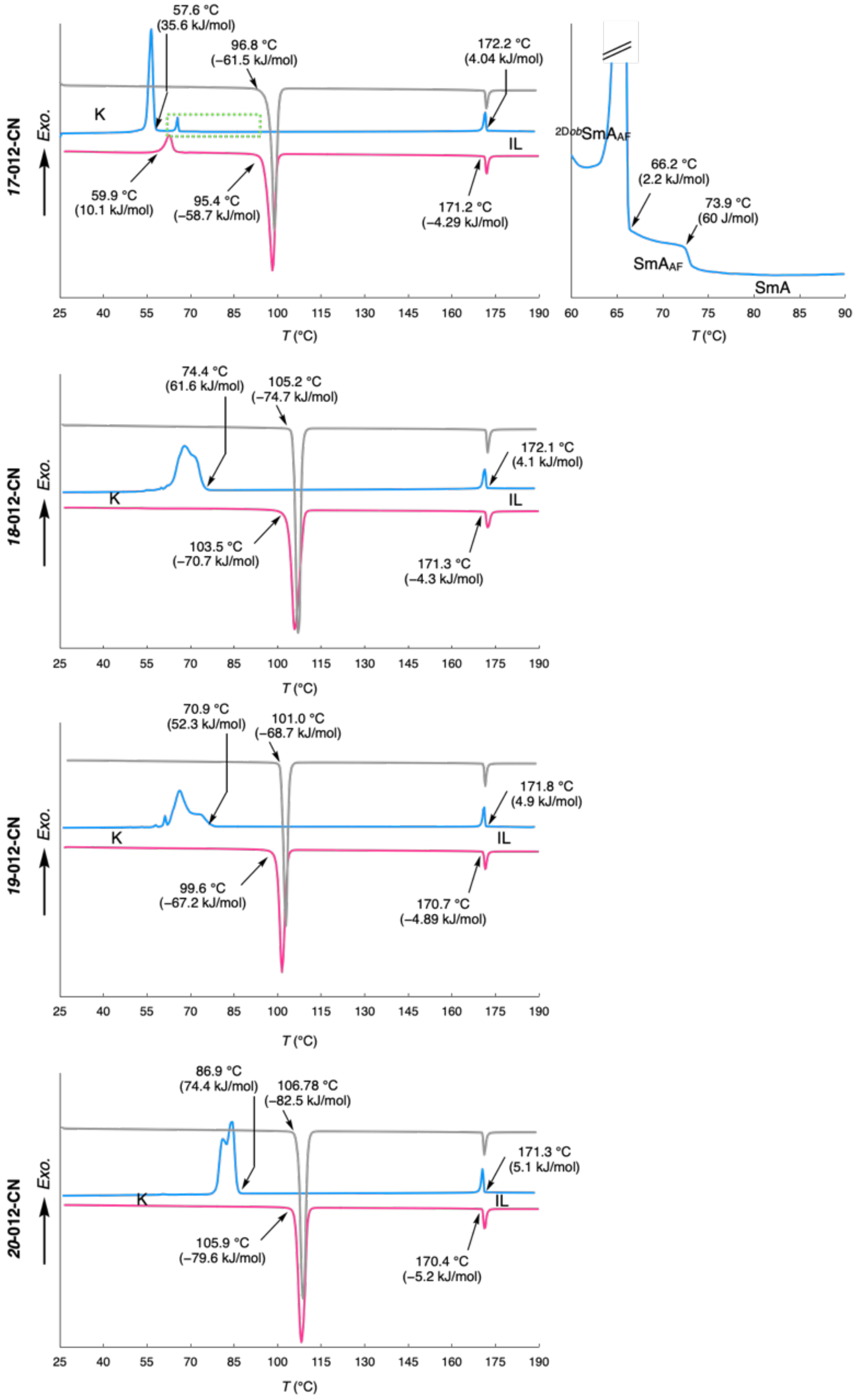

**Supplementary Fig. 11 |** DSC curves for ***n*-012-CN** (n = 17–20). The values in parentheses indicate the phase-transition enthalpies. The gray, blue, and magenta curves denote the first heating, first cooling, and second heating scans, respectively. Scan rate: 10 K min$^{-1}$.

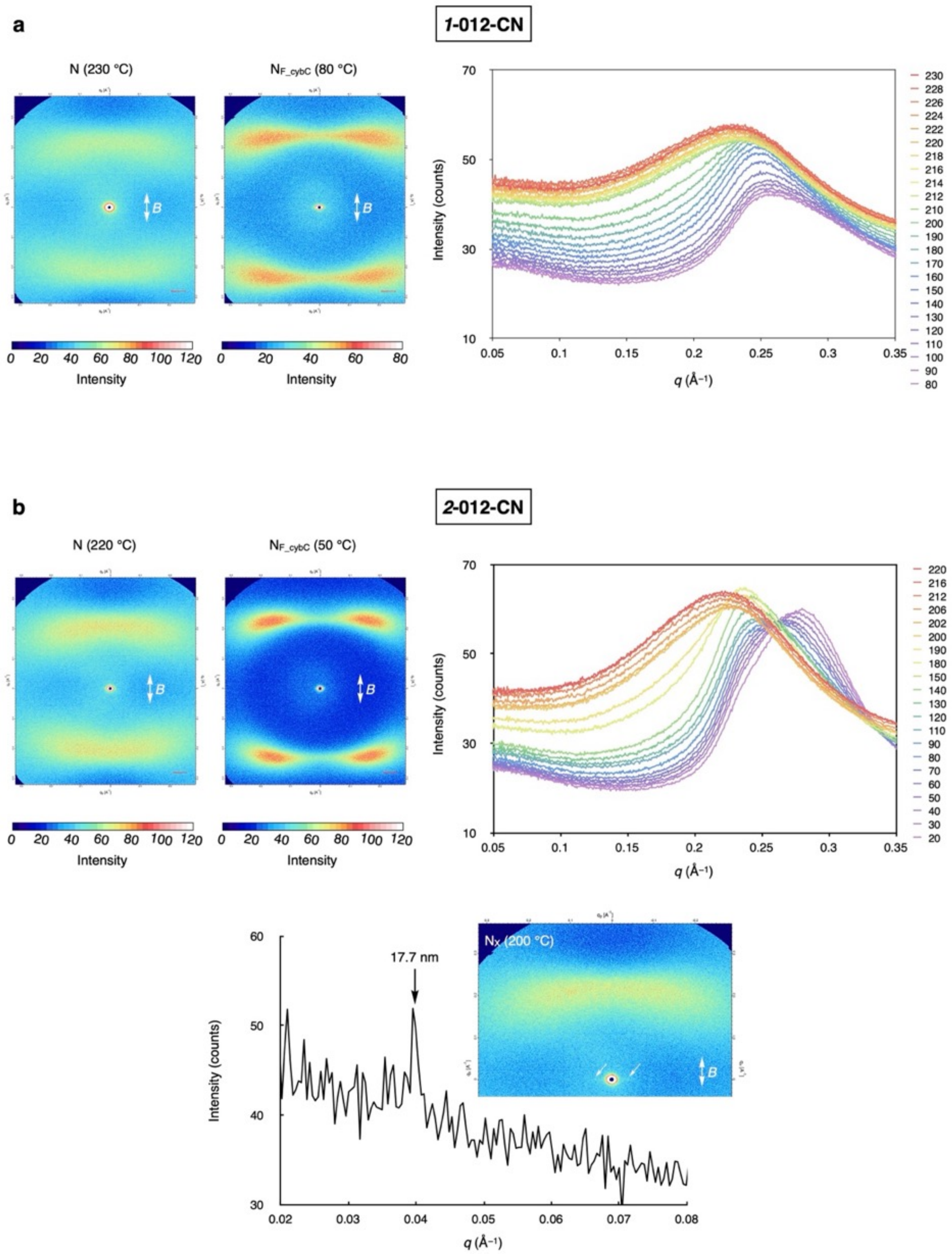

**Supplementary Fig. 12 |** 2D and corresponding 1D X-ray diffraction patterns at various temperatures under the magnetic field (~1 T, **B ∥ n**) for ***1*-012-CN** (a) and ***2*-012-CN** (b).

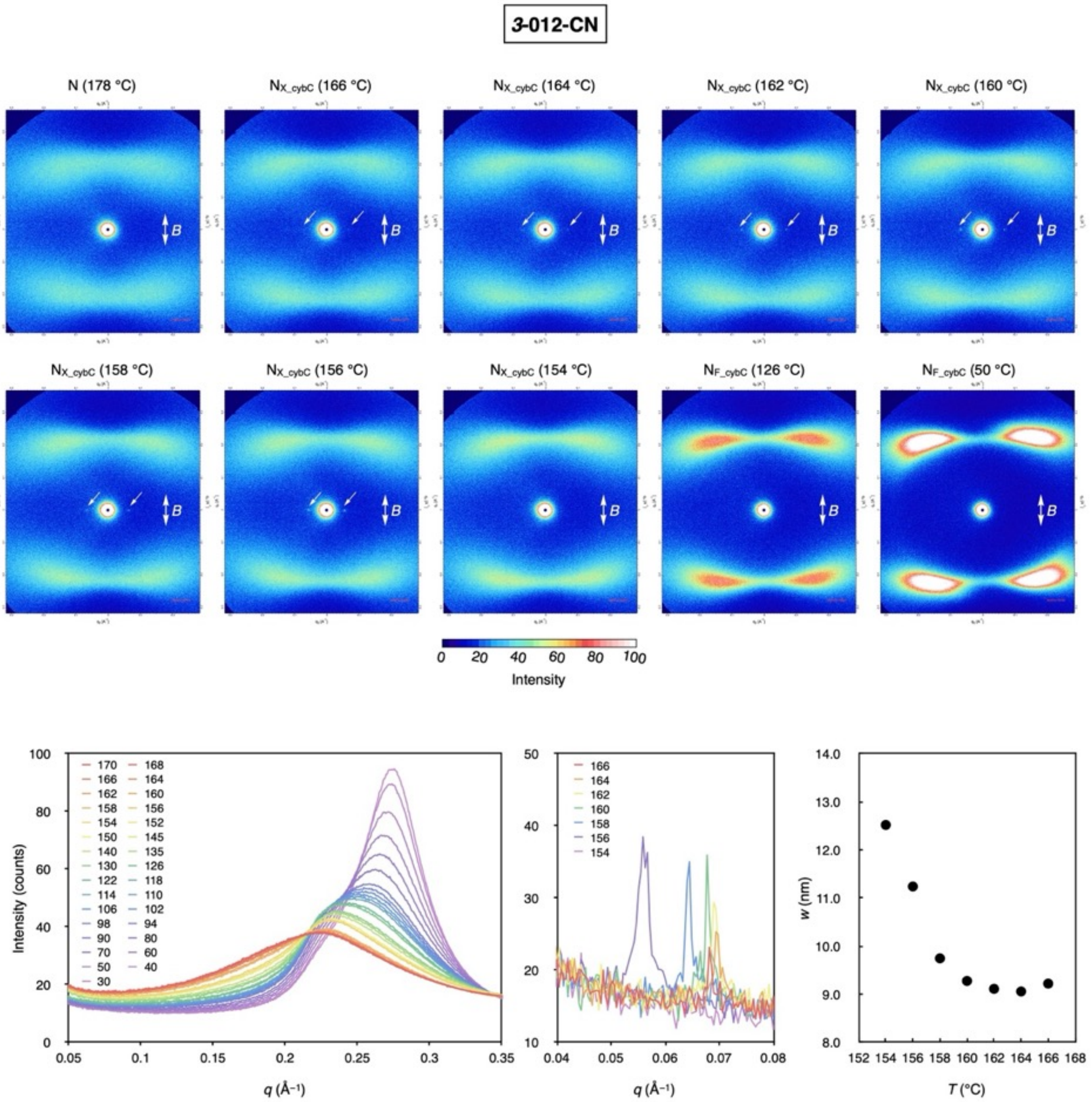

**Supplementary Fig. 13 |** 2D, corresponding 1D X-ray diffraction patterns and analyzed data at various temperatures under the magnetic field (~1 T, **B ∥ n**) for ***3*-012-CN**.

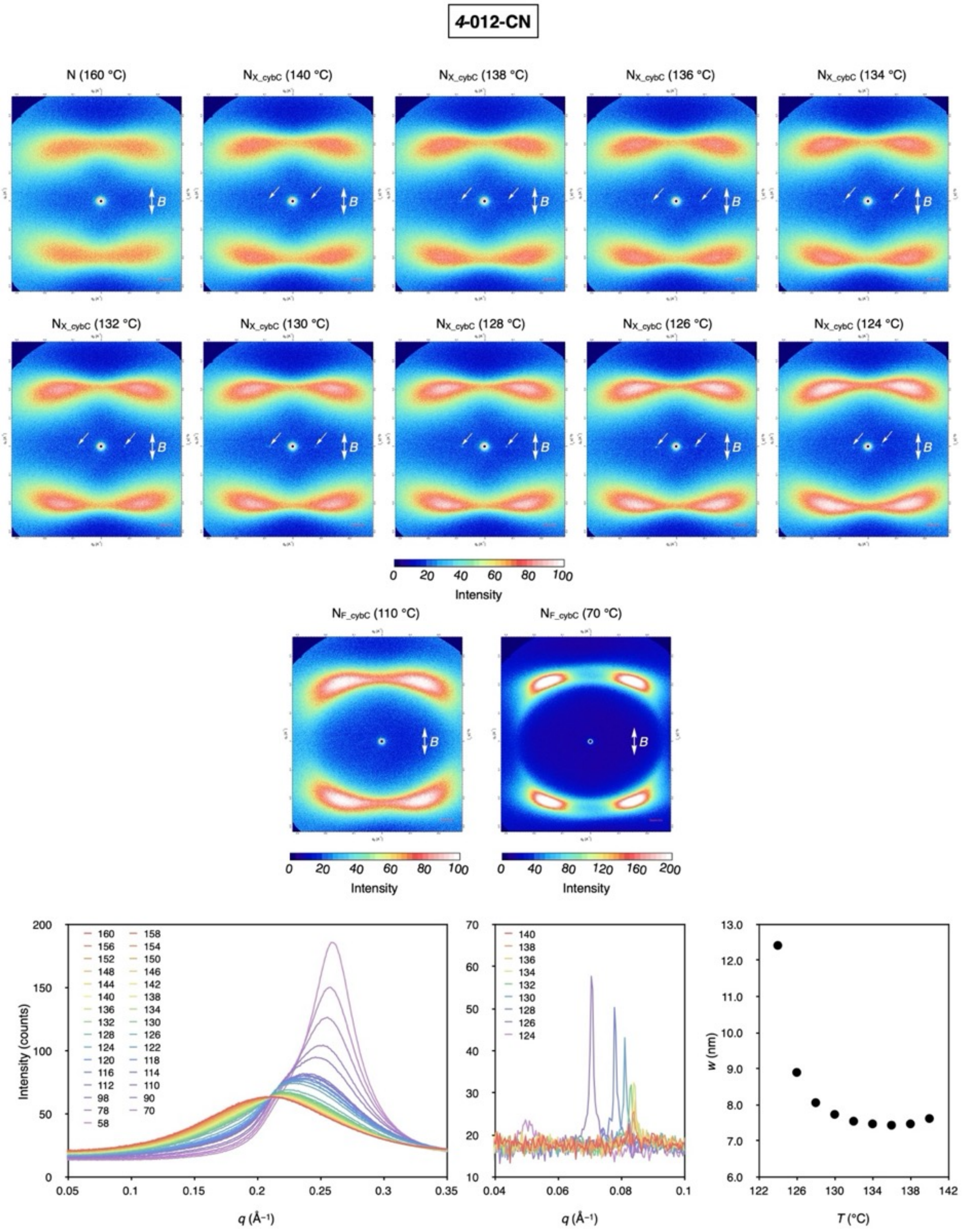

**Supplementary Fig. 14 |** 2D, corresponding 1D X-ray diffraction patterns and analyzed data at various temperatures under the magnetic field (~1 T, **B ‖ n**) for ***4*-012-CN**.

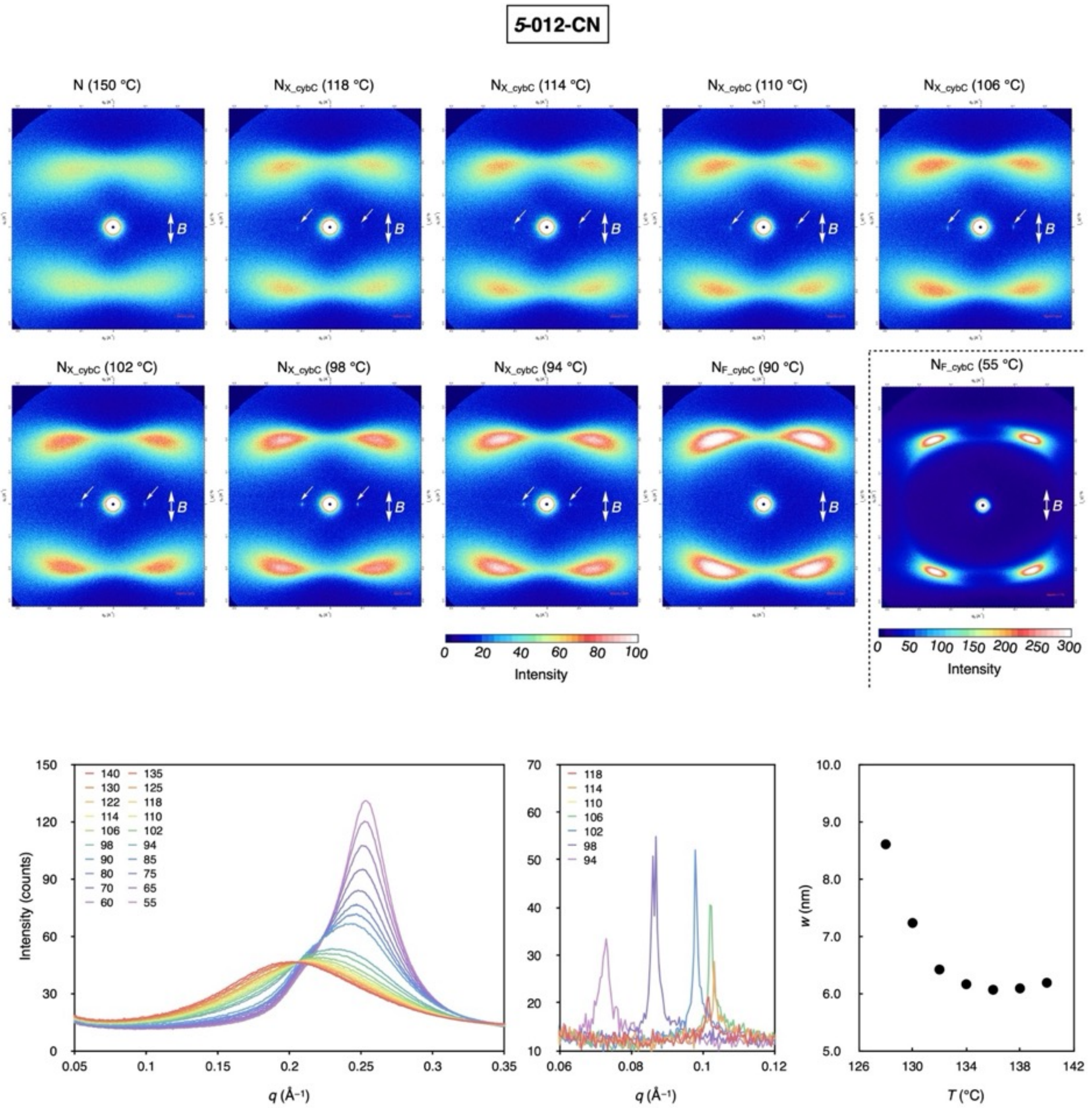

**Supplementary Fig. 15 |** 2D, corresponding 1D X-ray diffraction patterns and analyzed data at various temperatures under the magnetic field (~1 T, **B ∥ n**) for **5-012-CN**.

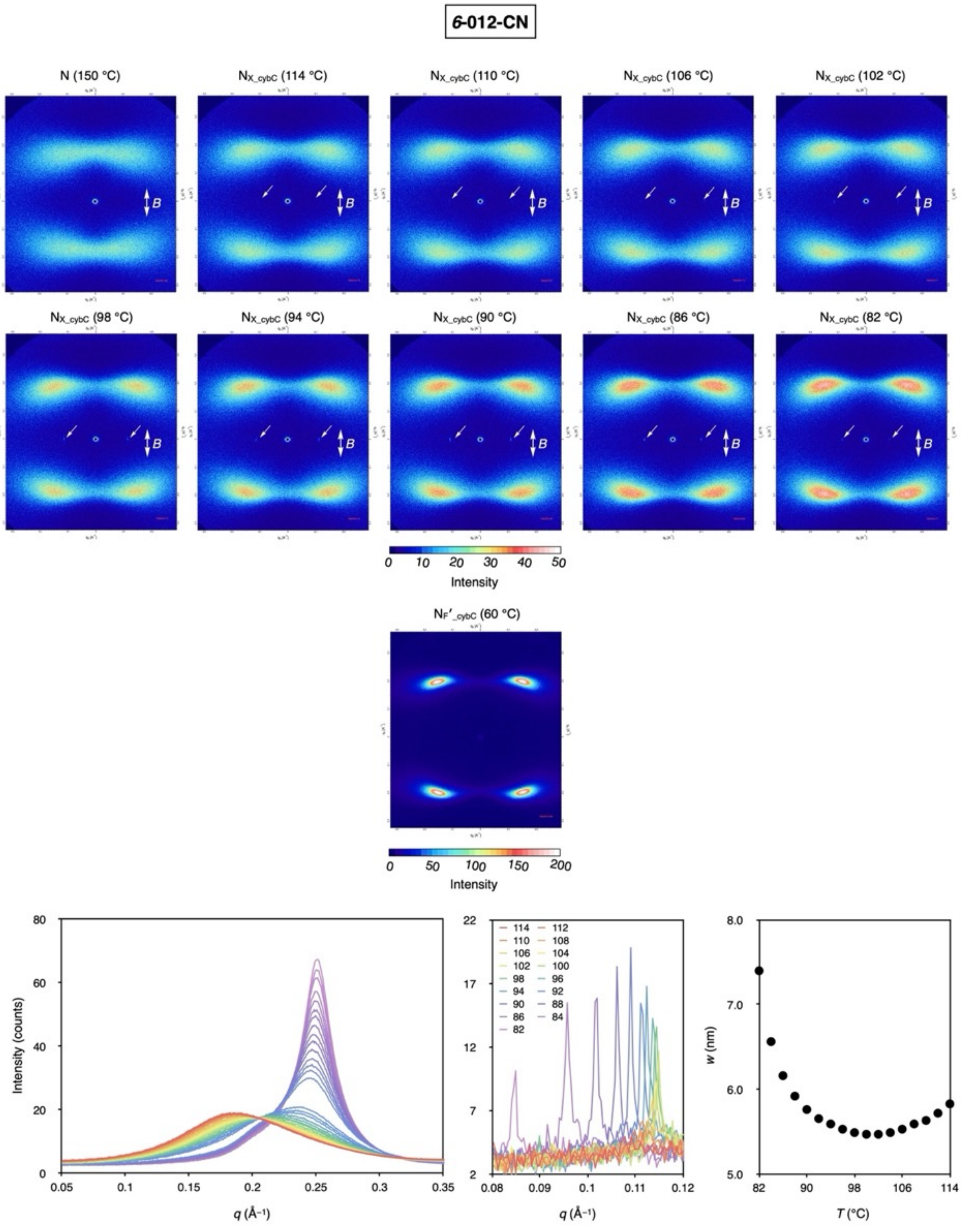

**Supplementary Fig. 16 |** 2D, corresponding 1D X-ray diffraction patterns and analyzed data at various temperatures under the magnetic field (~1 T, **B ∥ n**) for ***6*-012-CN**.

**7-012-CN**

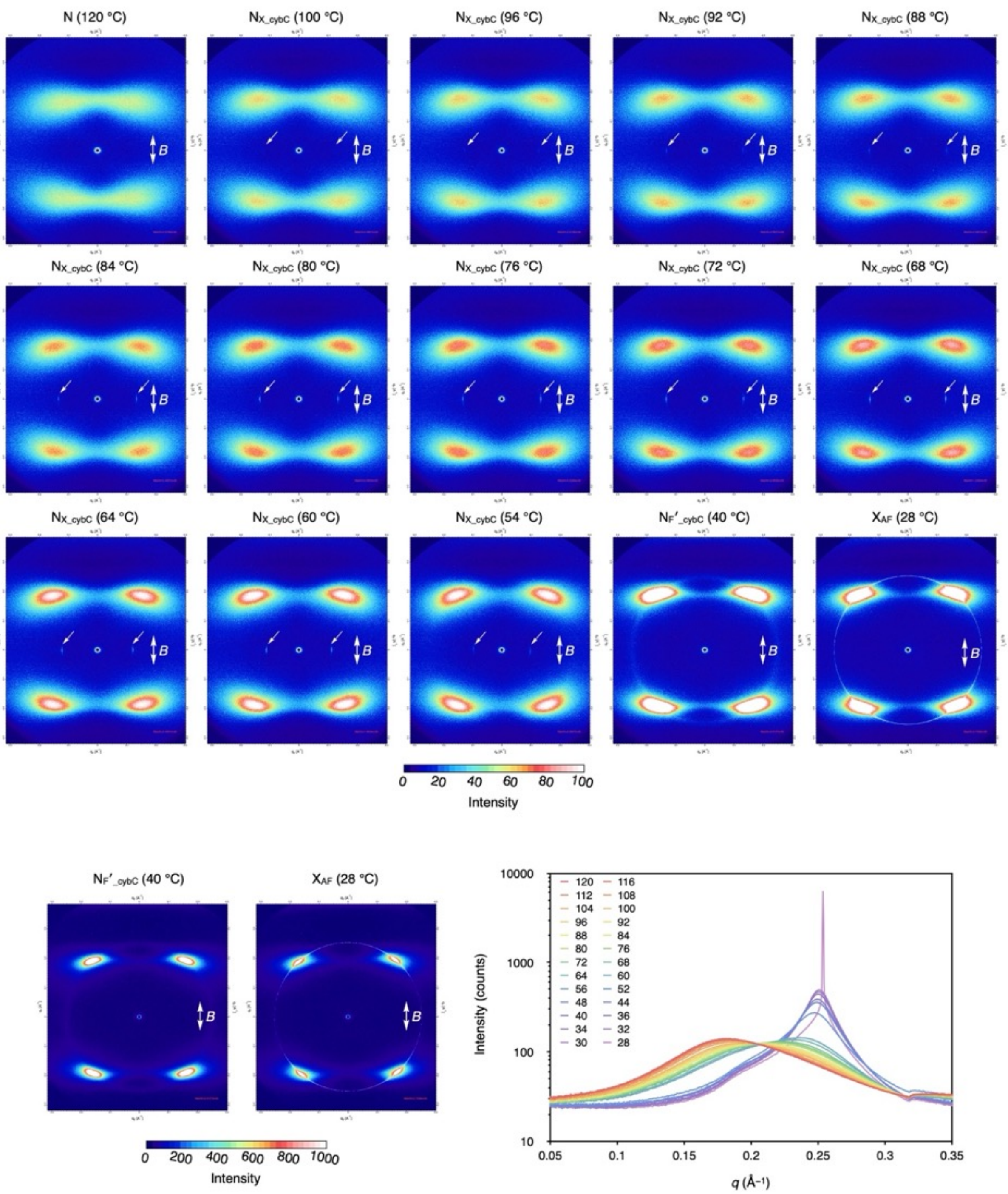

**Supplementary Fig. 17 |** 2D and corresponding 1D X-ray diffraction patterns and analyzed data at various temperatures under the magnetic field (~1 T, **B ∥ n**) for ***7*-012-CN**.

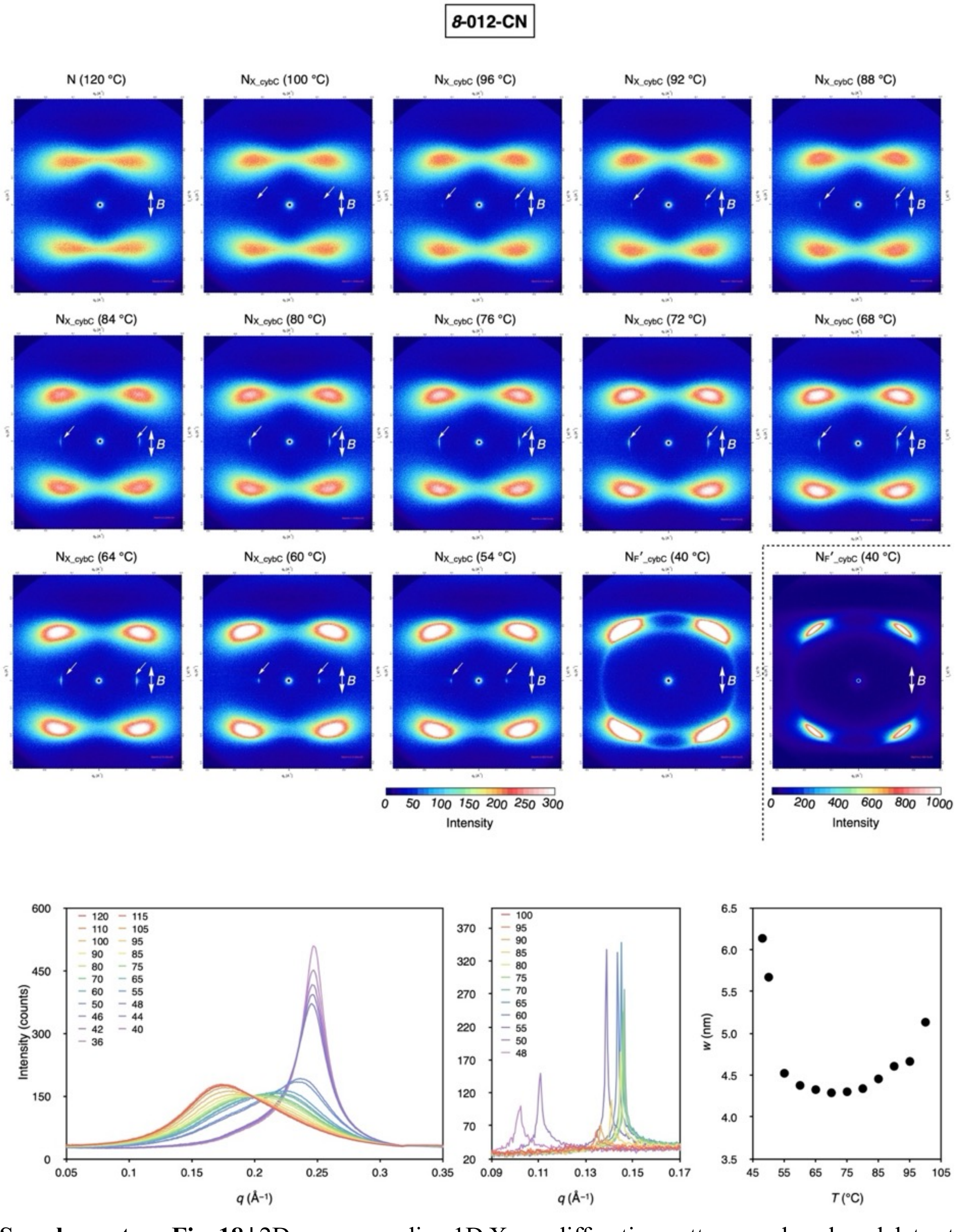

**Supplementary Fig. 18 |** 2D, corresponding 1D X-ray diffraction patterns and analyzed data at various temperatures under the magnetic field (~1 T, **B ∥ n**) for ***8*-012-CN**.

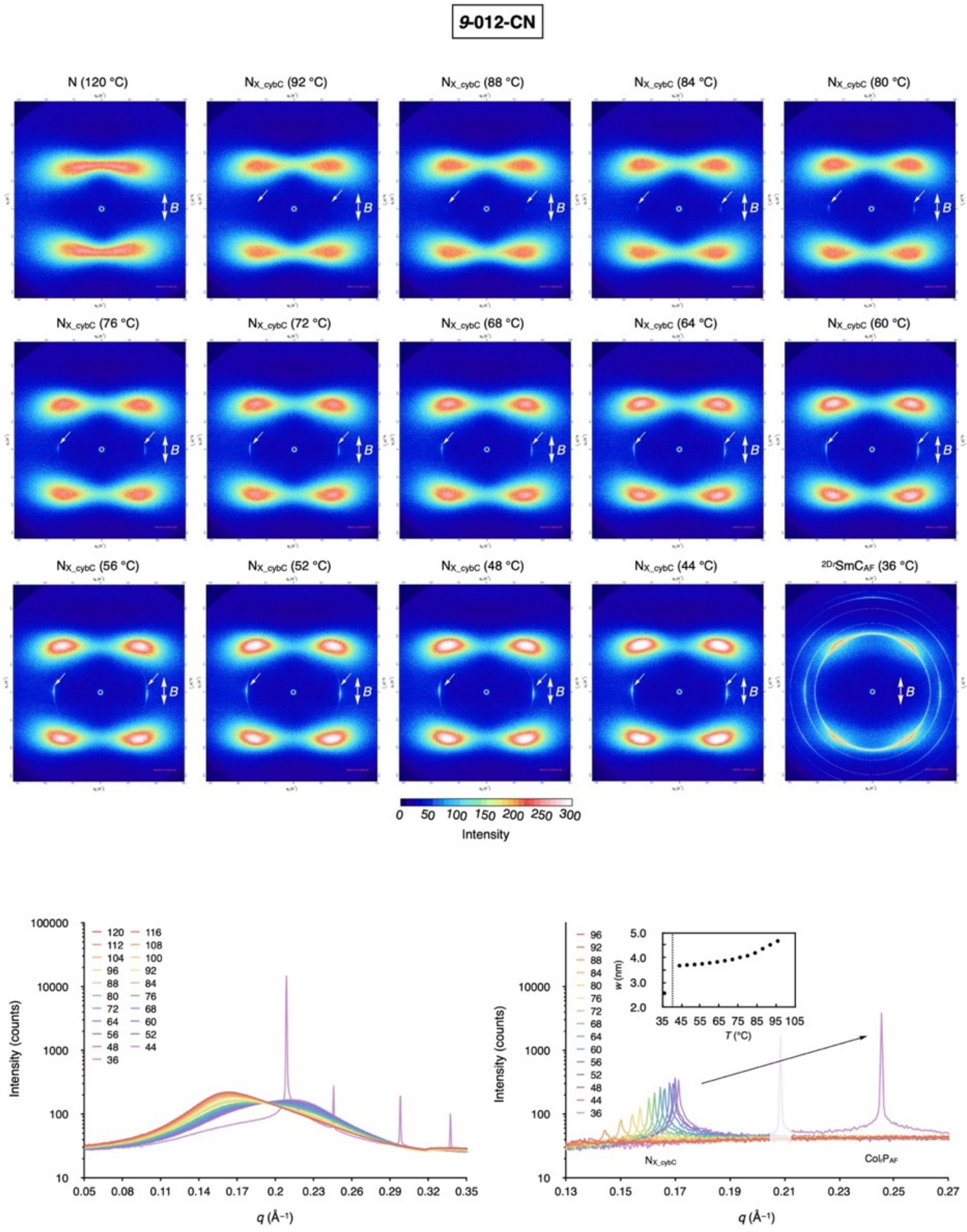

**Supplementary Fig. 19 |** 2D, corresponding 1D X-ray diffraction patterns and analyzed data at various temperatures under the magnetic field (~1 T, **B ∥ n**) for ***9*-012-CN**.

**9-012-CN**

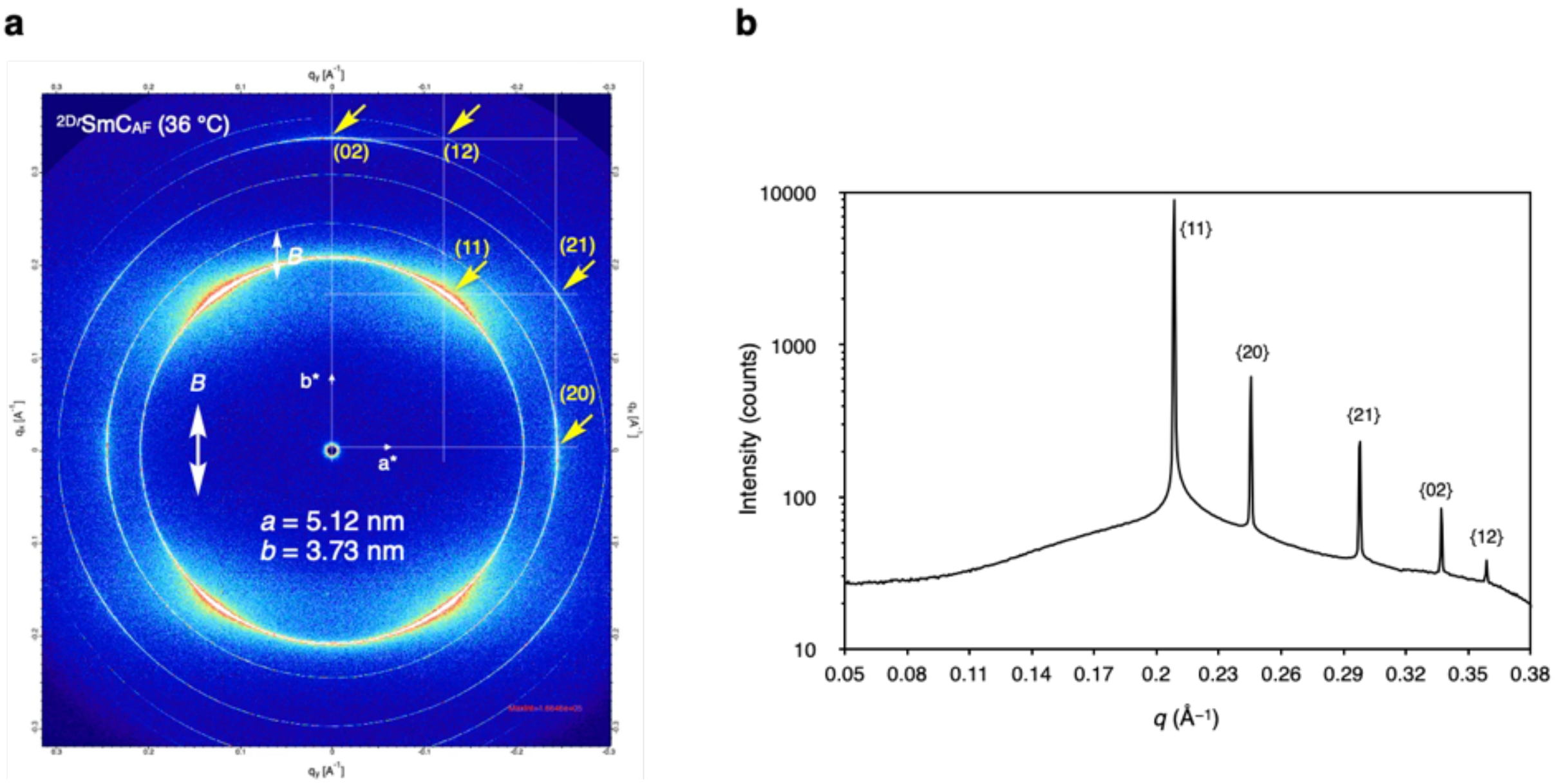

**Supplementary Fig. 20 |** XRD data at various temperatures under the magnetic field (~1 T, **B** ∥ **n**) for ***9*-012-CN**: 2D (a) and corresponding 1D (b) X-ray diffraction patterns.

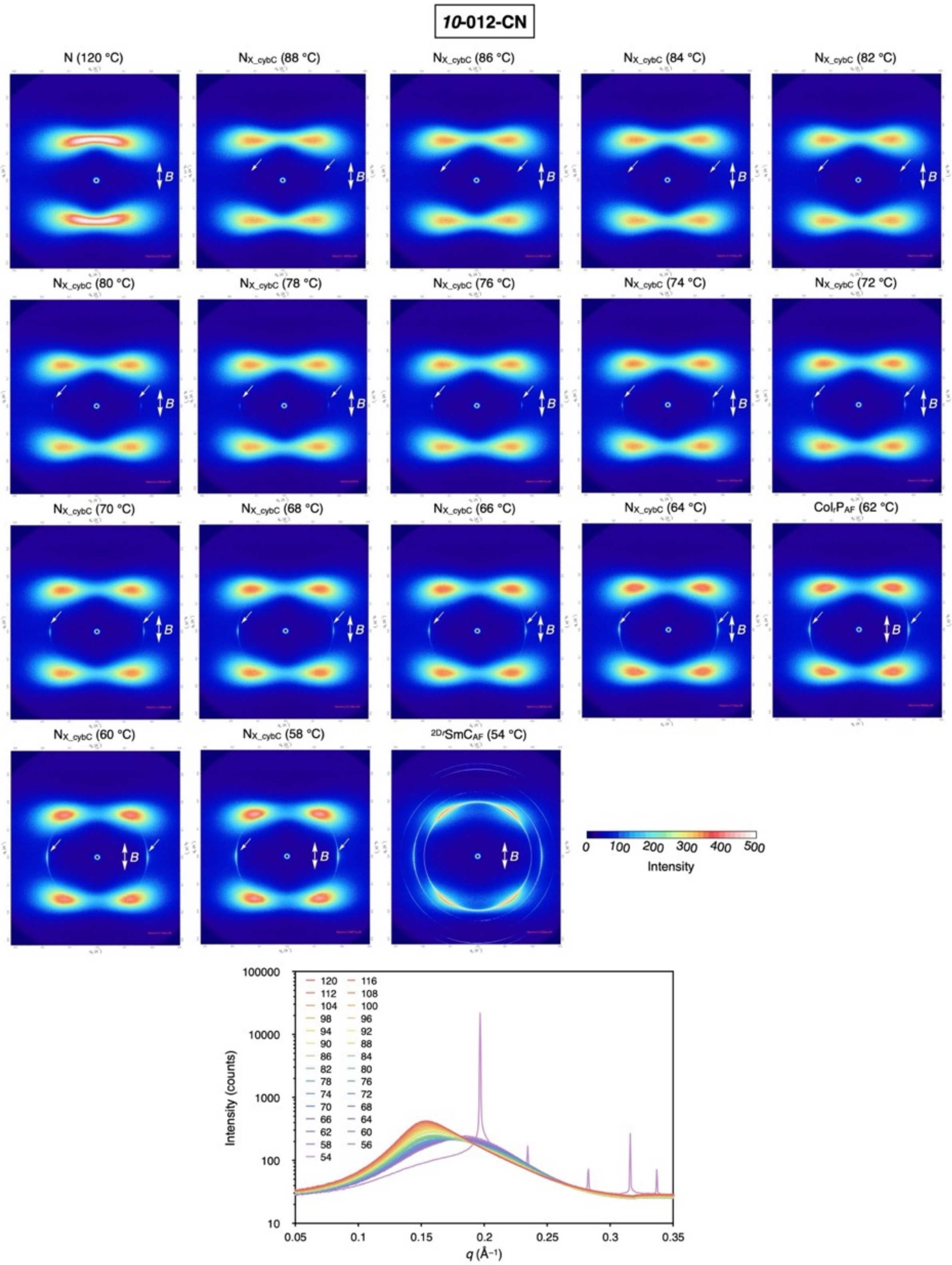

**Supplementary Fig. 21 |** 2D and corresponding 1D X-ray diffraction patterns at various temperatures under the magnetic field (~1 T, **B ∥ n**) for ***10*-012-CN**.

**10-012-CN**

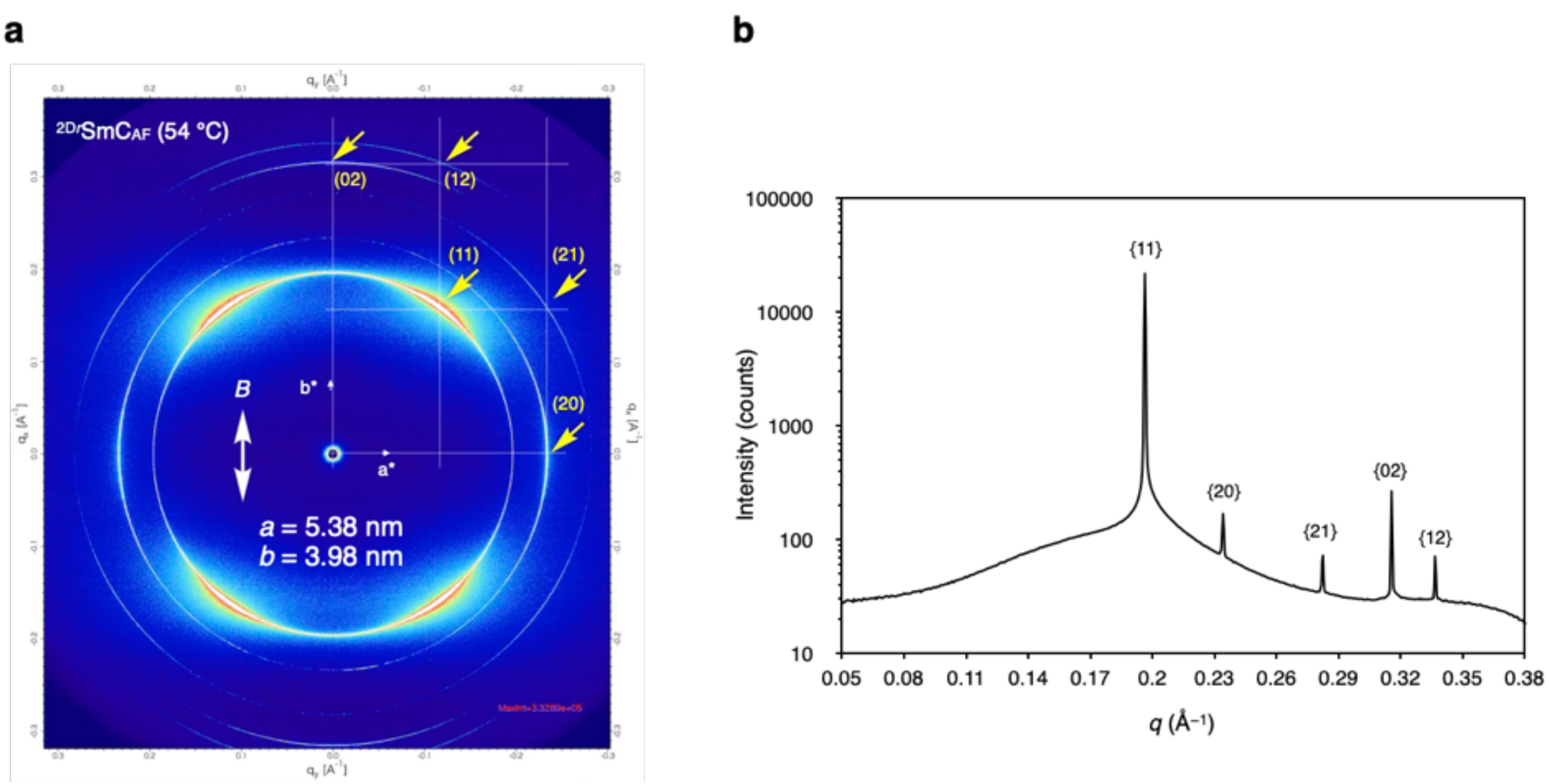

**Supplementary Fig. 22 |** XRD data at various temperatures under the magnetic field (~1 T, **B** ∥ **n**) for ***10*-012-CN**: 2D (a) and corresponding 1D (b) X-ray diffraction patterns.

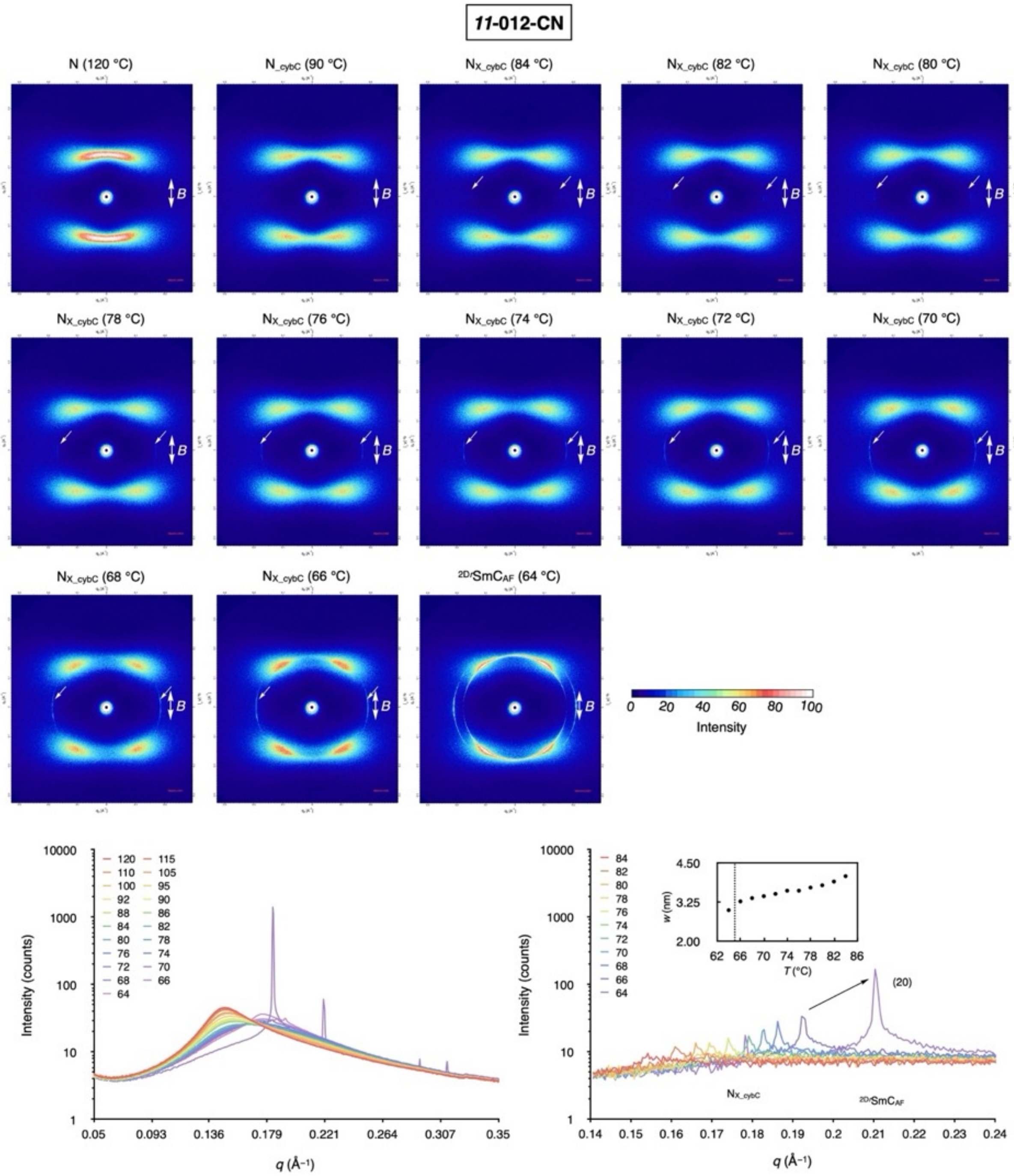

**Supplementary Fig. 23 |** 2D, corresponding 1D X-ray diffraction patterns and analyzed data at various temperatures under the magnetic field (~1 T, **B ∥ n**) for ***11*-012-CN**.

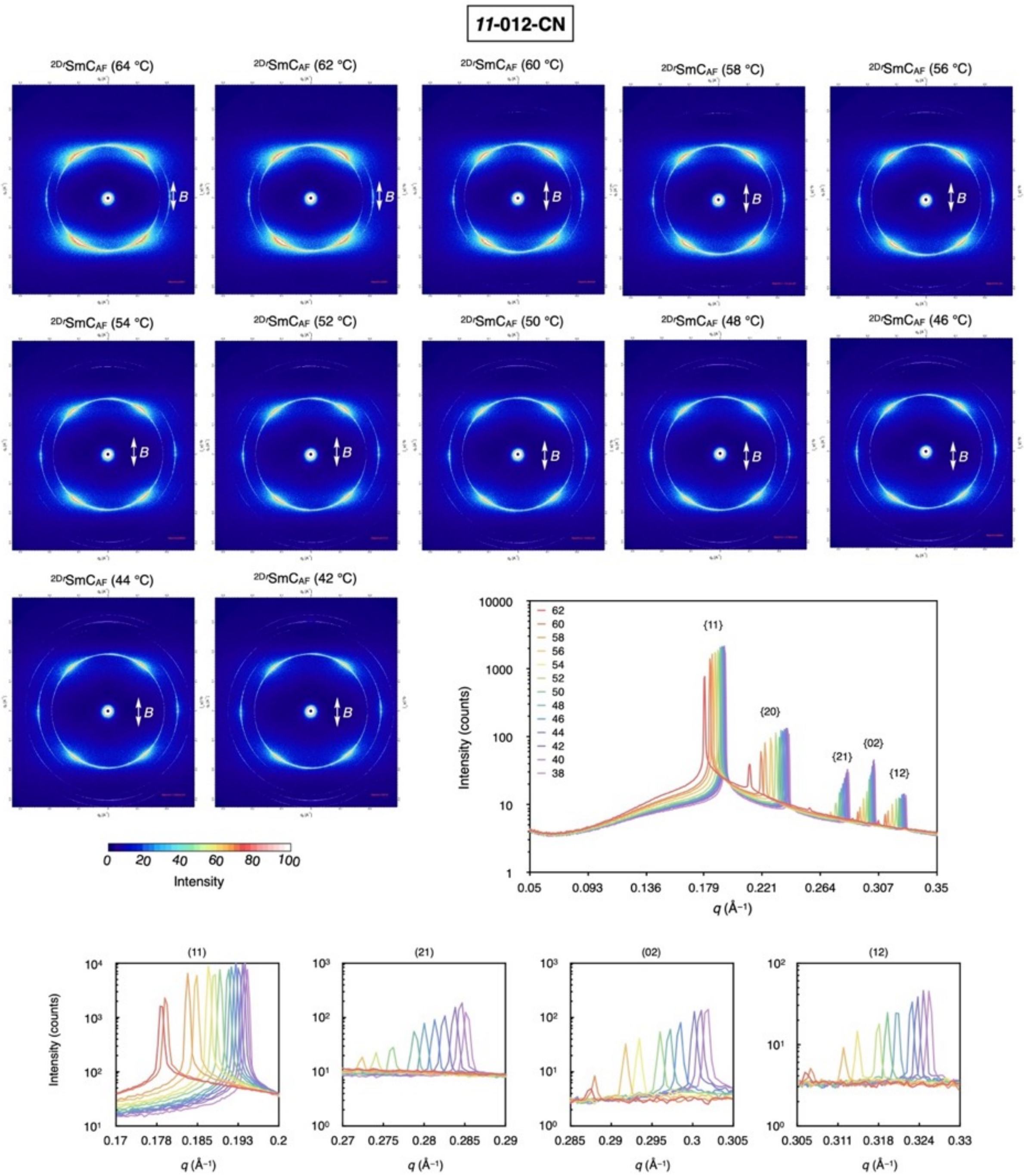

**Supplementary Fig. 24 |** 2D, corresponding 1D X-ray diffraction patterns and analyzed data at various temperatures under the magnetic field (~1 T, **B ∥ n**) for ***11*-012-CN**.

**11-012-CN**

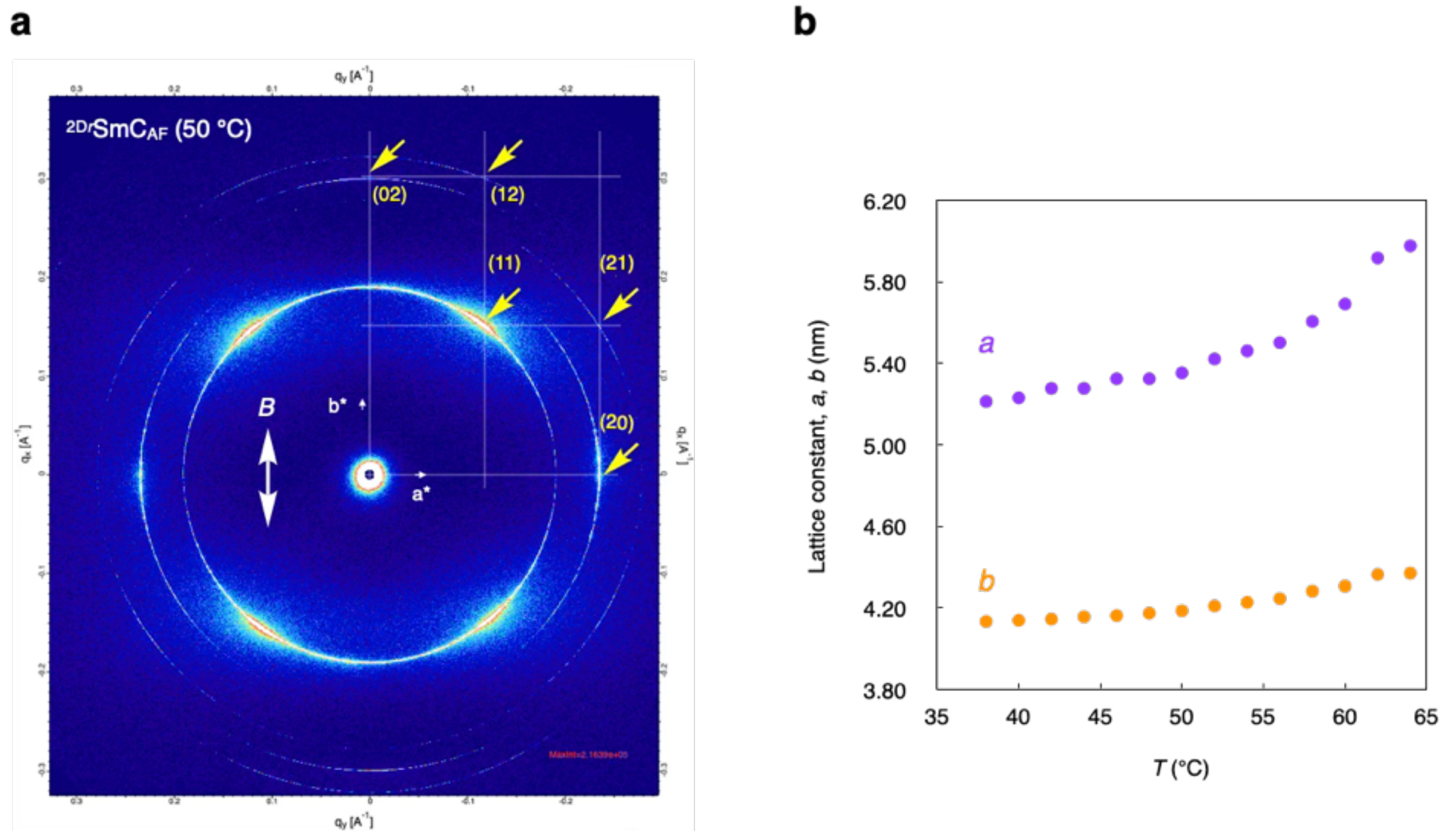

**Supplementary Fig. 25 |** XRD data at various temperatures under the magnetic field (~1 T, **B** ∥ **n**) for ***11*-012-CN**.

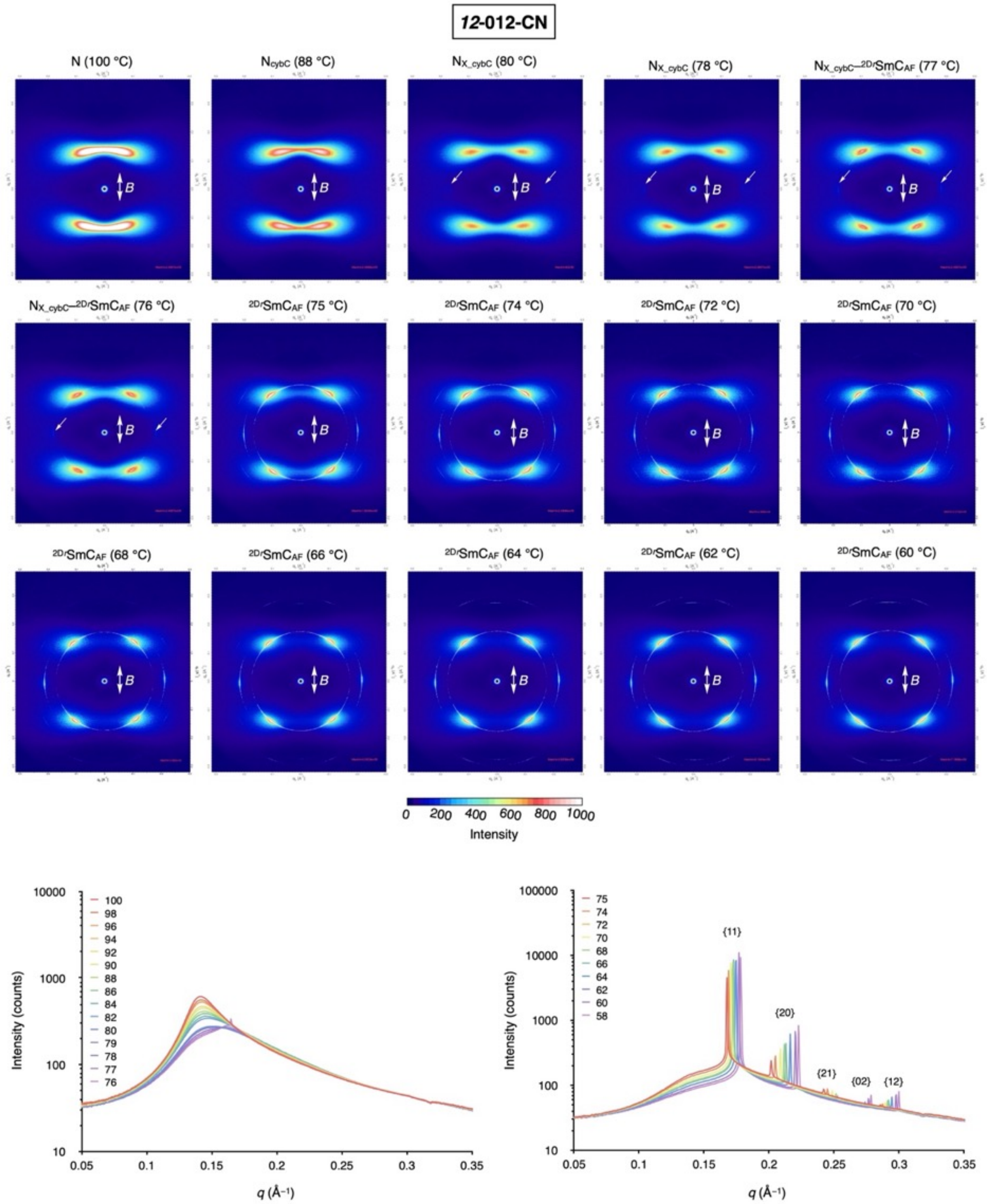

**Supplementary Fig. 26 |** 2D and corresponding 1D X-ray diffraction patterns at various temperatures under the magnetic field (~1 T, **B ∥ n**) for ***12*-012-CN**.

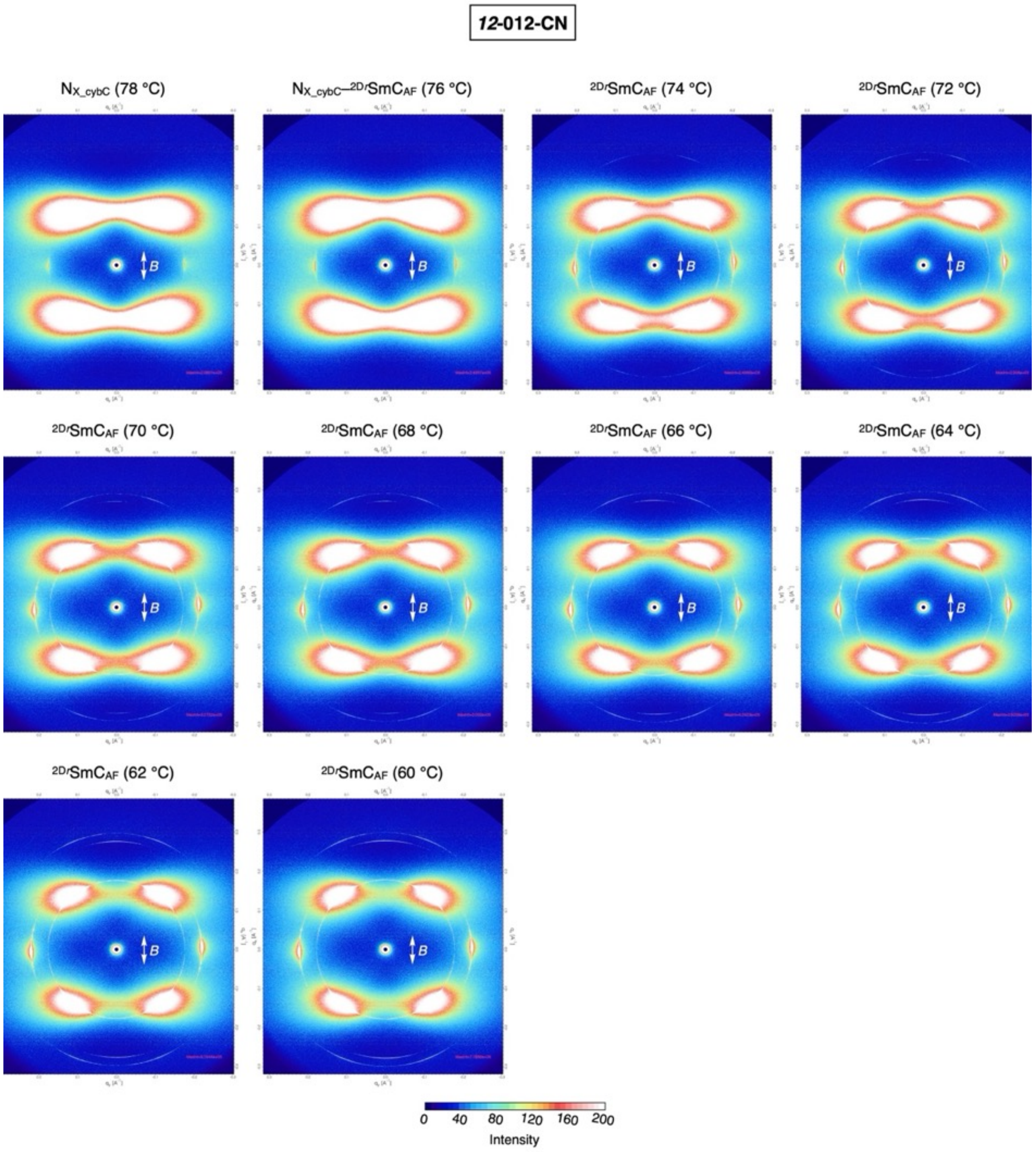

**Supplementary Fig. 27 |** 2D SAXS patterns at various temperatures under the magnetic field (~1 T, **B** ∥ **n**) for ***12*-012-CN**.

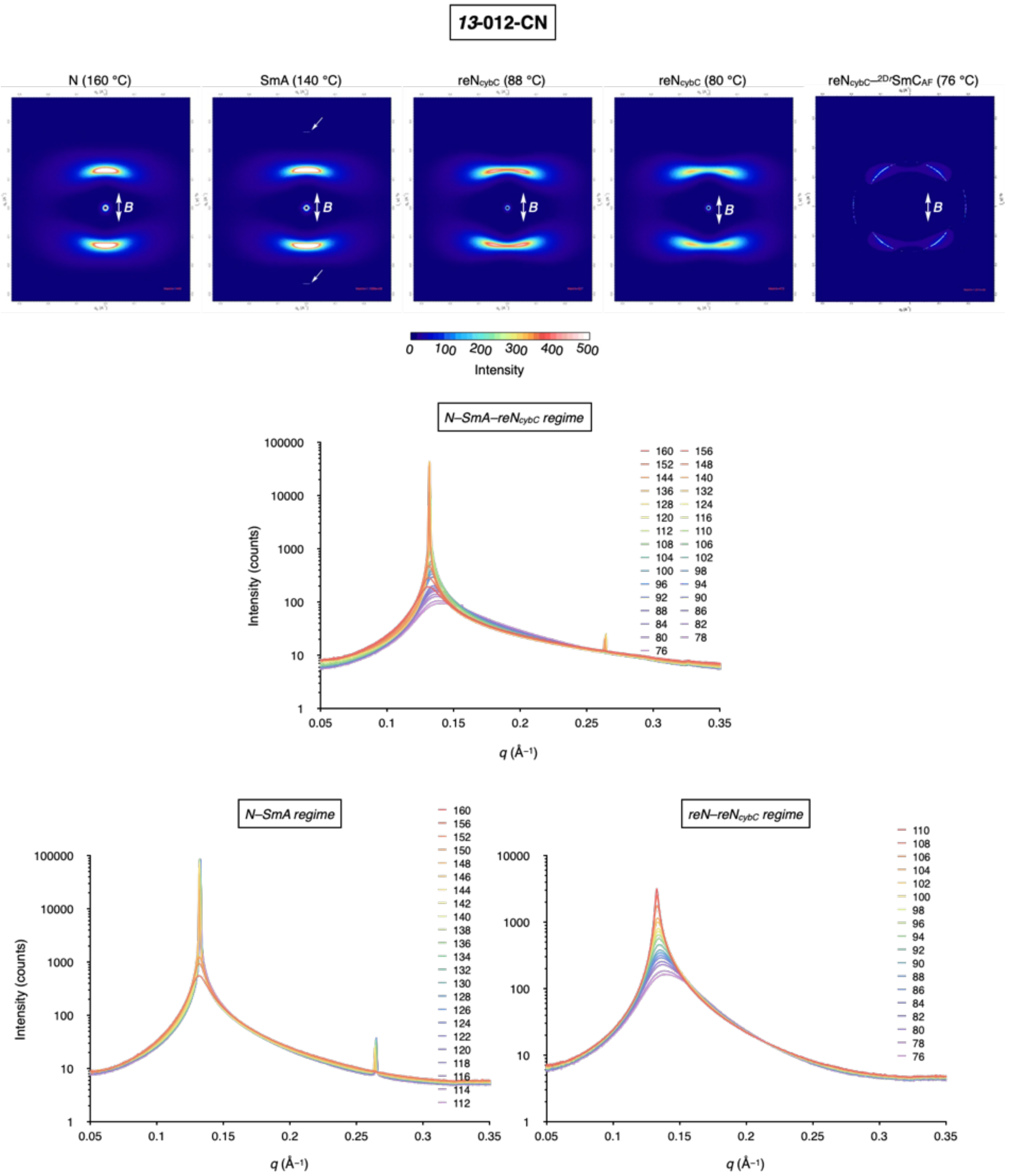

**Supplementary Fig. 28 |** 2D and corresponding 1D X-ray diffraction patterns at various temperatures under the magnetic field (~1 T, **B** ∥ **n**) for ***13*-012-CN**. In the SmA phase, white arrows indicate overtone peaks.

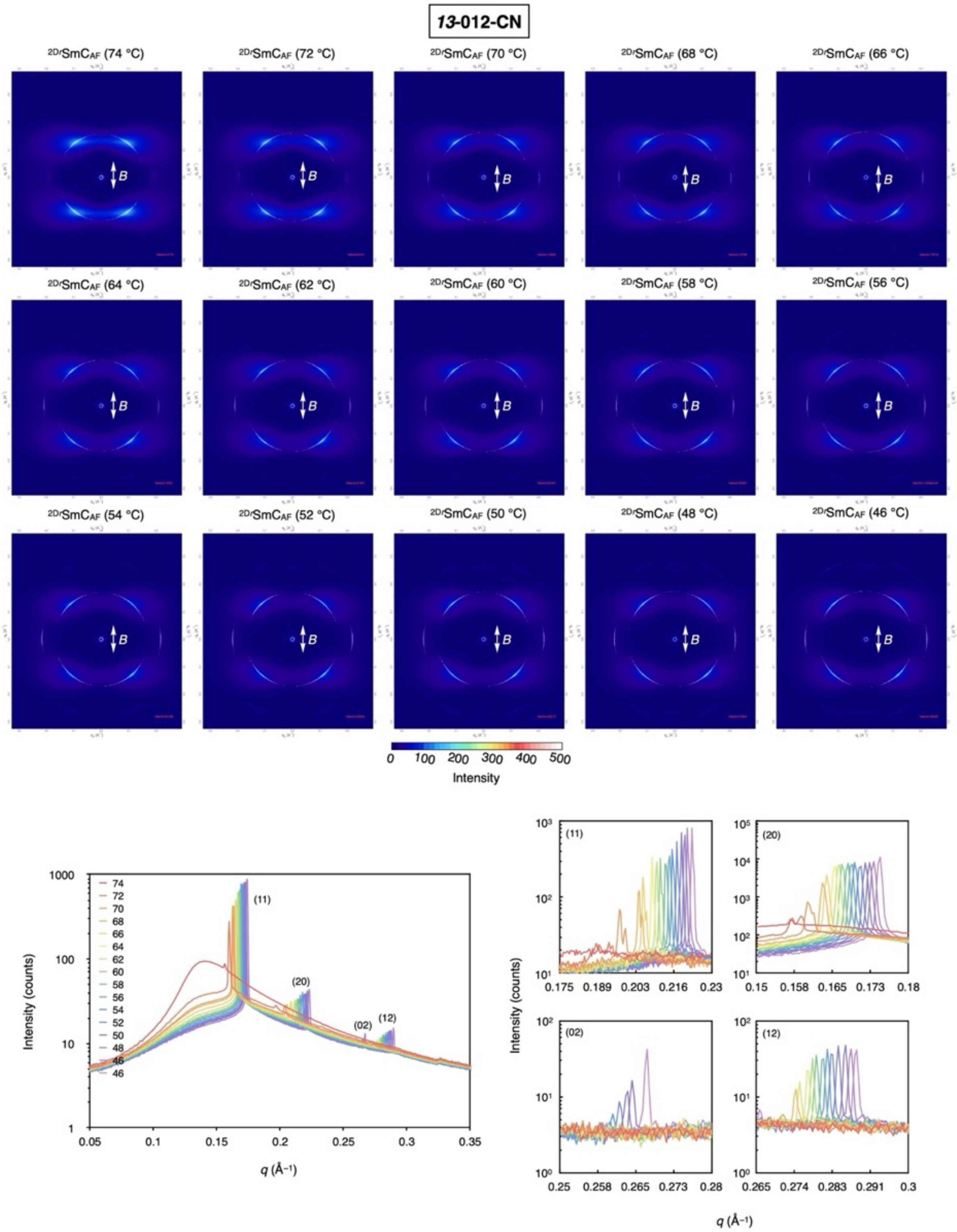

**Supplementary Fig. 29 |** 2D and corresponding 1D X-ray diffraction patterns at various temperatures under the magnetic field (~1 T, **B ∥ n**) for ***13*-012-CN**.

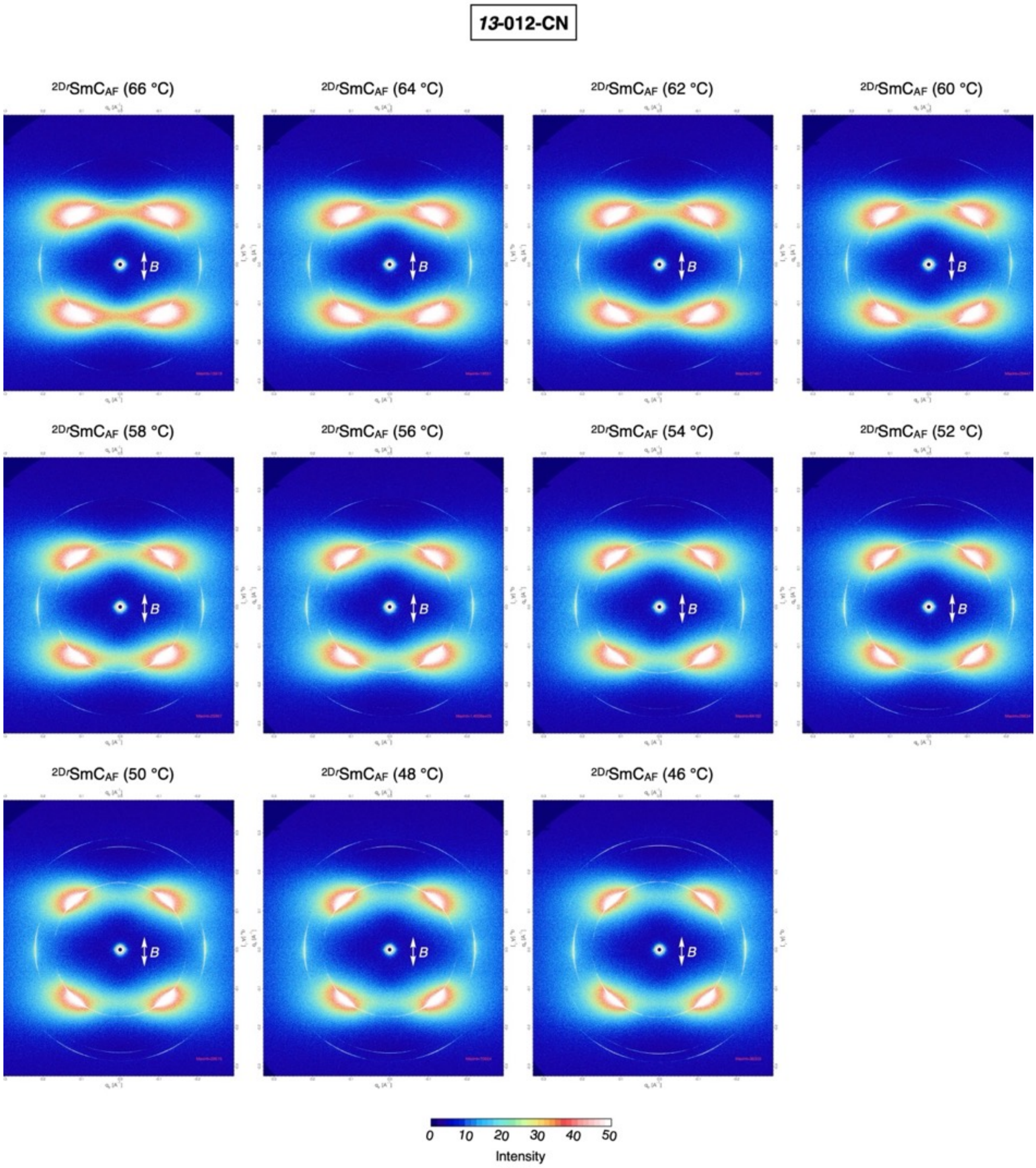

**Supplementary Fig. 30 |** 2D SAXS patterns at various temperatures under the magnetic field (~1 T, **B** ∥ **n**) for ***13*-012-CN**.

**_13_-012-CN**

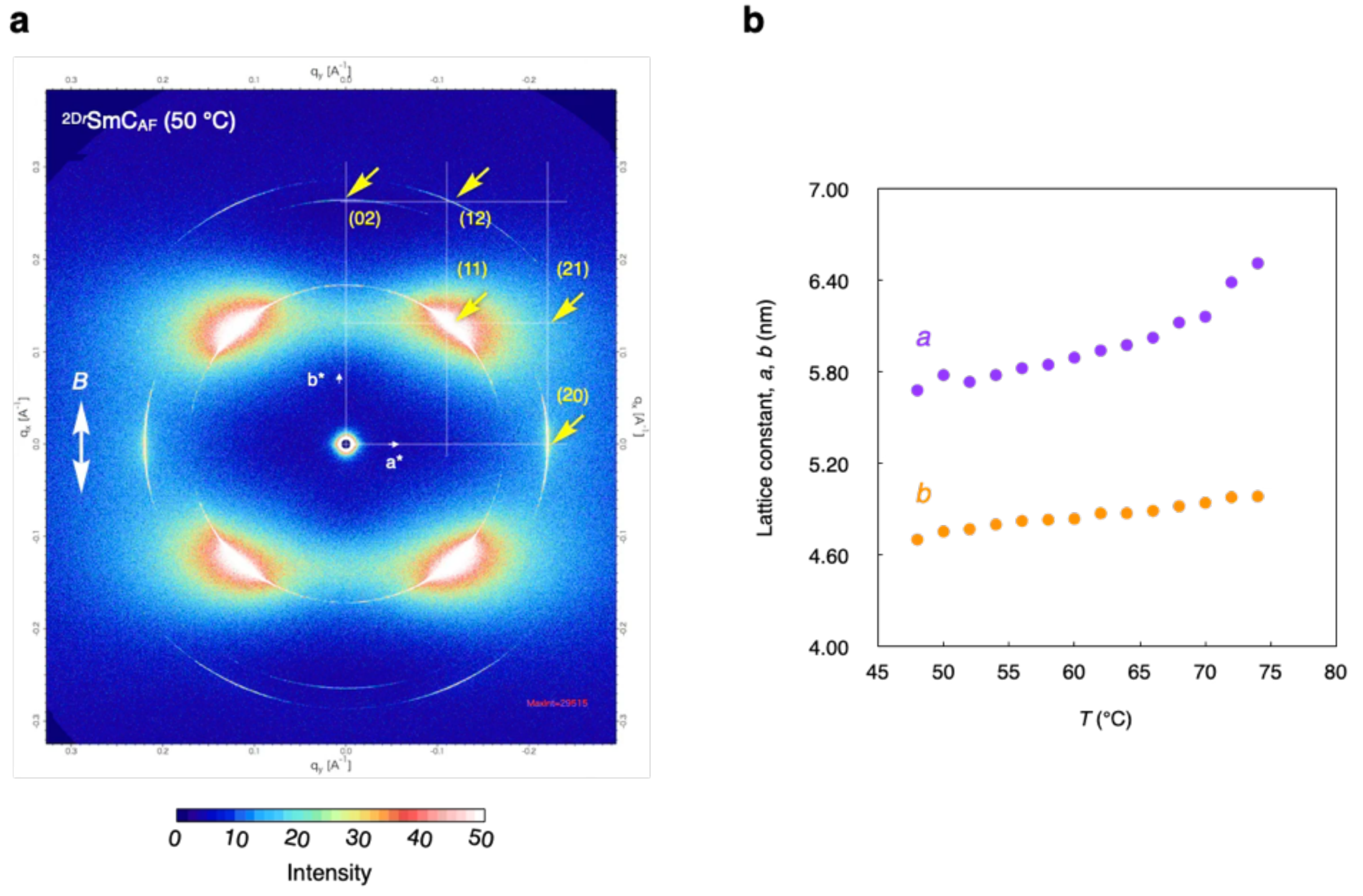

**Supplementary Fig. 31 |** XRD data at various temperatures under the magnetic field (~1 T, **B** ∥ **n**) for ***13*-012-CN**. (a) 2D SAXS patterns and (b) lattice constants *a*, *b* vs *T*.

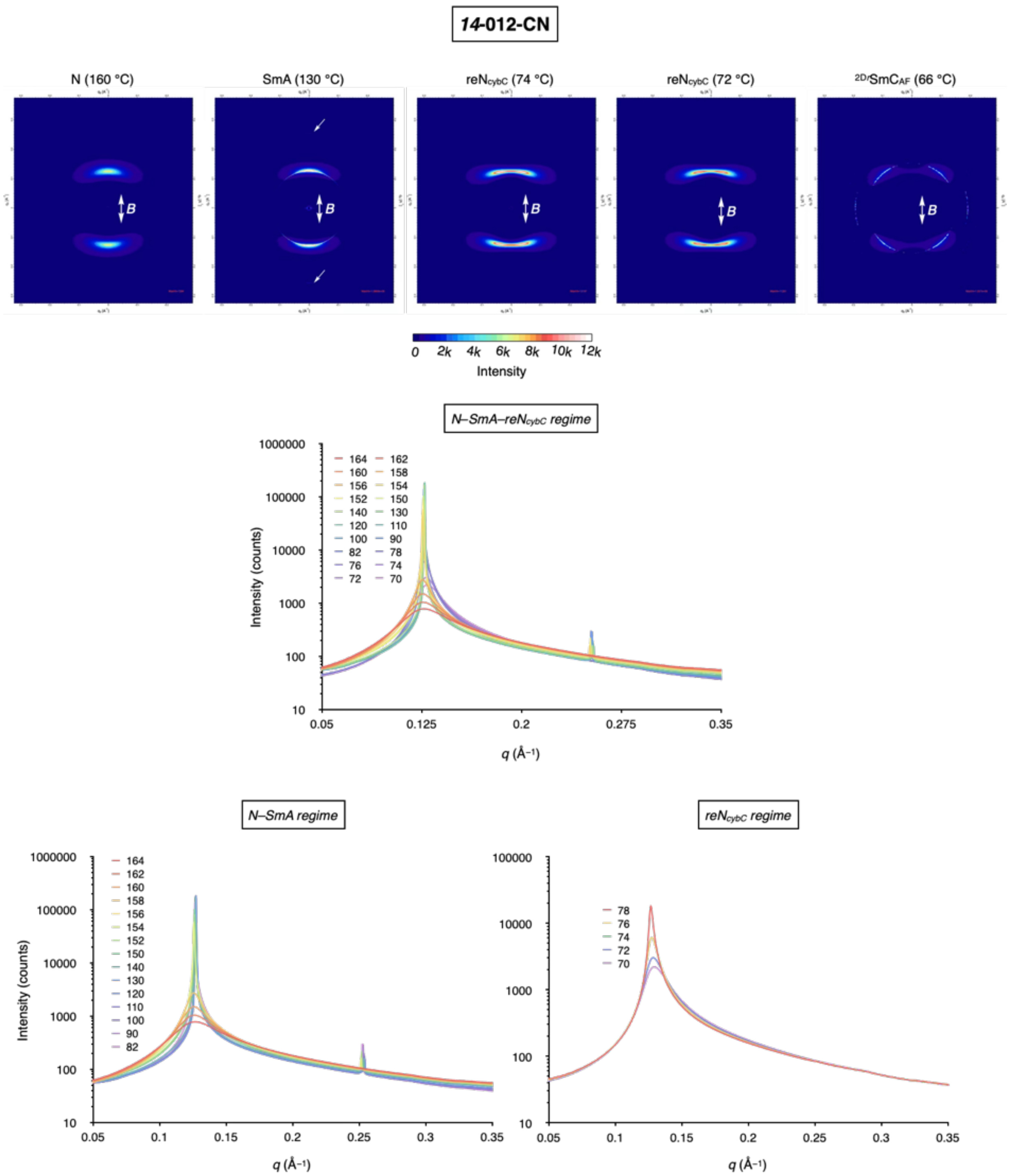

**Supplementary Fig. 32 |** 2D and corresponding 1D X-ray diffraction patterns at various temperatures under the magnetic field (~1 T, **B** ∥ **n**) for ***14*-012-CN**. In the SmA phase, white arrows indicate overtone peaks.

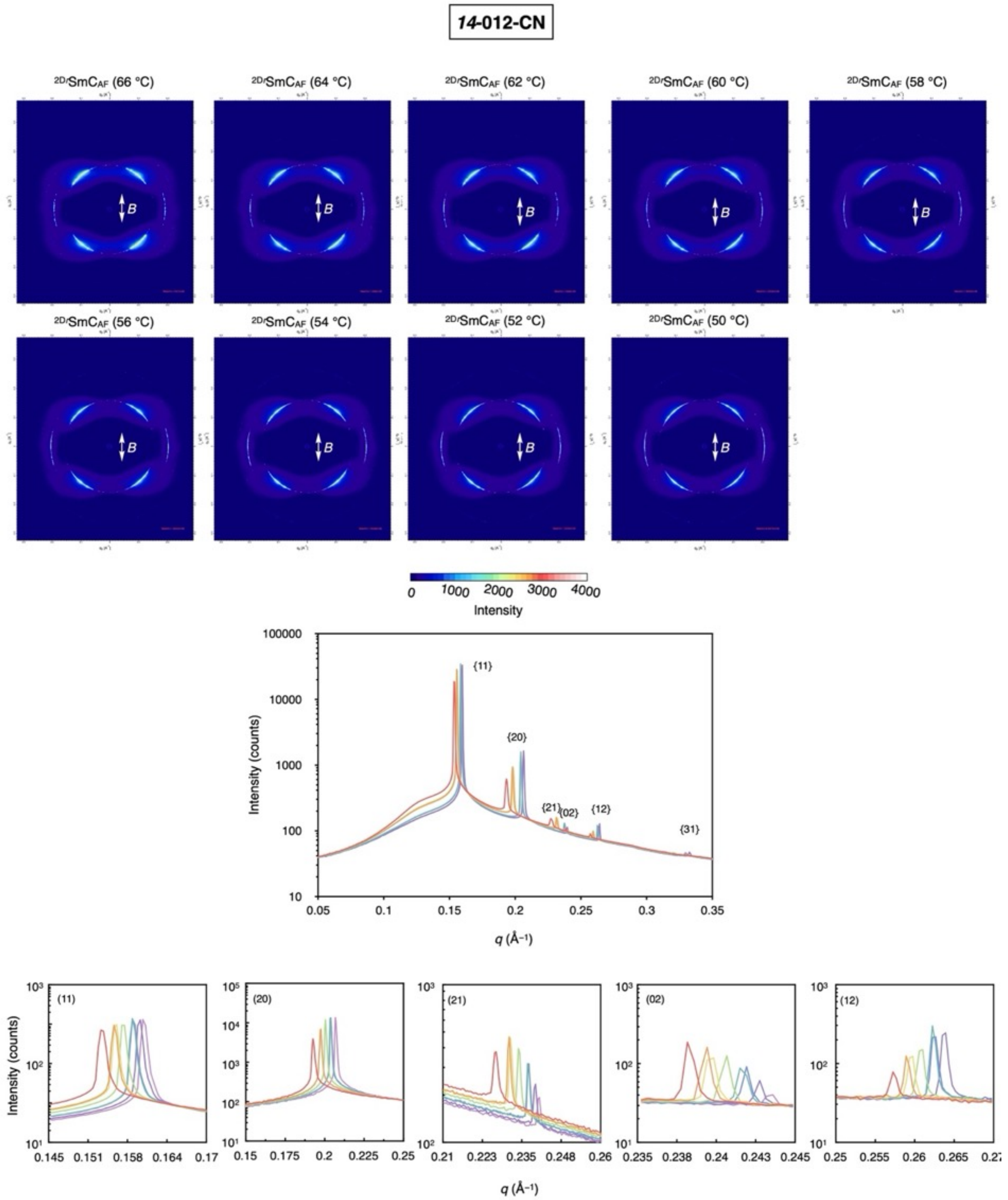

**Supplementary Fig. 33 |** 2D and corresponding 1D X-ray diffraction patterns at various temperatures under the magnetic field (~1 T, **B ∥ n**) for ***14*-012-CN**.

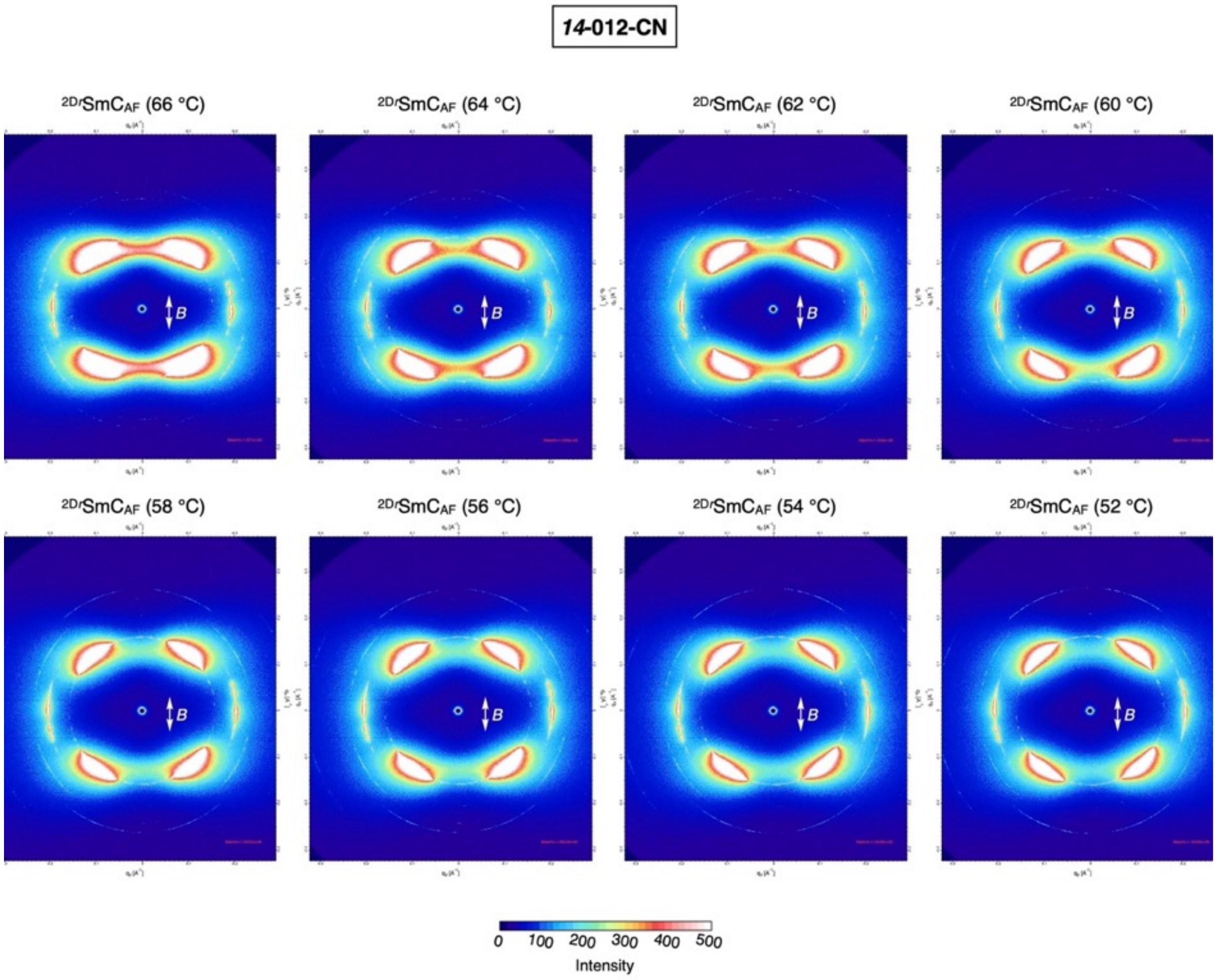

**Supplementary Fig. 34 |** XRD data at various temperatures under the magnetic field (~1 T, **B** ∥ **n**) for ***14*-012-CN**.

**14-012-CN**

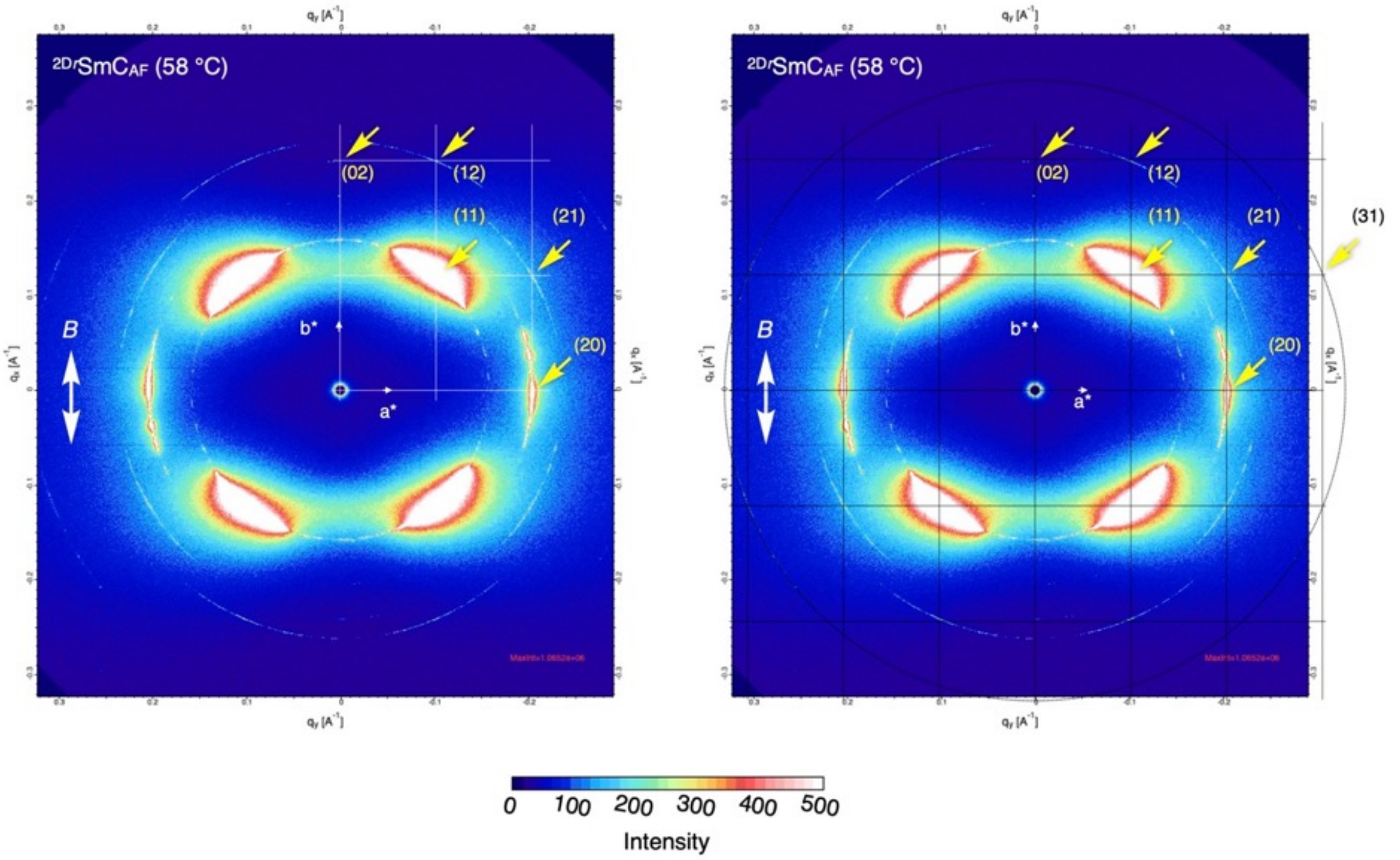

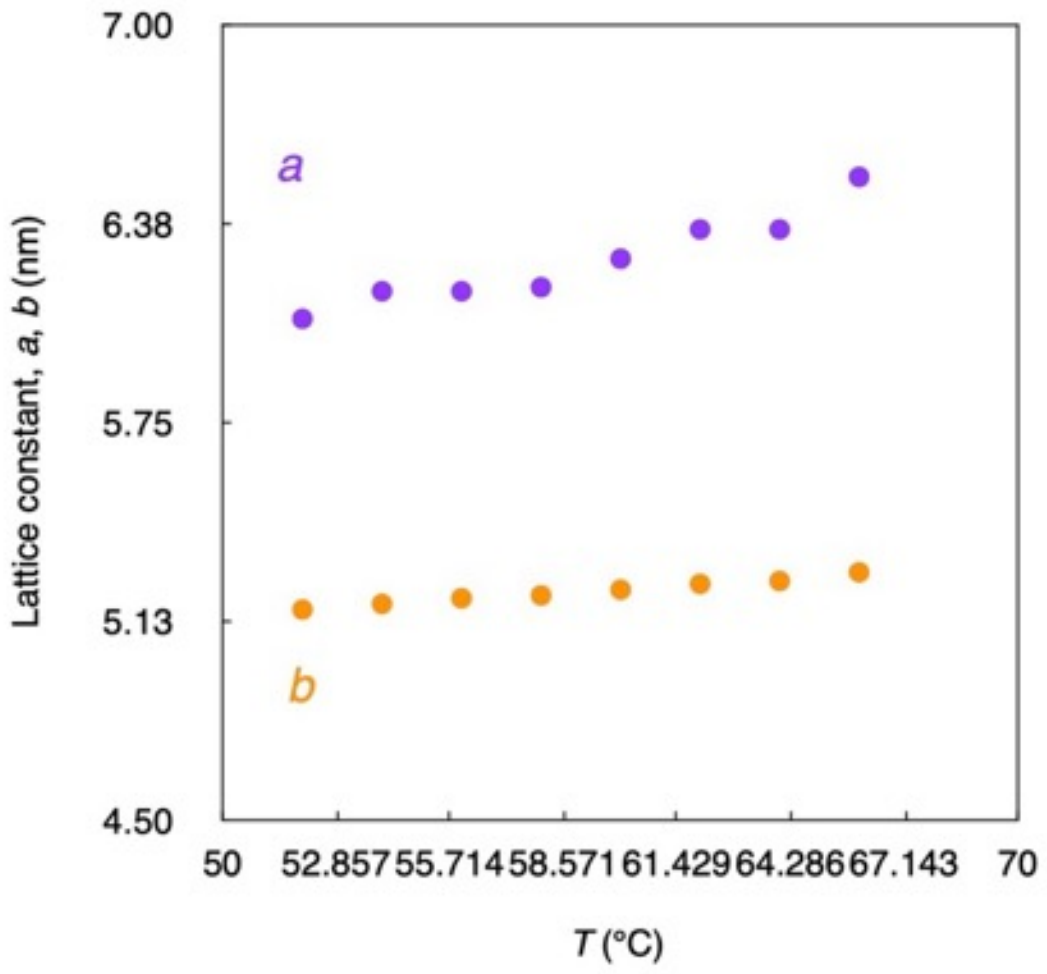

**Supplementary Fig. 35 |** (Upper) 2D SAXS patterns under the magnetic field (~1 T, **B ∥ n**) for ***14*-012-CN**. (Bottom) lattice constants *a*, *b* vs. *T*.

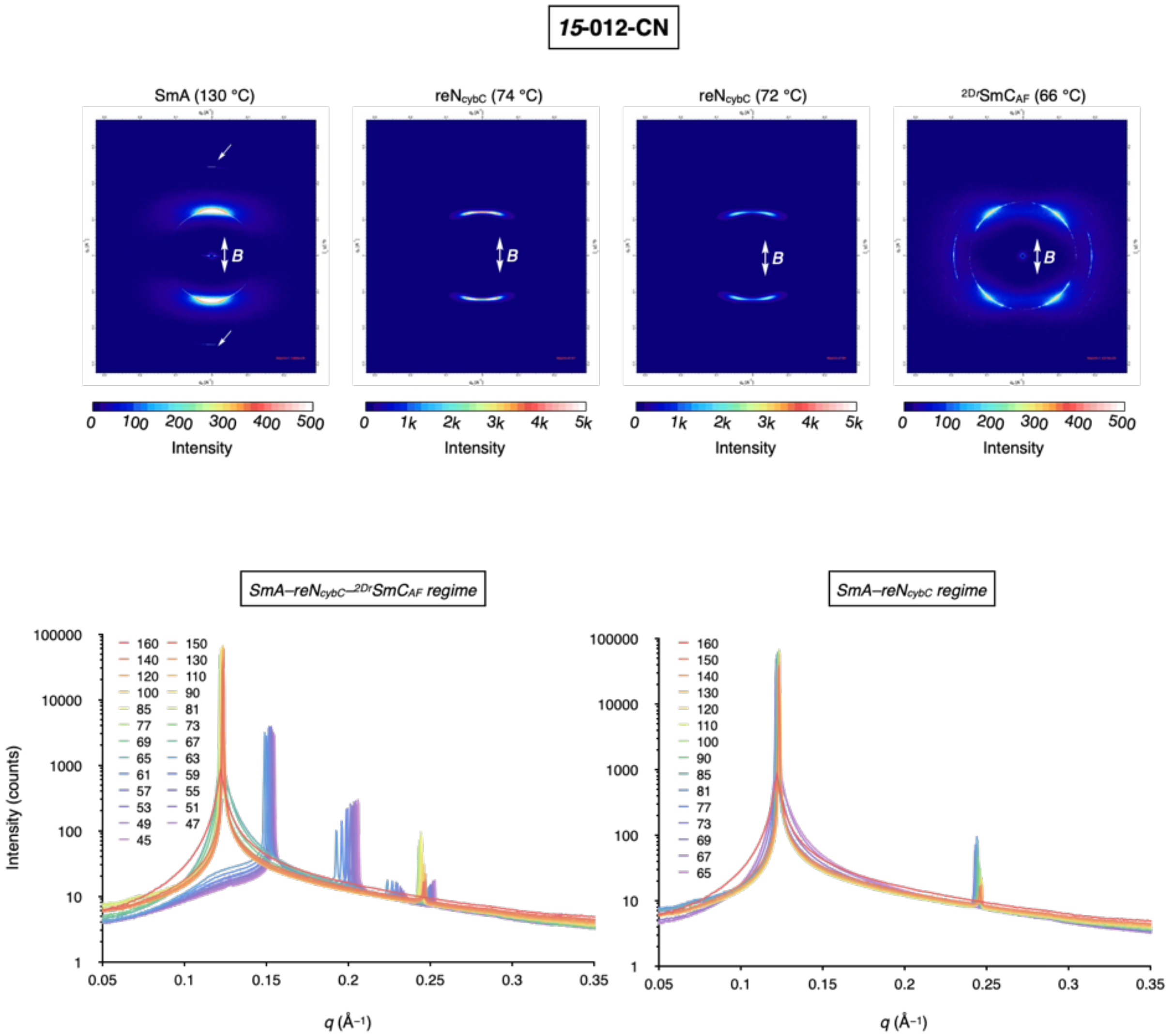

**Supplementary Fig. 36 |** 2D and corresponding 1D X-ray diffraction patterns at various temperatures under the magnetic field (~1 T, **B** ∥ **n**) for ***15*-012-CN**. In the SmA phase, white arrows indicate overtone peaks.

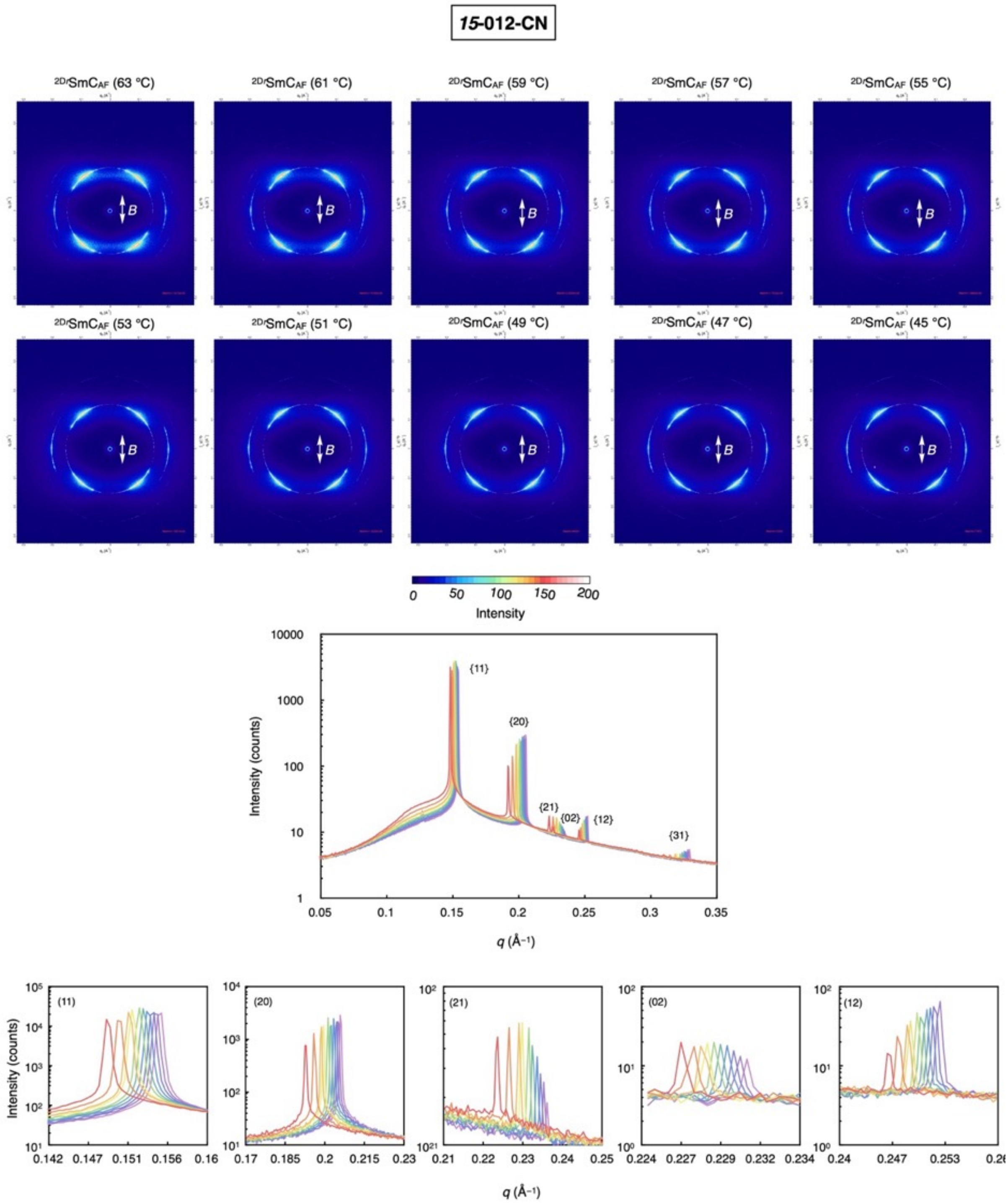

**Supplementary Fig. 37 |** 2D and corresponding 1D X-ray diffraction patterns at various temperatures under the magnetic field (~1 T, **B ∥ n**) for ***15*-012-CN**.

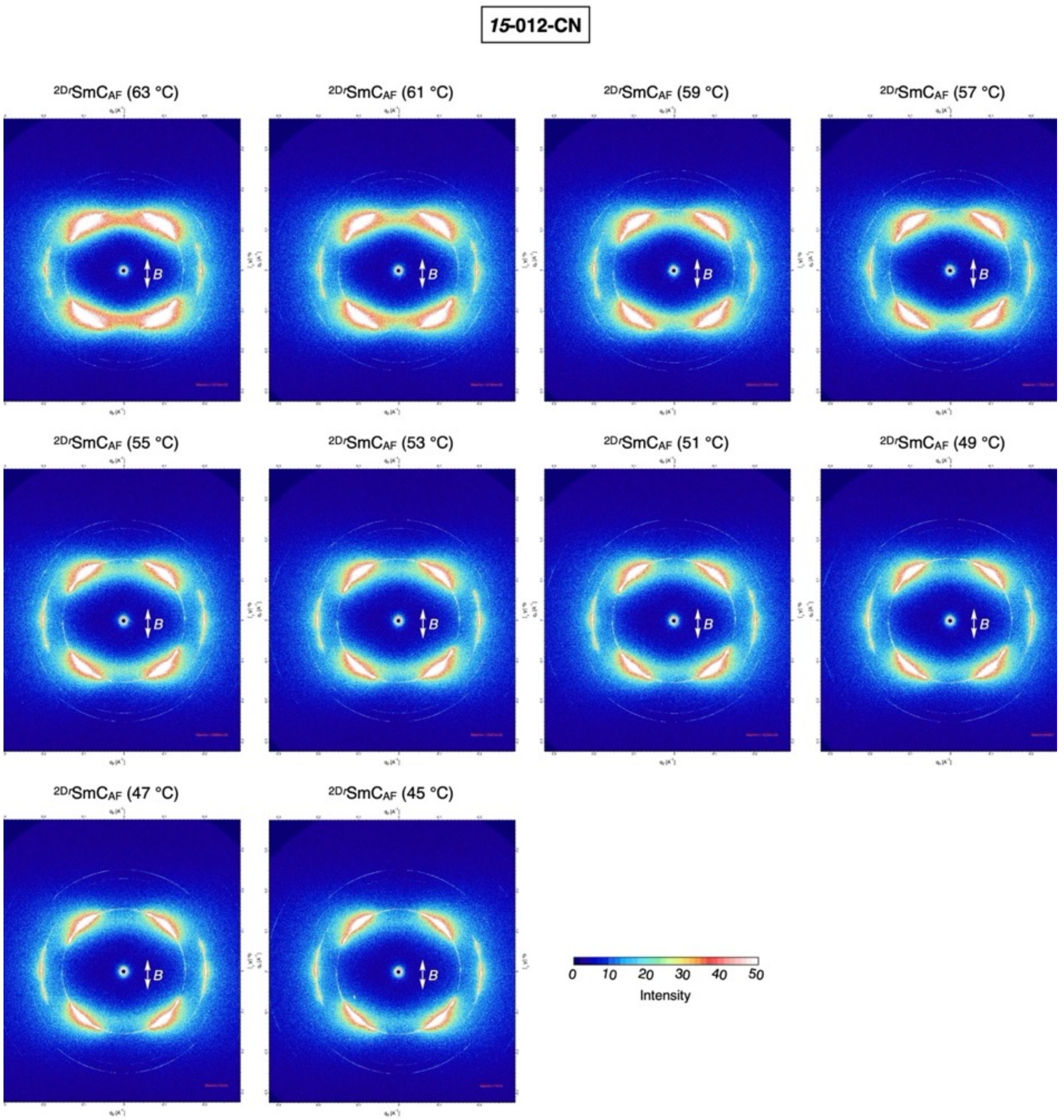

**Supplementary Fig. 38 |** 2D SAXS patterns at various temperatures under the magnetic field (~1 T, **B** ∥ **n**) for ***15*-012-CN**.

**_15_-012-CN**

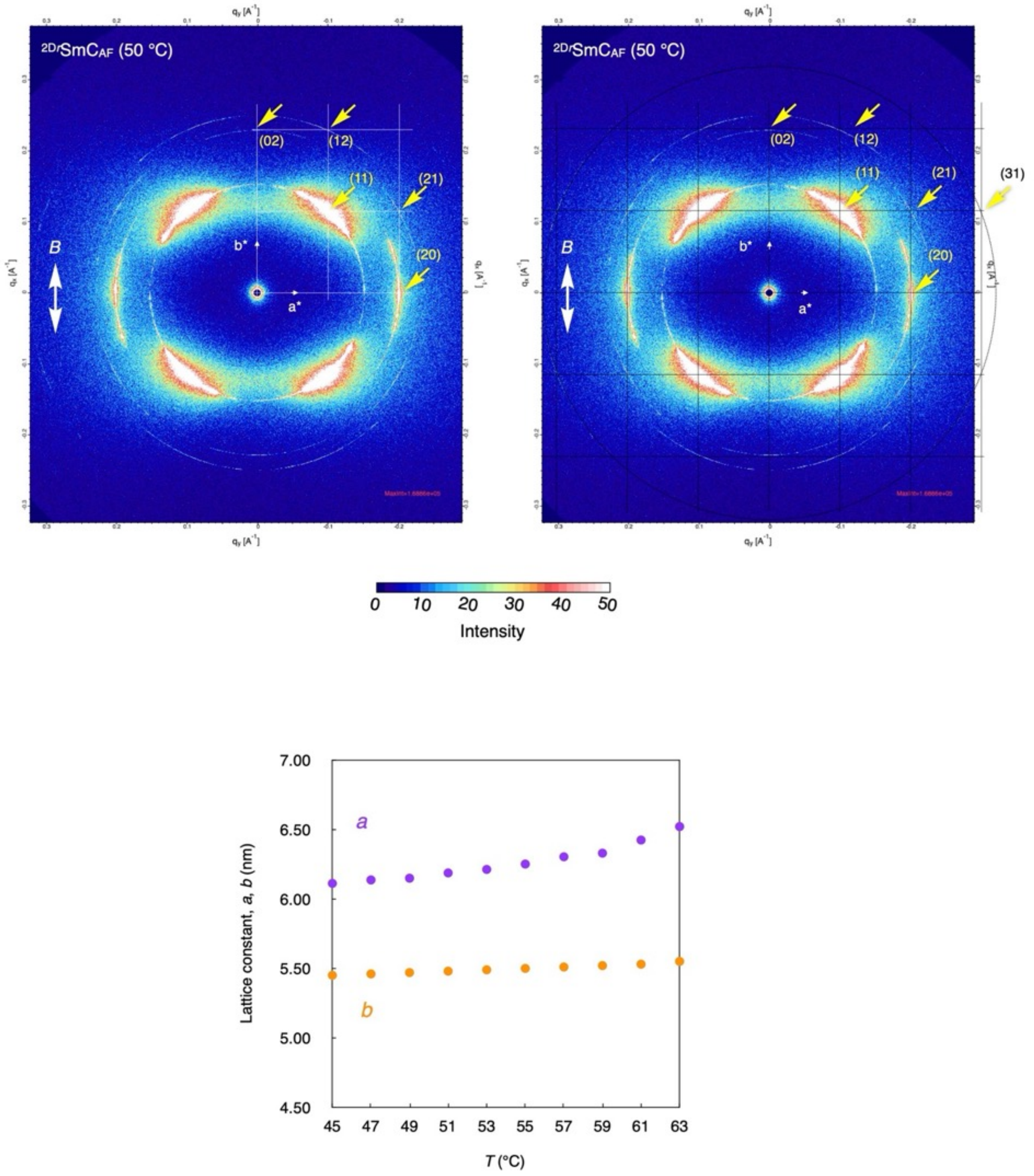

**Supplementary Fig. 39 |** (Upper) 2D SAXS patterns at various temperatures under the magnetic field (~1 T, **B ∥ n**) for ***15*-012-CN**. (Bottom) lattice constants *a*, *b* vs. *T*.

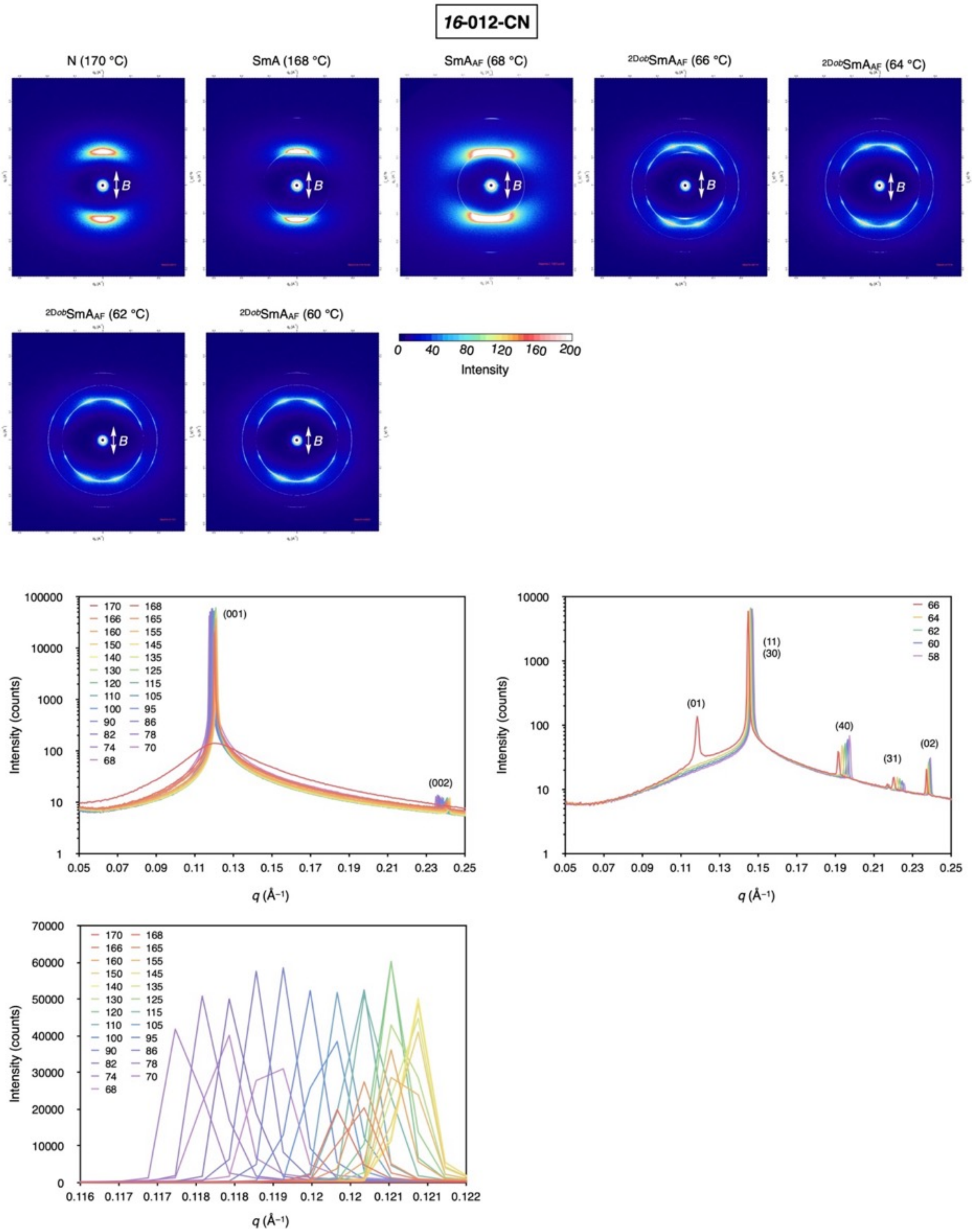

**Supplementary Fig. 40 |** 2D and corresponding 1D X-ray diffraction patterns at various temperatures under the magnetic field (~1 T, **B ∥ n**) for ***16*-012-CN**.

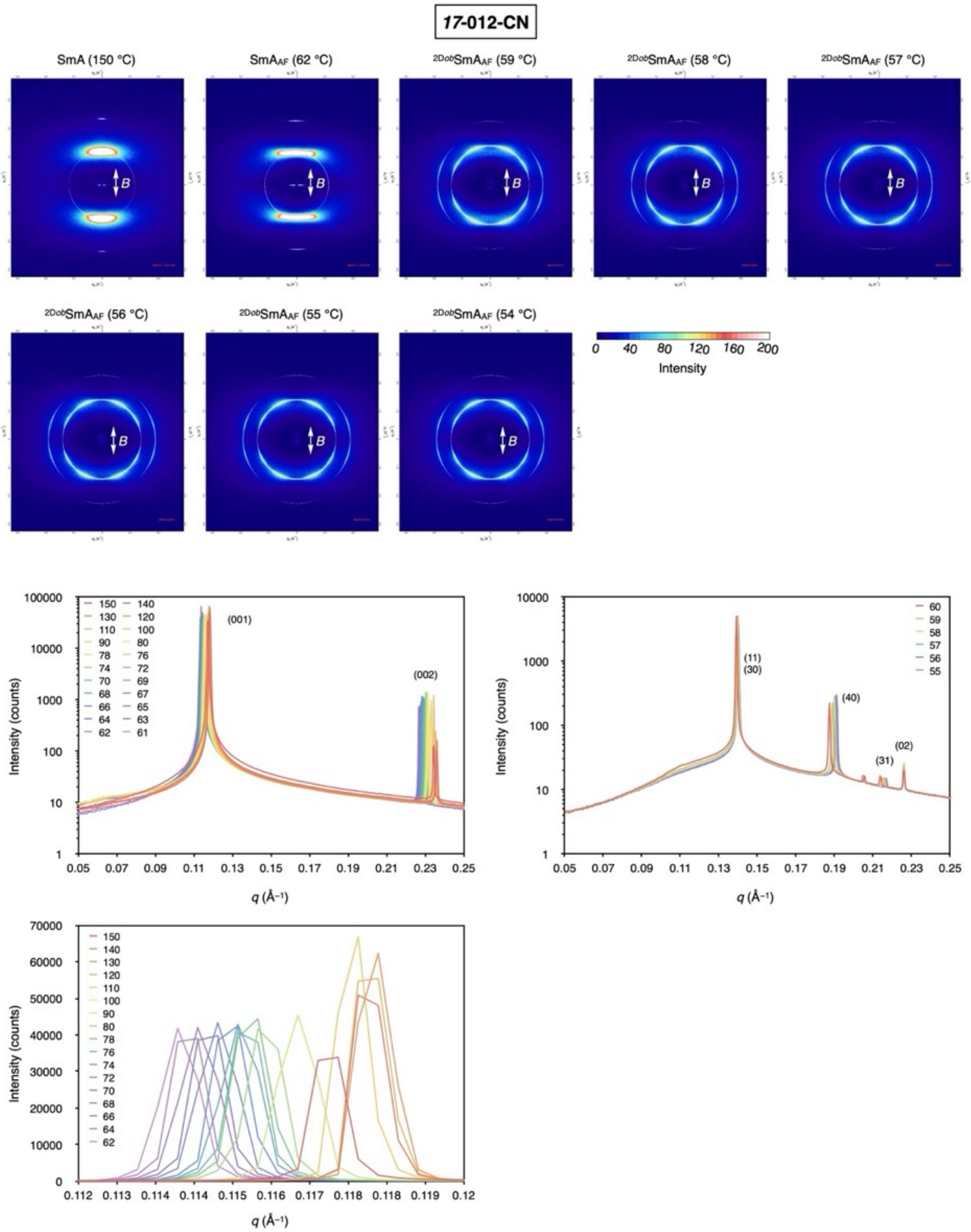

**Supplementary Fig. 41 |** 2D and corresponding 1D X-ray diffraction patterns at various temperatures under the magnetic field (~1 T, **B ∥ n**) for ***17*-012-CN**.

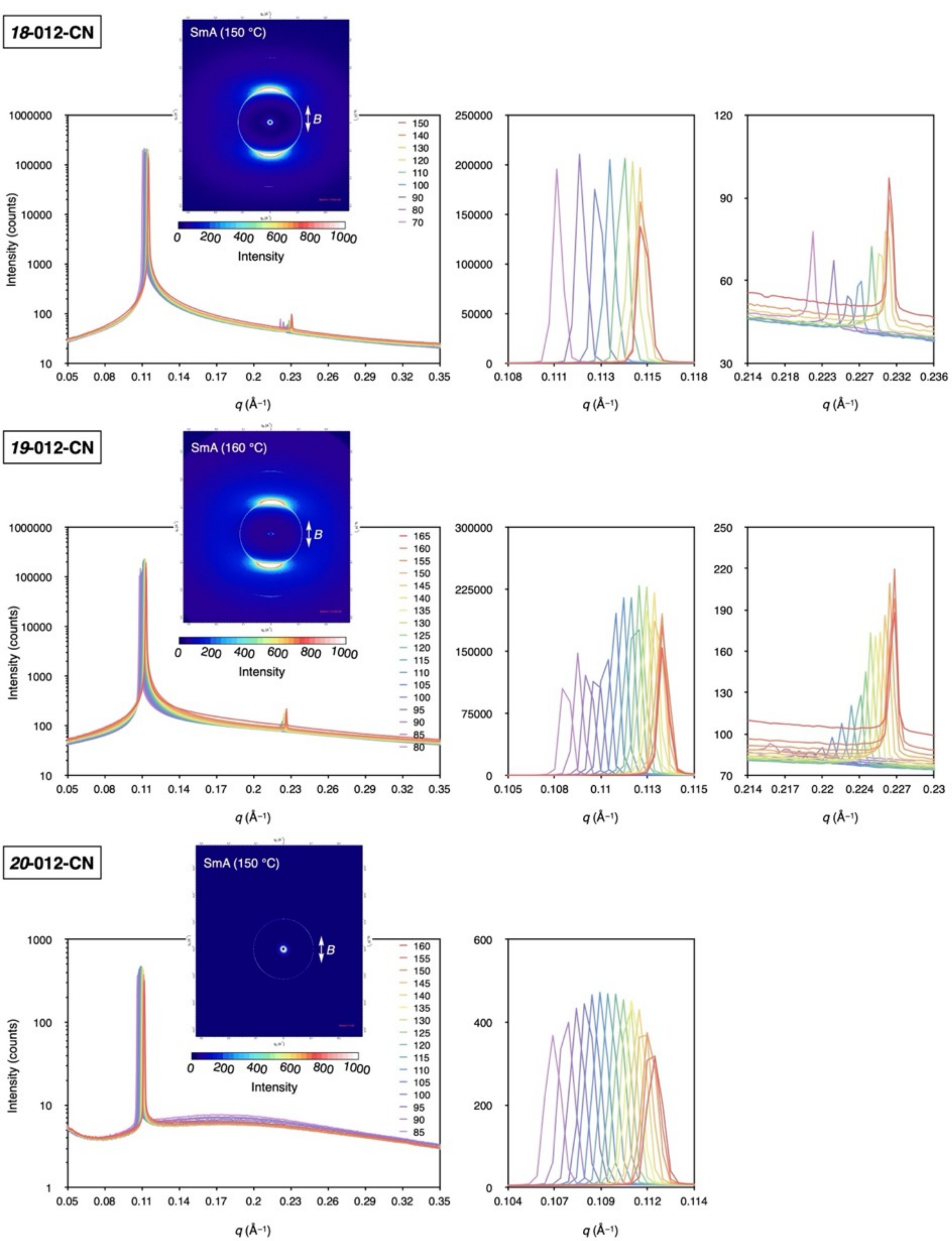

**Supplementary Fig. 42 |** XRD data at various temperatures under the magnetic field (~1 T, **B ∥ n**) for ***18*-012-CN** (upper), ***19*-012-CN** (middle) and ***20*-012-CN** (bottom). For ***20*-012-CN**, sufficient alignment could not be achieved under the magnetic field.

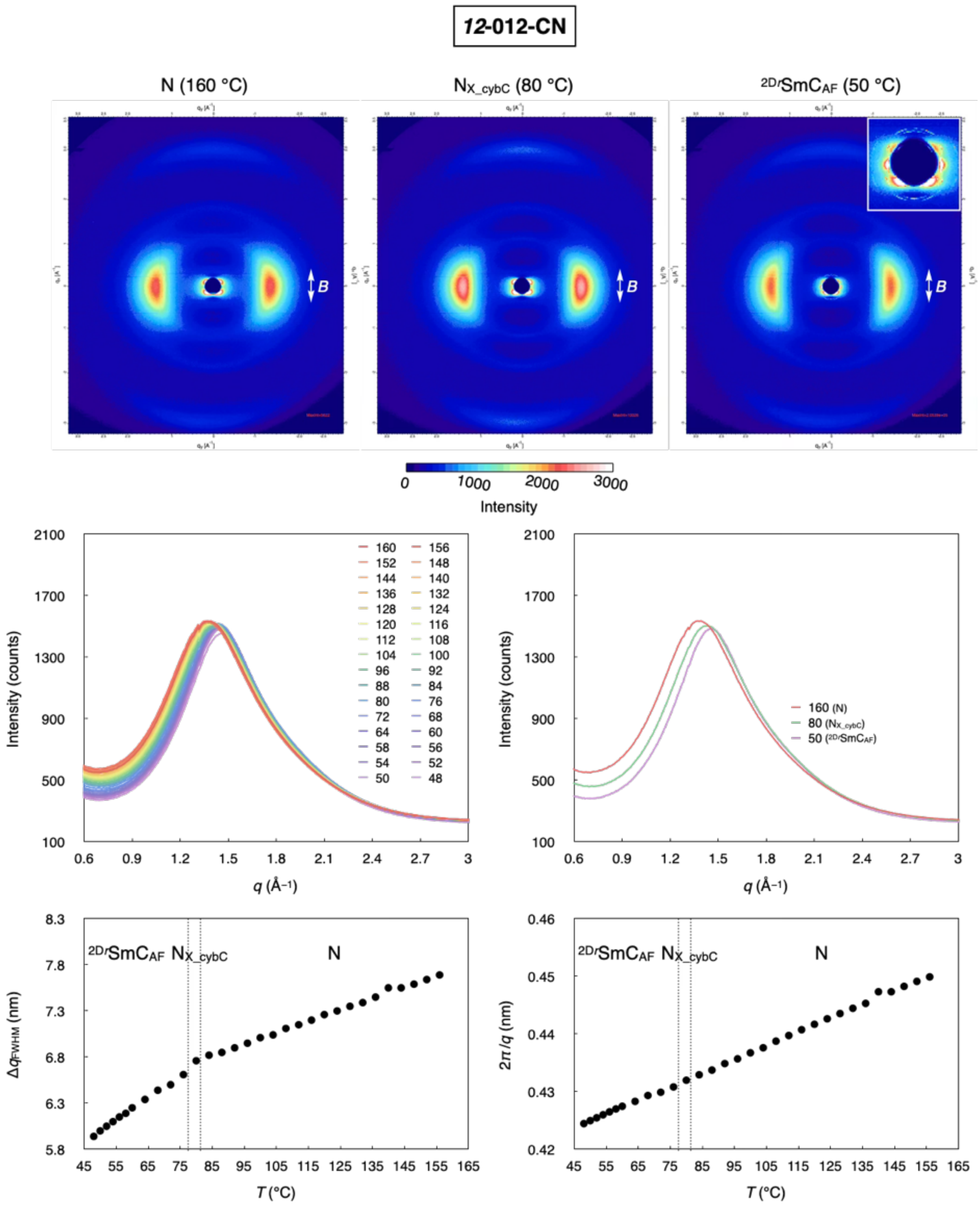

**Supplementary Fig. 43 |** WAXD data at various temperatures under the magnetic field (~1 T, **B** ∥ **n**) for ***12*-012-CN**.

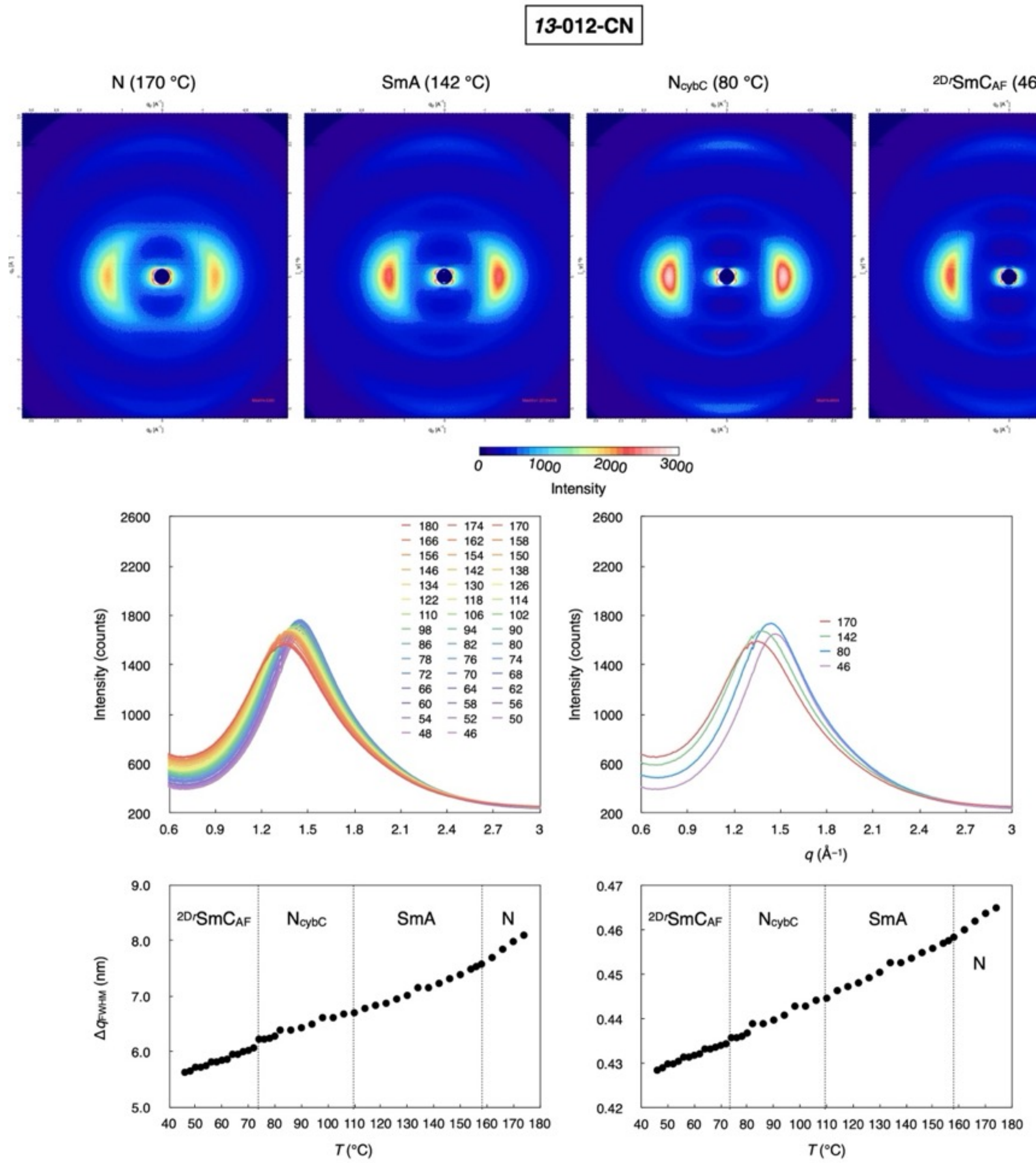

**Supplementary Fig. 44 |** WAXD data at various temperatures under the magnetic field (~1 T, **B ∥ n**) for ***13*-012-CN**.

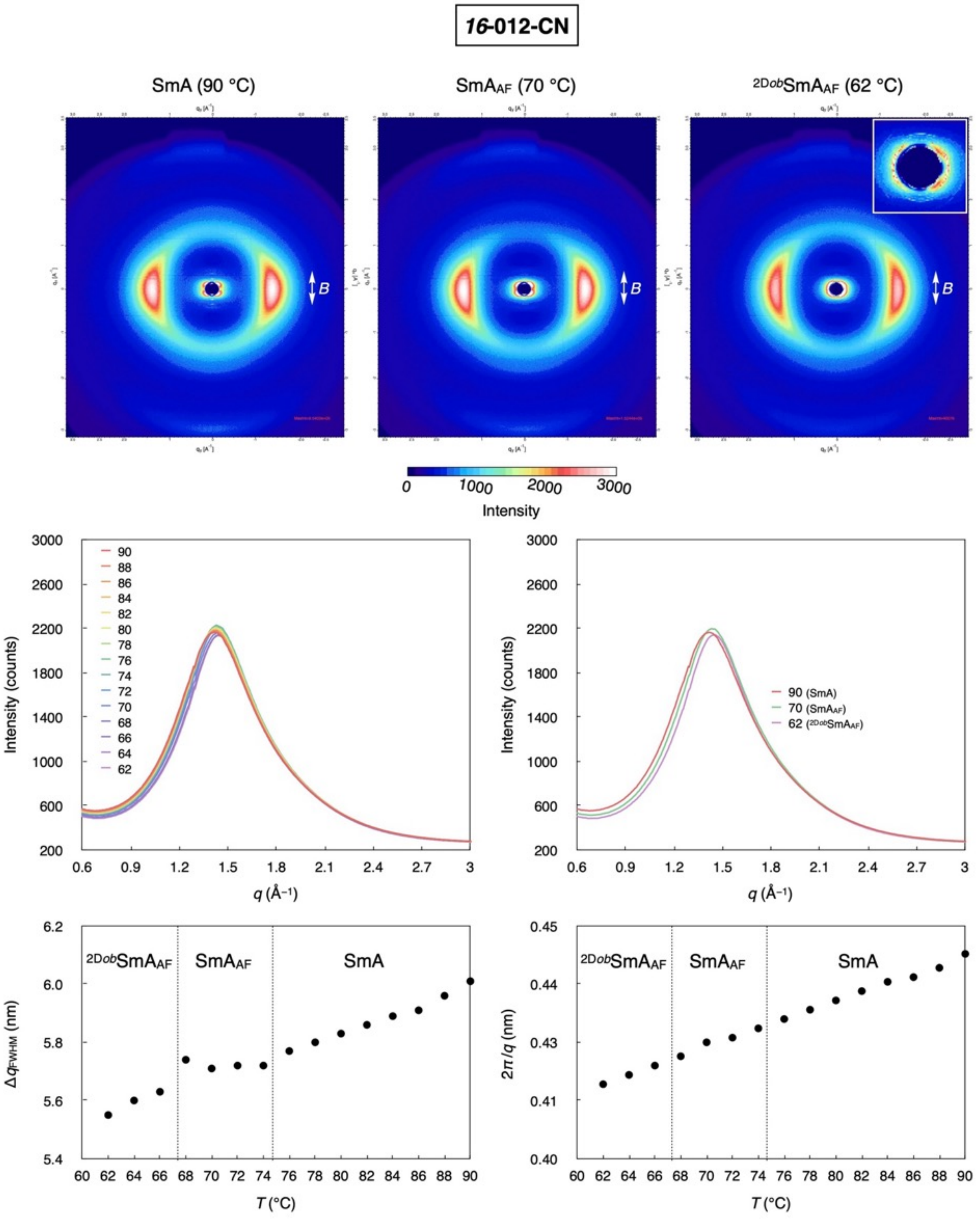

**Supplementary Fig. 45 |** WAXD data at various temperatures under the magnetic field (~1 T, **B** ∥ **n**) for ***16*-012-CN**.

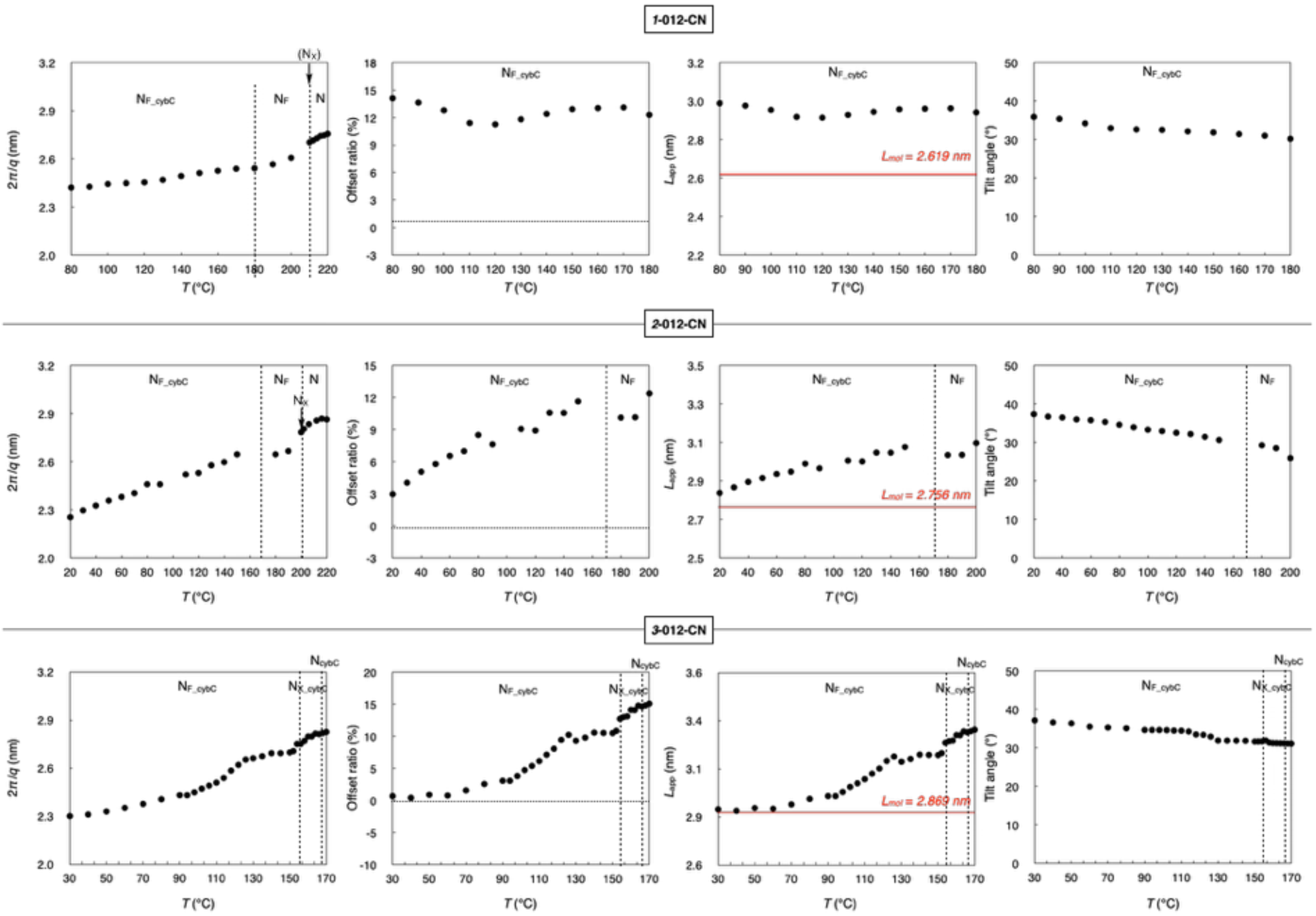

**Supplementary Fig. 46 |** XRD data set for ***n*-012-CN** (n = 1–3).

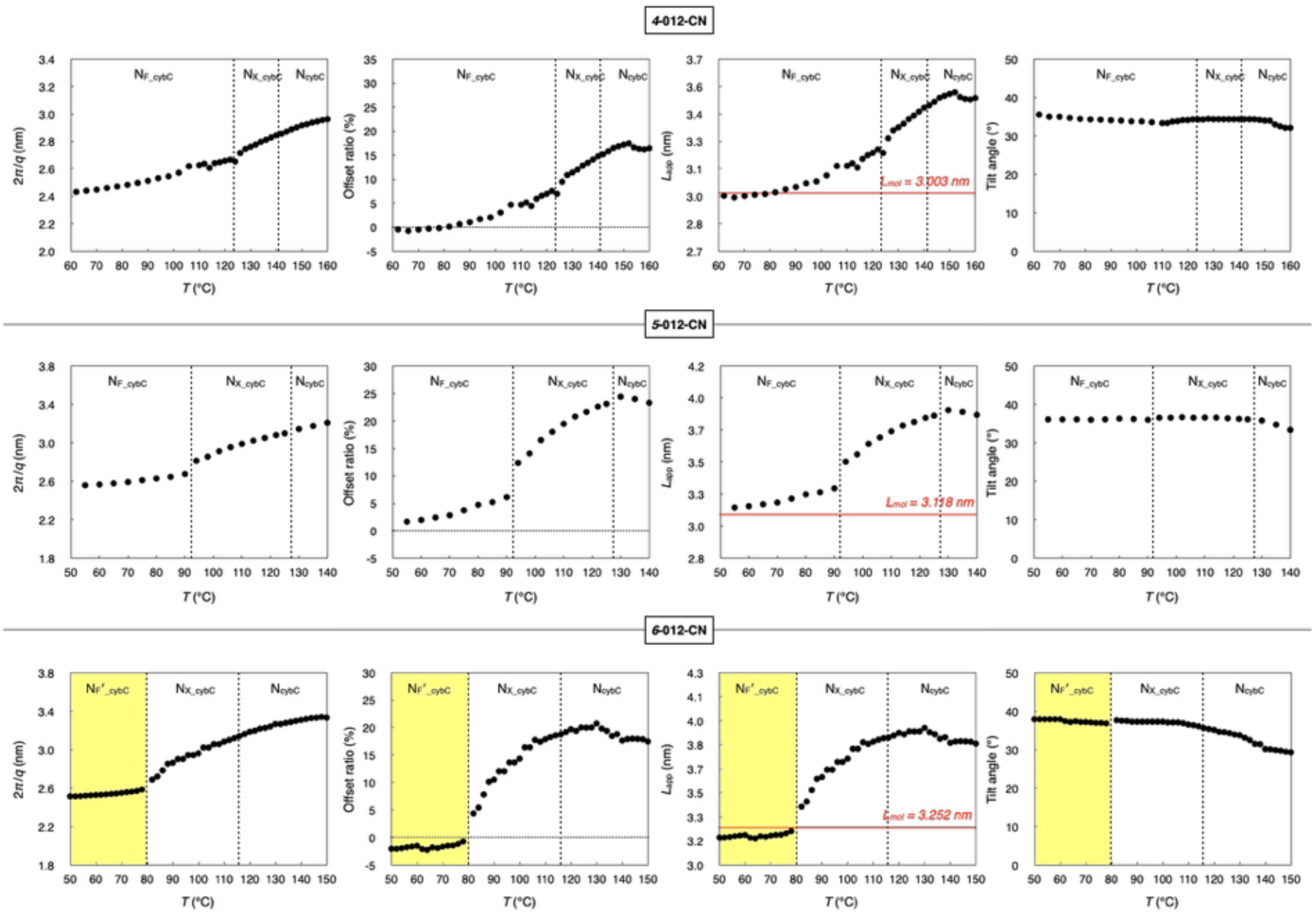

**Supplementary Fig. 47 |** XRD data set for ***4*-012-CN** (top), ***5*-012-CN** (middle) and ***6*-012-CN** (bottom).

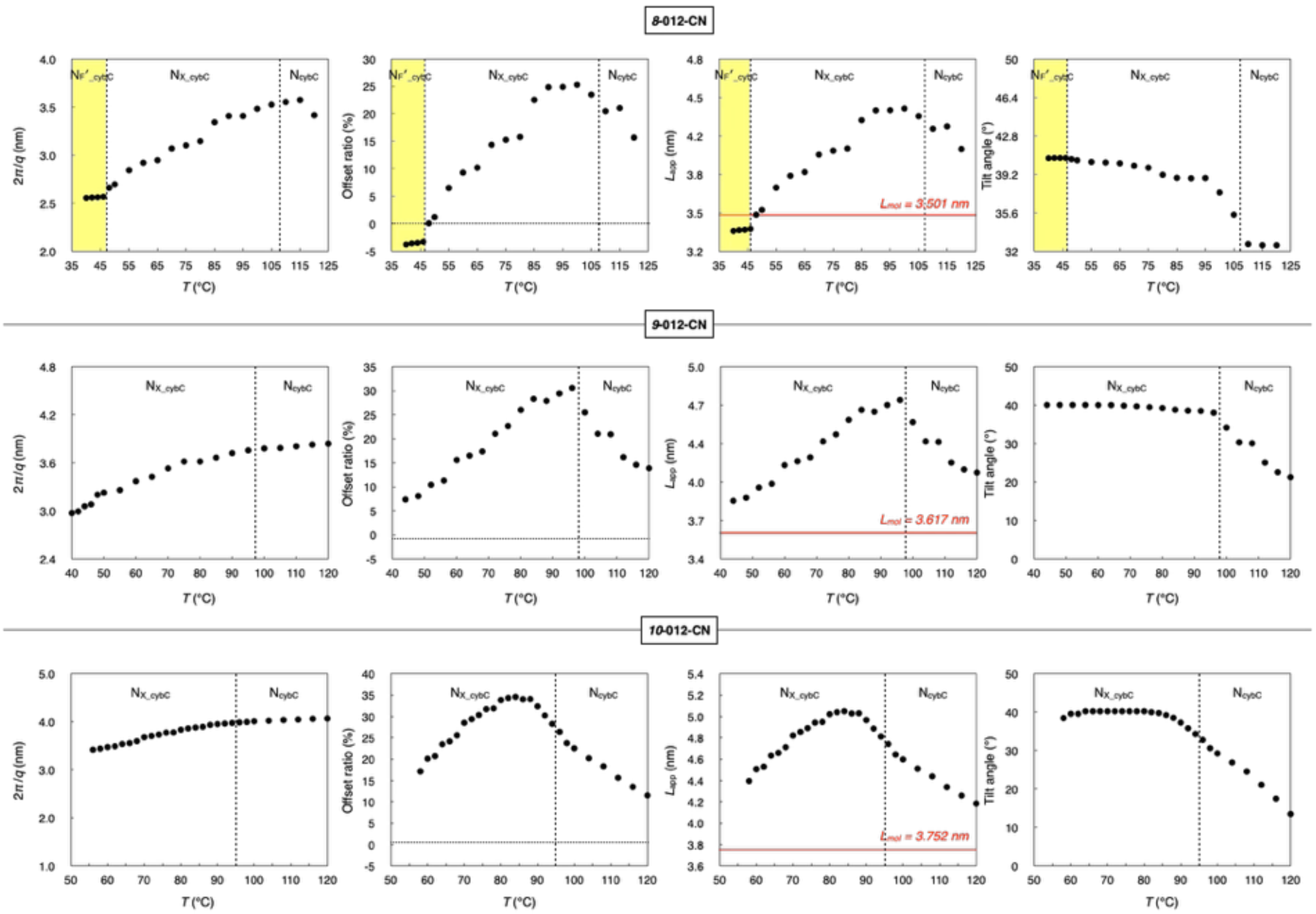

**Supplementary Fig. 48 |** XRD data set for ***8*-012-CN** (top), ***9*-012-CN** (middle) and ***10*-012-CN** (bottom).

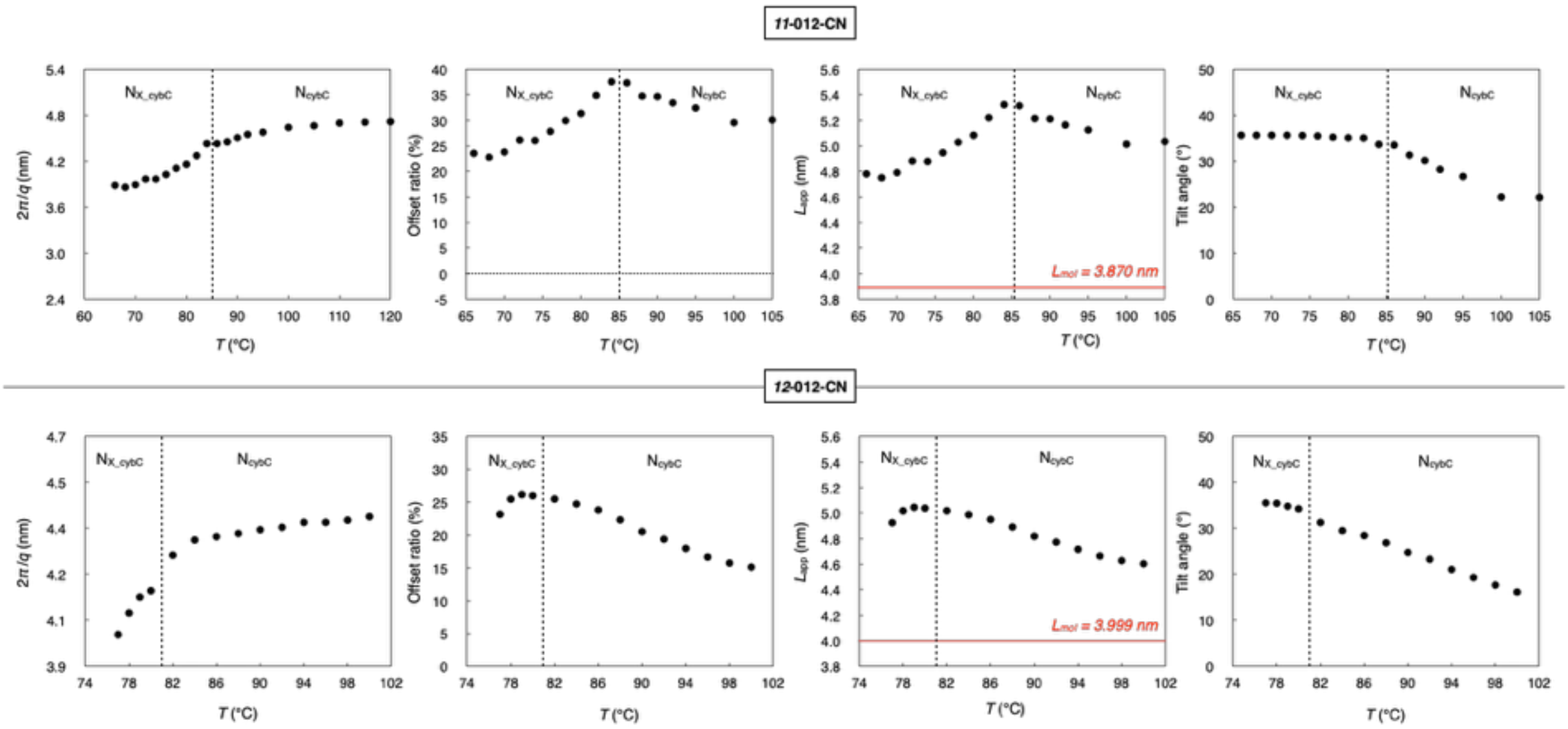

**Supplementary Fig. 49 |** XRD data set for ***11*-012-CN** (top) and ***12*-012-CN** (bottom).

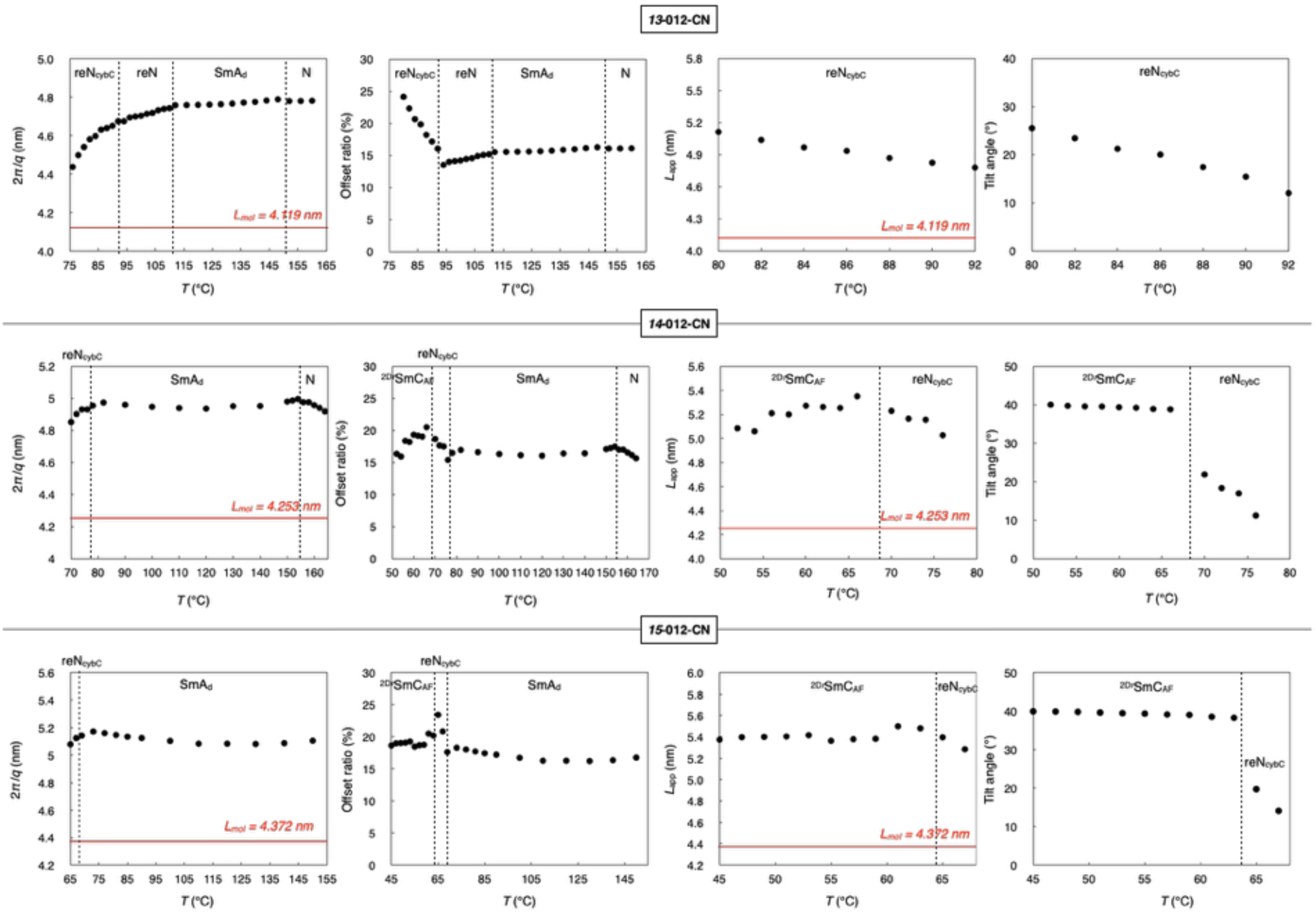

**Supplementary Fig. 50 |** XRD data set for ***13*-012-CN** (top), ***14*-012-CN** (middle) and ***15*-012-CN** (bottom).

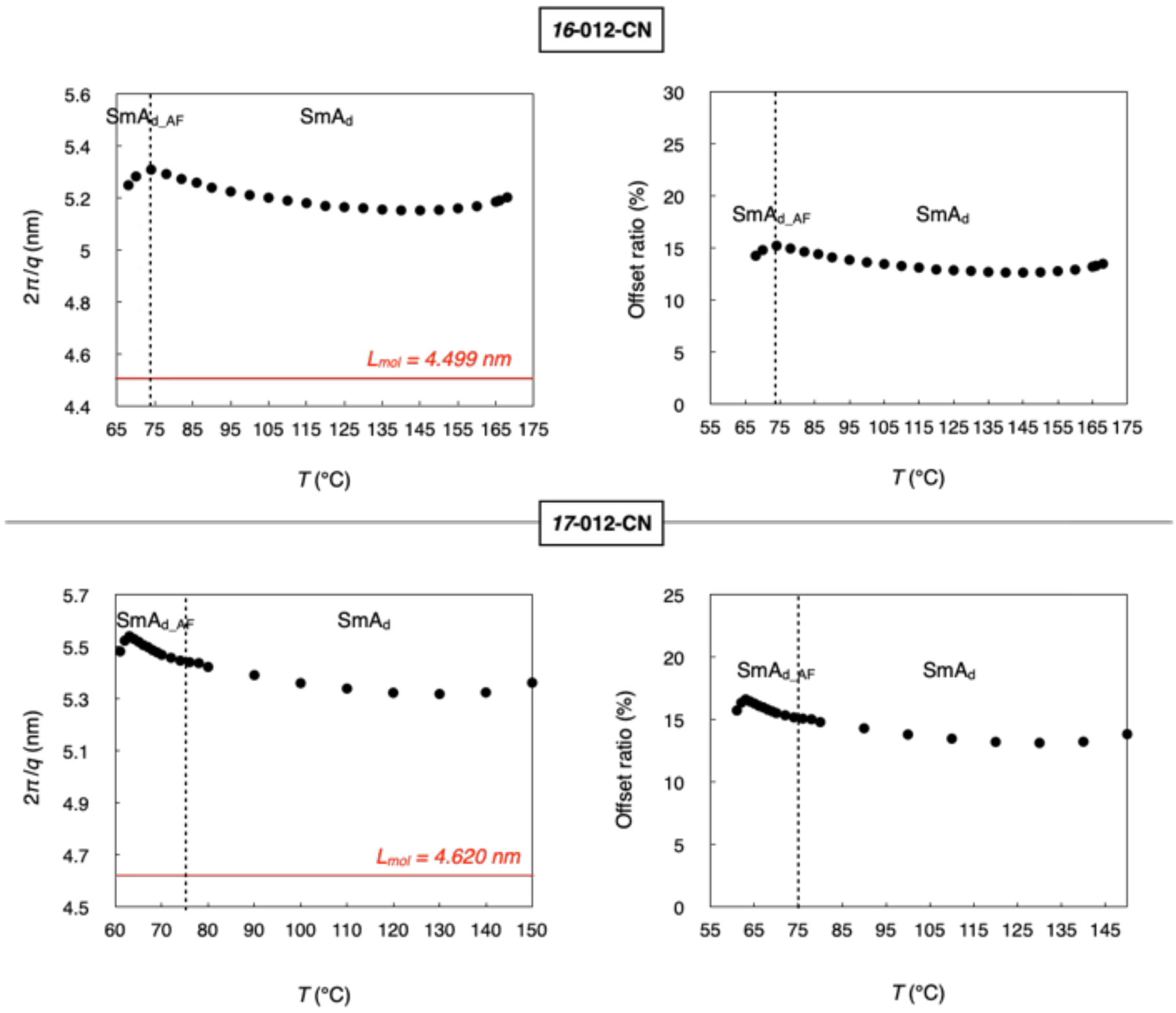

**Supplementary Fig. 51 |** XRD data set for ***16*-012-CN** (top) and ***17*-012-CN** (bottom).

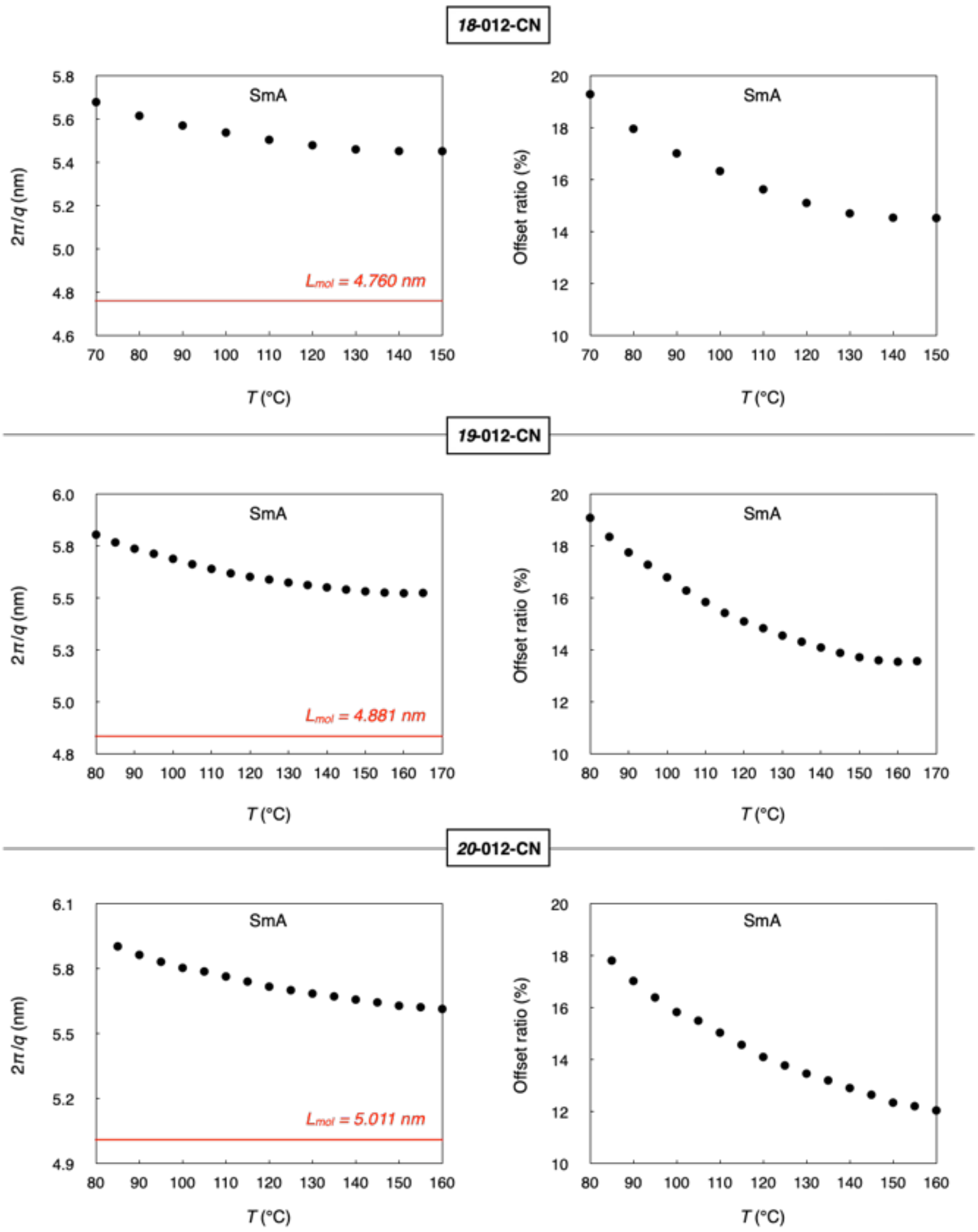

**Supplementary Fig. 52 |** XRD data set for ***18*-012-CN** (top), ***19*-012-CN** (middle) and ***20*-012-CN** (bottom).

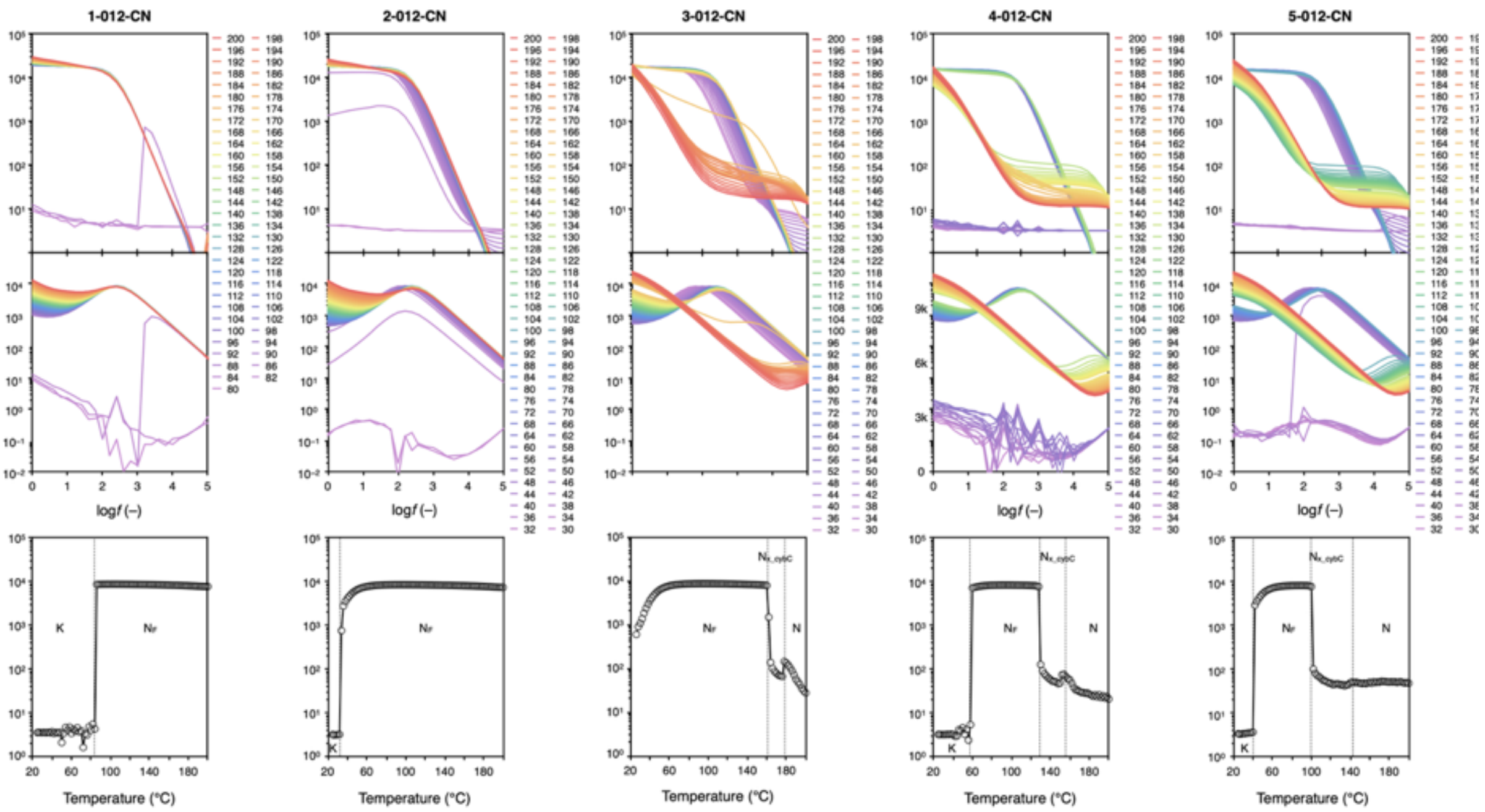

**Supplementary Fig. 53 |** BDS spectra and relative dielectric permittivity ($\varepsilon'_{\mathrm{app}}$) vs. $T$ ($f$ = 400 Hz) for ***n*-012-CN** (n = 1–5).

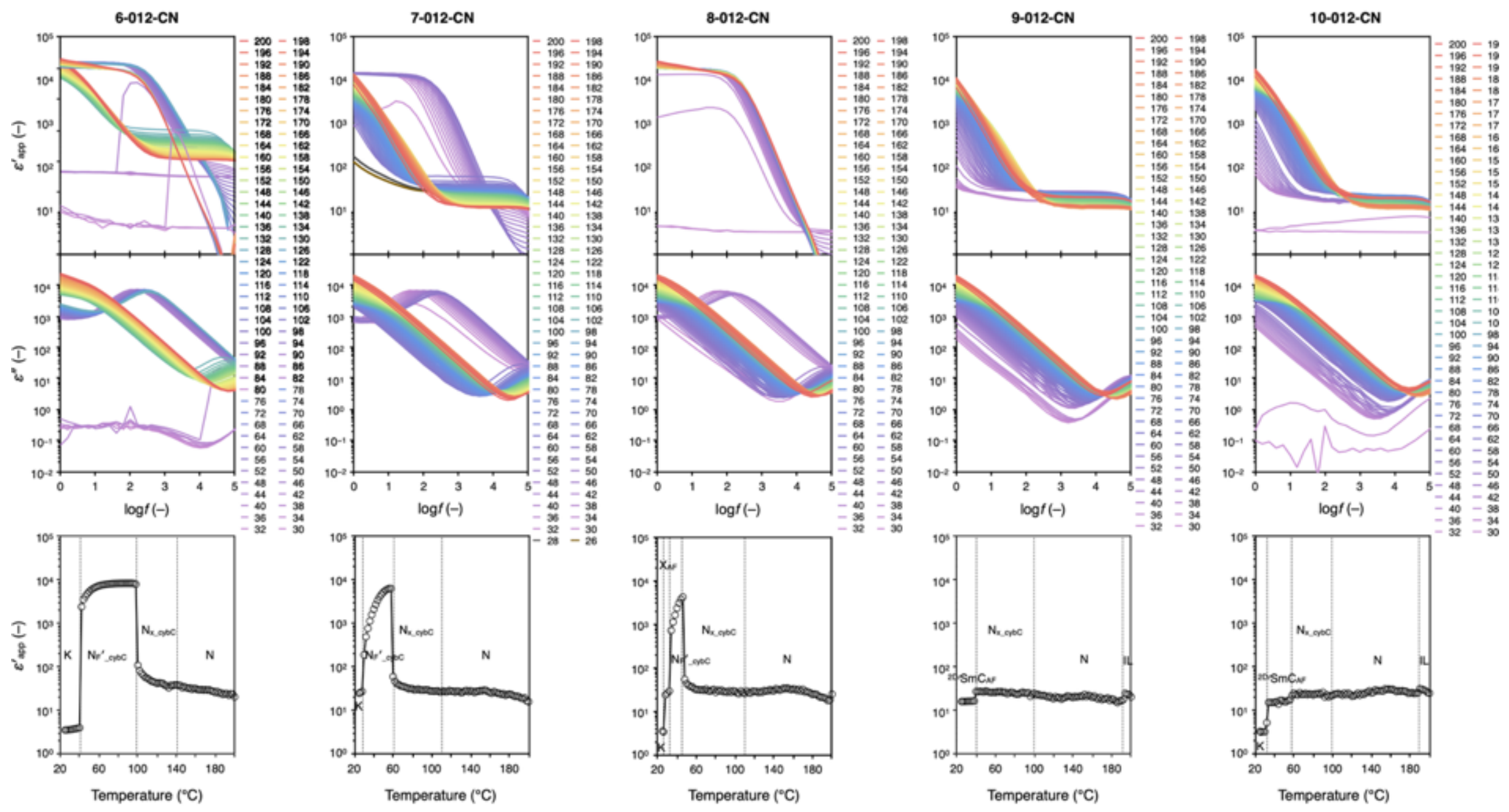

**Supplementary Fig. 54 |** BDS spectra and relative dielectric permittivity ($\varepsilon'_{\mathrm{app}}$) vs. $T$ ($f$ = 400 Hz) for ***n*-012-CN** (n = 6–10).

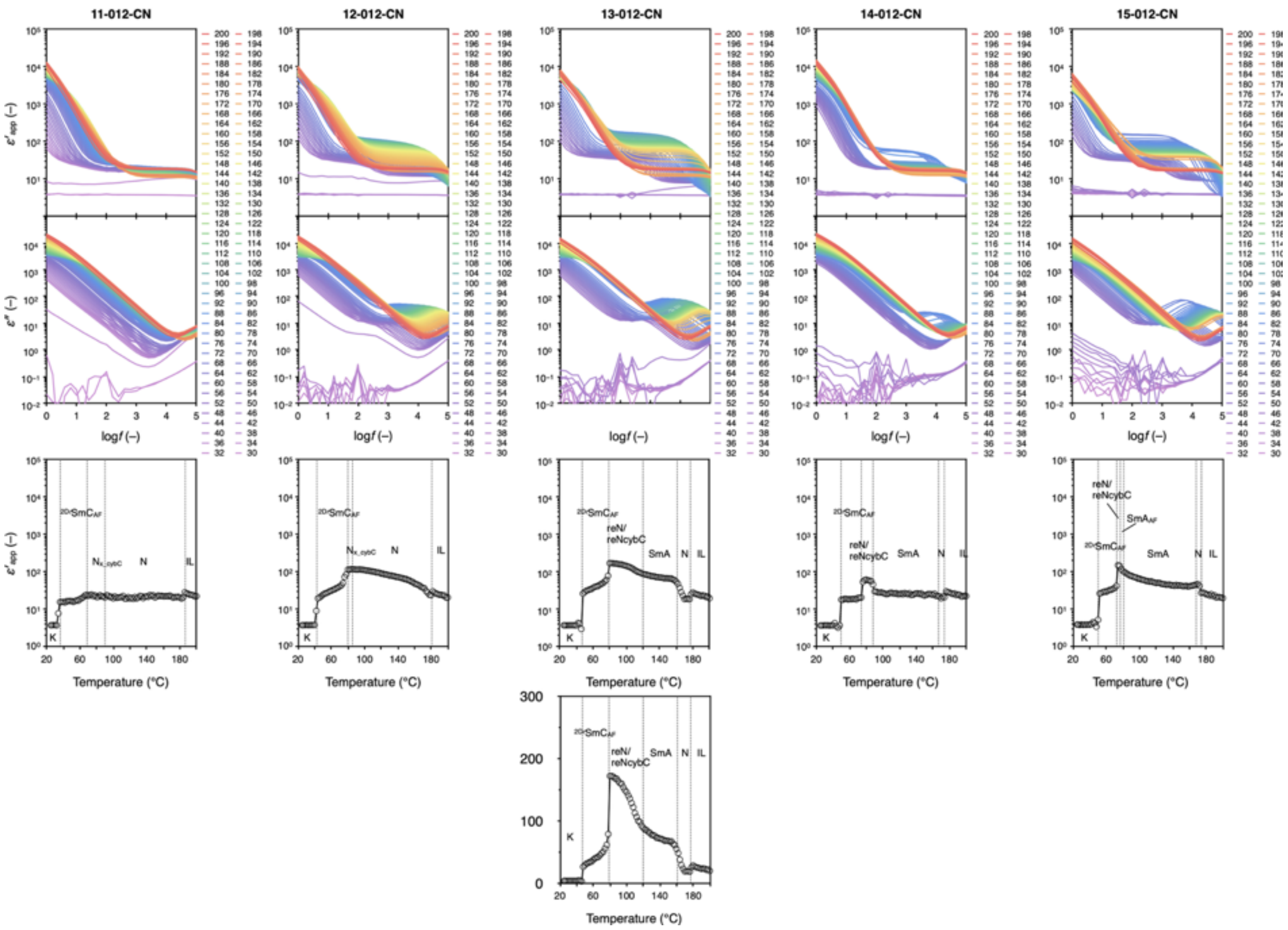

**Supplementary Fig. 55 |** BDS spectra and relative dielectric permittivity ($\varepsilon'_{\mathrm{app}}$) vs. $T$ ($f$ = 400 Hz) for ***n*-012-CN** (n = 11–15).

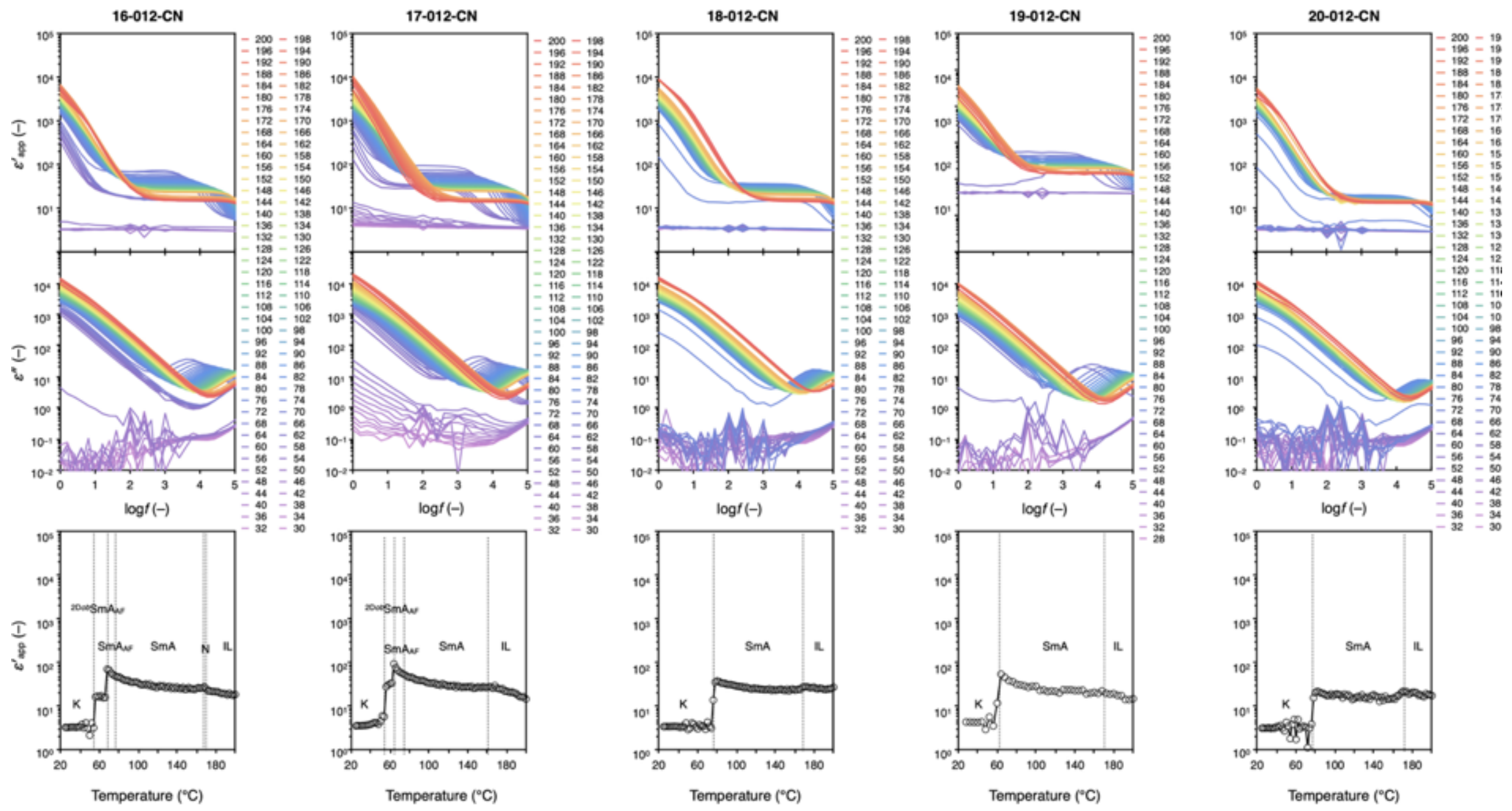

**Supplementary Fig. 56 |** BDS spectra and relative dielectric permittivity ($\varepsilon'_{\mathrm{app}}$) vs. $T$ ($f$ = 400 Hz) for ***n*-012-CN** (n = 16–20).

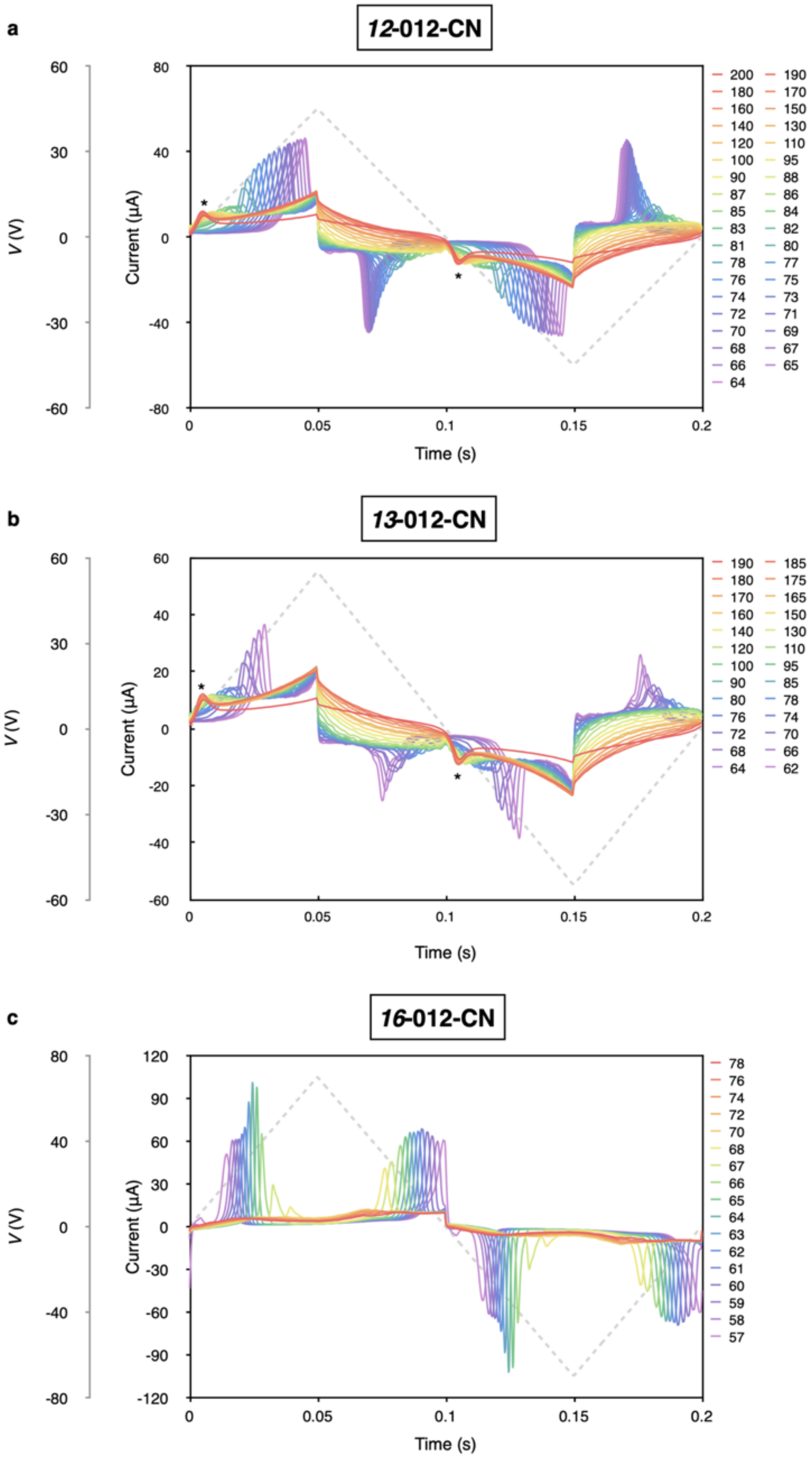

**Supplementary Fig. 57 |** PRC data for ***12*-012-CN** (a) and ***13*-012-CN** (b) and ***16*-012-CN** (c). Note: The peaks marked by asterisk symbols may be due to the ionic migration because it appeared from the isotropic state and then remained to the $\mathrm{reN_{cybC}}$ phase.

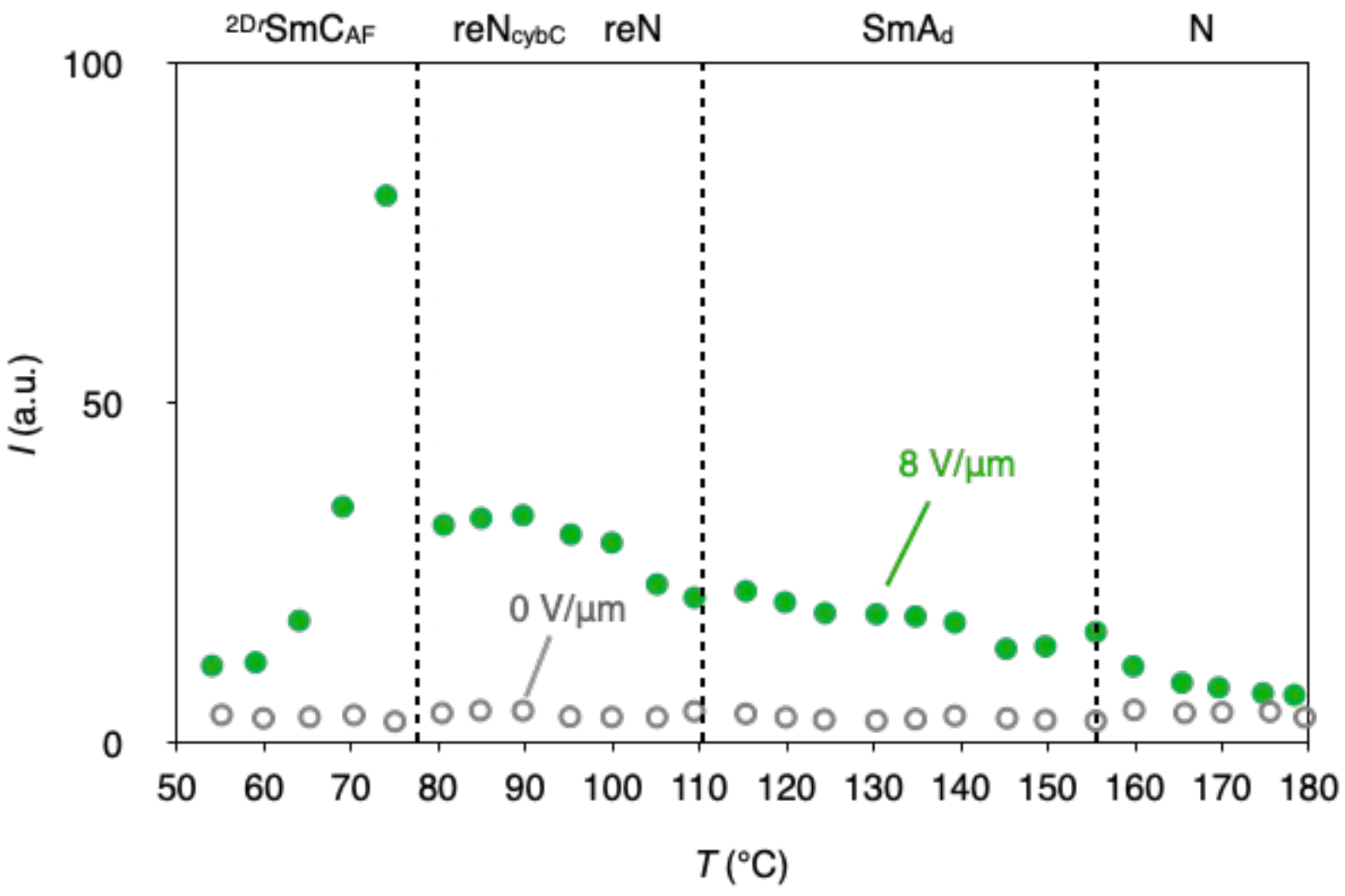

**Supplementary Fig. 58 |** Temperature-dependent SHG under zero and $E$-field (8.0 V µm$^{-1}$) for ***13*-012-CN**. The field-induced SHG signal decreased in the $^{2Dr}$SmC$_{AF}$ region because the applied field was insufficient to induce effective polarization switching at low temperatures. Because higher fields caused sample breakdown, the field strength was fixed at 8 V µm$^{-1}$ (176 V / 22 µm).

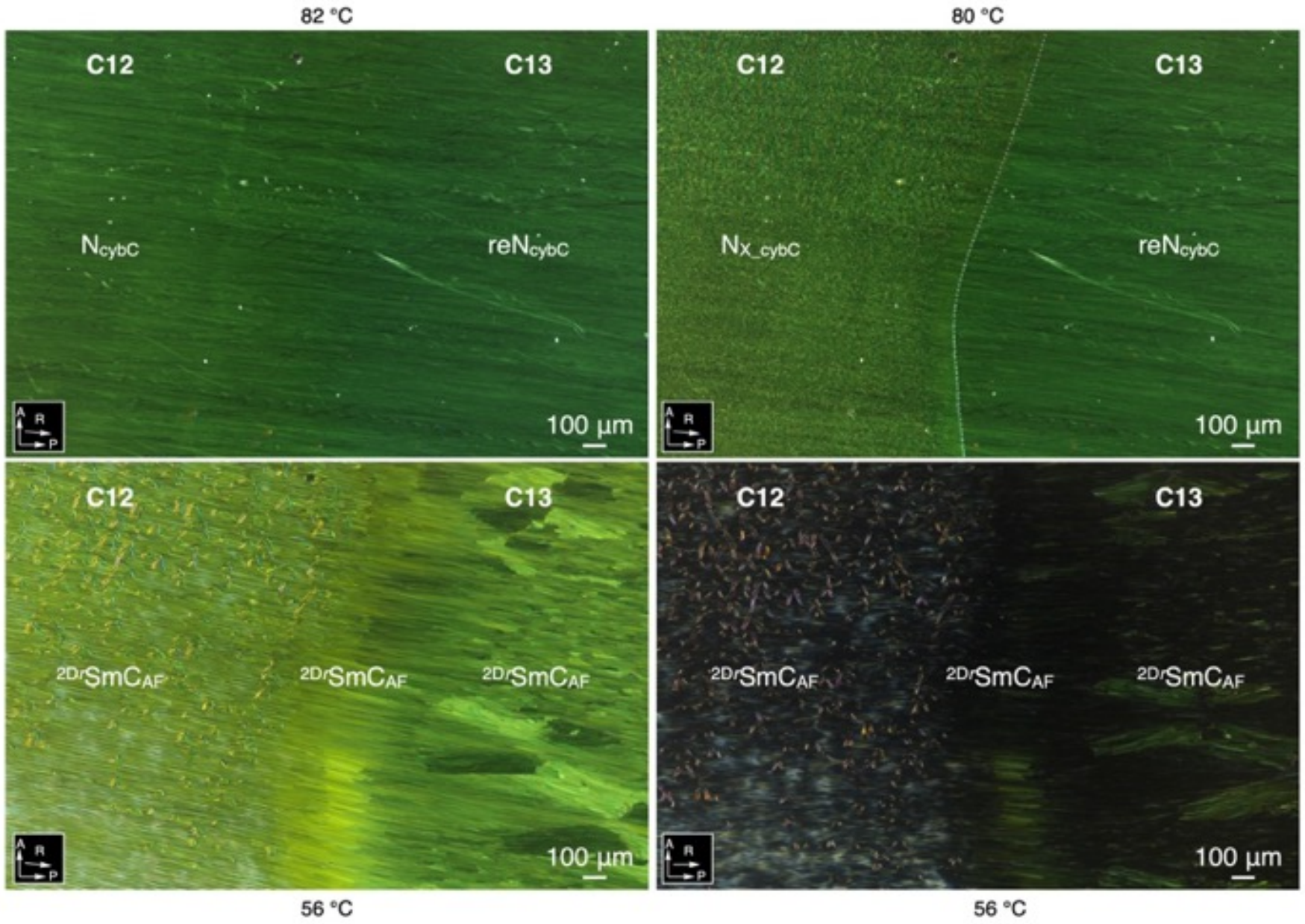

**Supplementary Fig. 59 |** Contact test in the LC cell between ***12*-012-CN** and ***13*-012-CN**.

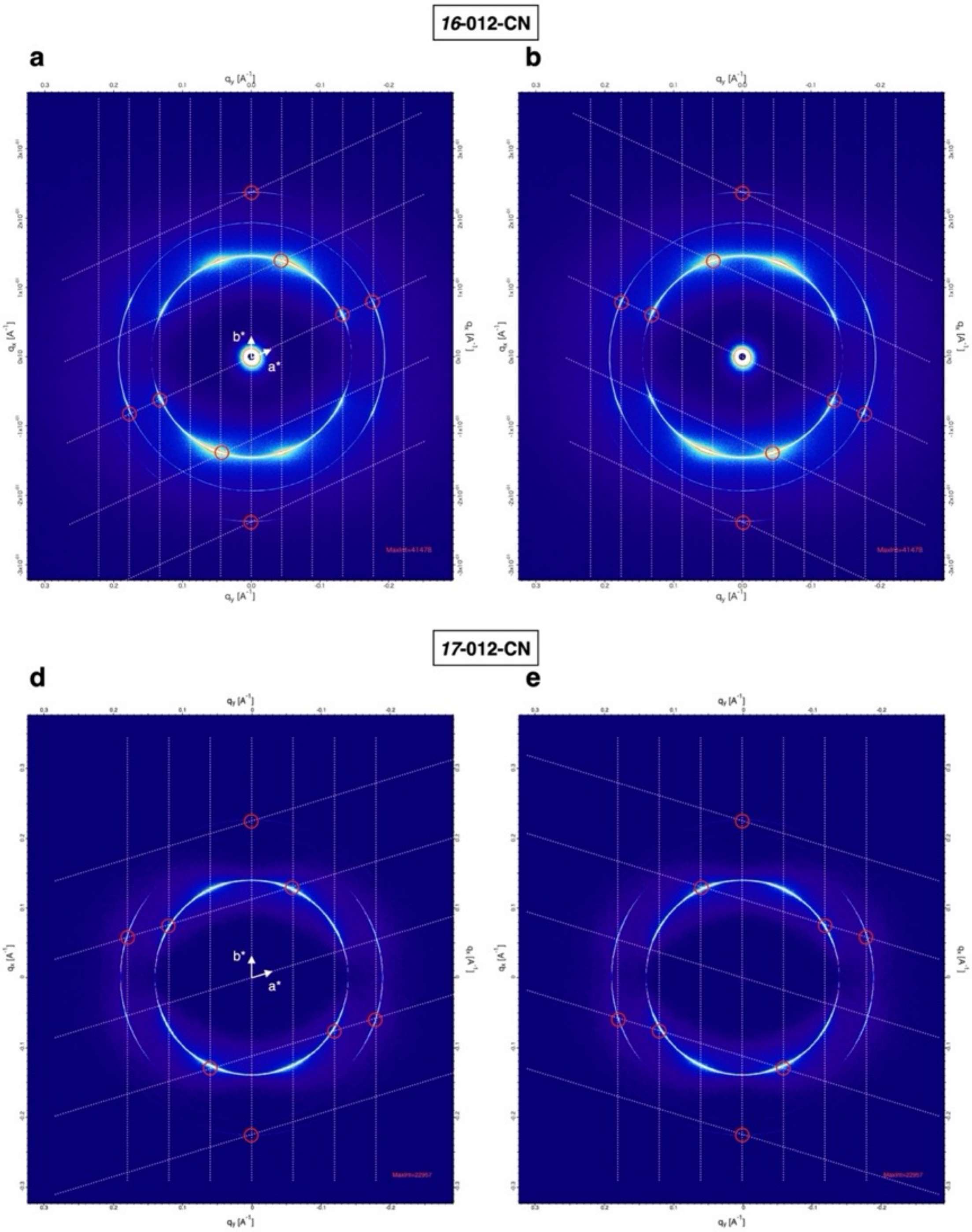

**Supplementary Fig. 60 |** Domain-resolved indexing of mirror-related oblique 2D lattices in ***16*-012-CN** and ***17*-012-CN**. (a,b) Magnetically aligned 2D SAXS patterns of ***16*-012-CN** at 64 °C, showing the reciprocal lattices associated with two mirror-related orientation domains. (d,e) Corresponding analysis for ***17*-012-CN** at 59 °C. The same diffraction pattern is displayed for each pair of panels, with the indexed reflections and reciprocal-lattice guides shown separately for each domain. Reflections from the two orientation domains were treated independently to avoid mixing their indices within a single lattice.

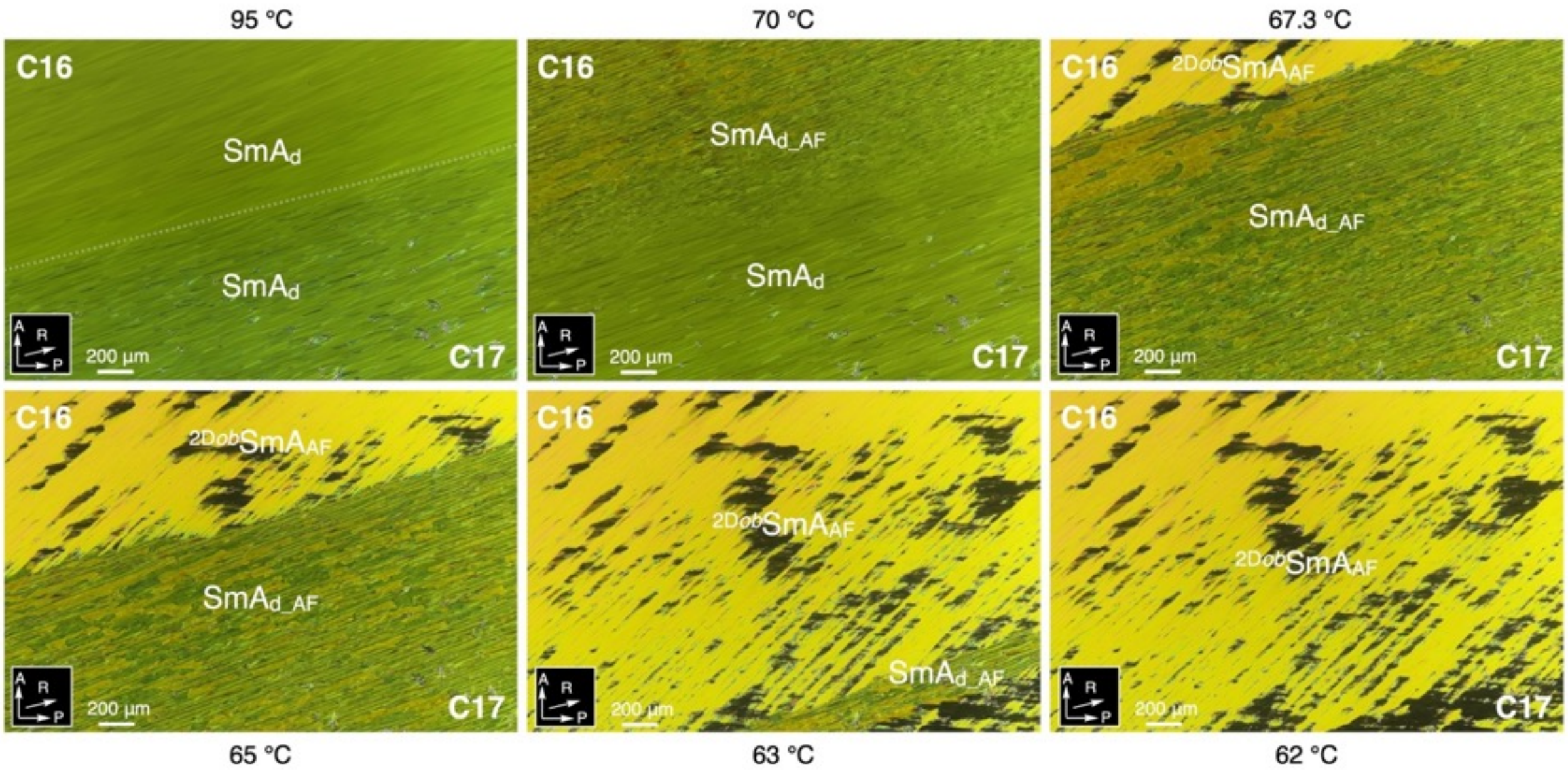

**Supplementary Fig. 61 |** Contact test in the LC cell between ***16*-012-CN** and ***17*-012-CN**.

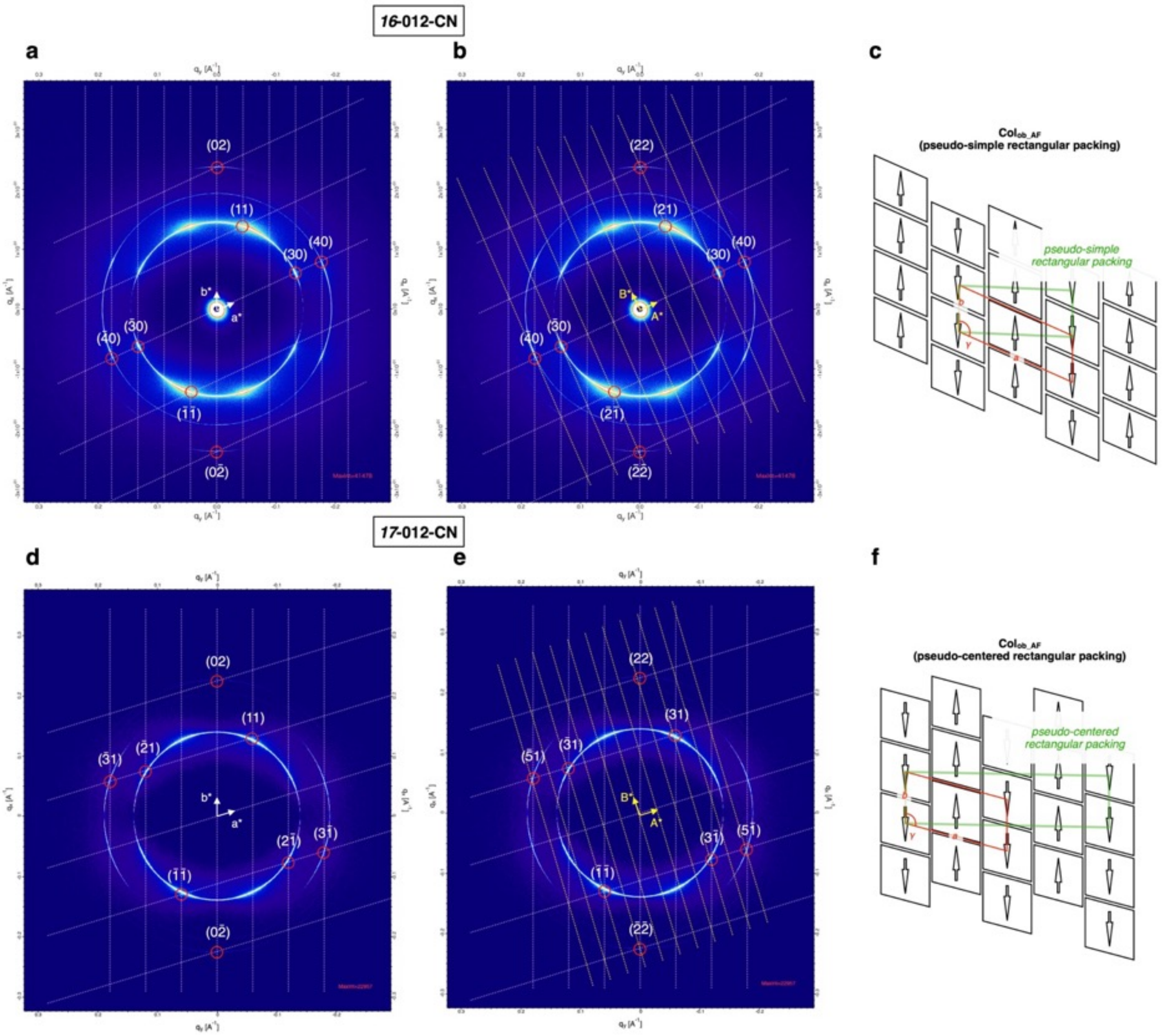

**Supplementary Fig. 62 |** Pseudo-rectangular representations of the oblique $Col_{ob_AF}$ lattices in ***16*-012-CN** and ***17*-012-CN.** (a,d) Magnetically aligned 2D SAXS patterns indexed using the original oblique unit cells for ***16*-012-CN** at 64 °C and ***17*-012-CN** at 59 °C, respectively. (b,e) Corresponding patterns reindexed using nearly orthogonal unit-cell representations obtained through basis transformations. The transformed lattice angles are $\Gamma \approx 90.0°$ for ***16*-012-CN** and $\Gamma \approx 90.0°$ for ***17*-012-CN**, corresponding to pseudo-simple rectangular and pseudo-centered rectangular representations, respectively. (c,f) Proposed polar-column packing models illustrating the two lattice representations. The transformation equations, lattice parameters, and reflection-index correspondences are provided in Supplementary Note 2.

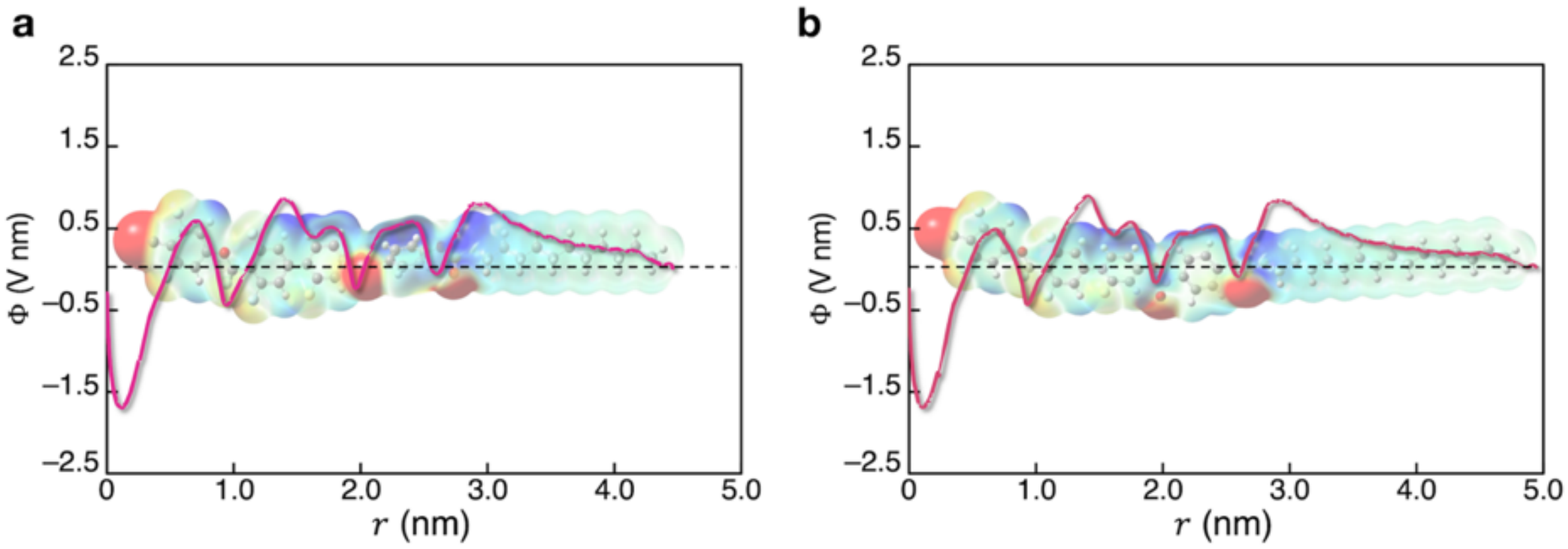

**Supplementary Fig. 63 |** 3D ESP surface and resulting 1D longitudinal ESP profile for ***12*-012-CN** (a) and ***16*-012-CN** (b).

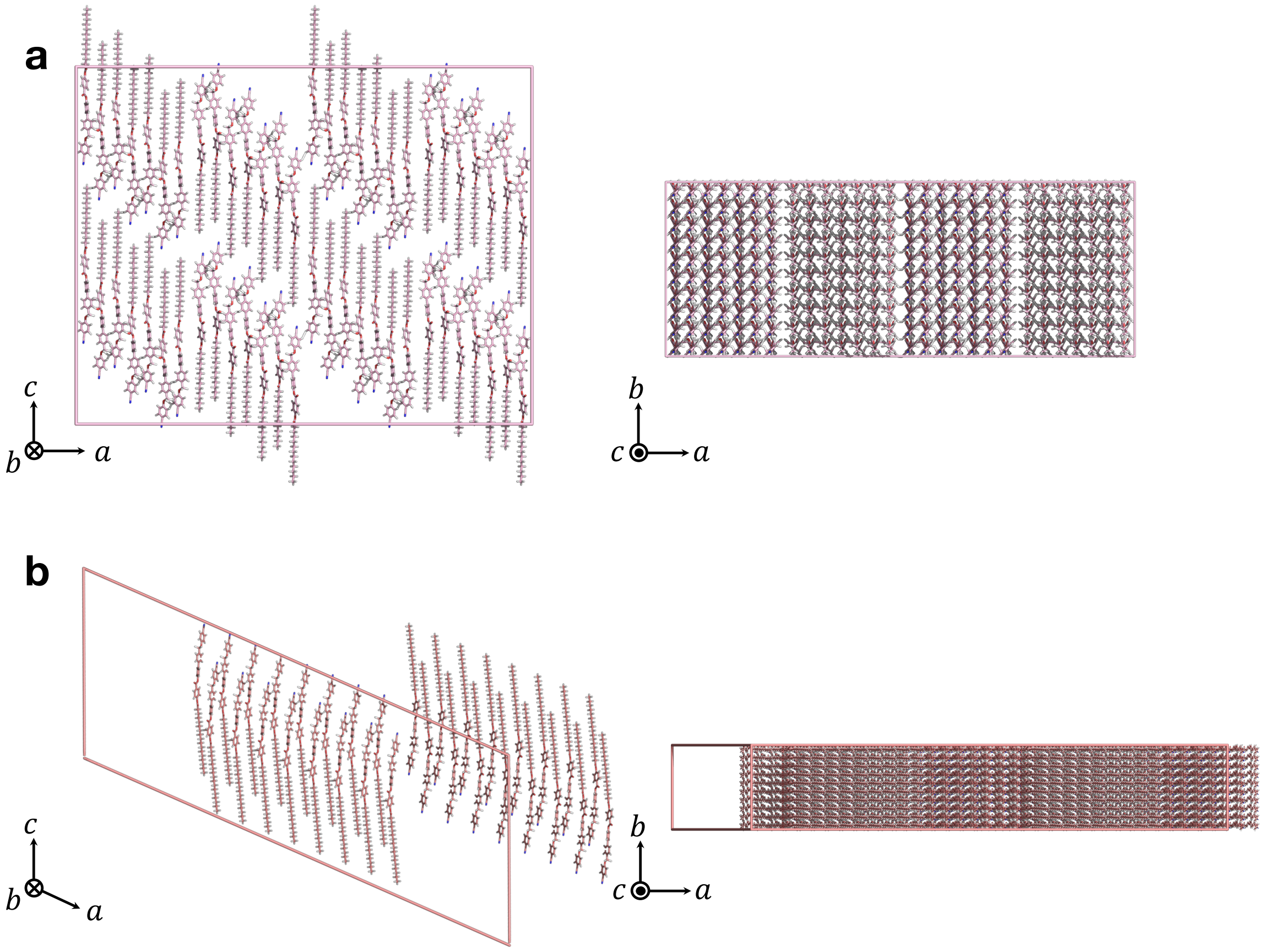

**Supplementary Fig. 64 |** Initial configurations for the all-atom MD simulations of ***12*-012-CN** (a) and ***16*-012-CN** (b). The initial polar columns with initial SmC-type packing contain 7 molecules per column segment with $\beta$ = 38° for ***12*-012-CN** and 16 molecules per column segment with $\beta$ = 19° for ***16*-012-CN**.

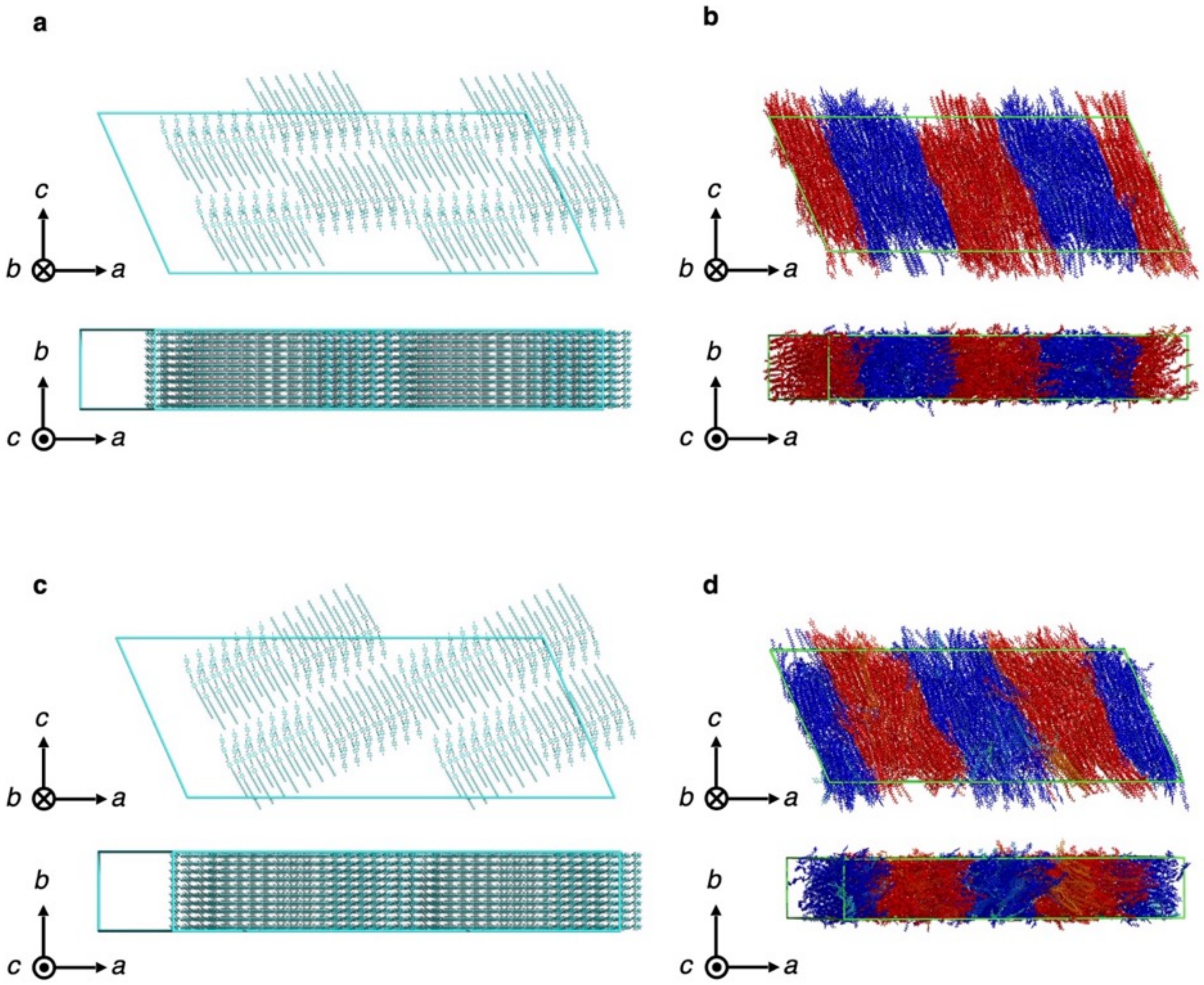

**Supplementary Fig. 65 |** Effect of initial molecular packing on the relaxed polar-column arrangements of ***16*-012-CN**. Initial configurations containing 14 molecules per column segment, with SmC-type ($\beta$ = 19°) (a) and SmA-type ($\beta$ = 0°) (c) packing, respectively. (b,d) Corresponding MD snapshots after 100 ns of equilibration at 335 K. Red and blue indicate oppositely polarized columns.

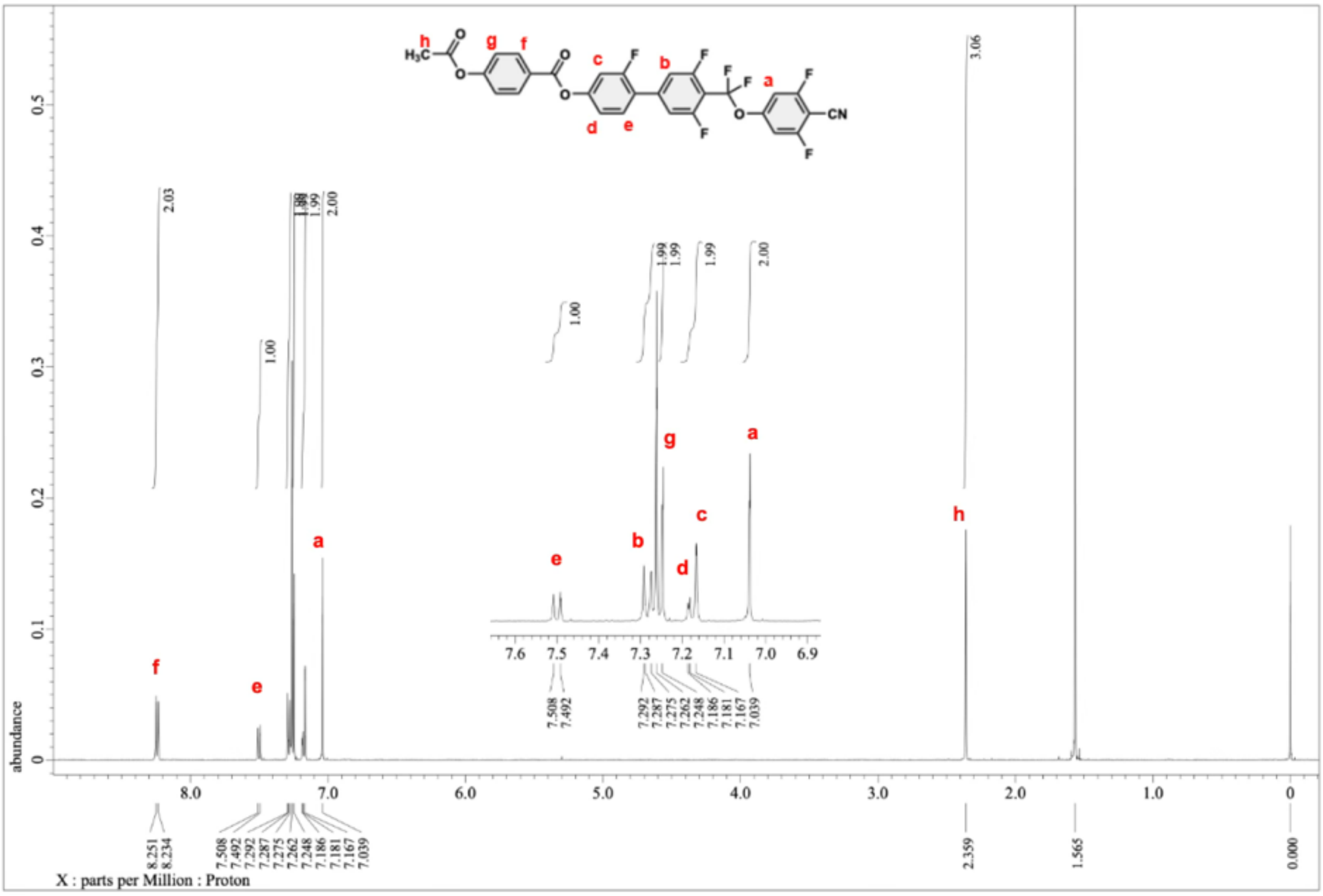

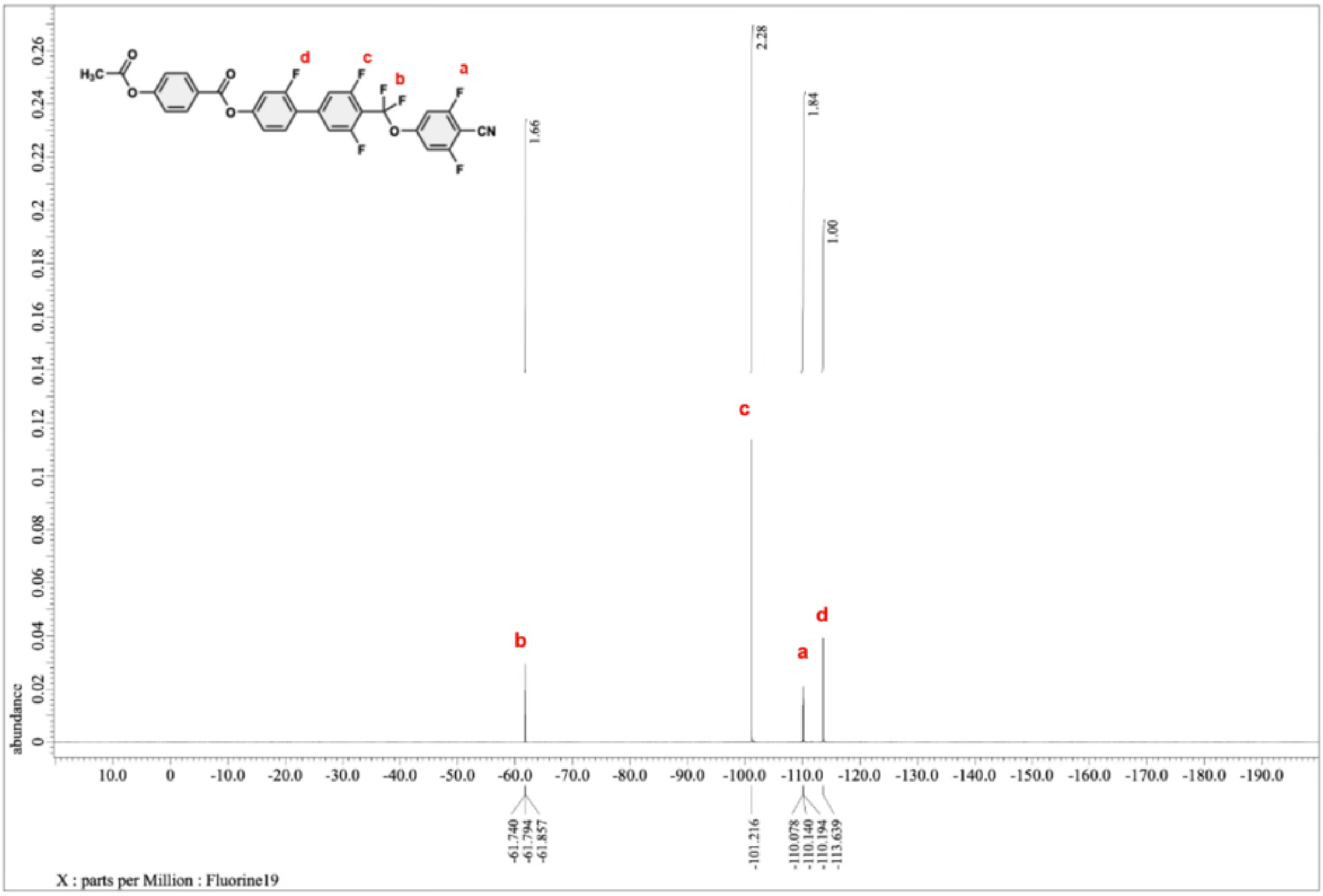

**Supplementary Fig. 66 |** $^{1}H\{^{19}F\}$ (upper) and $^{19}F\{^{1}H\}$ (bottom) NMR spectra for ***1*-012-CN**.

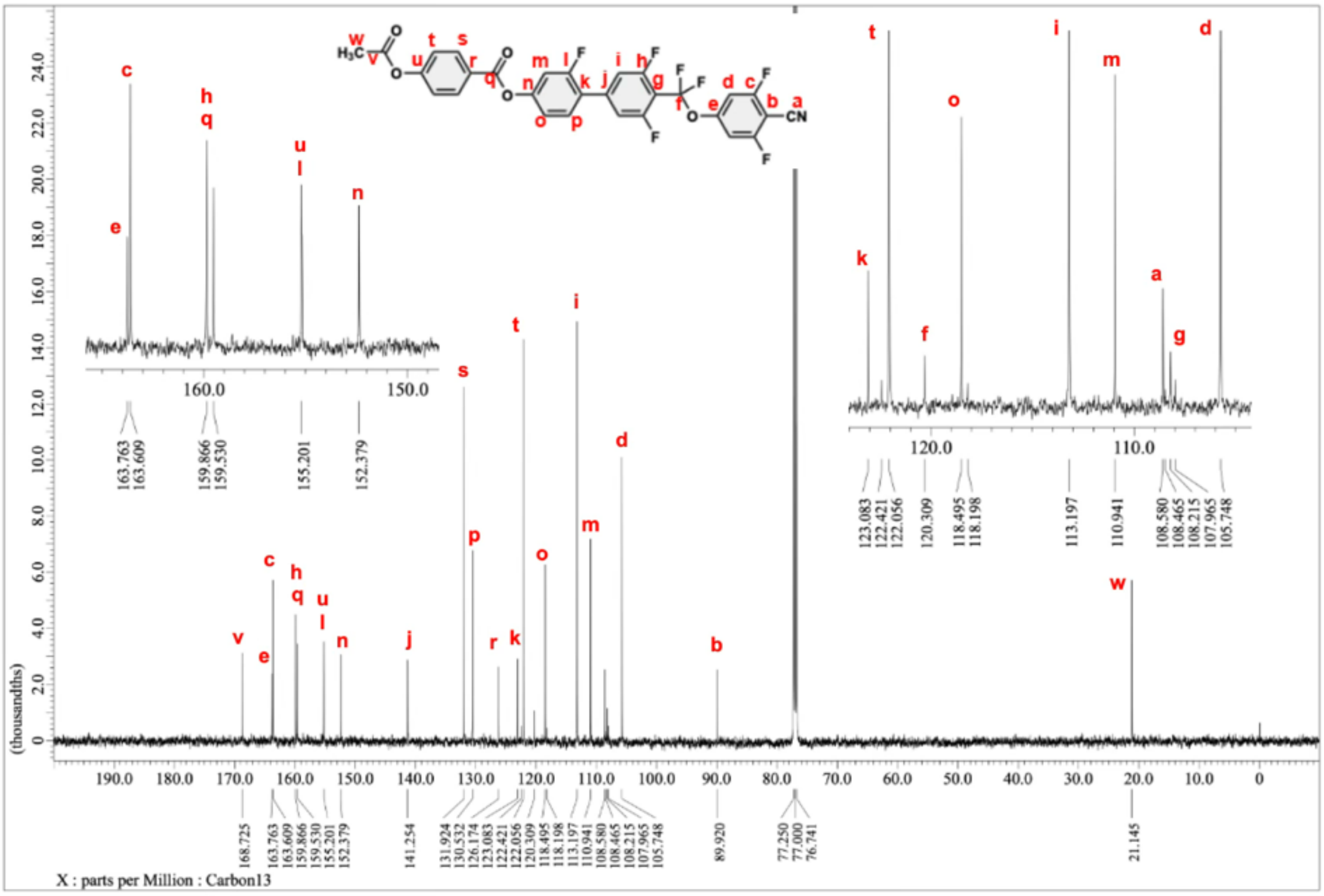

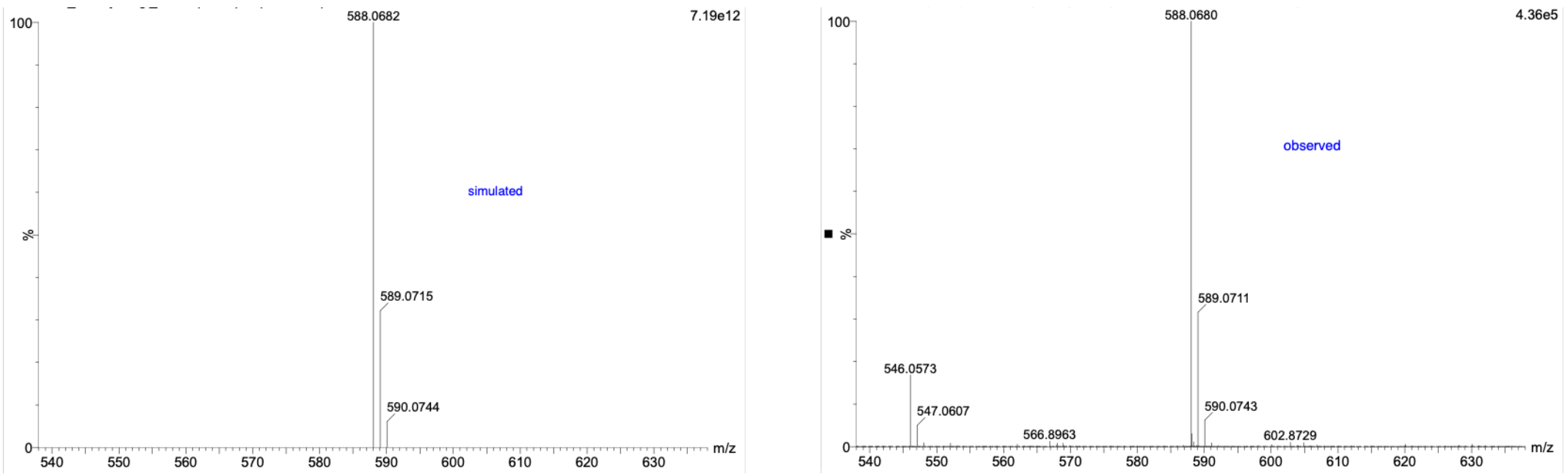

**Supplementary Fig. 67 |** $^{13}C\{^1H,^{19}F\}$NMR (upper) and HR-ESI-TOF-MS spectra (bottom) for ***1*-012-CN**.

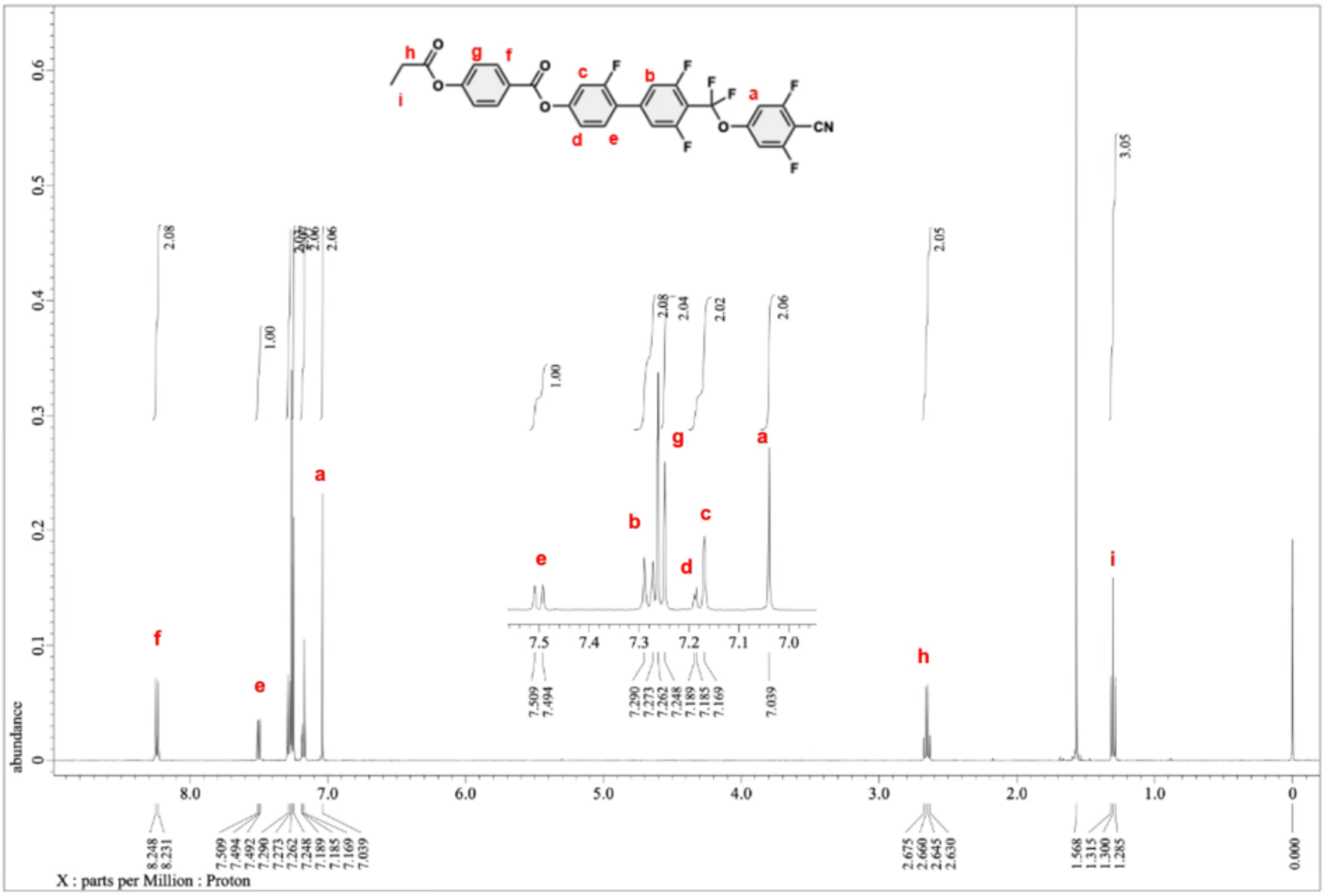

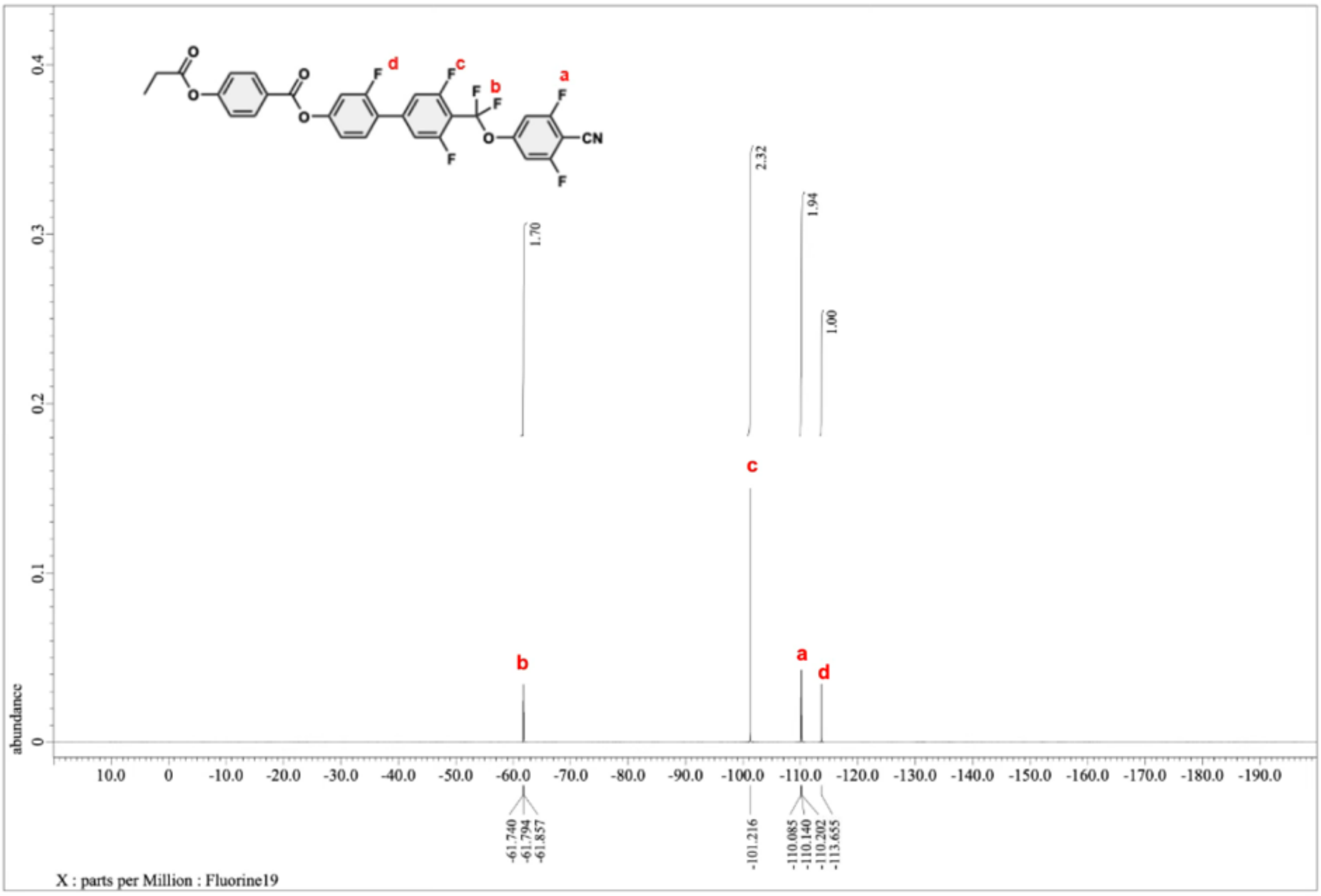

**Supplementary Fig. 68 |** $^{1}H\{^{19}F\}$ (upper) and $^{19}F\{^{1}H\}$ (bottom) NMR spectra for ***2*-012-CN**.

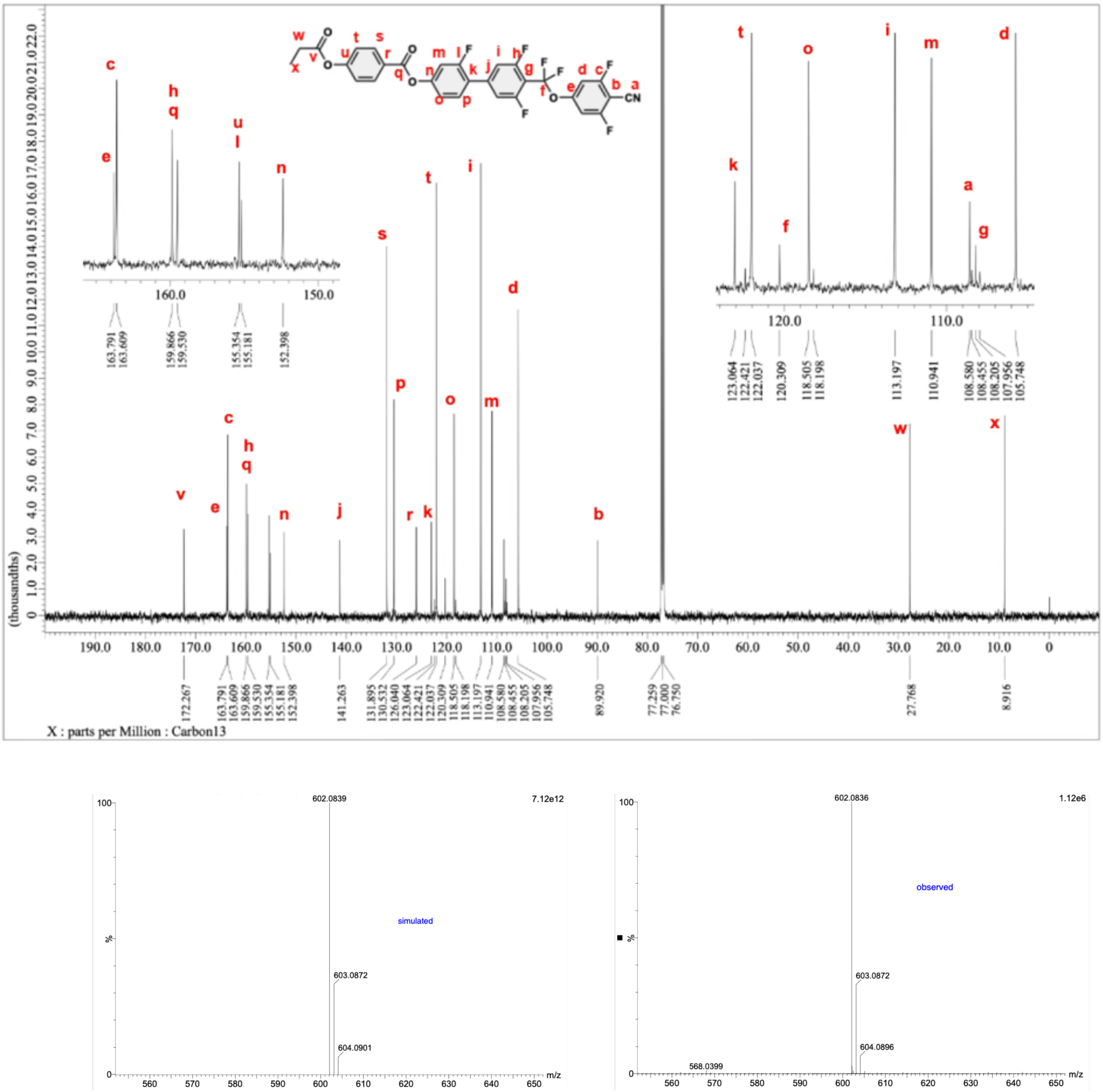

**Supplementary Fig. 69 |** $^{13}C\{^1H,^{19}F\}$NMR (upper) and HR-ESI-TOF-MS spectra (bottom) for ***2*-012-CN**.

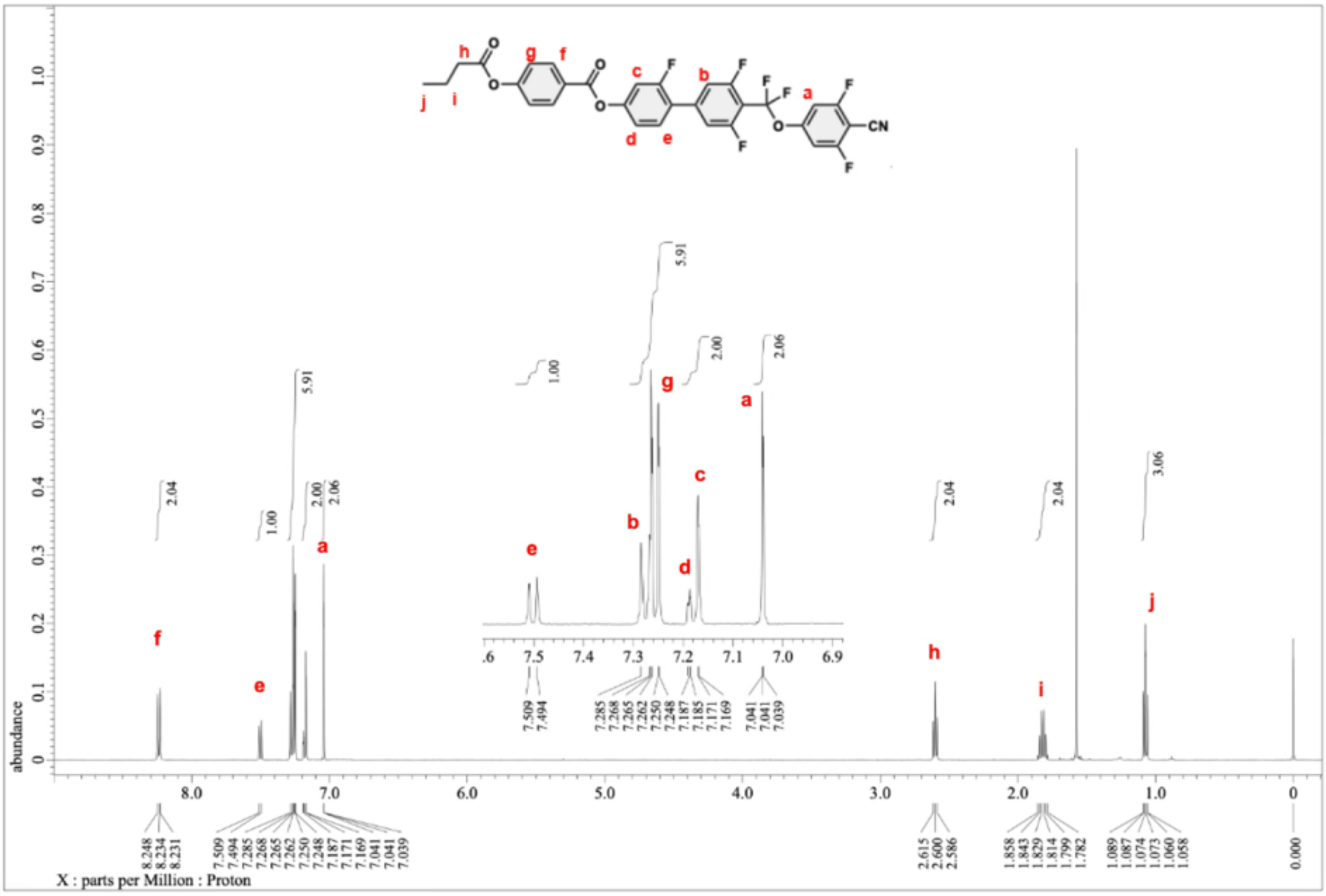

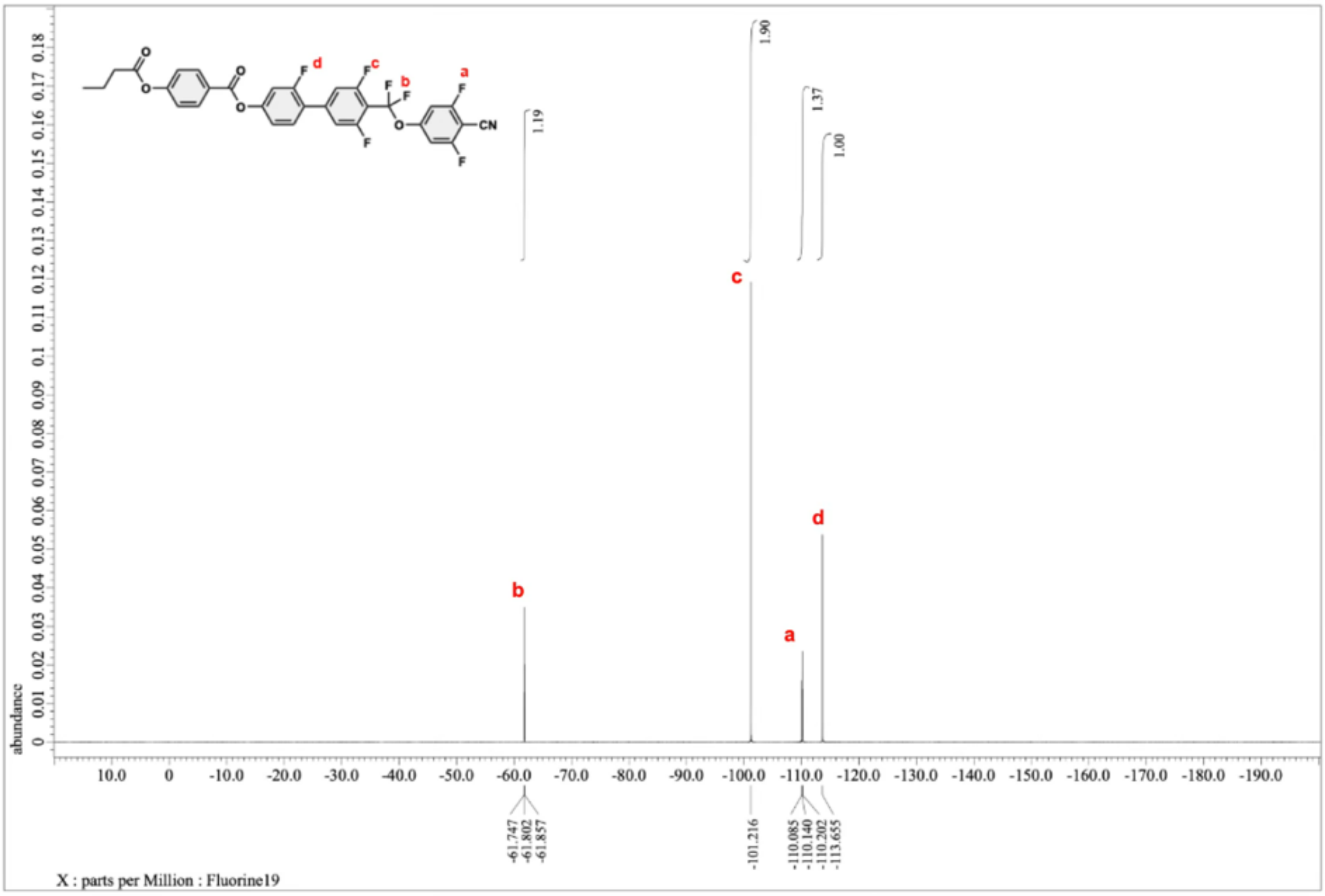

**Supplementary Fig. 70 |** $^{1}H\{^{19}F\}$ (upper) and $^{19}F\{^{1}H\}$ (bottom) NMR spectra for ***3*-012-CN**.

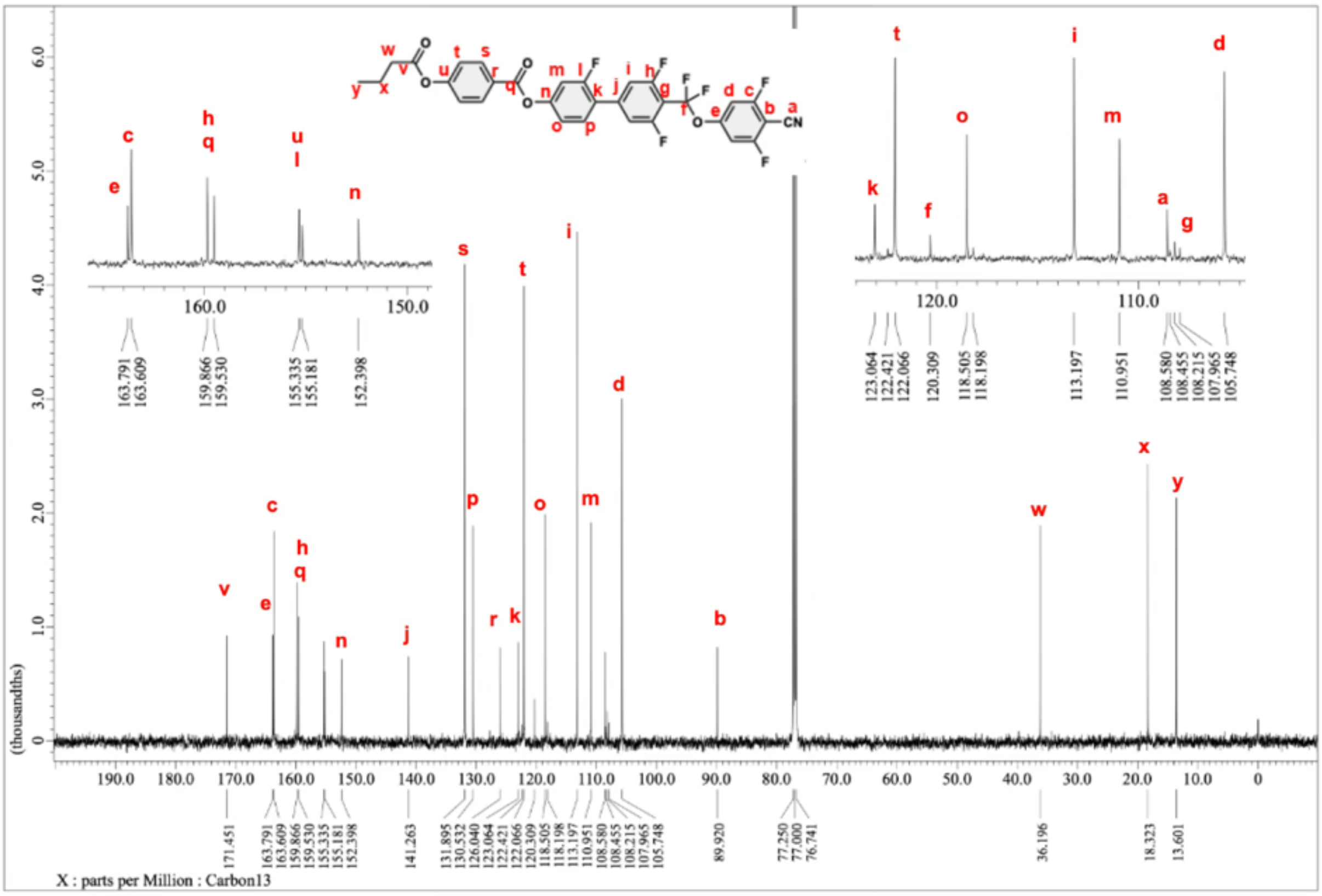

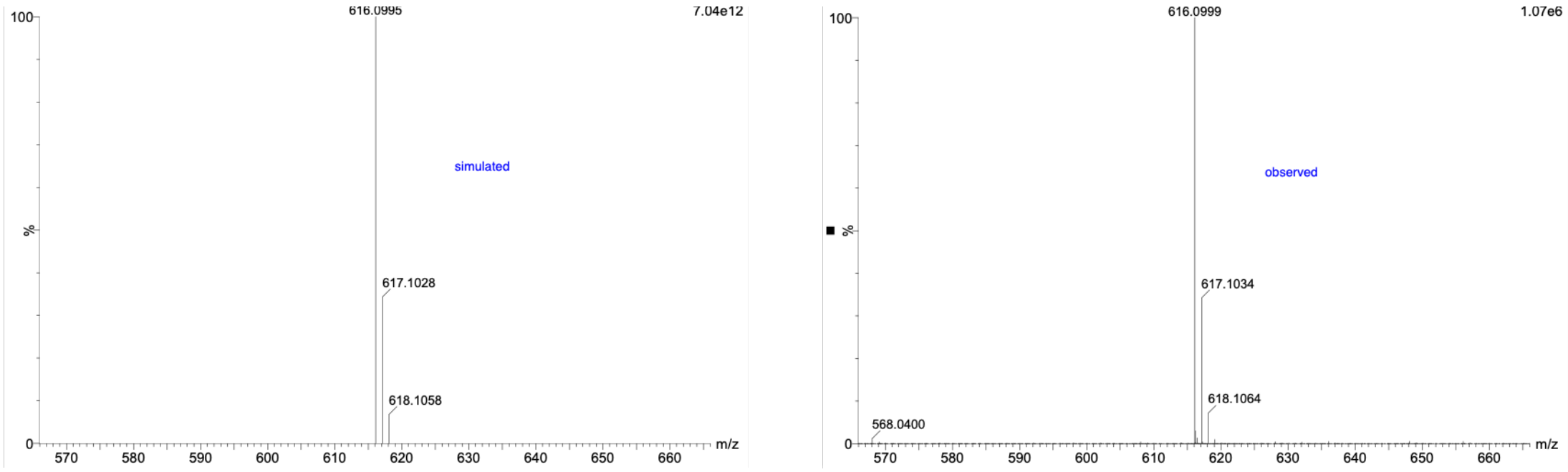

**Supplementary Fig. 71 |** $^{13}C\{^1H,^{19}F\}$NMR (upper) and HR-ESI-TOF-MS spectra (bottom) for ***3*-012-CN**.

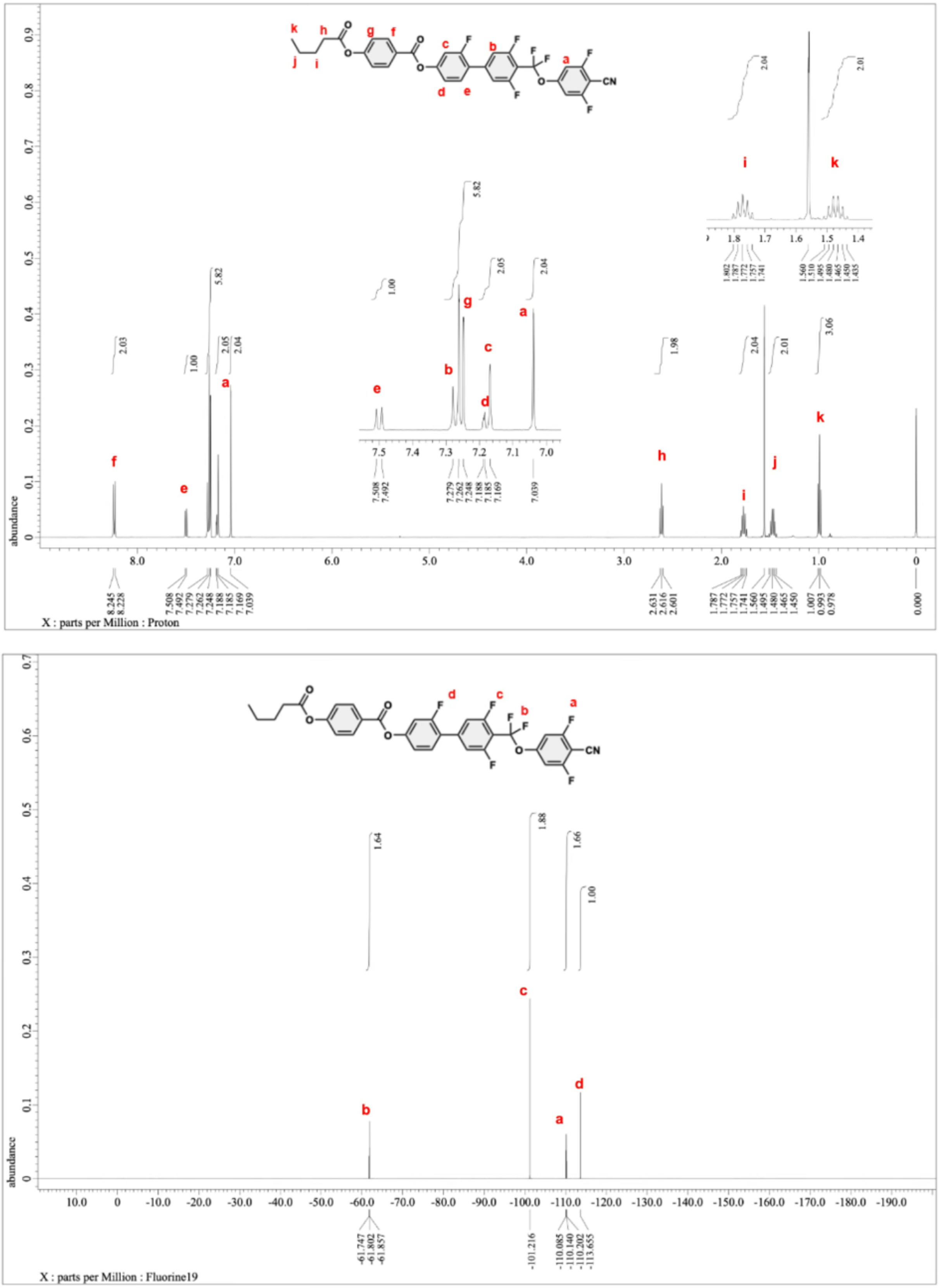

**Supplementary Fig. 72 |** $^{1}H\{^{19}F\}$ (upper) and $^{19}F\{^{1}H\}$ (bottom) NMR spectra for ***4*-012-CN**.

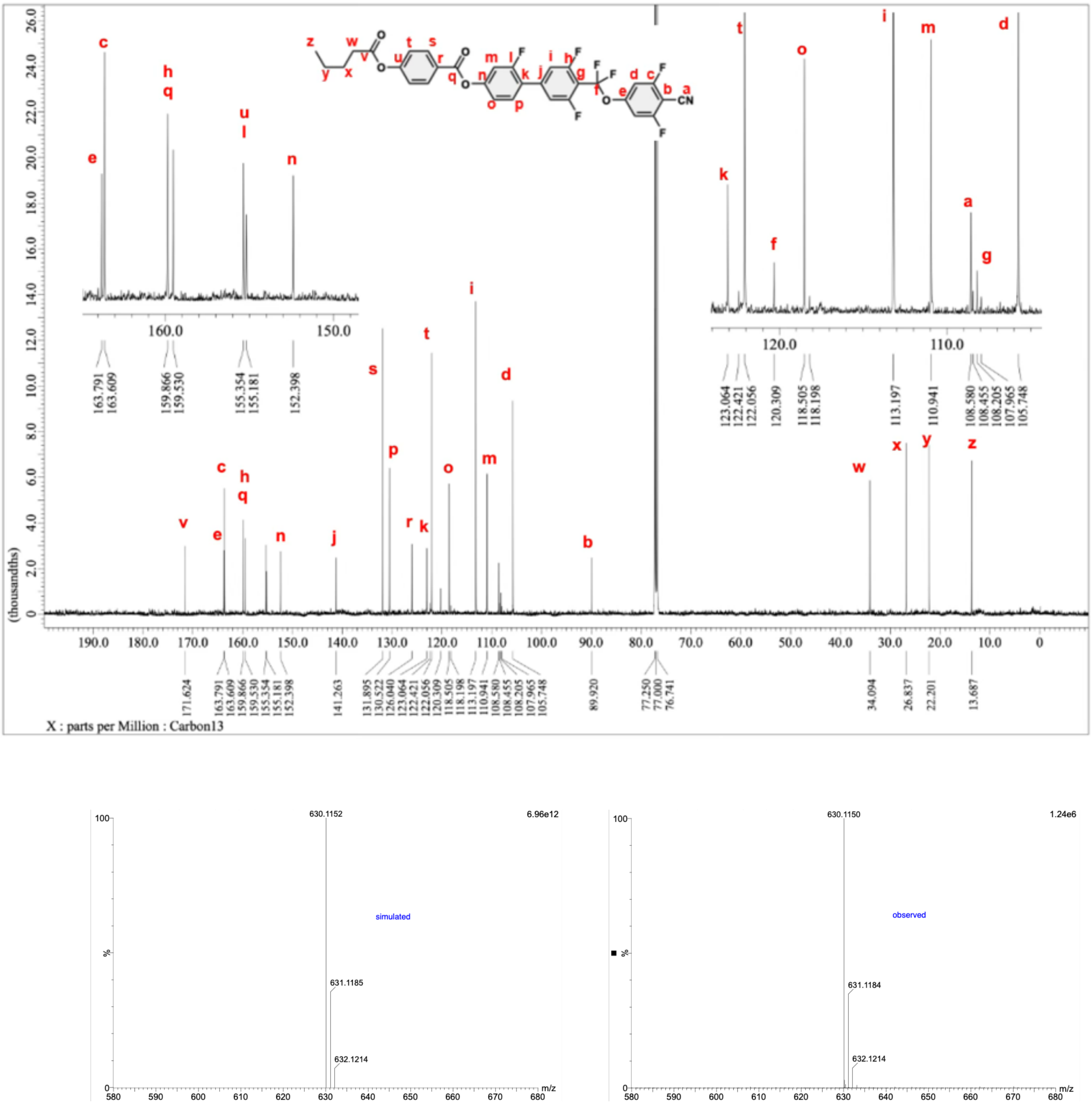

**Supplementary Fig. 73 |** $^{13}C\{^{1}H,^{19}F\}$NMR (upper) and HR-ESI-TOF-MS spectra (bottom) for ***4*-012-CN**.

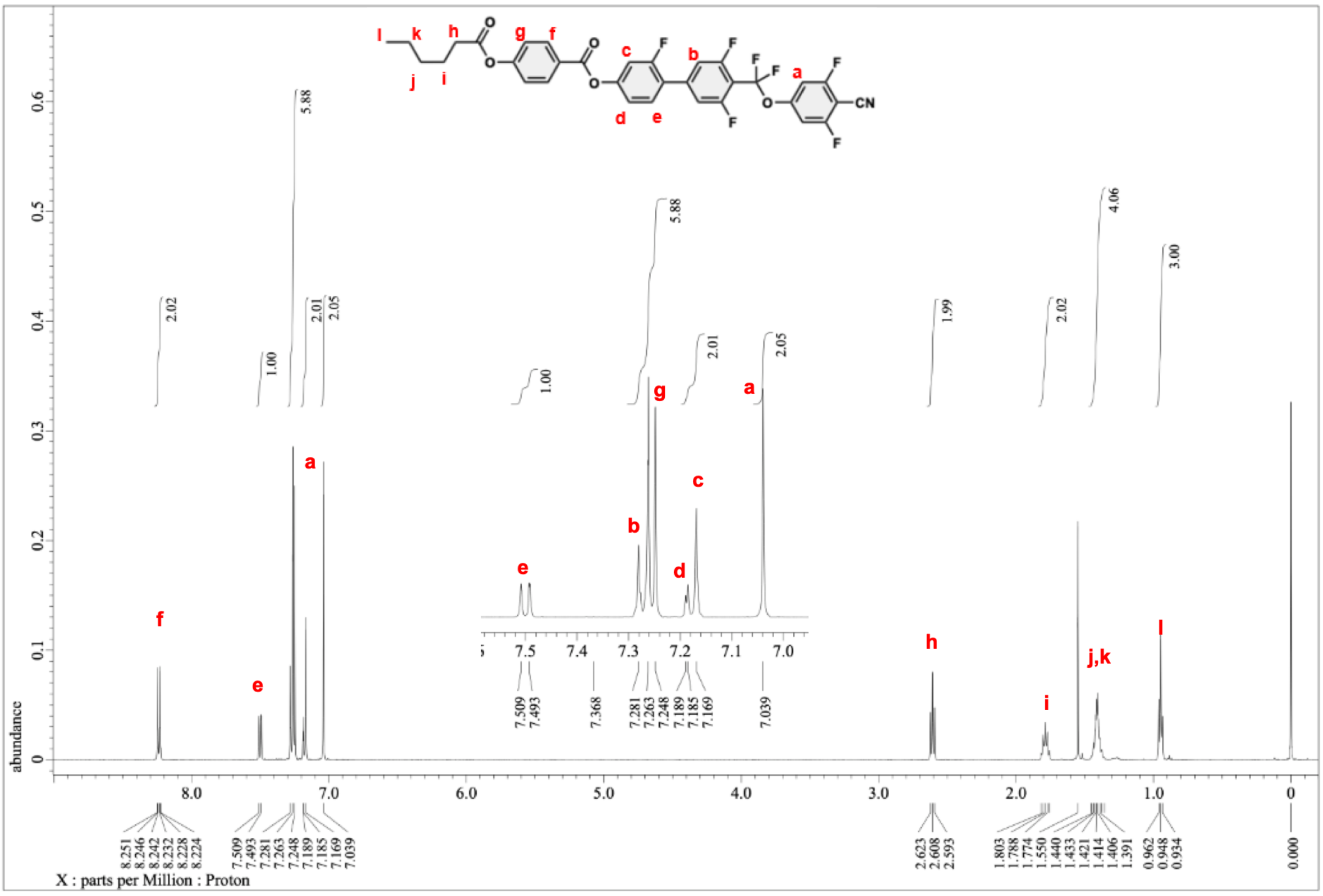

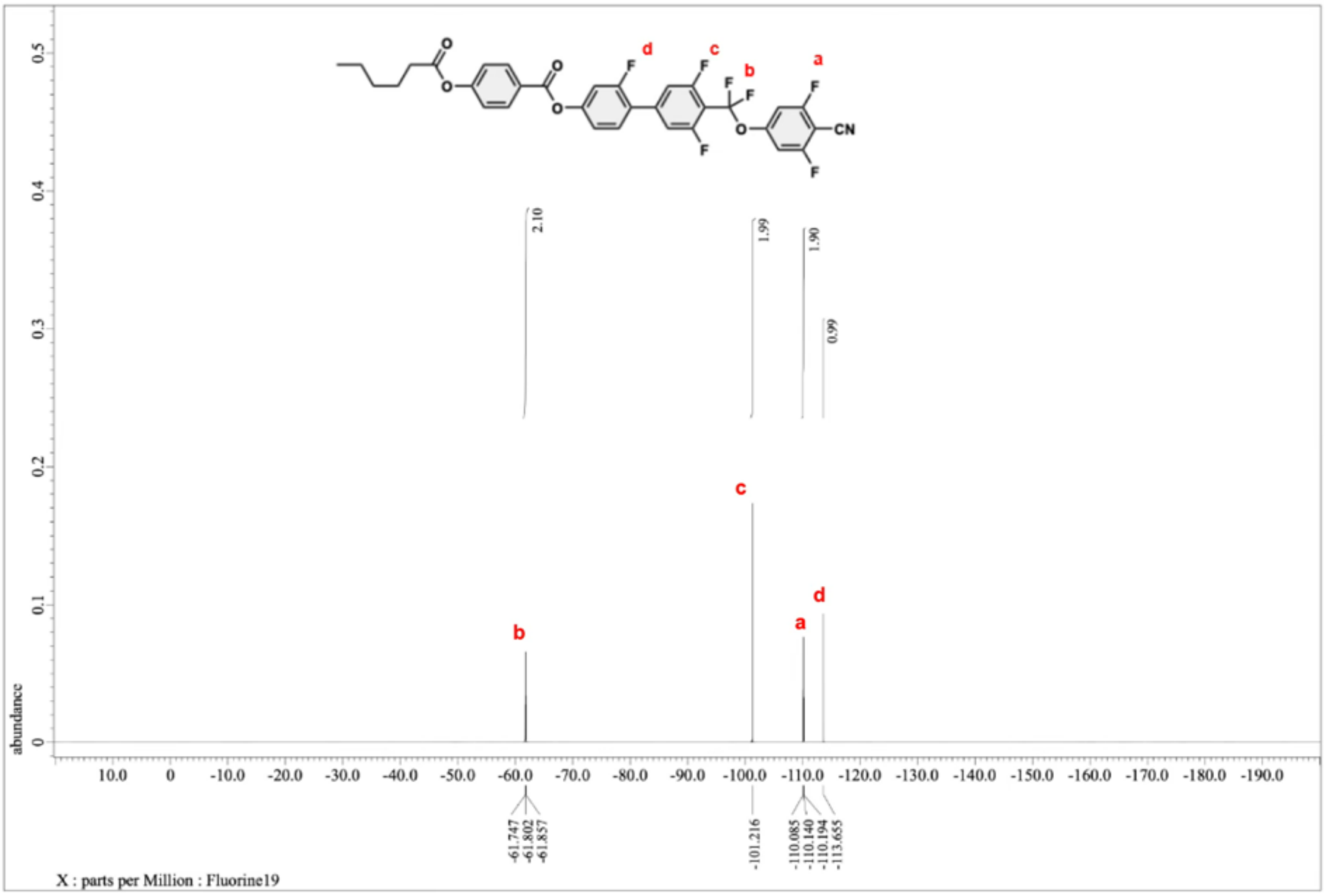

**Supplementary Fig. 74 |** $^{1}H\{^{19}F\}$ (upper) and $^{19}F\{^{1}H\}$ (bottom) NMR spectra for **5-012-CN**.

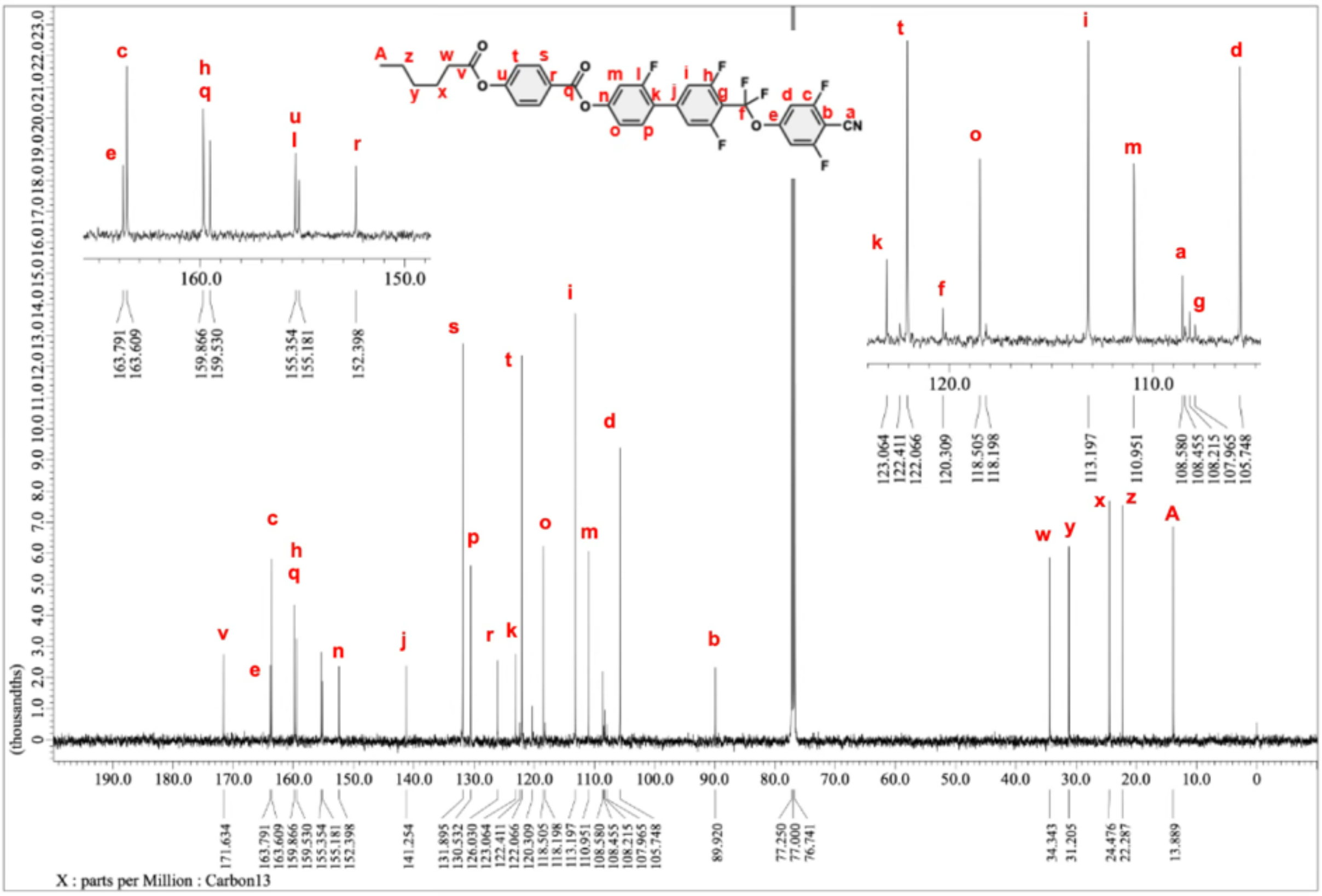

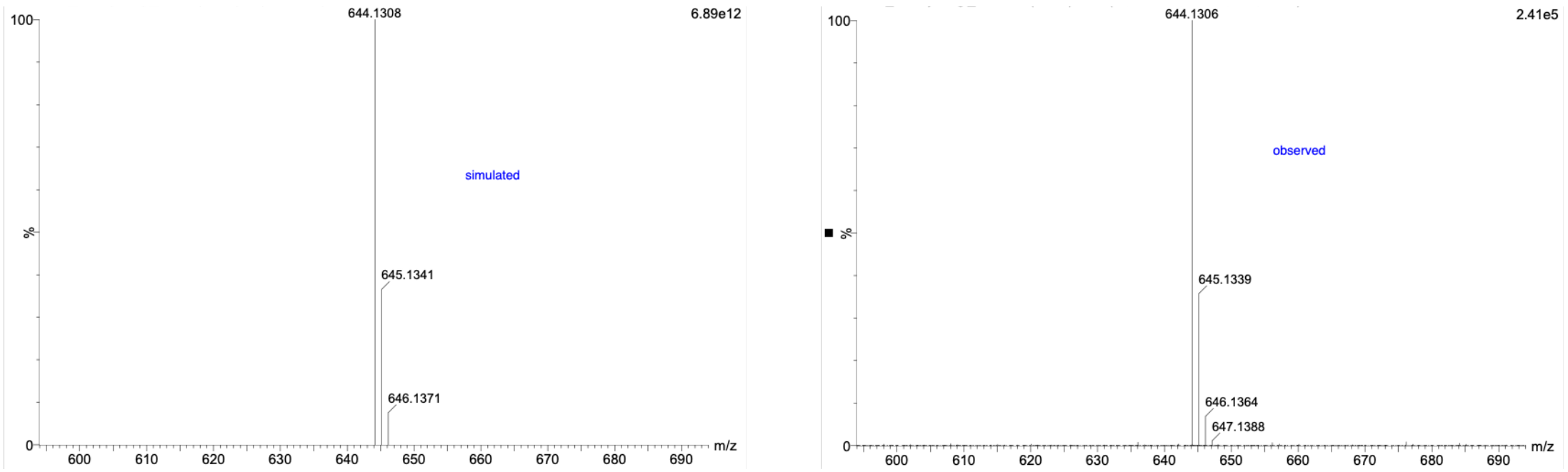

**Supplementary Fig. 75 |** $^{13}C\{^1H,^{19}F\}$NMR (upper) and HR-ESI-TOF-MS spectra (bottom) for **5-012-CN**.

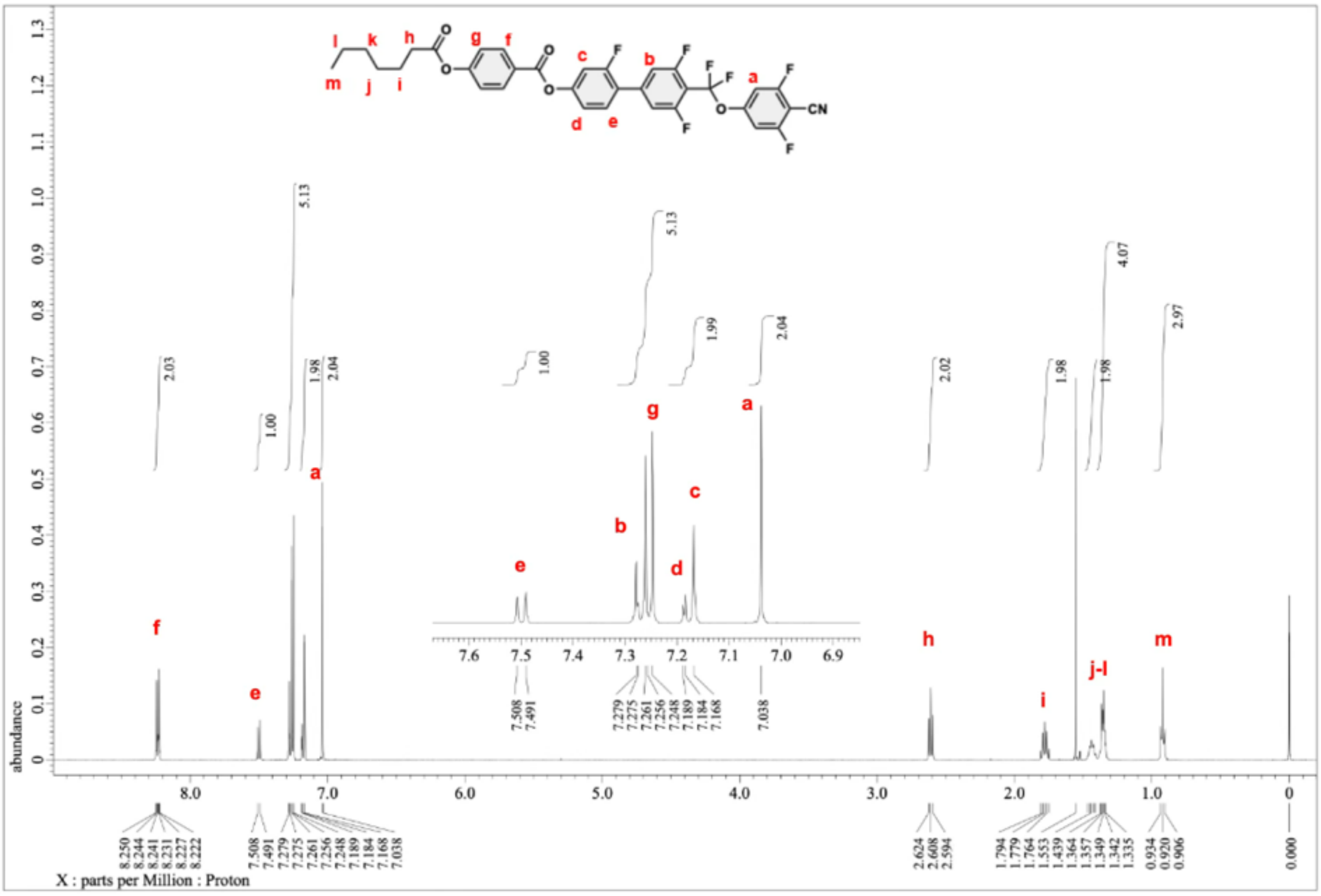

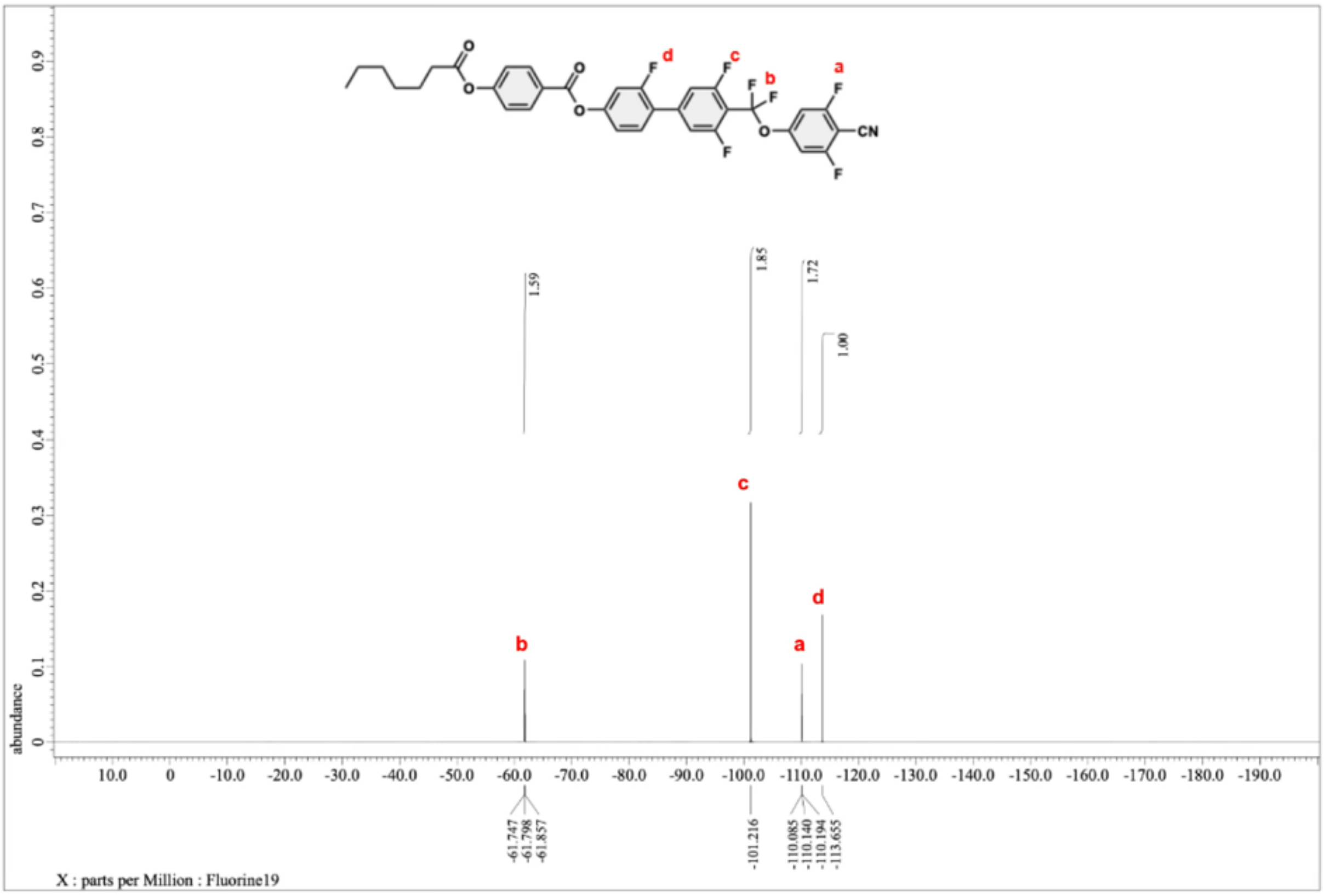

**Supplementary Fig. 76 |** $^{1}H\{^{19}F\}$ (upper) and $^{19}F\{^{1}H\}$ (bottom) NMR spectra for ***6*-012-CN**.

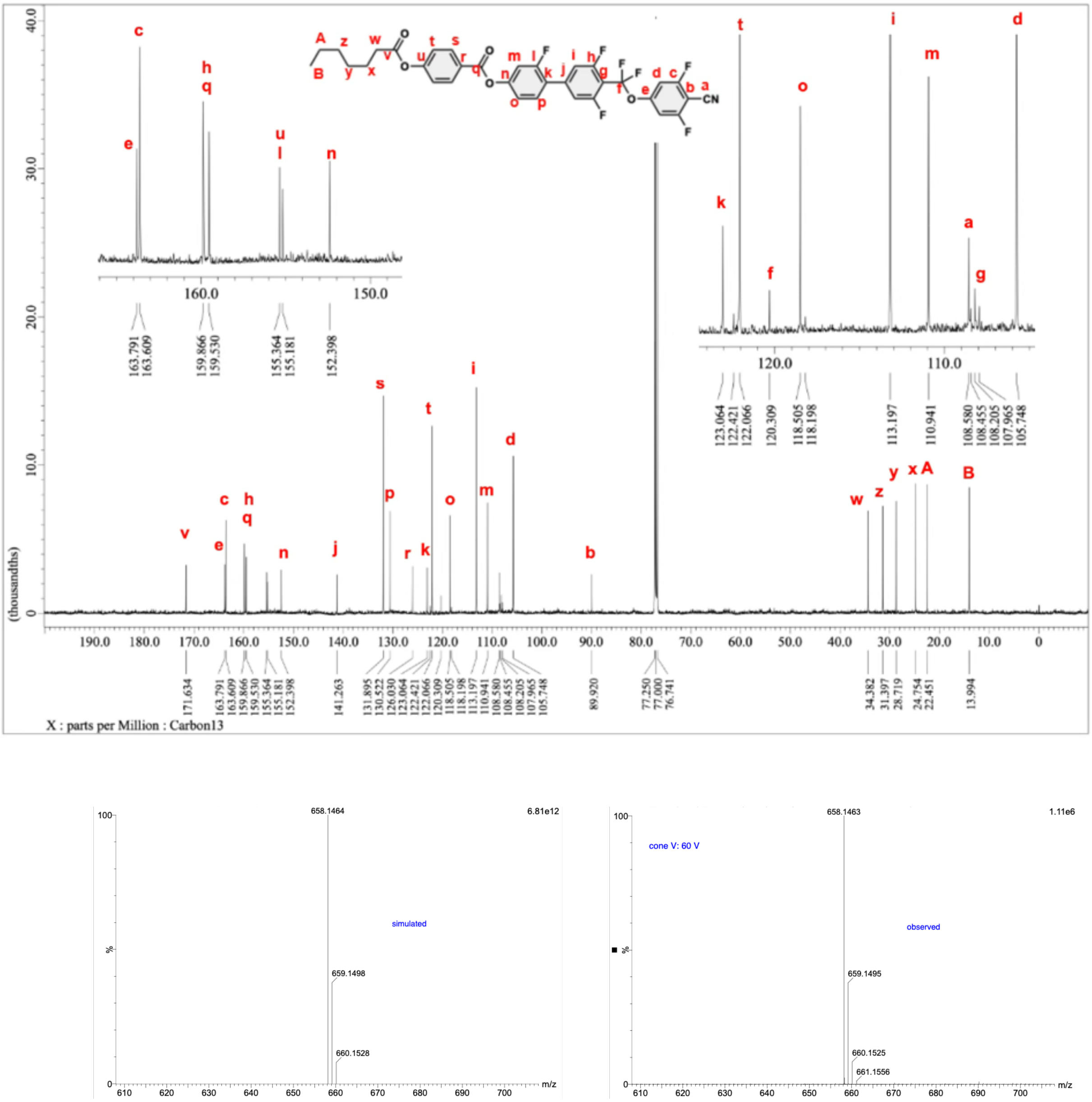

**Supplementary Fig. 77 |** $^{13}$C{1H,$^{19}$F}NMR (upper) and HR-ESI-TOF-MS spectra (bottom) for ***6*-012-CN**.

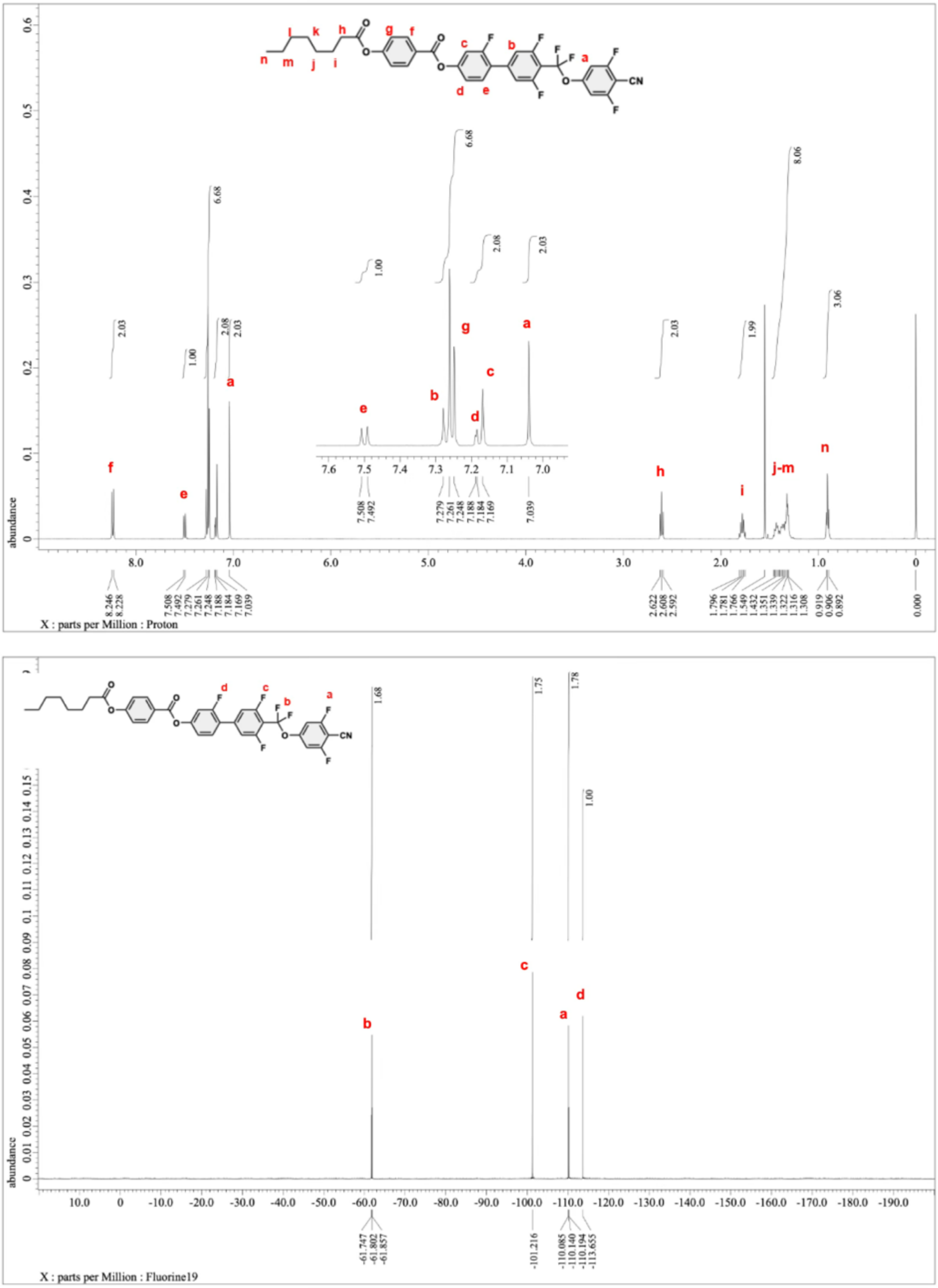

**Supplementary Fig. 78 |** $^{1}H\{^{19}F\}$ (upper) and $^{19}F\{^{1}H\}$ (bottom) NMR spectra for ***7*-012-CN**.

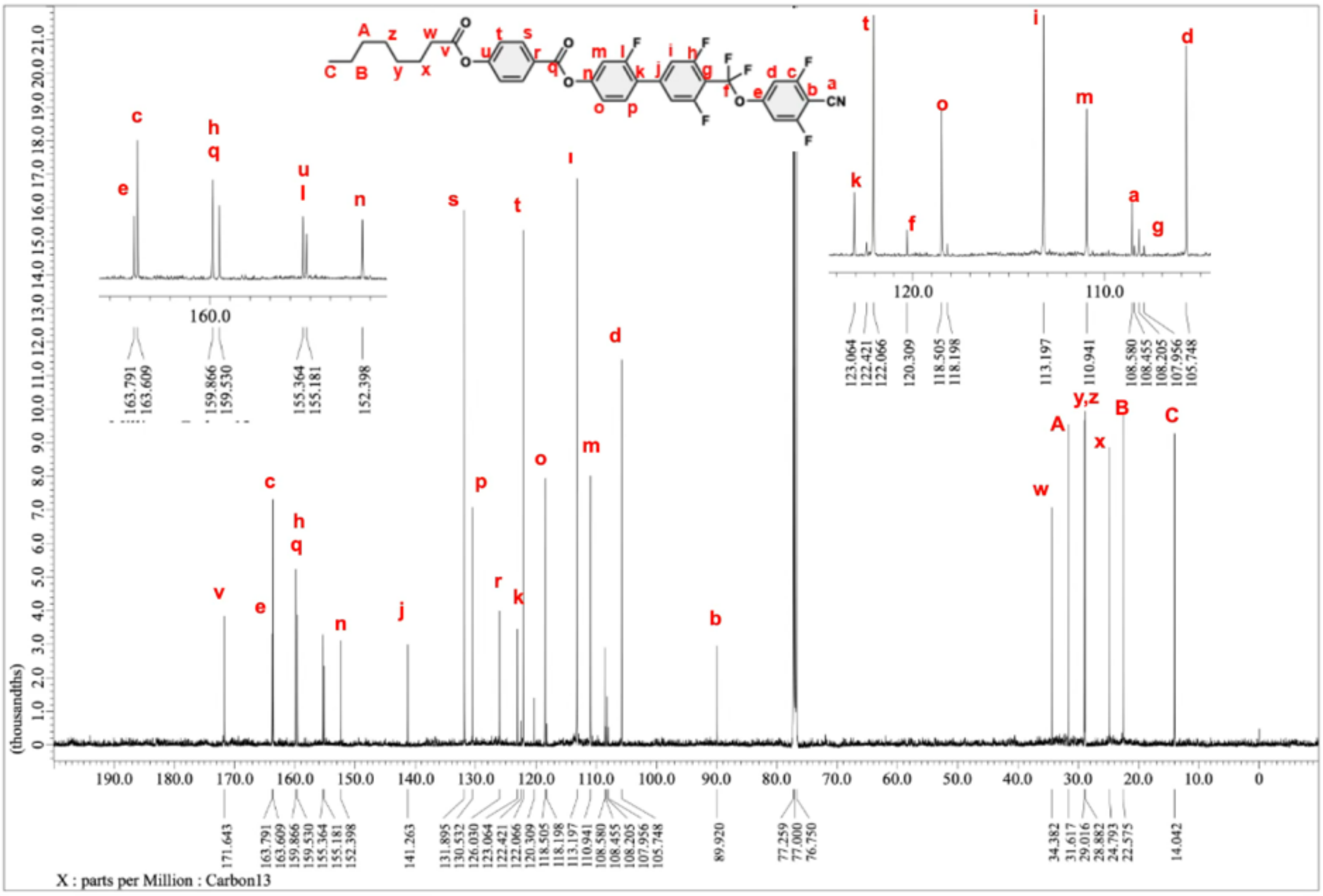

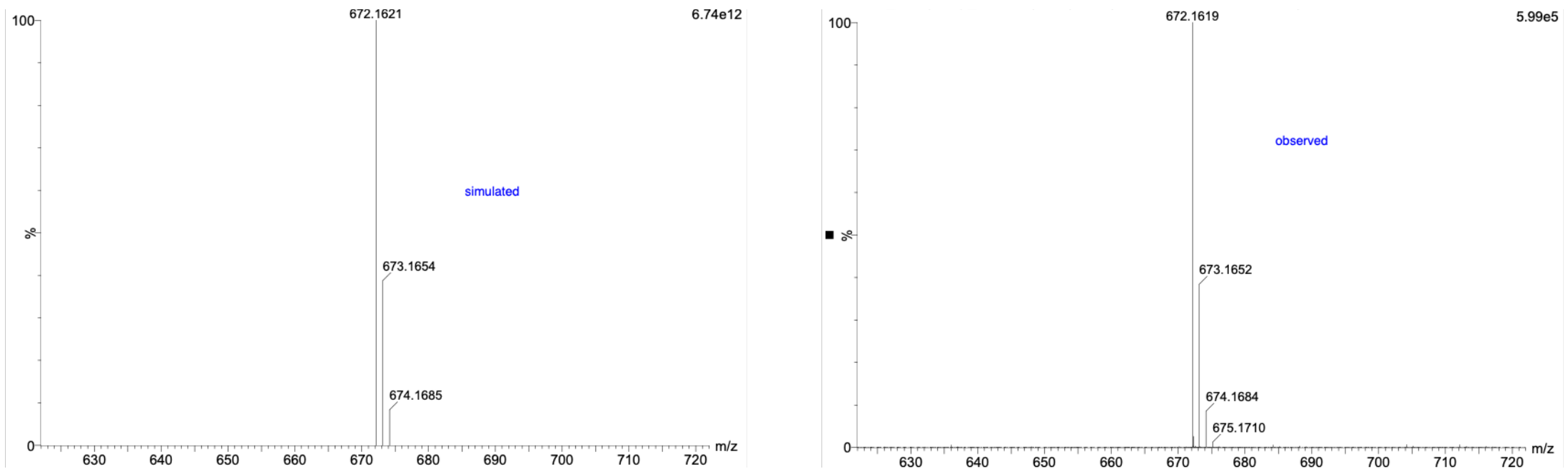

**Supplementary Fig. 79 |** $^{13}C\{1H,^{19}F\}$NMR (upper) and HR-ESI-TOF-MS spectra (bottom) for ***7*-012-CN**.

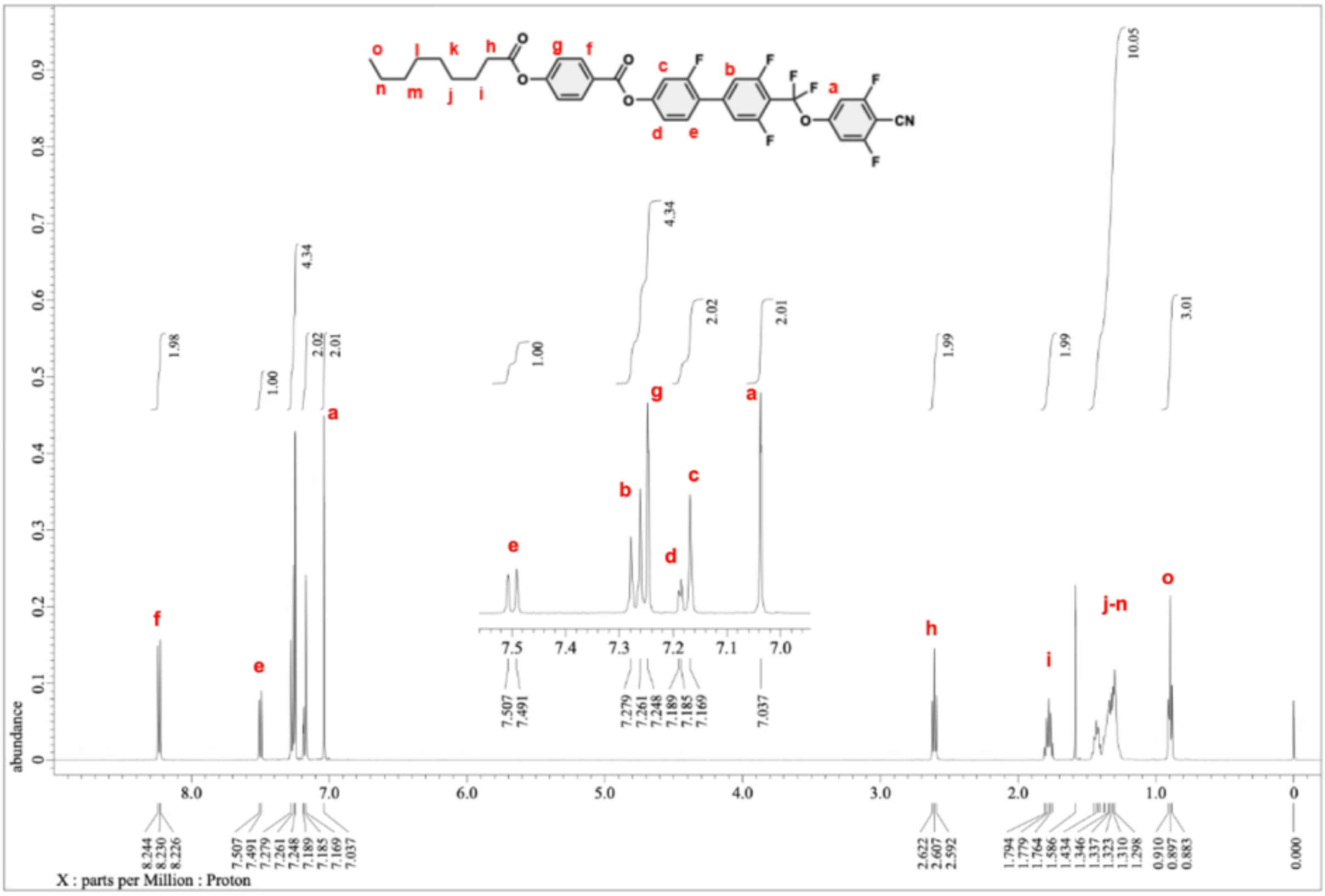

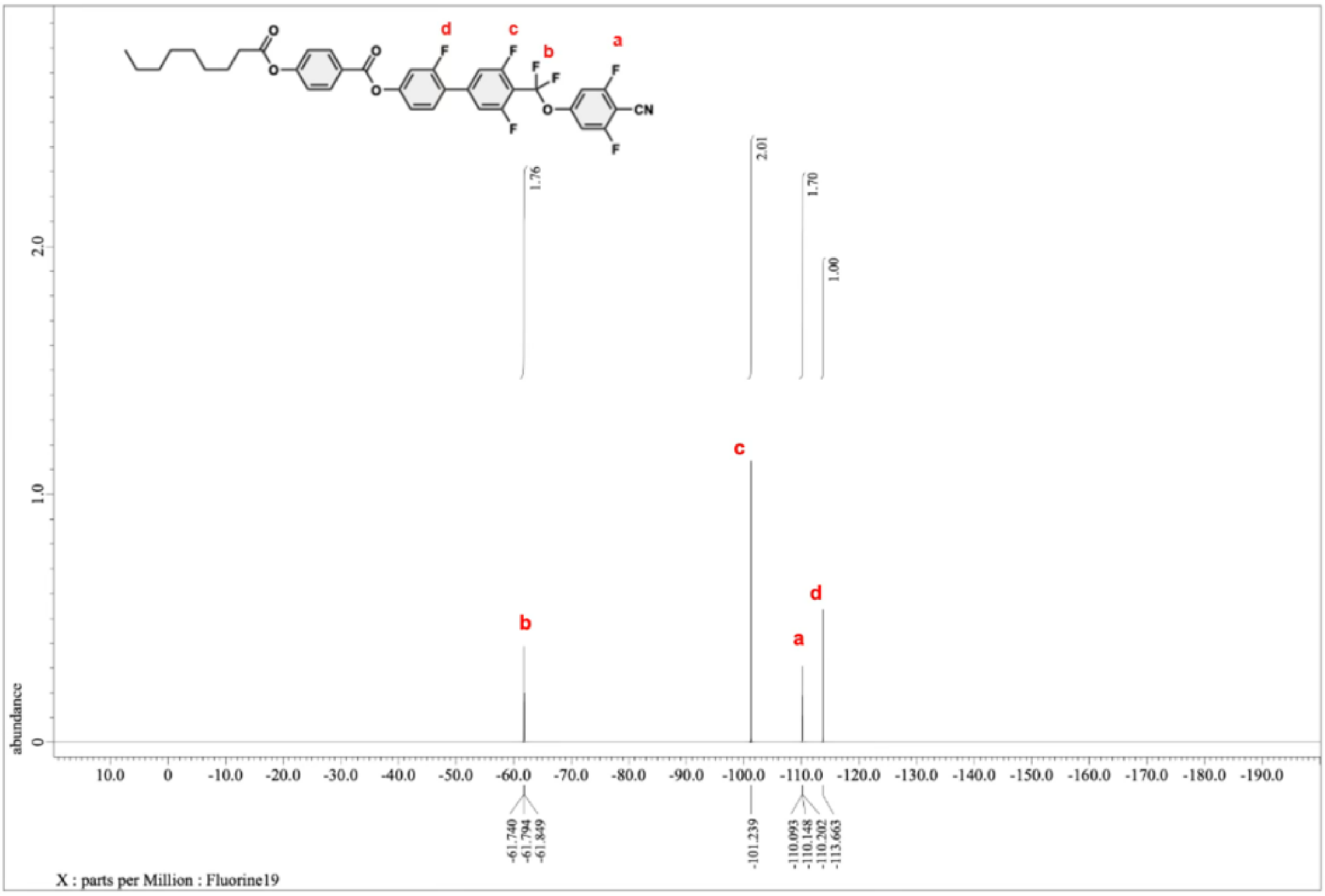

**Supplementary Fig. 80 |** $^{1}H\{^{19}F\}$ (upper) and $^{19}F\{^{1}H\}$ (bottom) NMR spectra for ***8*-012-CN**.

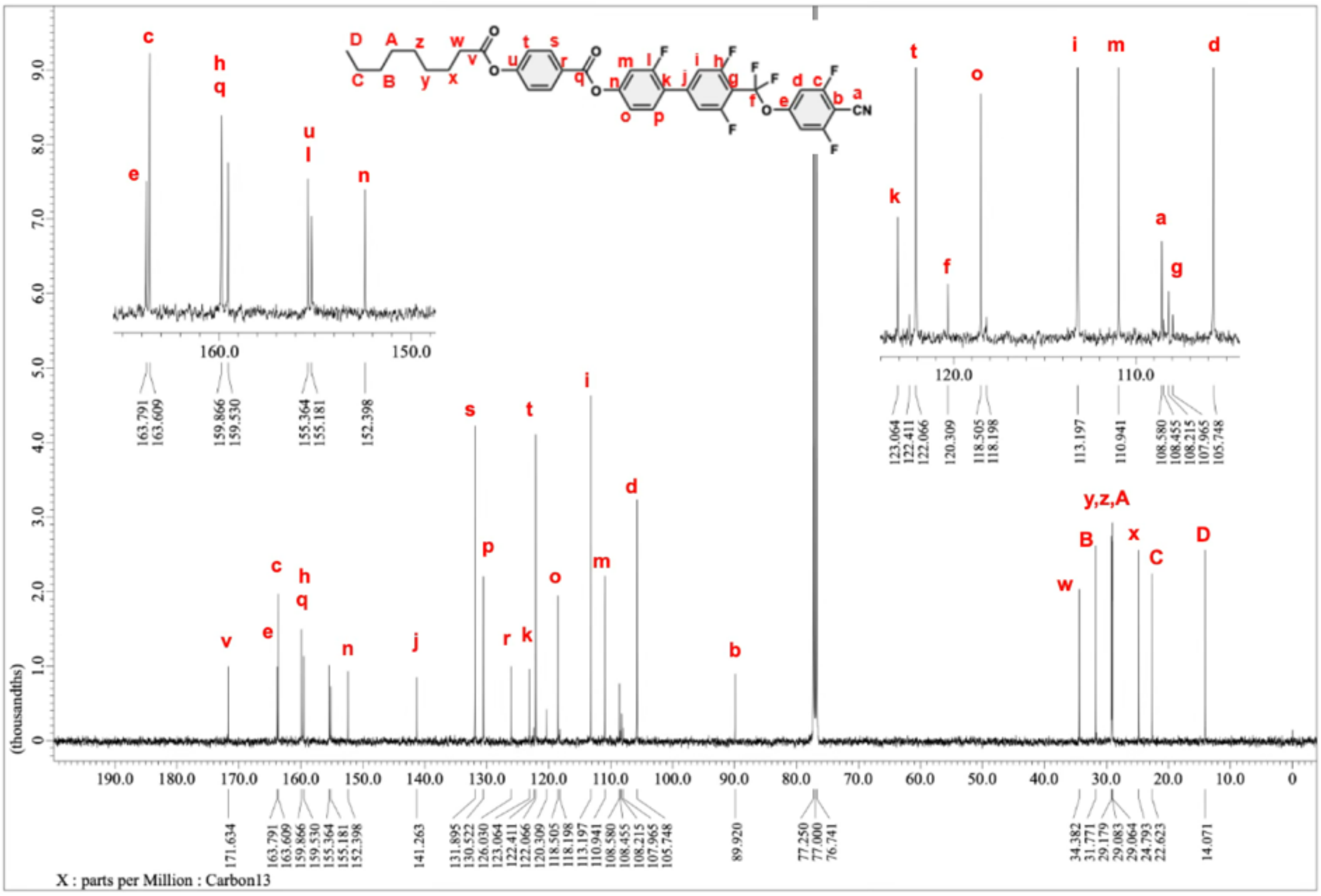

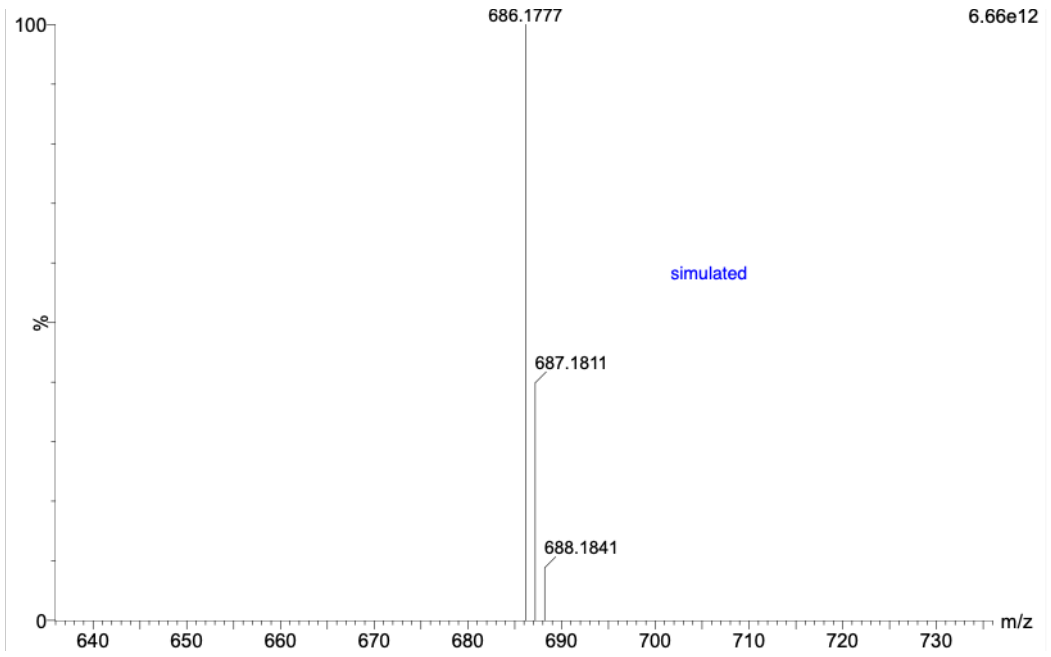

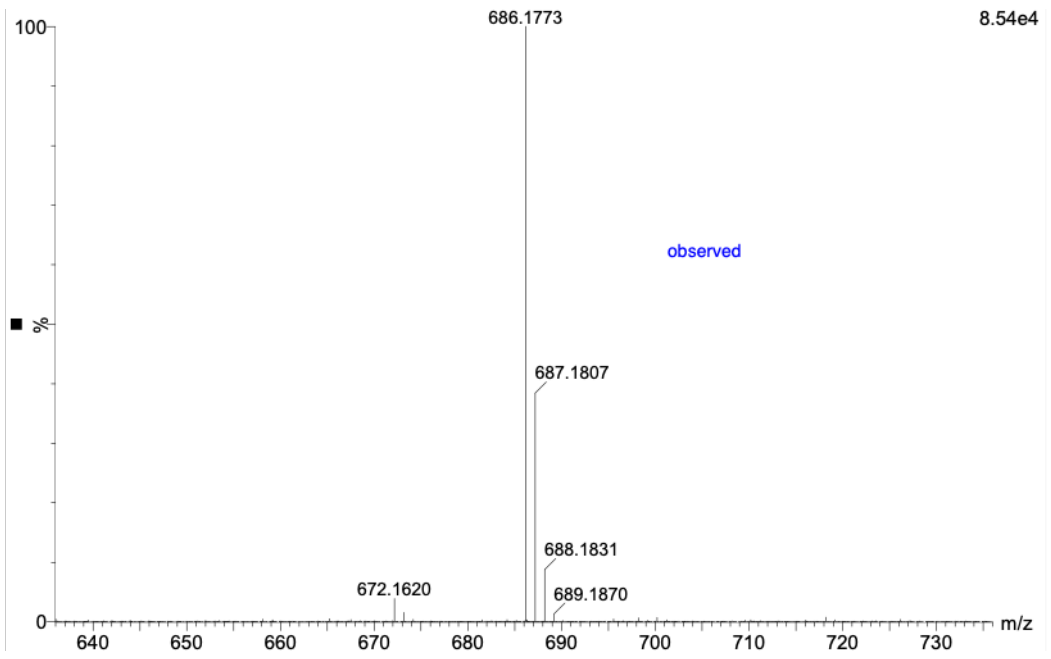

**Supplementary Fig. 81 |** $^{13}C\{1H,^{19}F\}$NMR (upper) and HR-ESI-TOF-MS spectra (bottom) for ***8*-012-CN**.

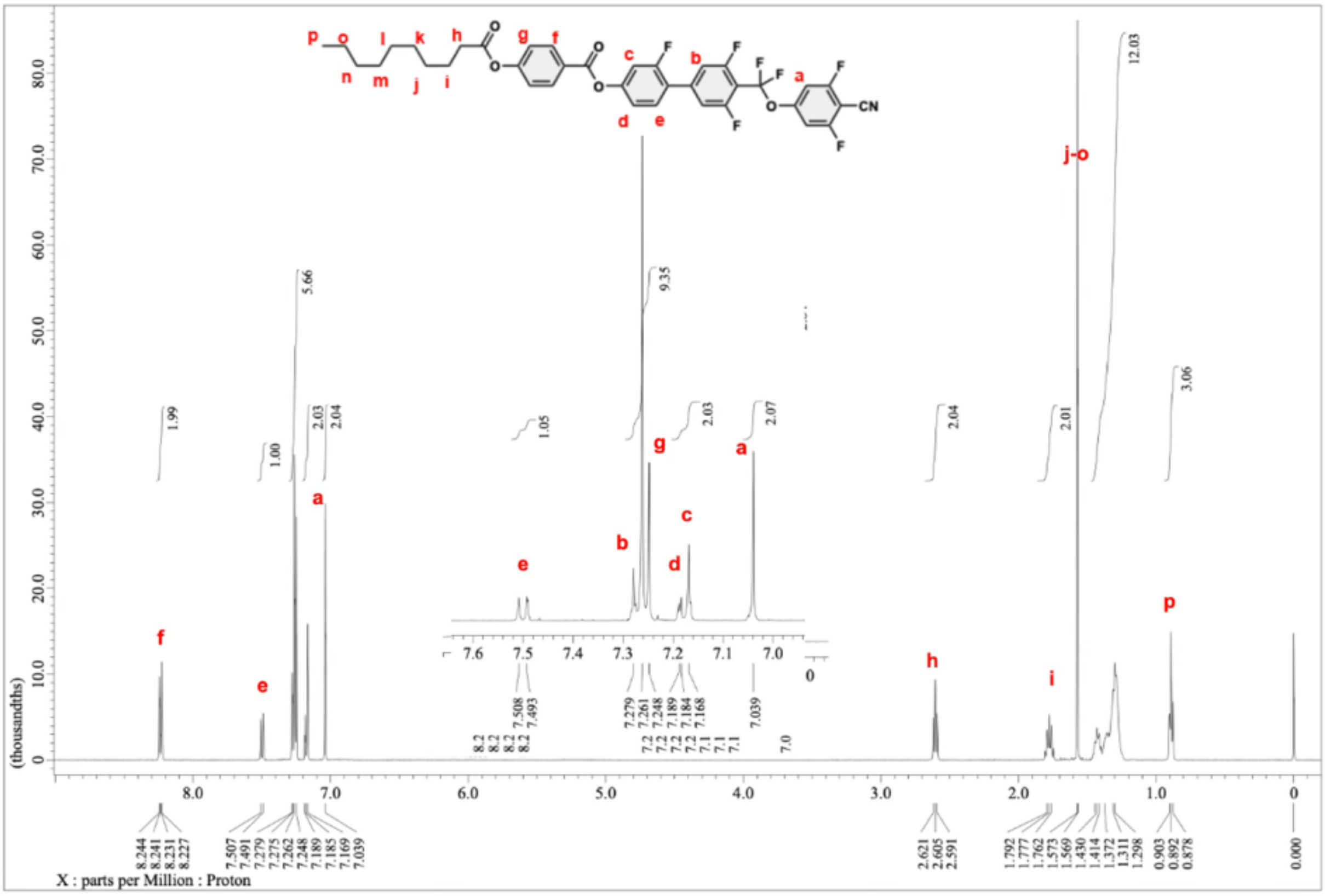

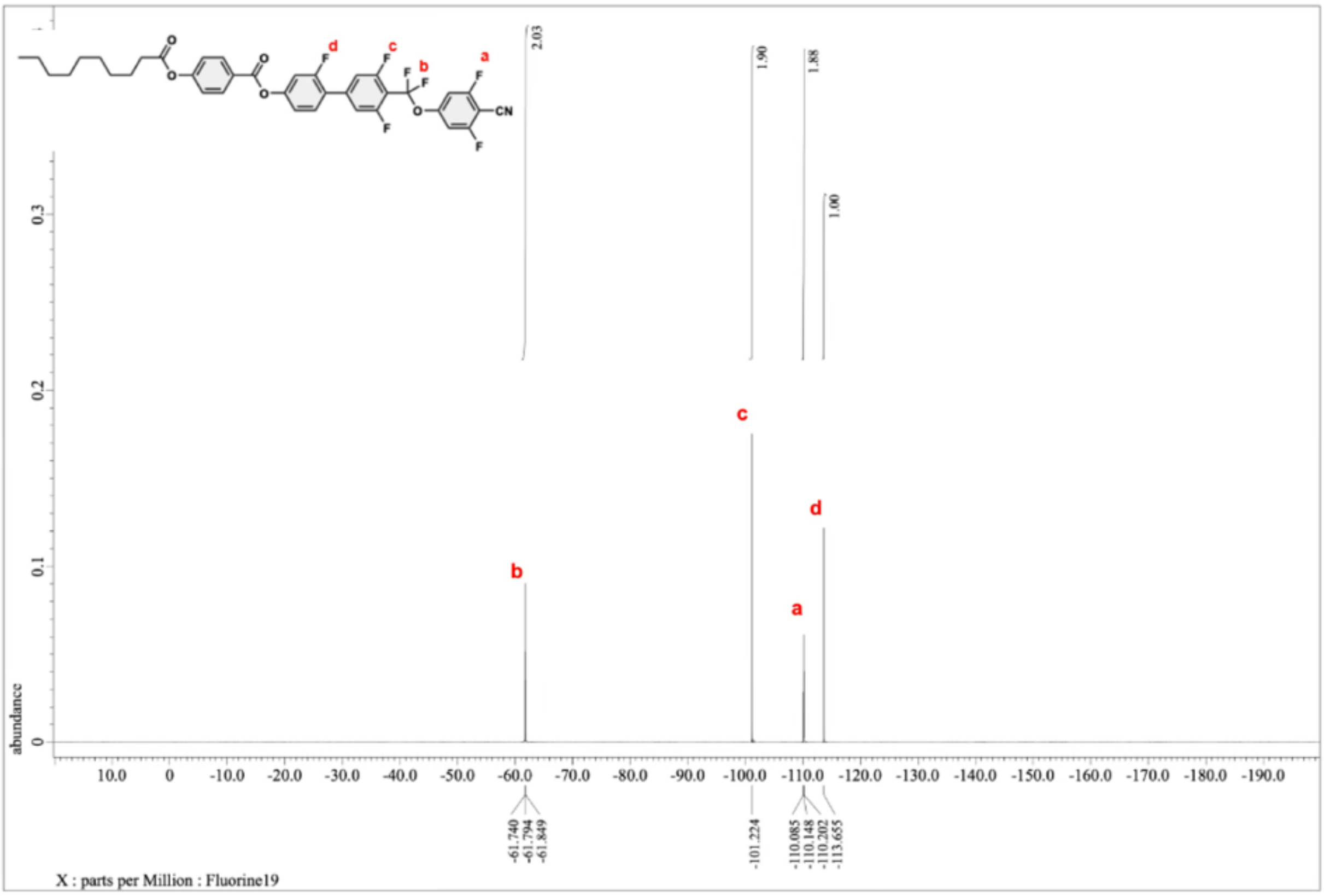

**Supplementary Fig. 82 |** $^{1}H\{^{19}F\}$ (upper) and $^{19}F\{^{1}H\}$ (bottom) NMR spectra for ***9*-012-CN**.

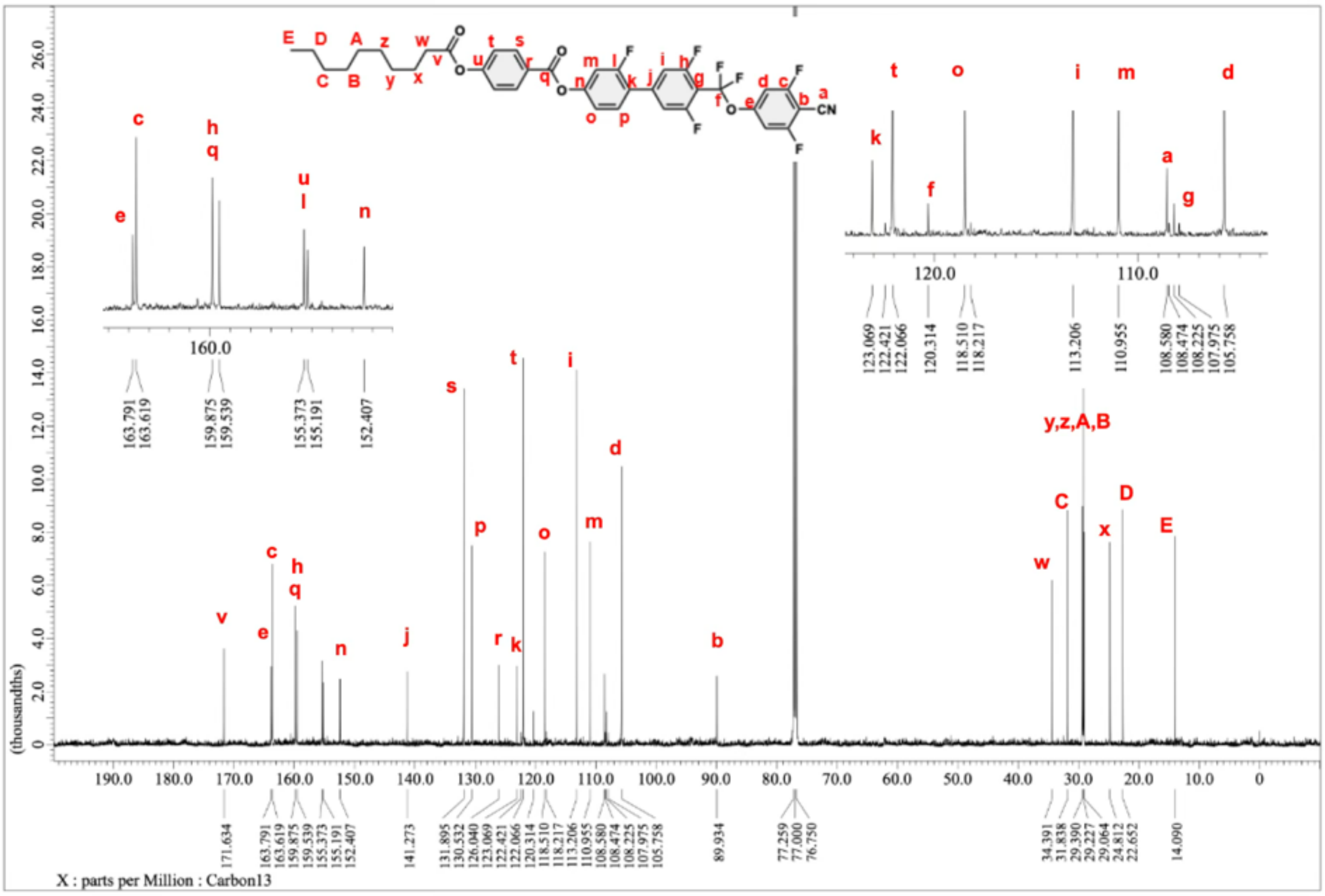

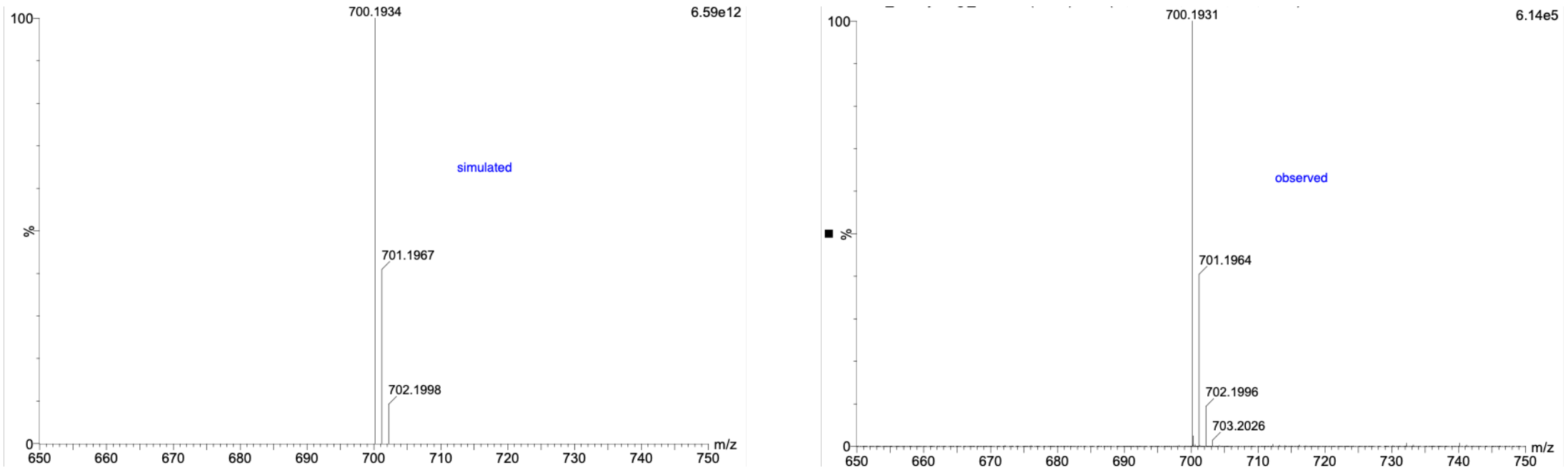

**Supplementary Fig. 83 |** $^{13}C\{1H,^{19}F\}$NMR (upper) and HR-ESI-TOF-MS spectra (bottom) for ***9*-012-CN**.

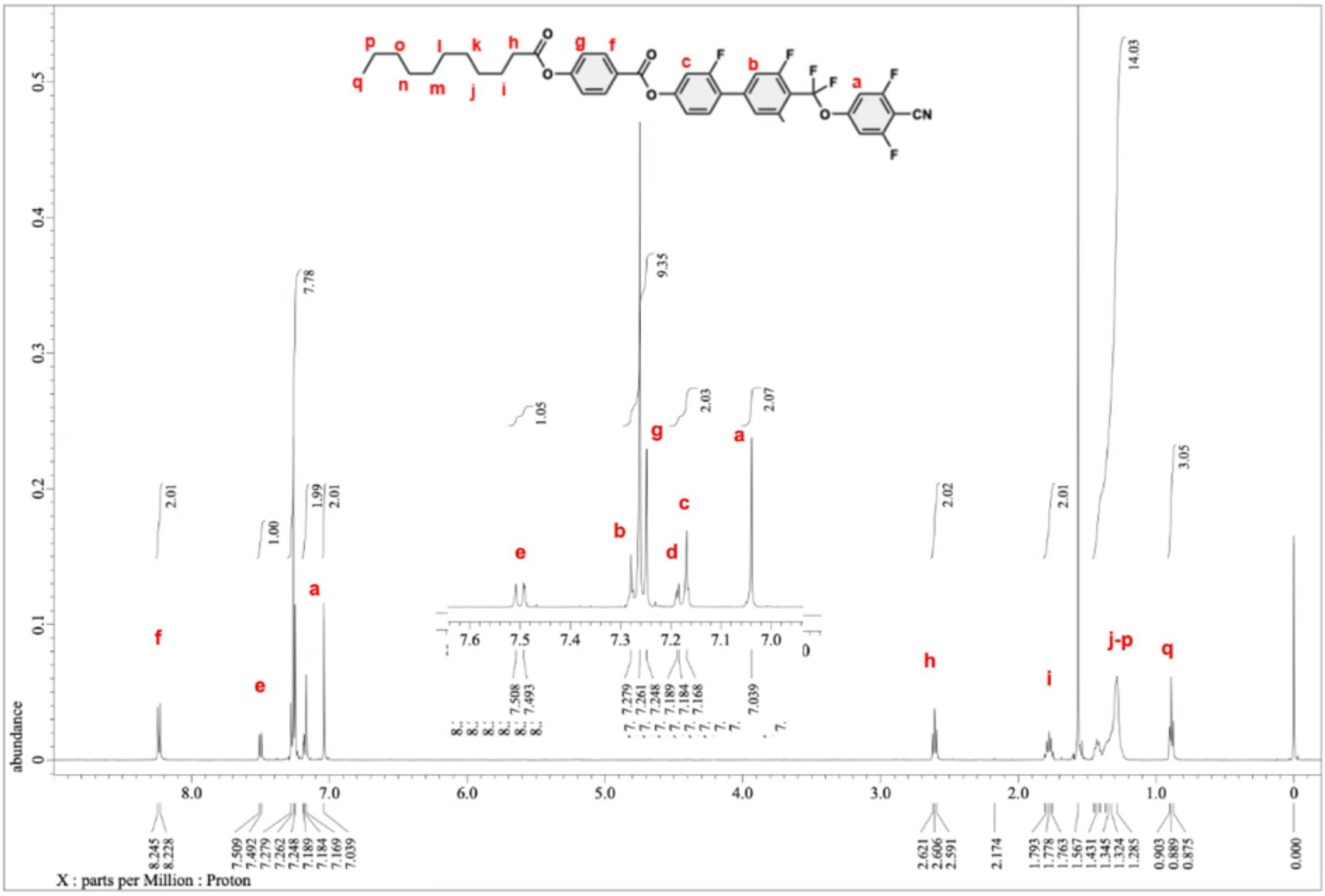

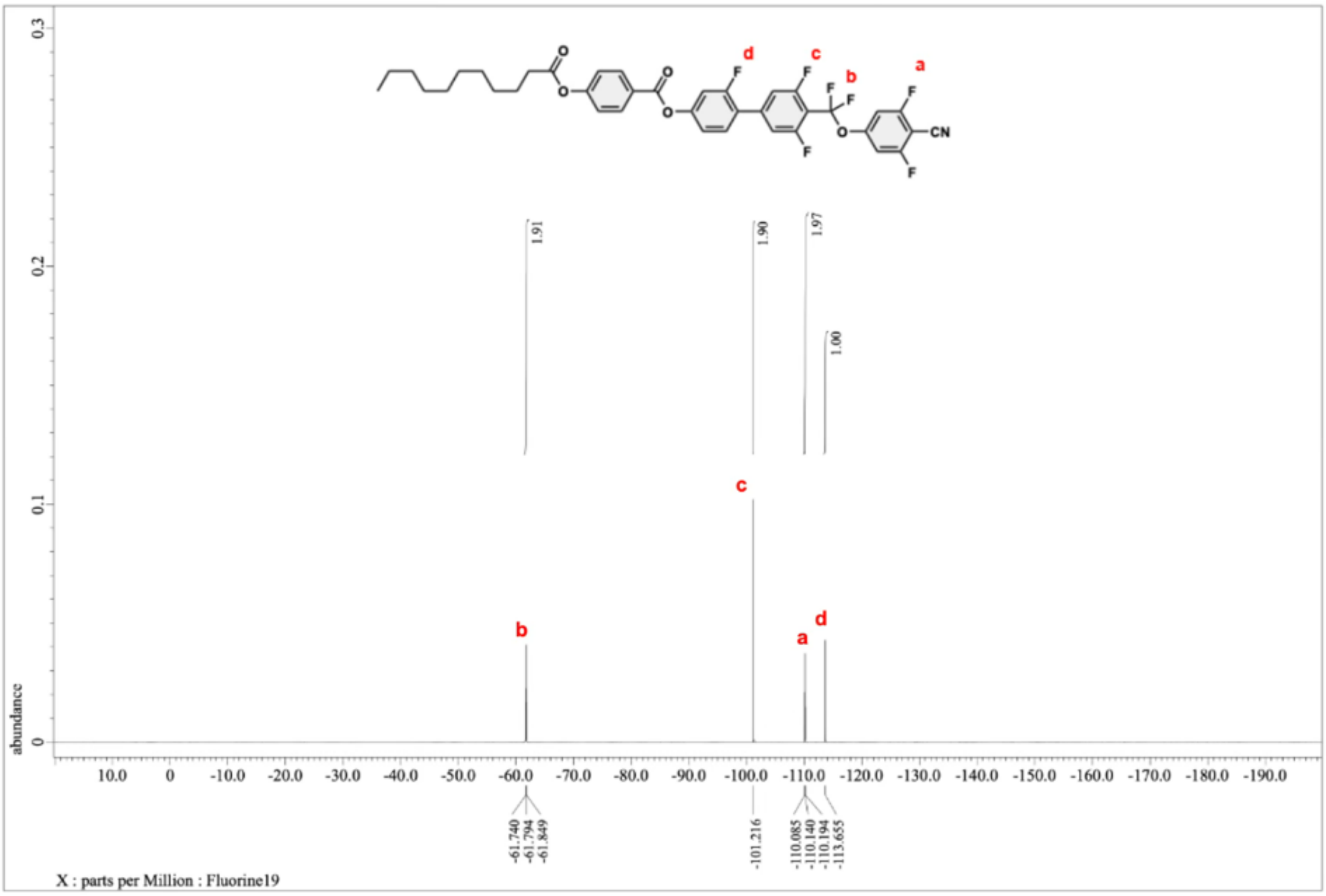

**Supplementary Fig. 84 |** $^{1}H\{^{19}F\}$ (upper) and $^{19}F\{^{1}H\}$ (bottom) NMR spectra for ***10*-012-CN**.

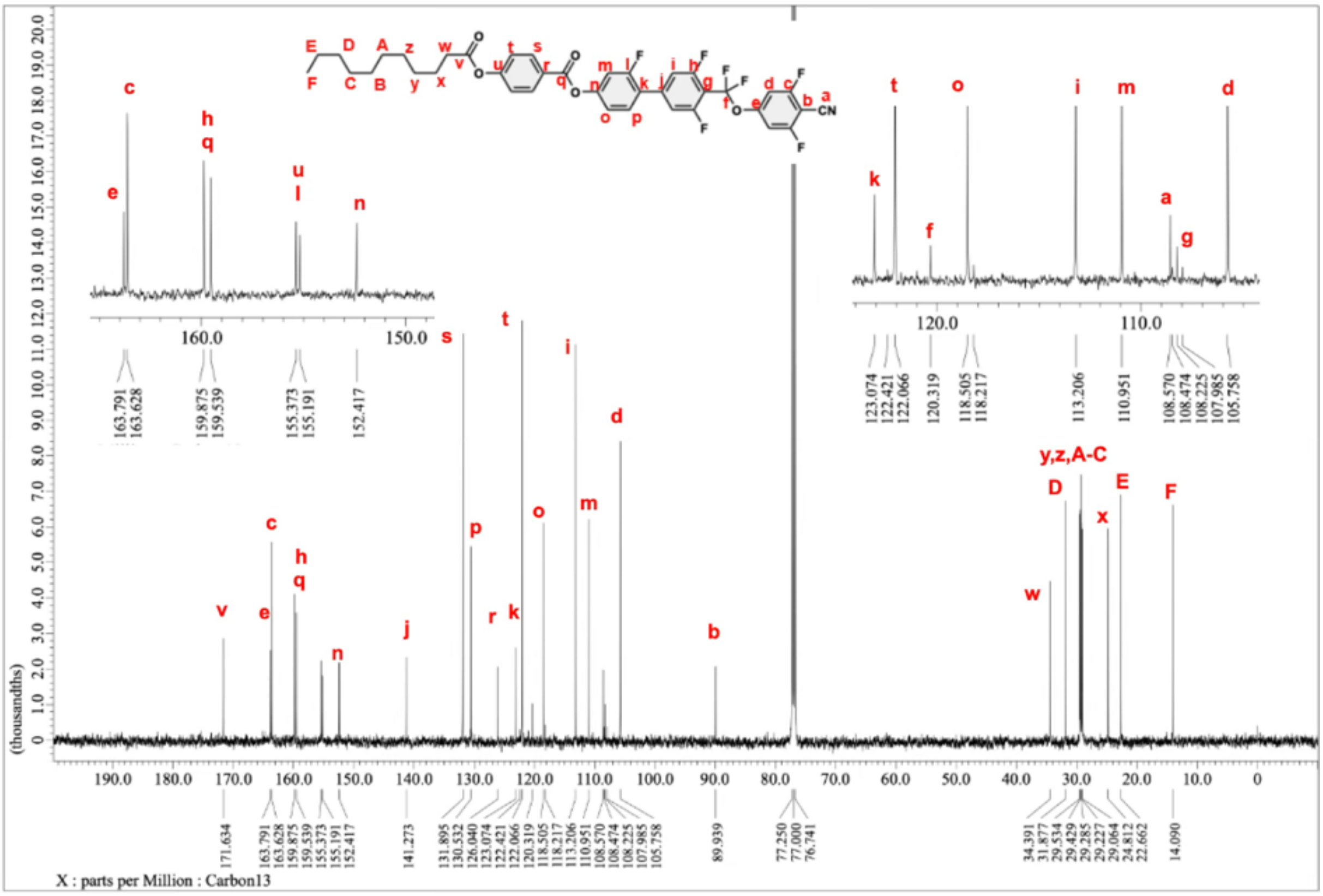

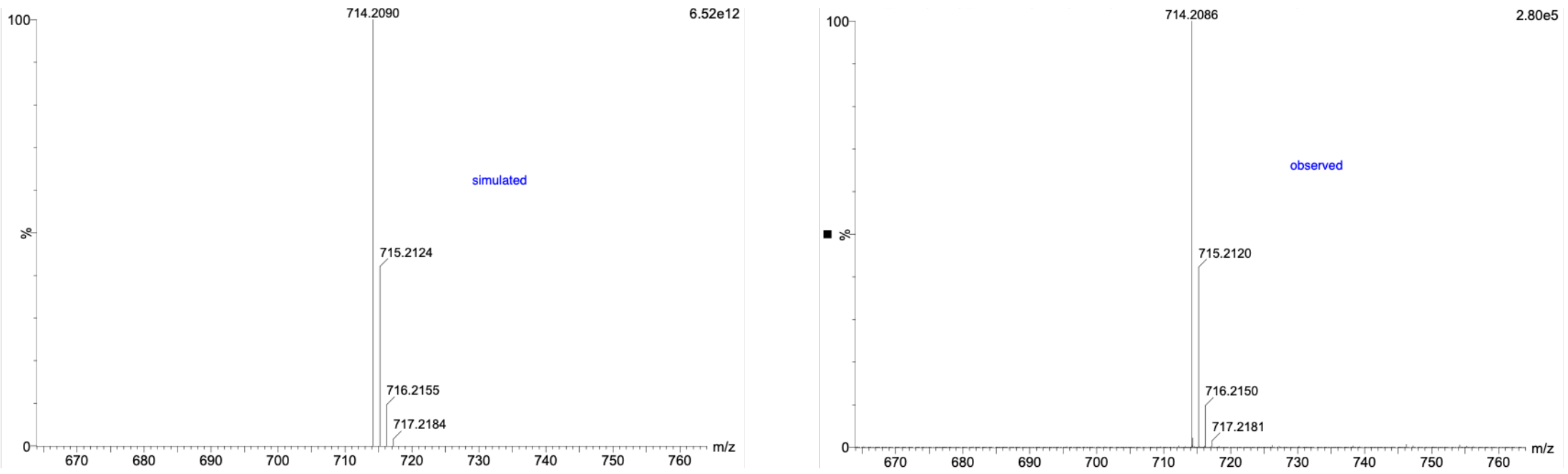

**Supplementary Fig. 85 |** $^{13}C\{^1H,^{19}F\}$NMR (upper) and HR-ESI-TOF-MS spectra (bottom) for ***10*-012-CN**.

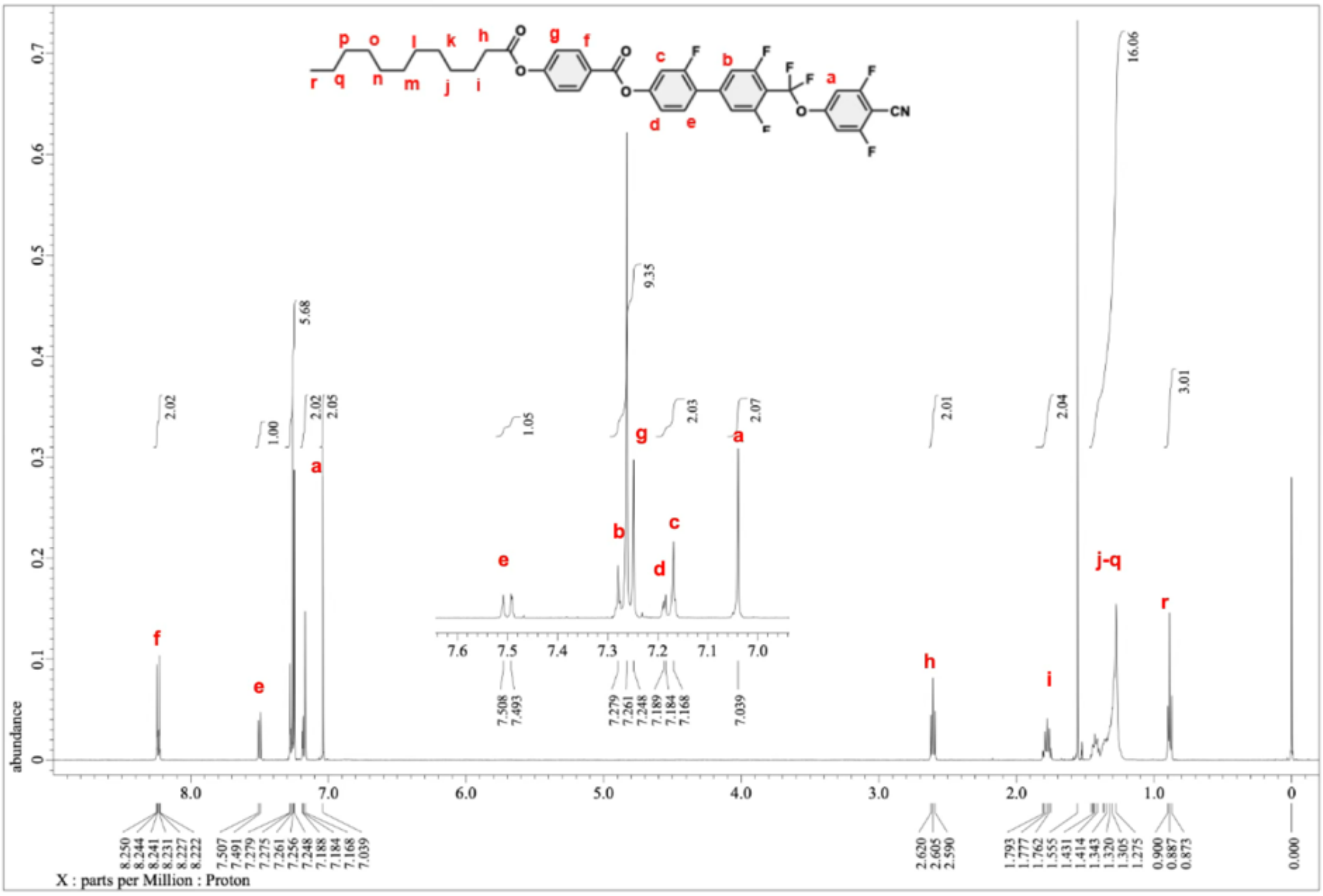

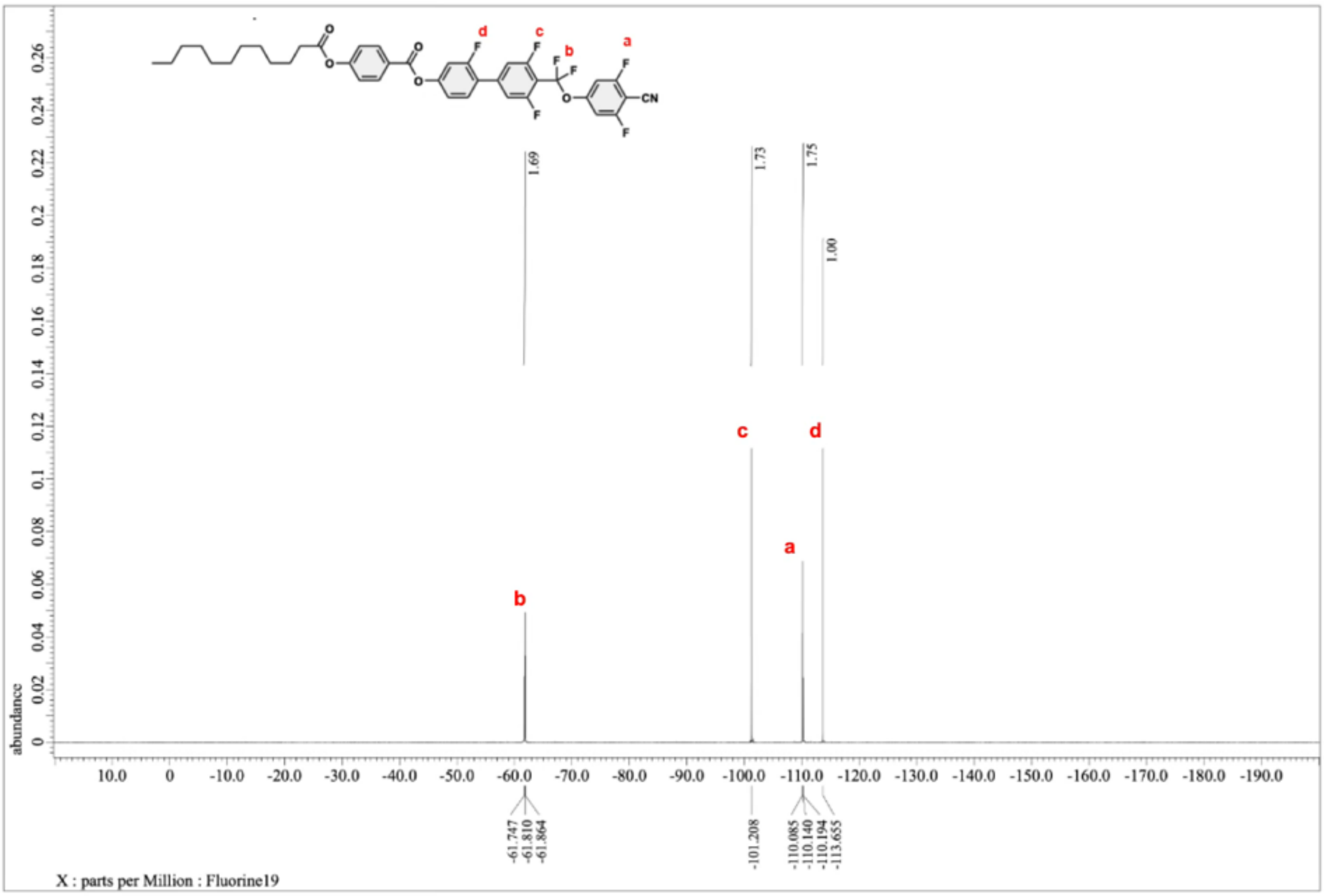

**Supplementary Fig. 86 |** $^{1}H\{^{19}F\}$ (upper) and $^{19}F\{^{1}H\}$ (bottom) NMR spectra for ***11*-012-CN**.

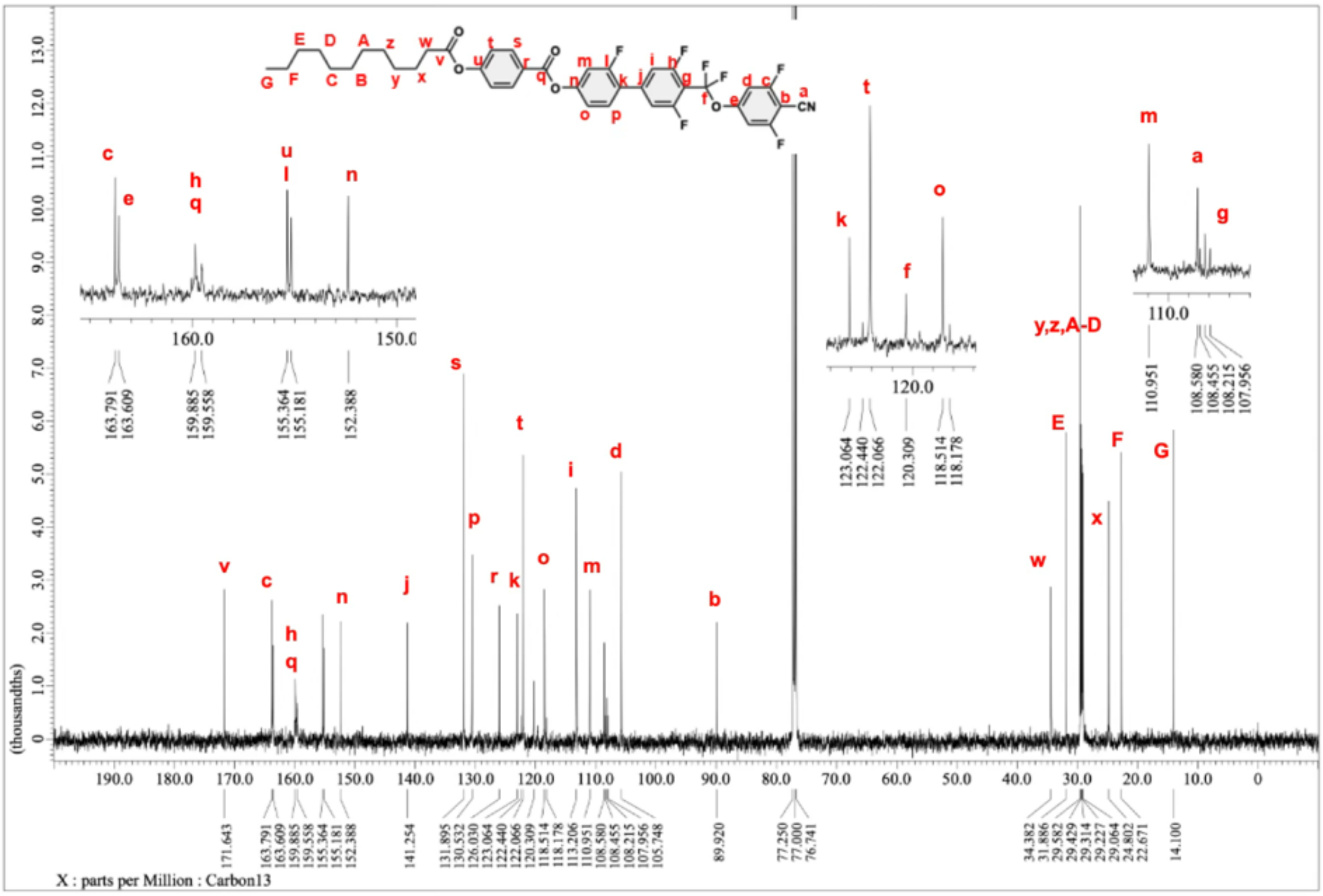

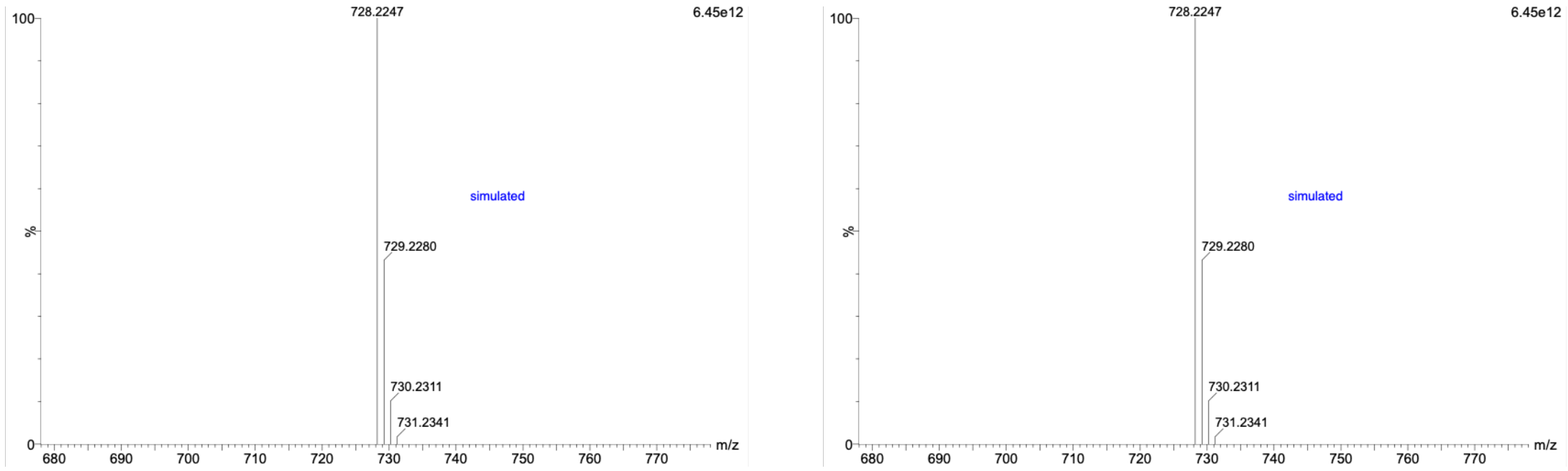

Supplementary Fig. 87 | $^{13}C\{^1H,^{19}F\}$NMR (upper) and HR-ESI-TOF-MS spectra (bottom) for ***11*-012-CN**.

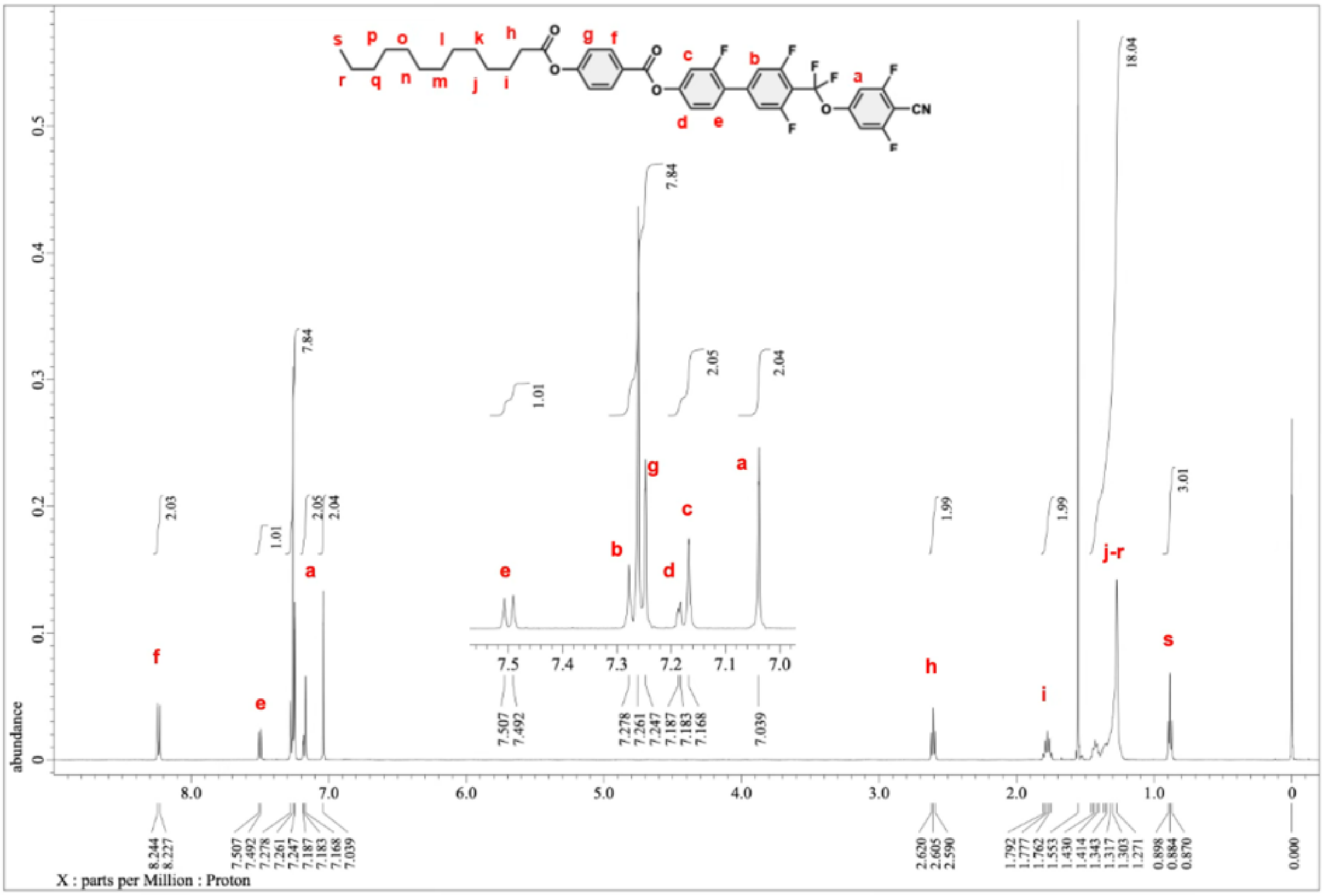

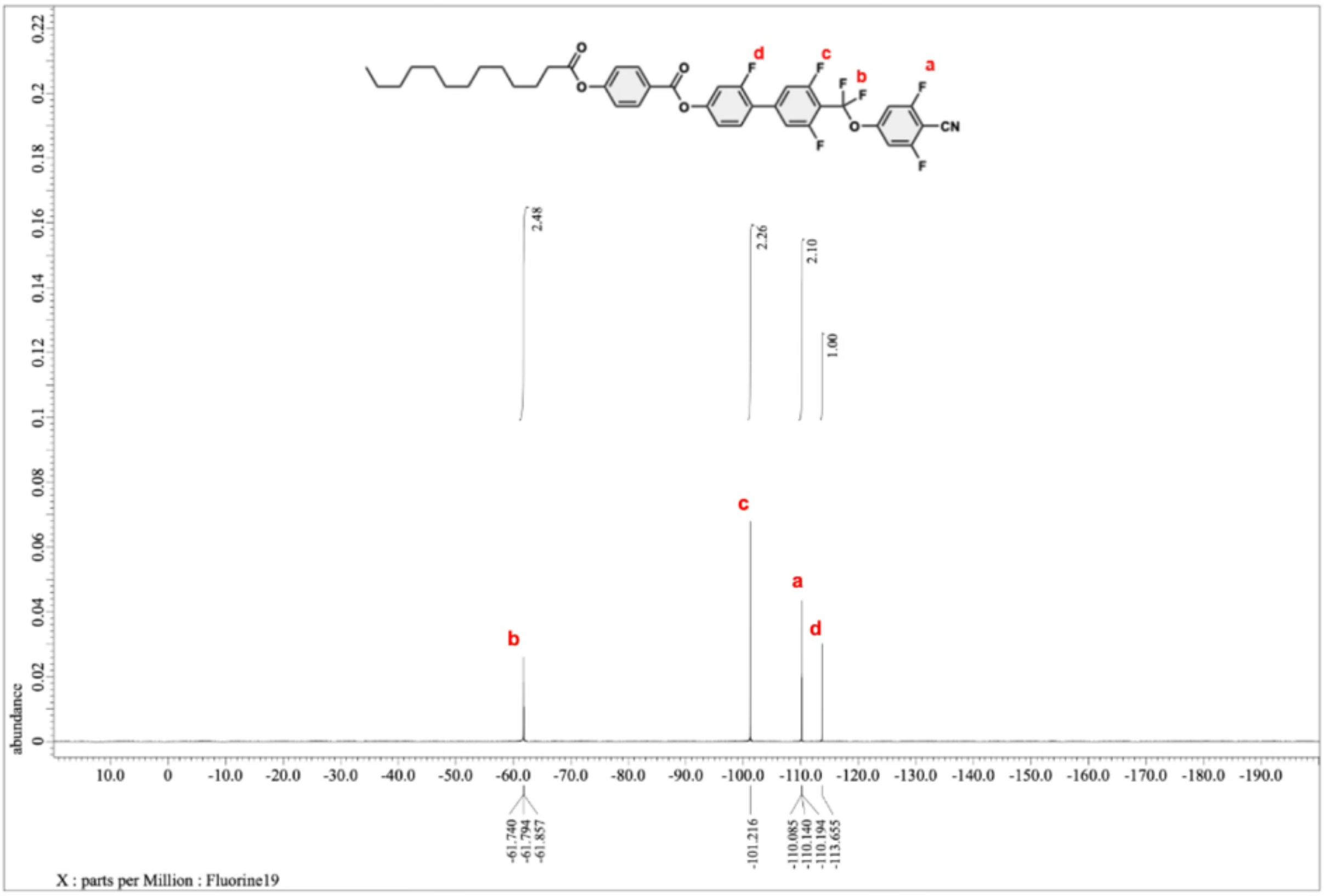

**Supplementary Fig. 88 |** $^{1}H\{^{19}F\}$ (upper) and $^{19}F\{^{1}H\}$ (bottom) NMR spectra for ***12*-012-CN**.

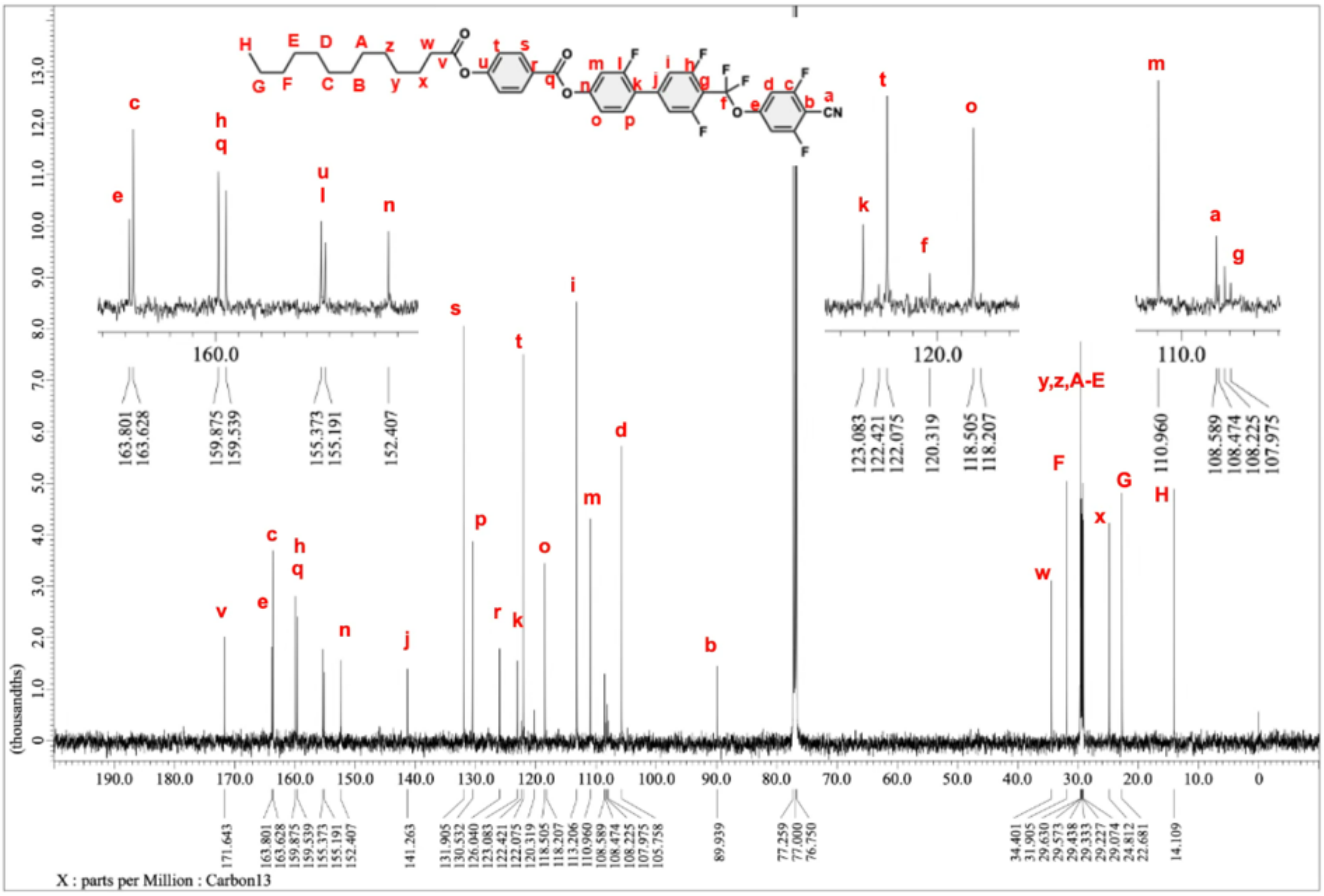

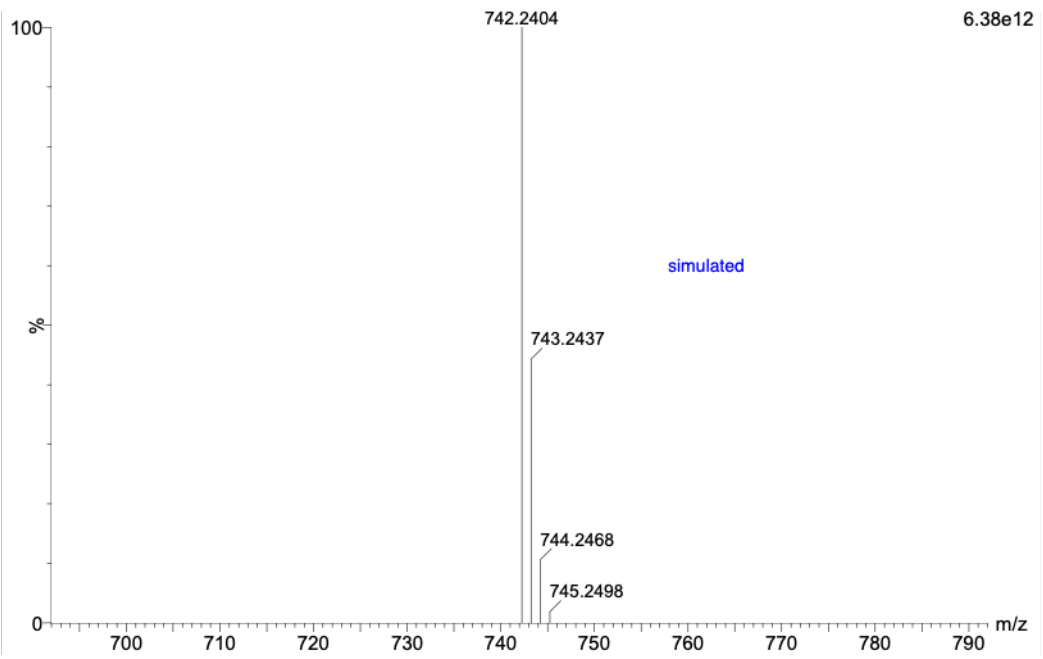

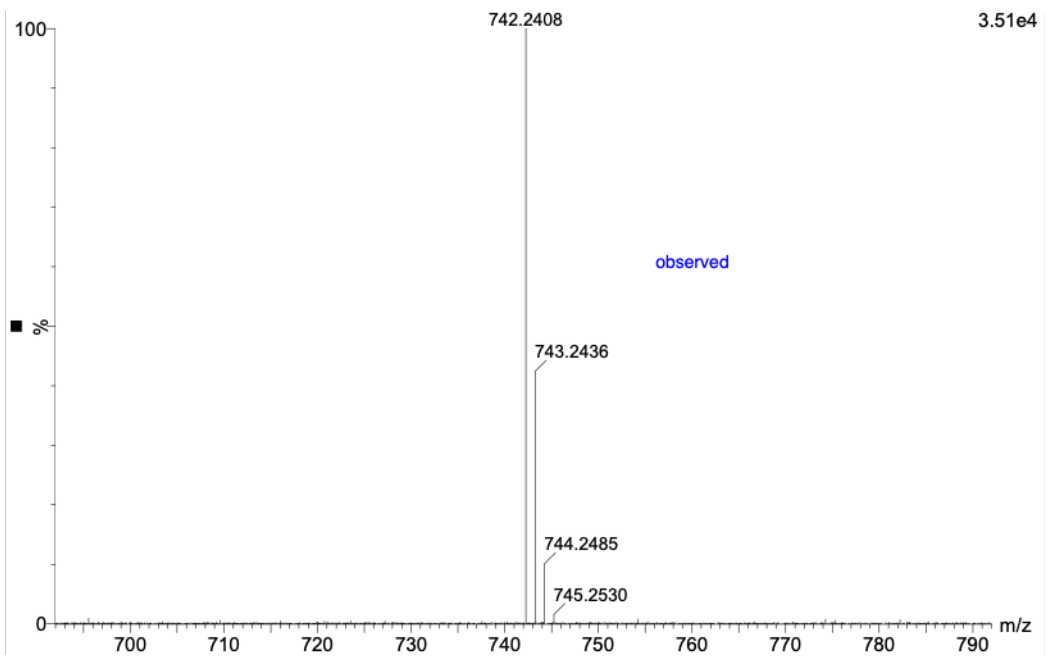

**Supplementary Fig. 89 |** $^{13}C\{^1H,^{19}F\}$NMR (upper) and HR-ESI-TOF-MS spectra (bottom) for ***12*-012-CN**.

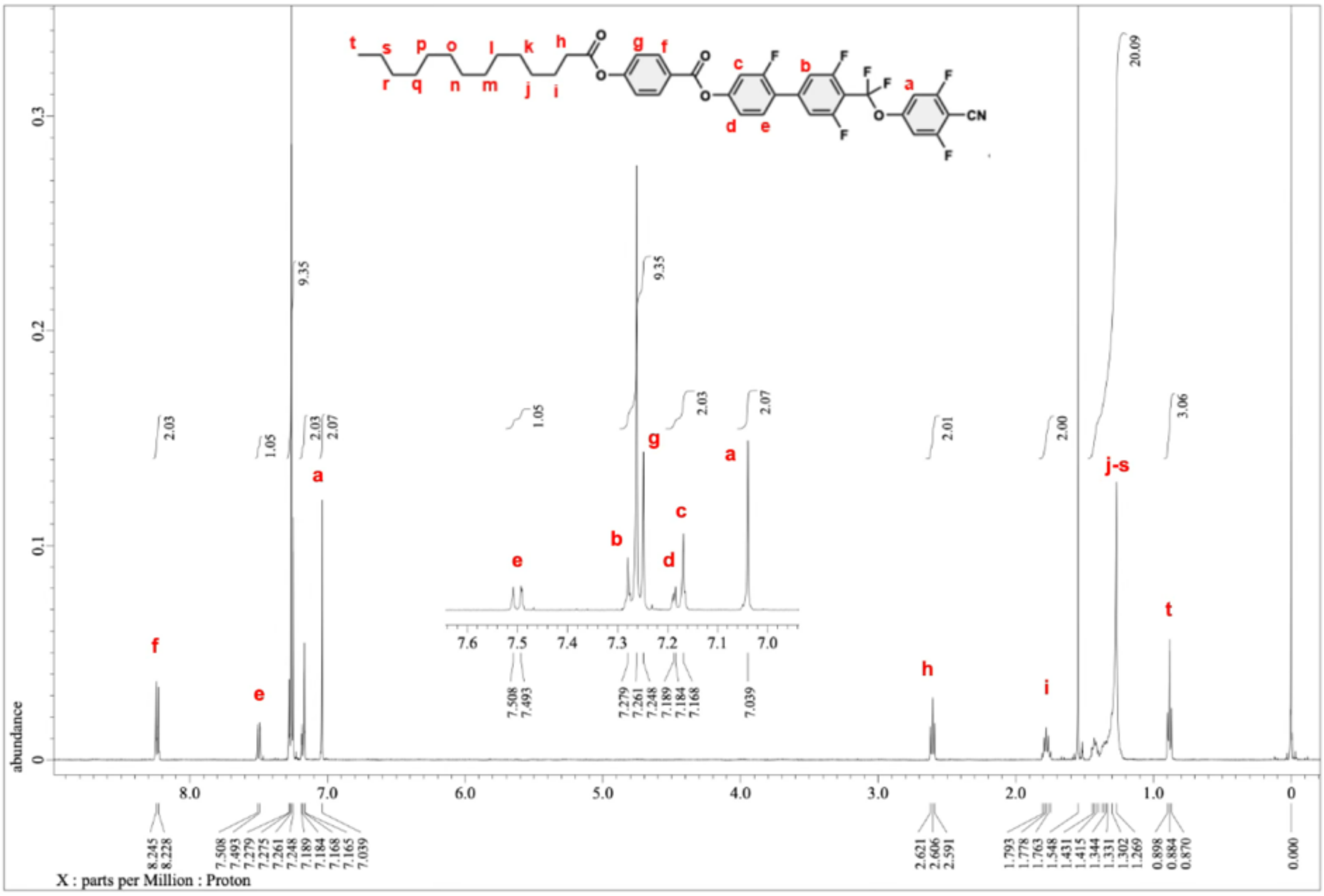

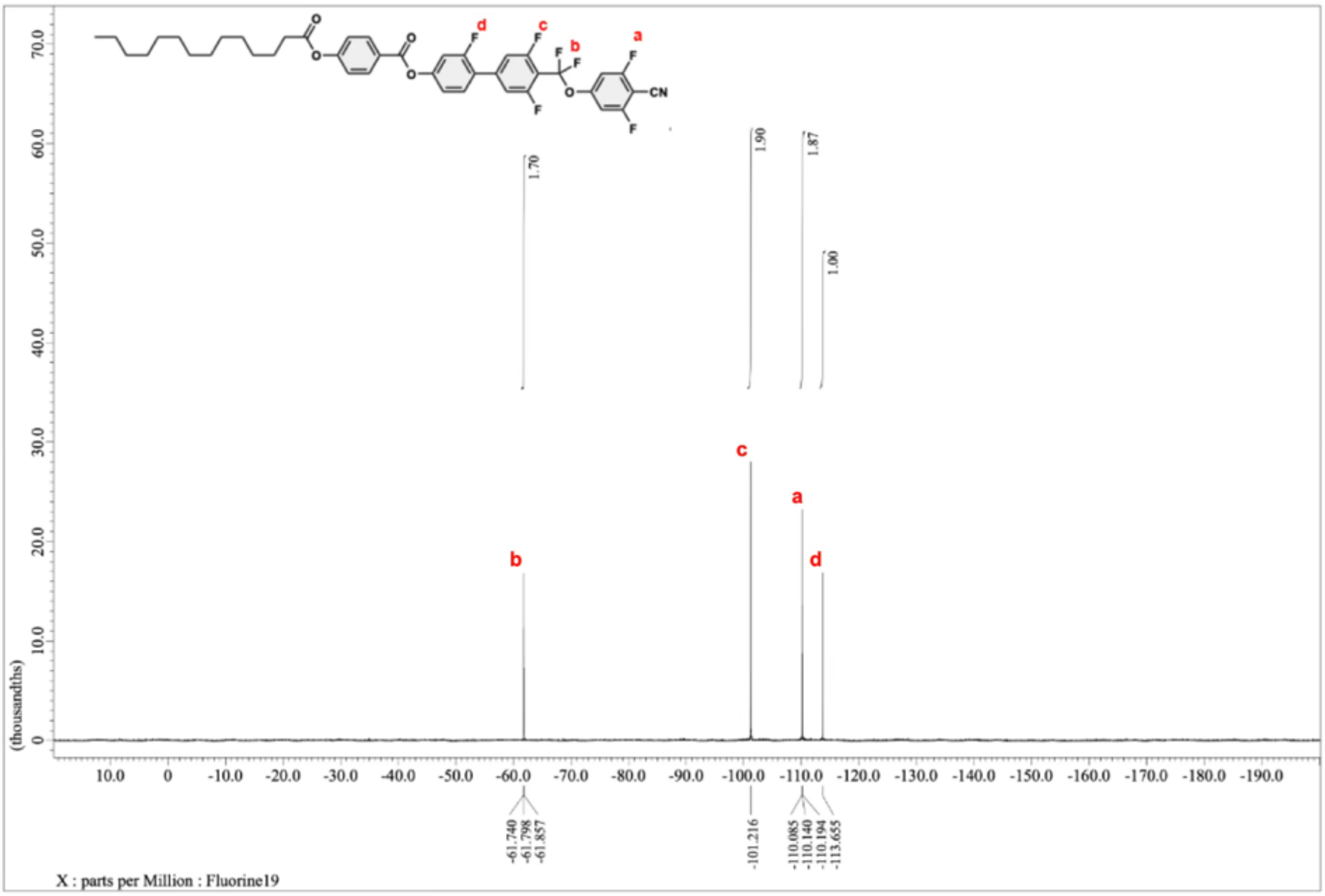

**Supplementary Fig. 90 |** $^{1}H\{^{19}F\}$ (upper) and $^{19}F\{^{1}H\}$ (bottom) NMR spectra for ***13*-012-CN**.

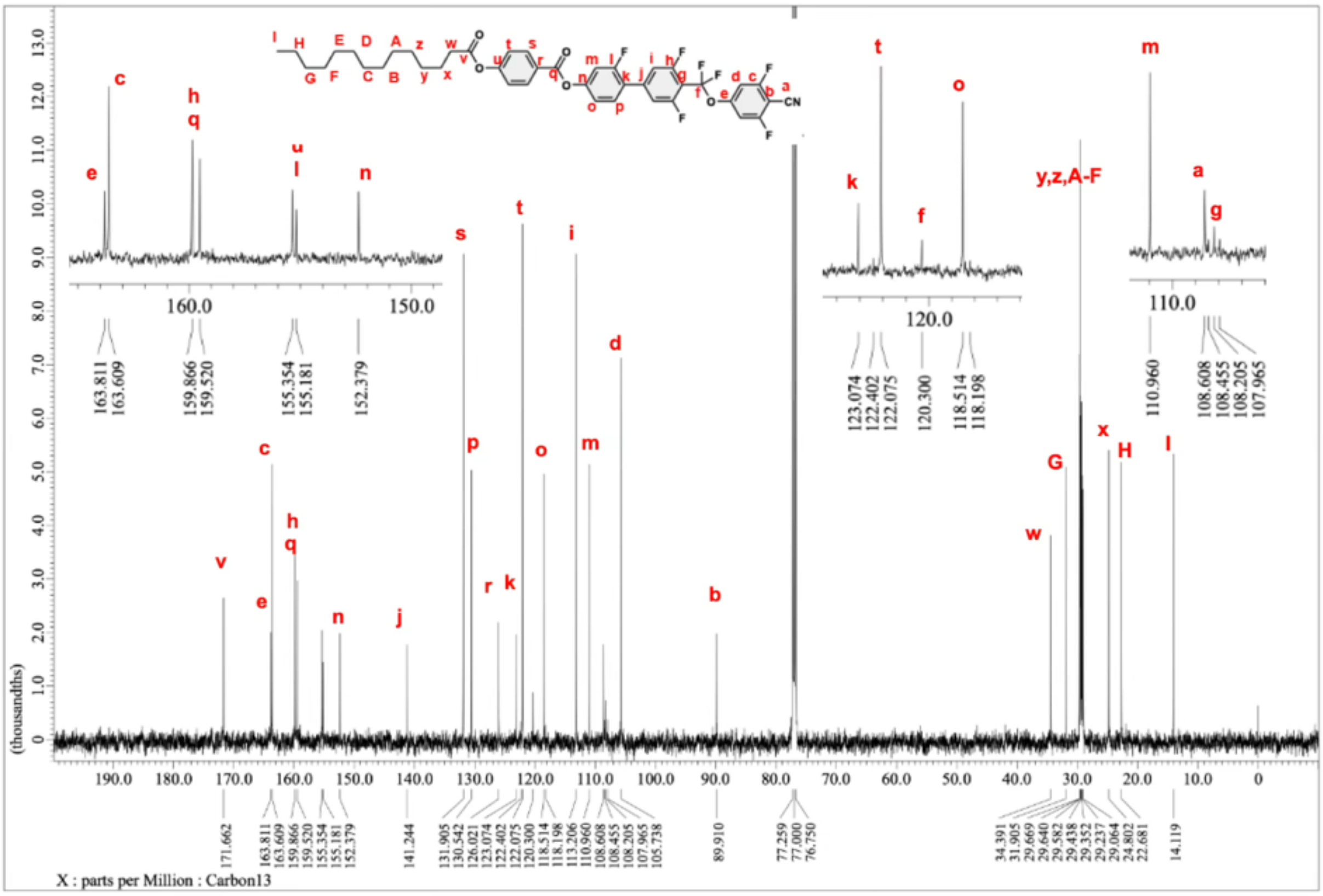

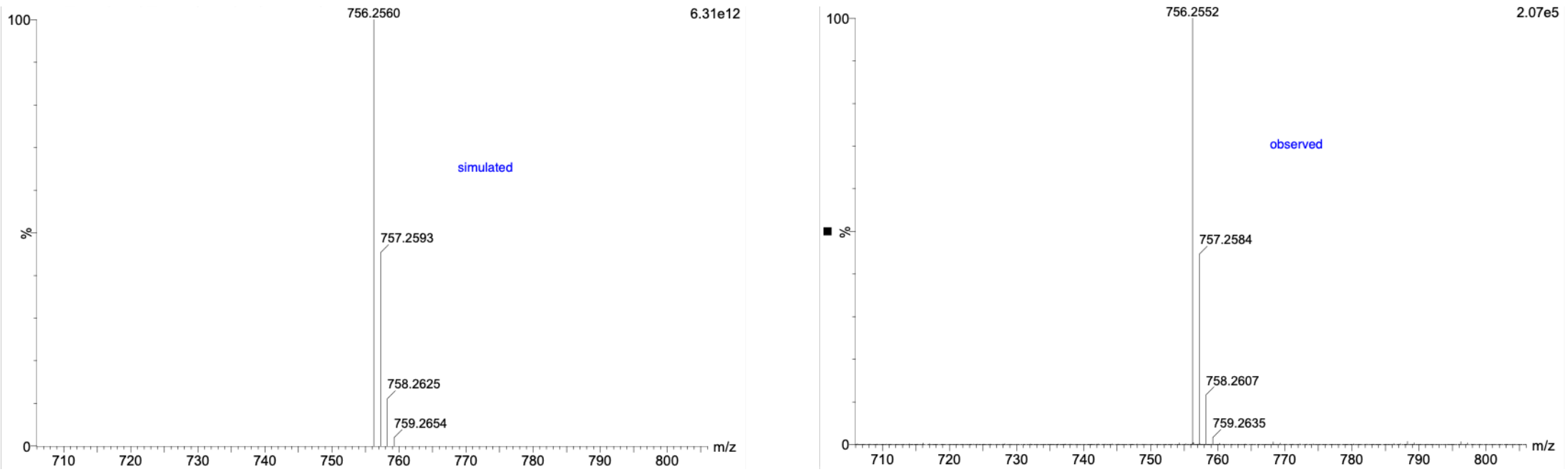

**Supplementary Fig. 91 |** $^{13}C\{1H,^{19}F\}$NMR (upper) and HR-ESI-TOF-MS spectra (bottom) for ***13*-012-CN**.

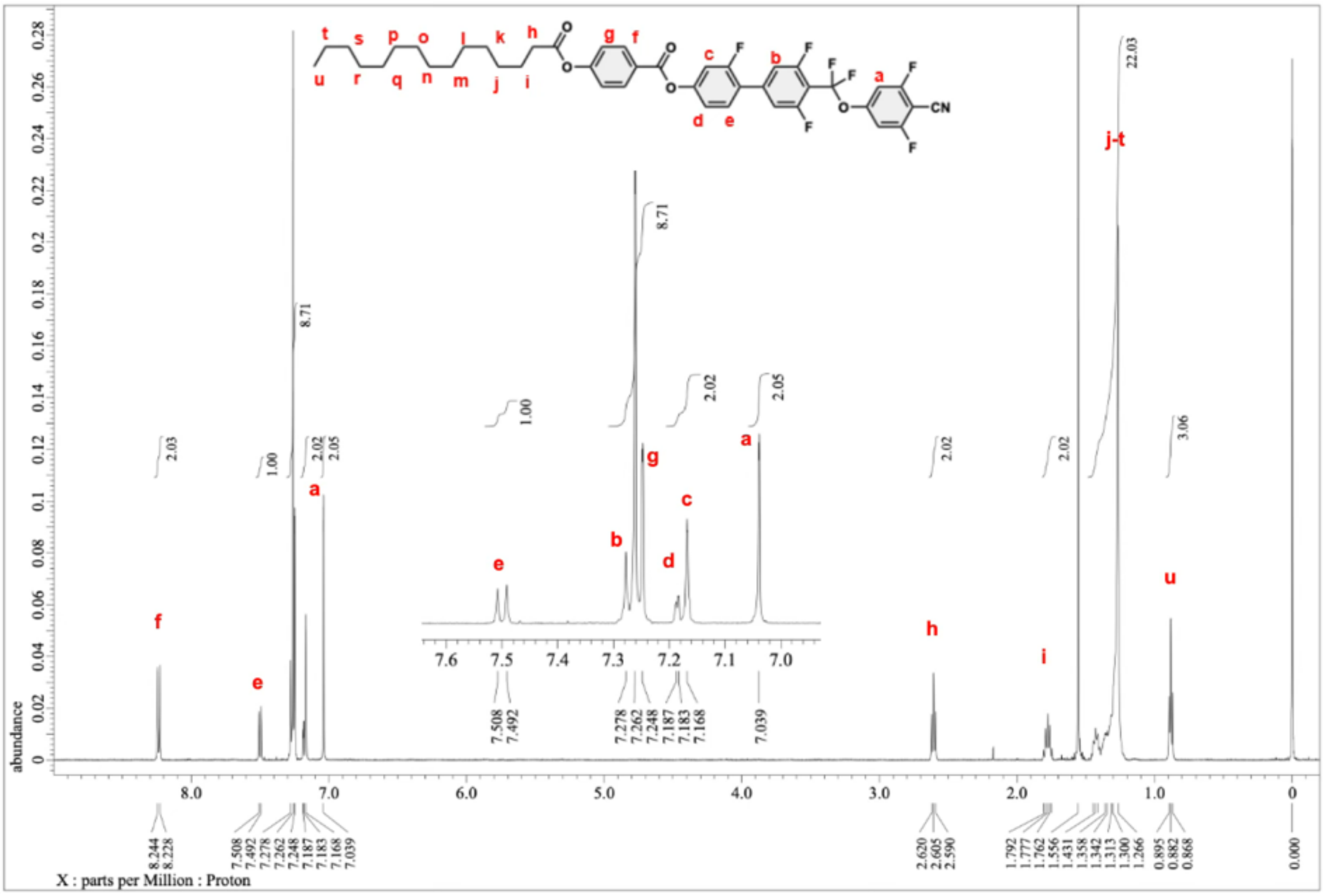

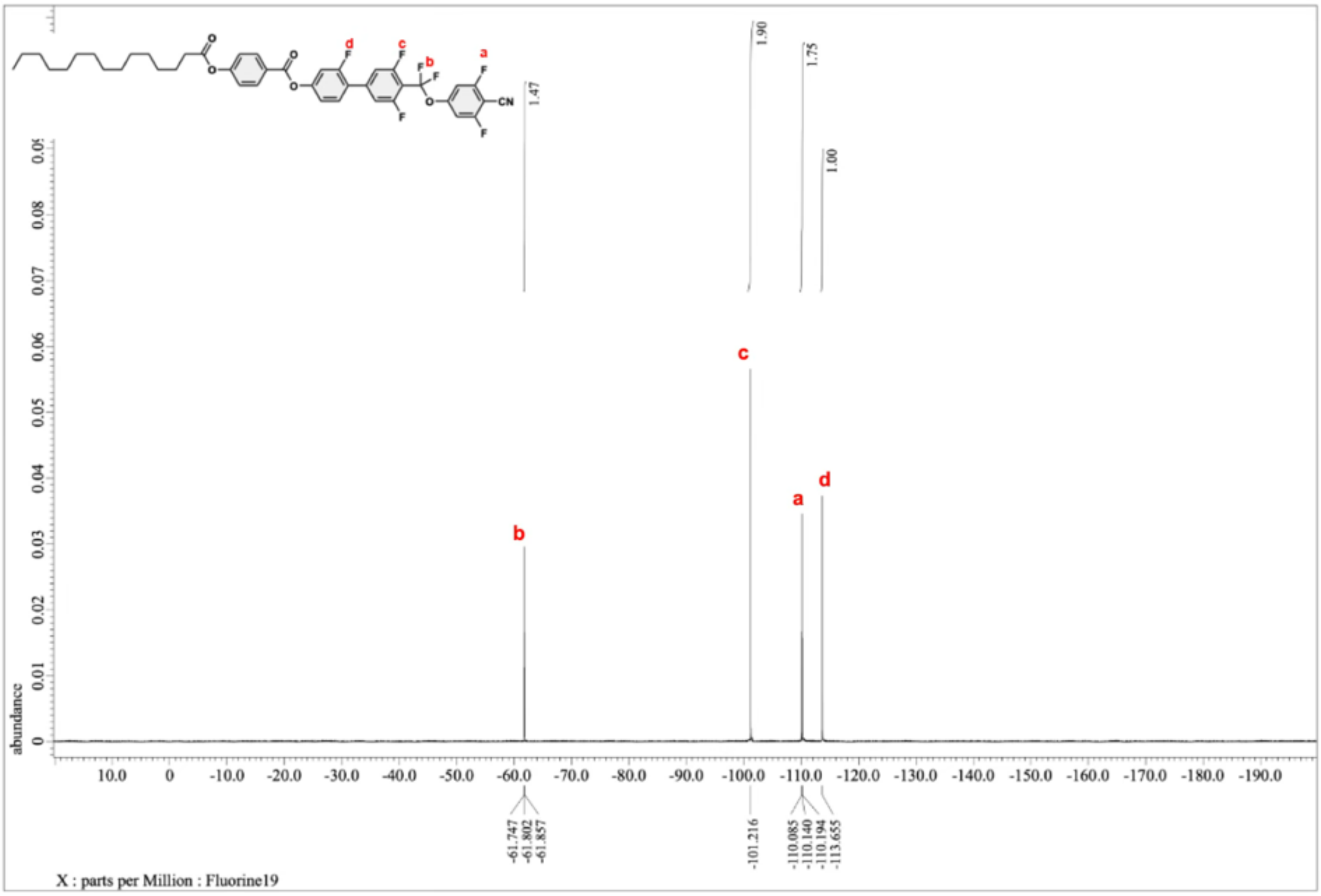

**Supplementary Fig. 92 |** $^{1}H\{^{19}F\}$ (upper) and $^{19}F\{^{1}H\}$ (bottom) NMR spectra for ***14*-012-CN**.

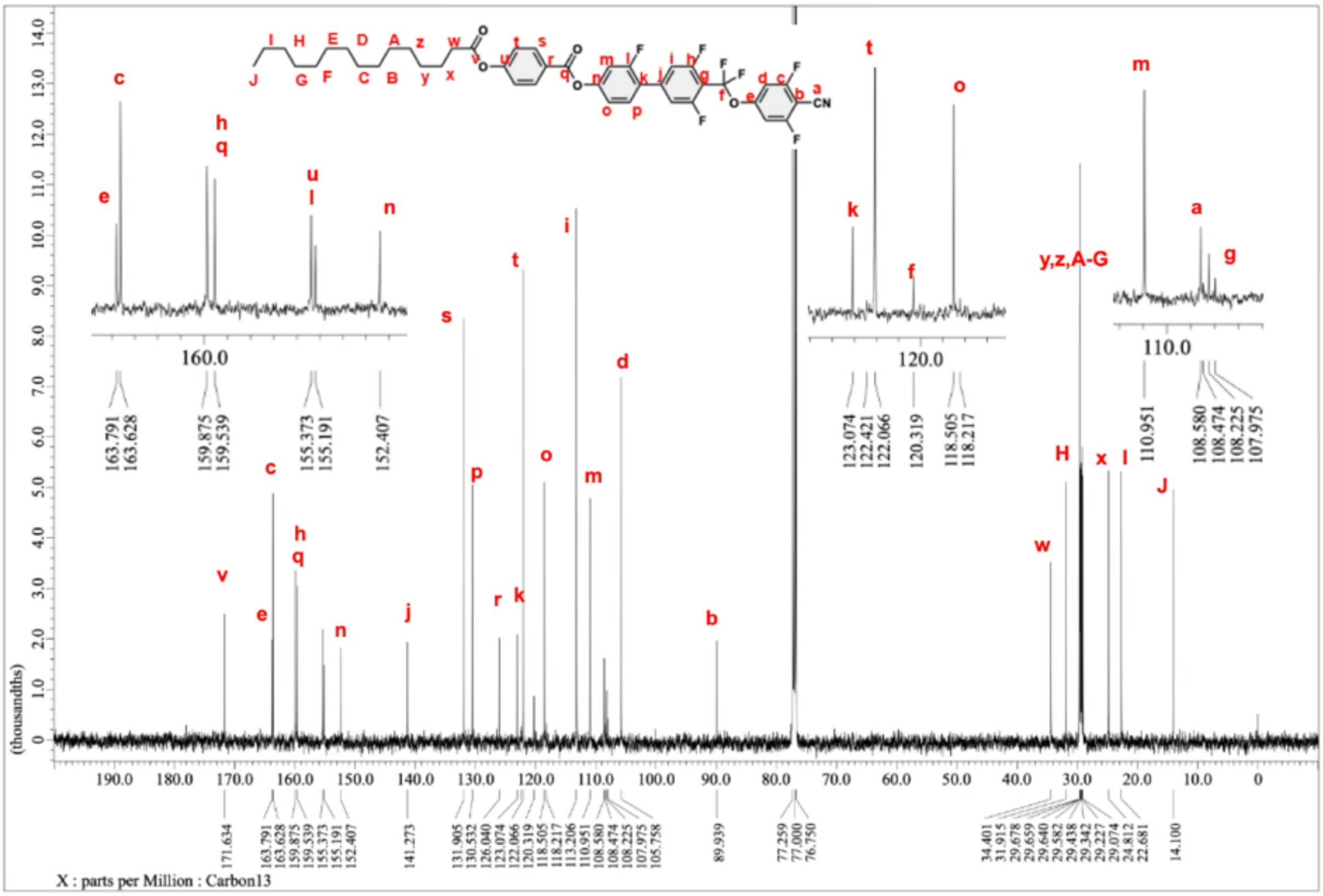

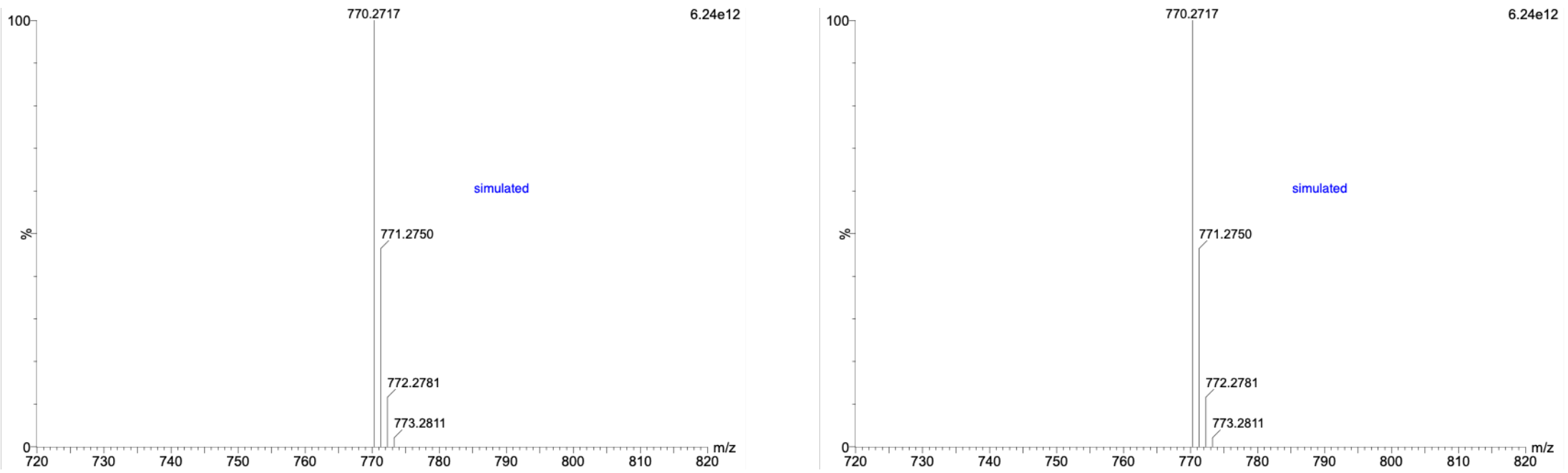

**Supplementary Fig. 93 |** $^{13}C\{1H,^{19}F\}$NMR (upper) and HR-ESI-TOF-MS spectra (bottom) for ***14*-012-CN**.

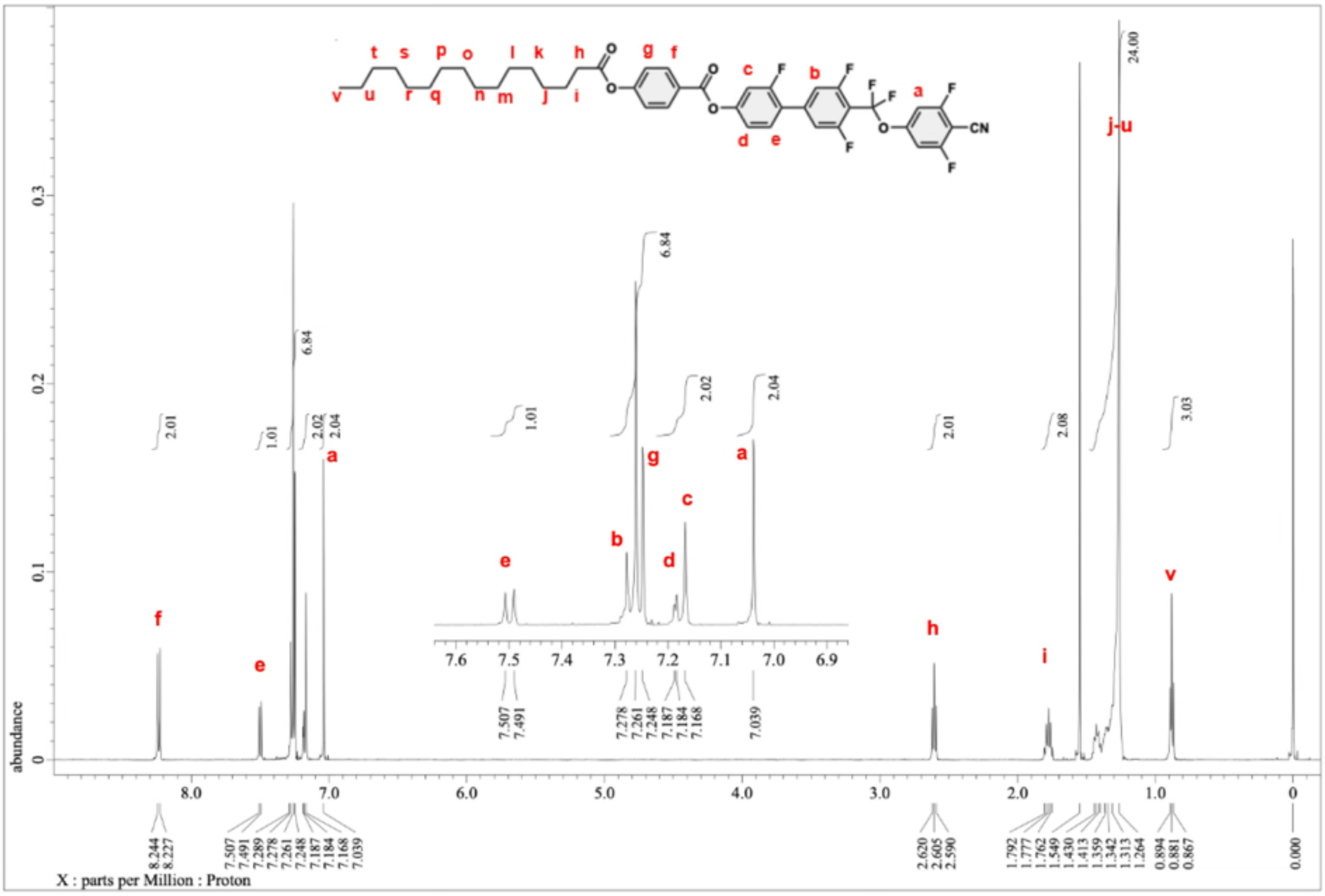

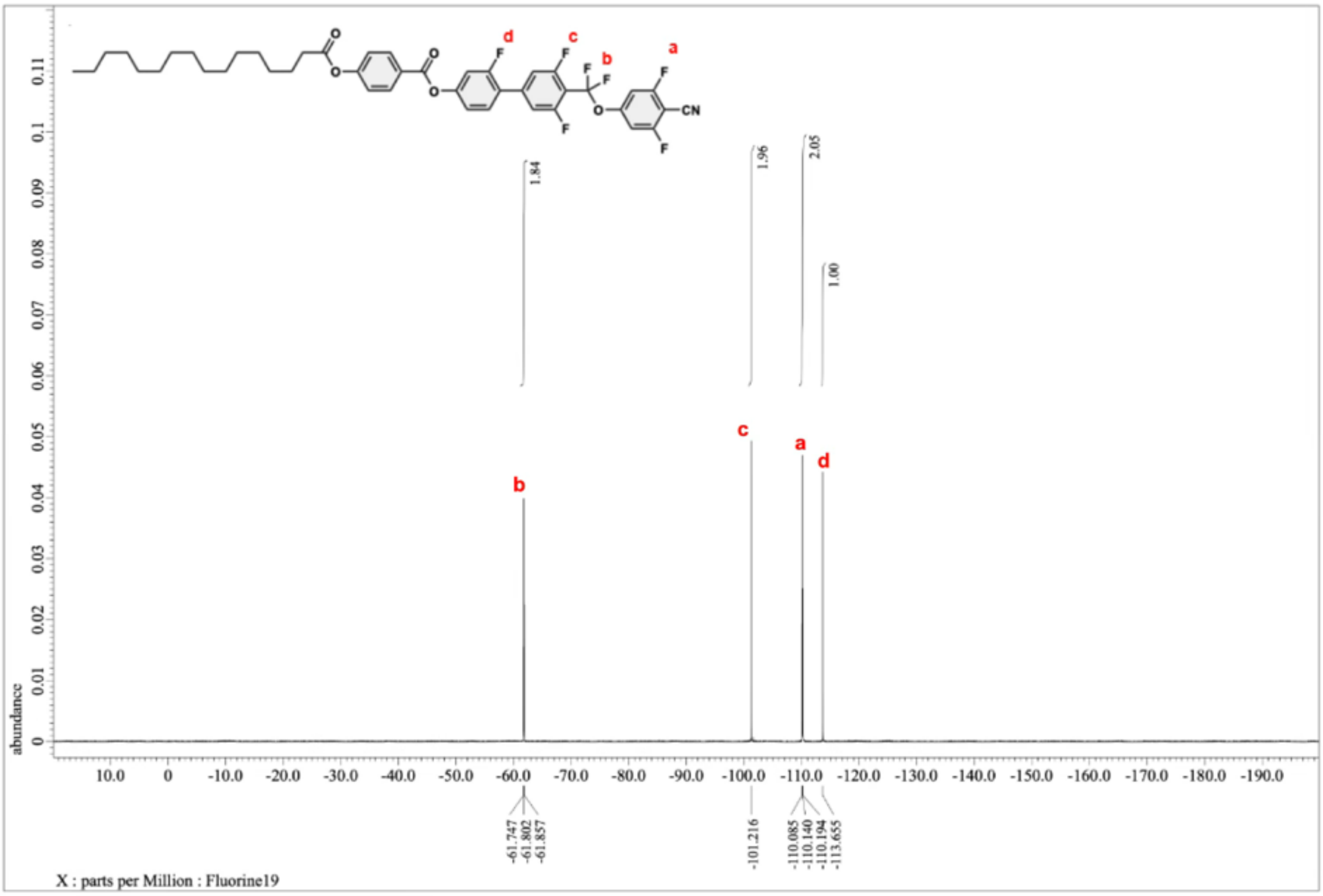

**Supplementary Fig. 94 |** $^{1}H\{^{19}F\}$ (upper) and $^{19}F\{^{1}H\}$ (bottom) NMR spectra for ***15*-012-CN**.

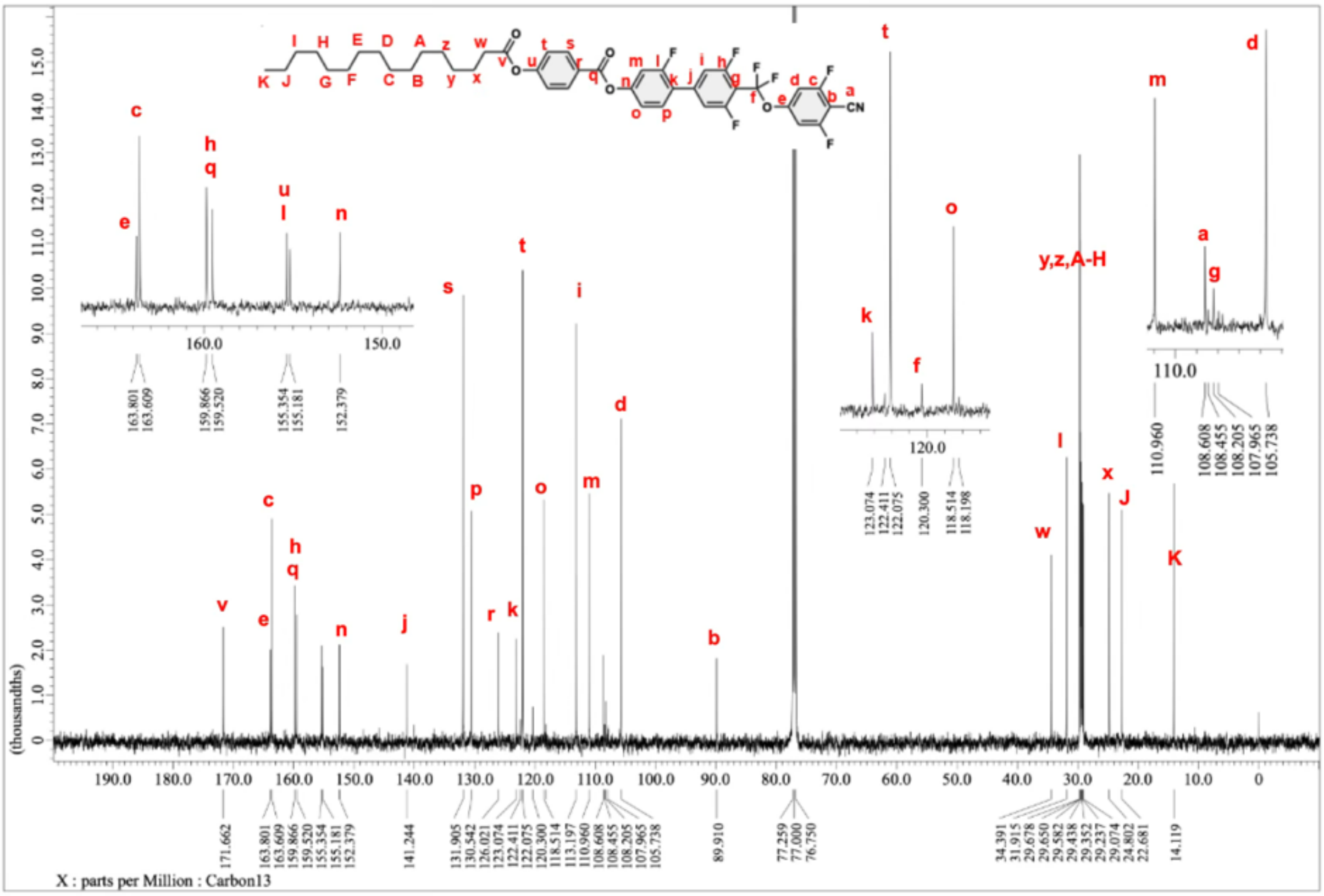

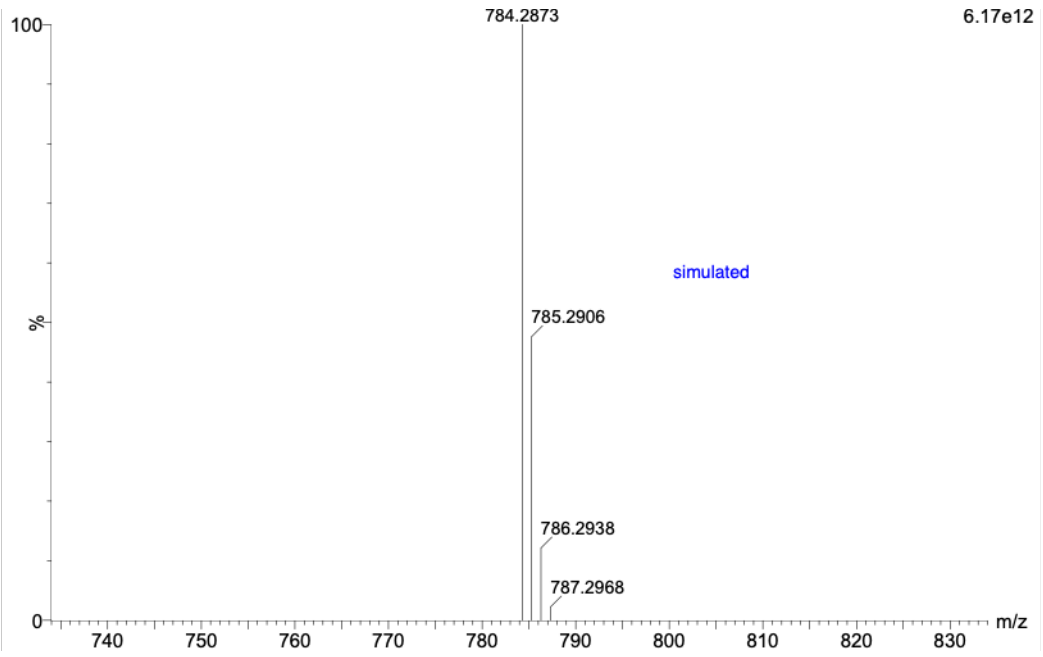

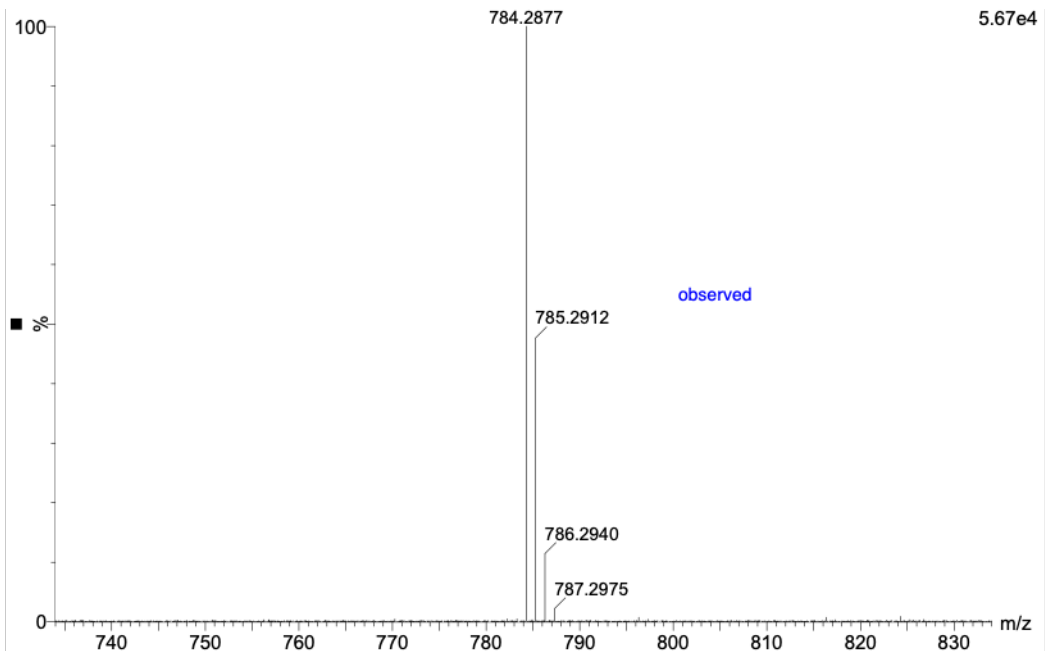

**Supplementary Fig. 95 |** $^{13}C\{^{1}H,^{19}F\}$NMR (upper) and HR-ESI-TOF-MS spectra (bottom) for ***15*-012-CN**.

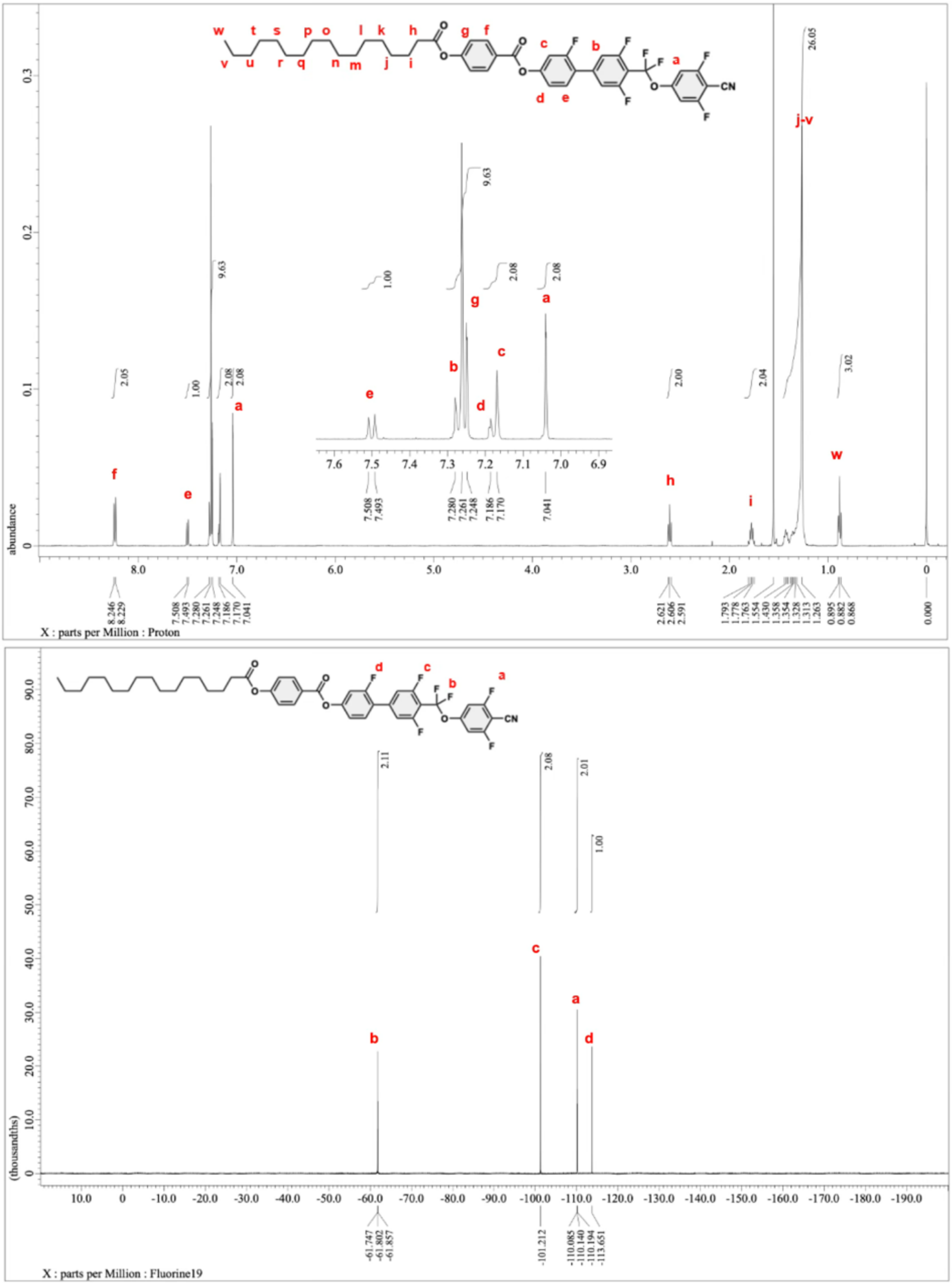

**Supplementary Fig. 96 |** $^{1}H\{^{19}F\}$ (upper) and $^{19}F\{^{1}H\}$ (bottom) NMR spectra for ***16*-012-CN**.

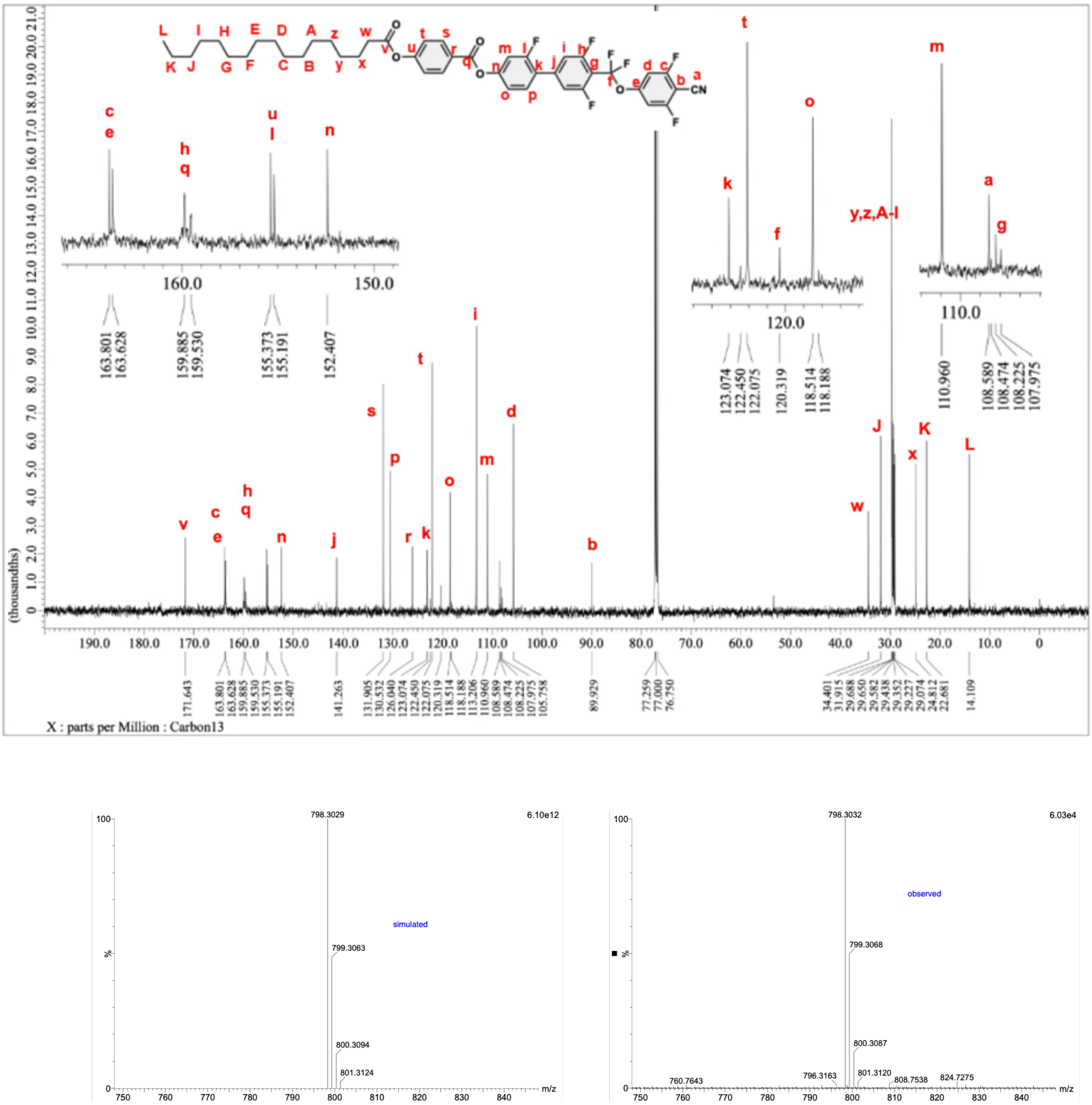

**Supplementary Fig. 97 |** $^{13}C\{1H,^{19}F\}$NMR (upper) and HR-ESI-TOF-MS spectra (bottom) for ***16*-012-CN**.

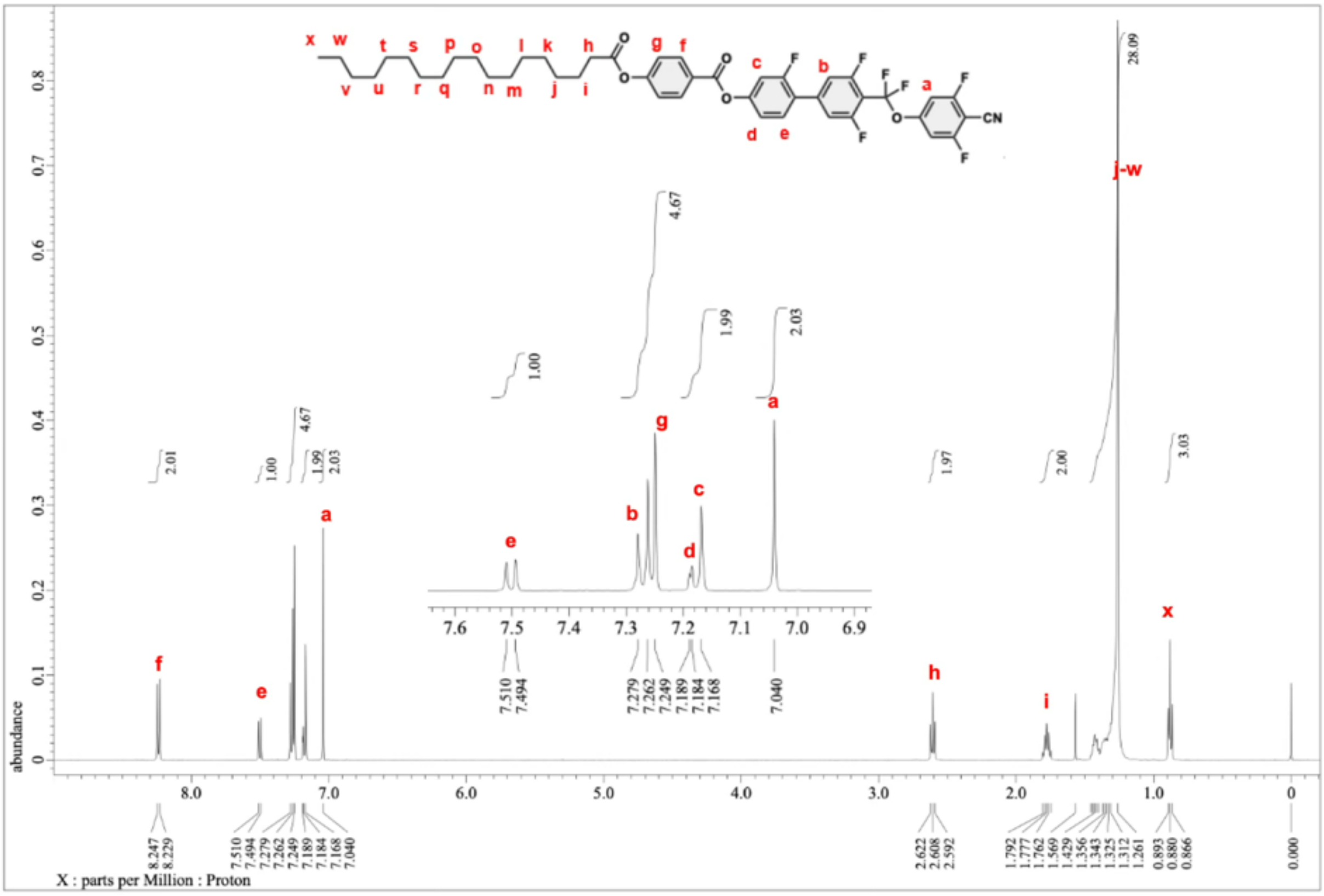

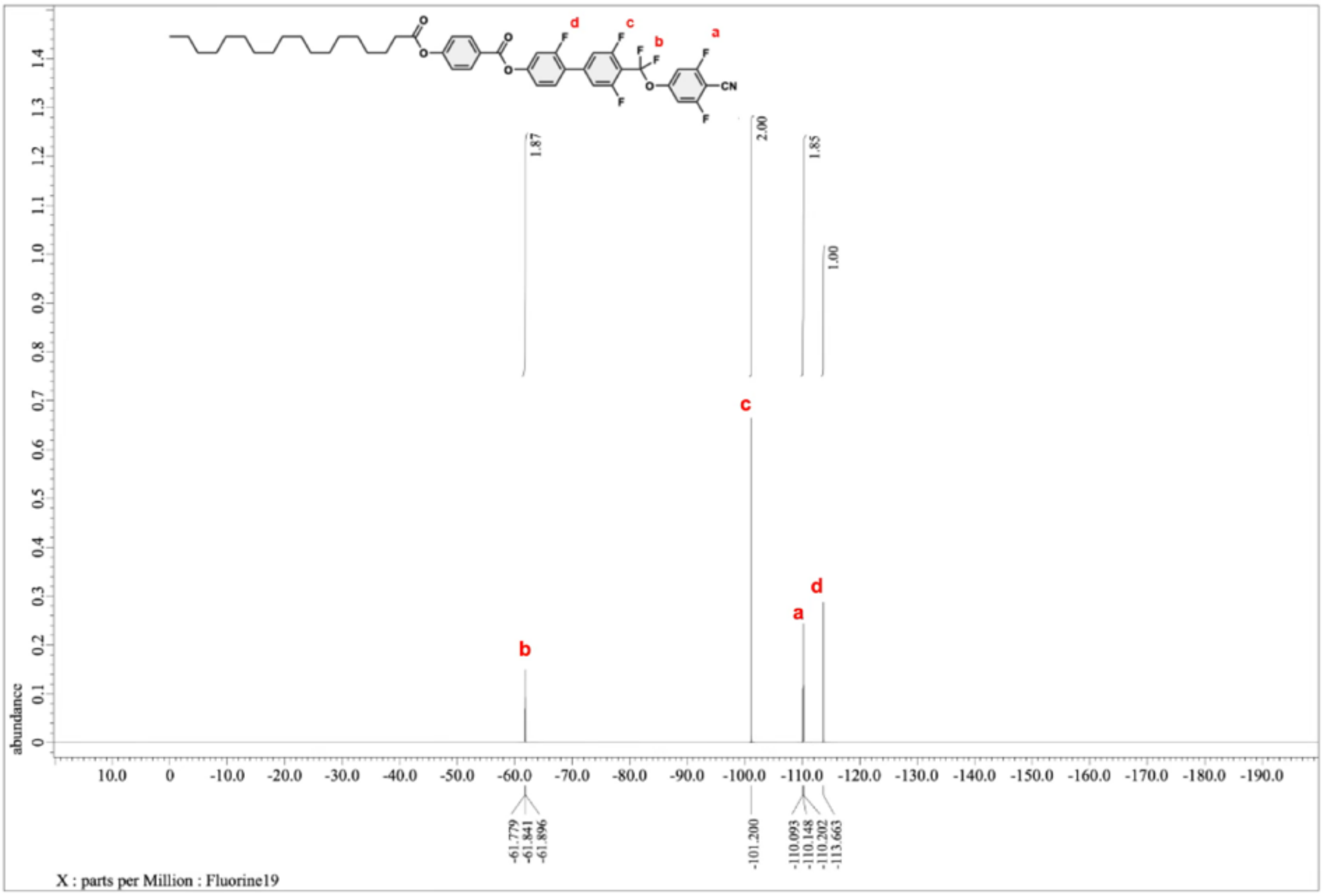

**Supplementary Fig. 98 |** $^{1}H\{^{19}F\}$ (upper) and $^{19}F\{^{1}H\}$ (bottom) NMR spectra for ***17*-012-CN**.

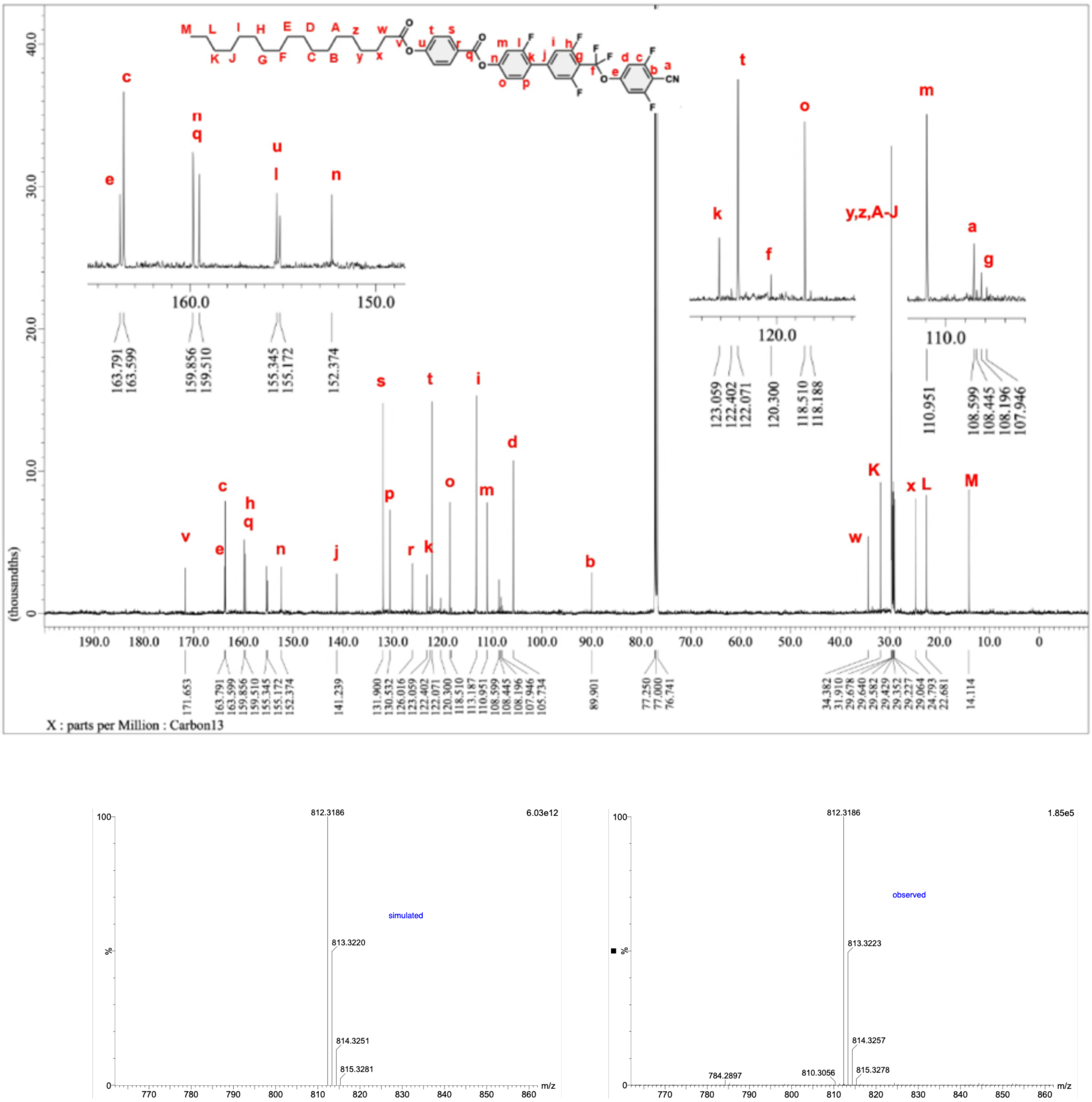

**Supplementary Fig. 99 |** $^{13}$C{1H,$^{19}$F}NMR (upper) and HR-ESI-TOF-MS spectra (bottom) for ***17*-012-CN**.

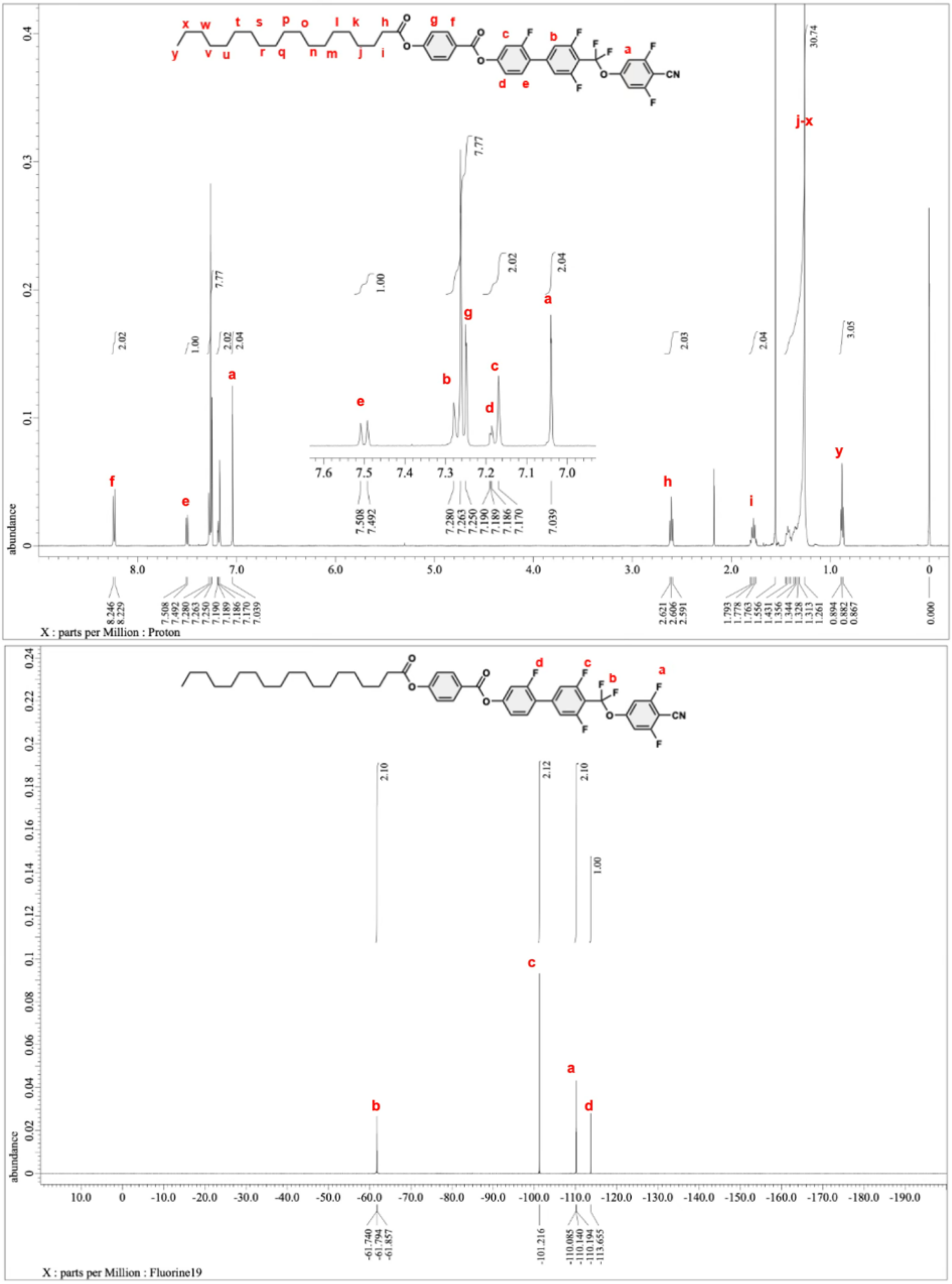

**Supplementary Fig. 100 |** $^{1}H\{^{19}F\}$ (upper) and $^{19}F\{^{1}H\}$ (bottom) NMR spectra for ***18*-012-CN**.

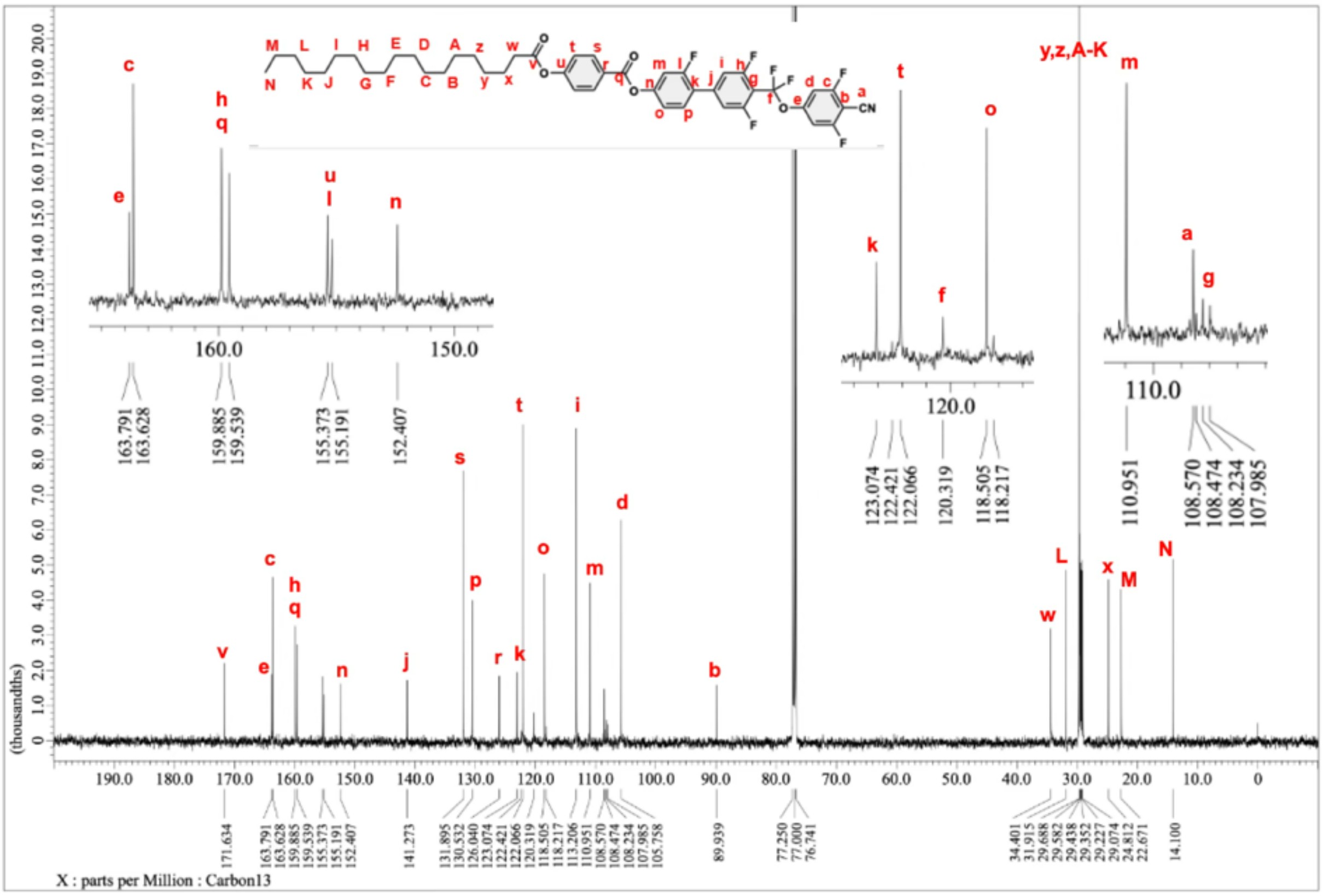

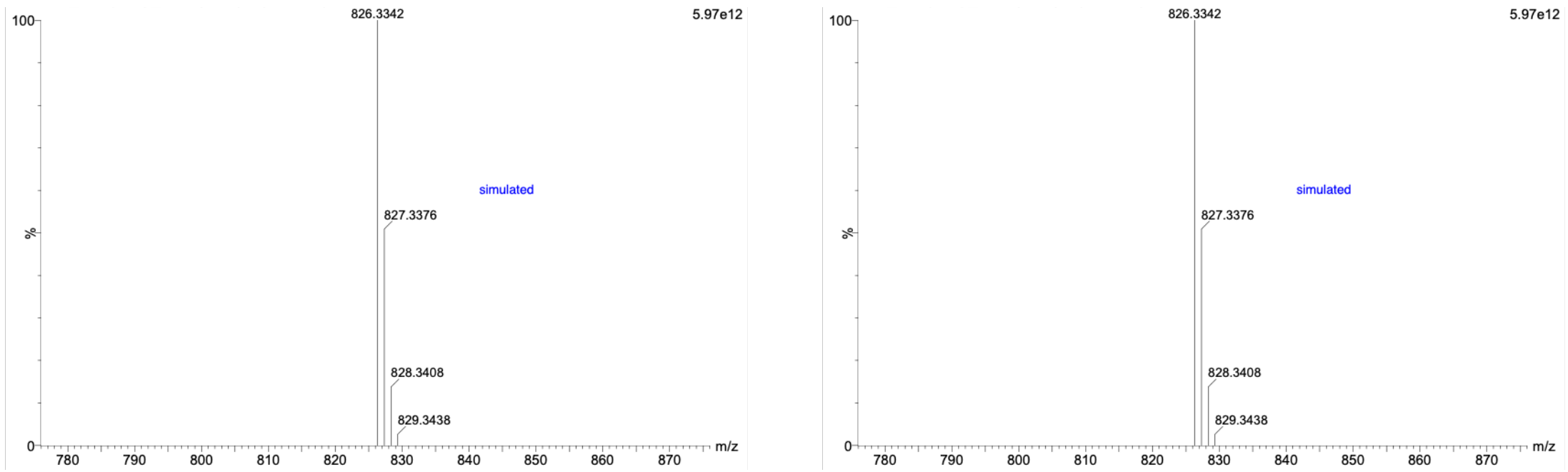

**Supplementary Fig. 101 |** $^{13}C\{1H,^{19}F\}$NMR (upper) and HR-ESI-TOF-MS spectra (bottom) for ***18*-012-CN**.

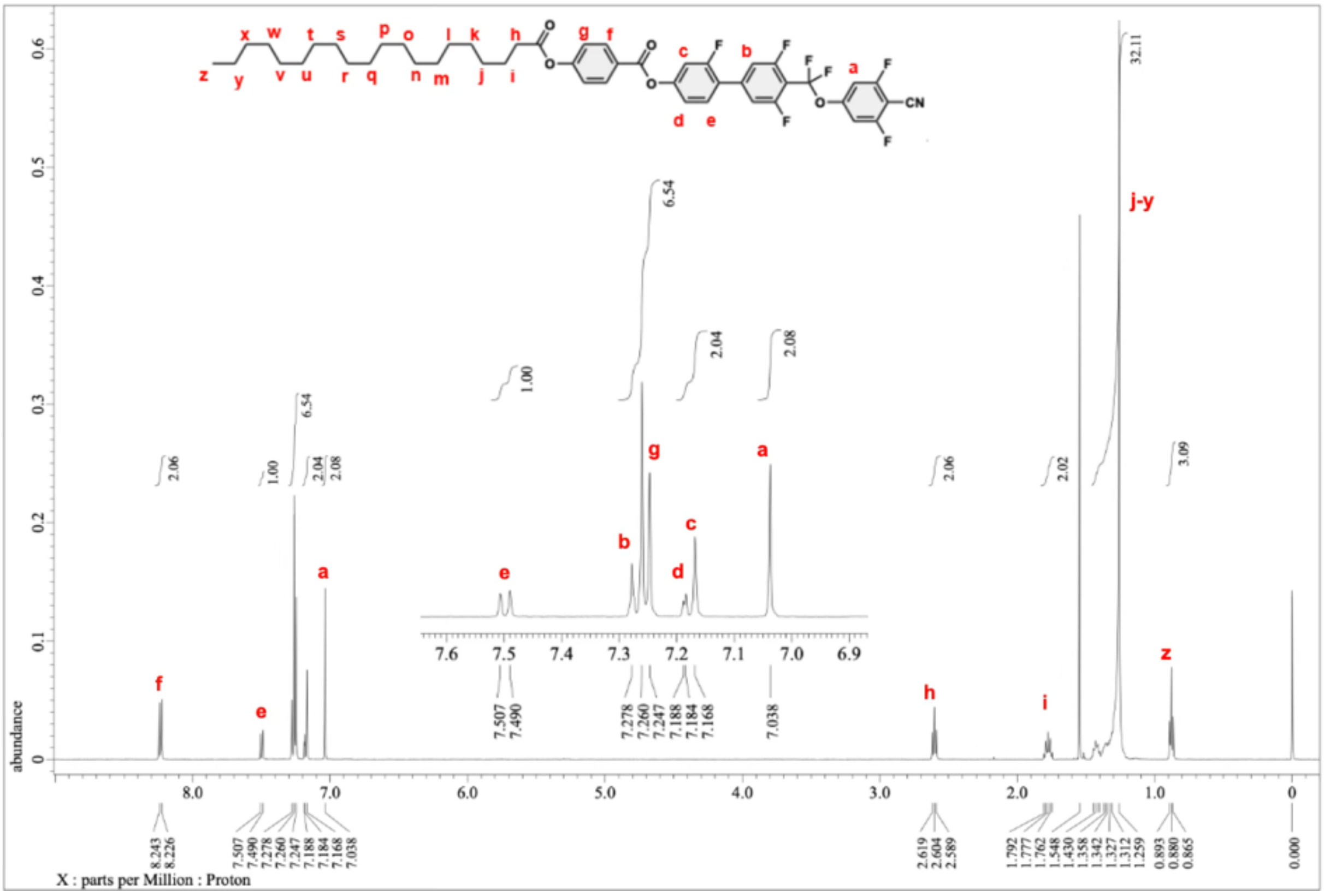

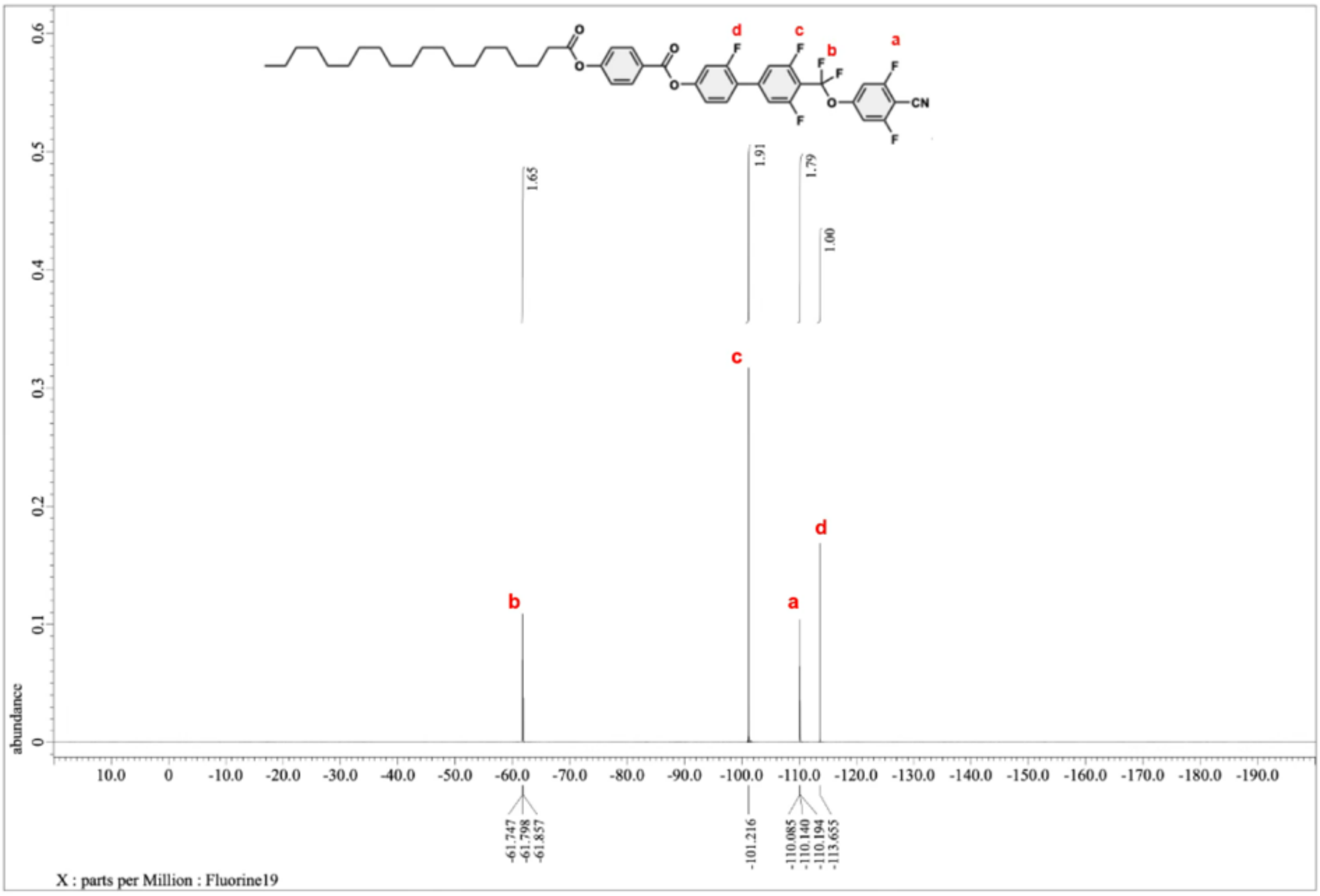

**Supplementary Fig. 102 |** $^{1}H\{^{19}F\}$ (upper) and $^{19}F\{^{1}H\}$ (bottom) NMR spectra for ***19*-012-CN**.

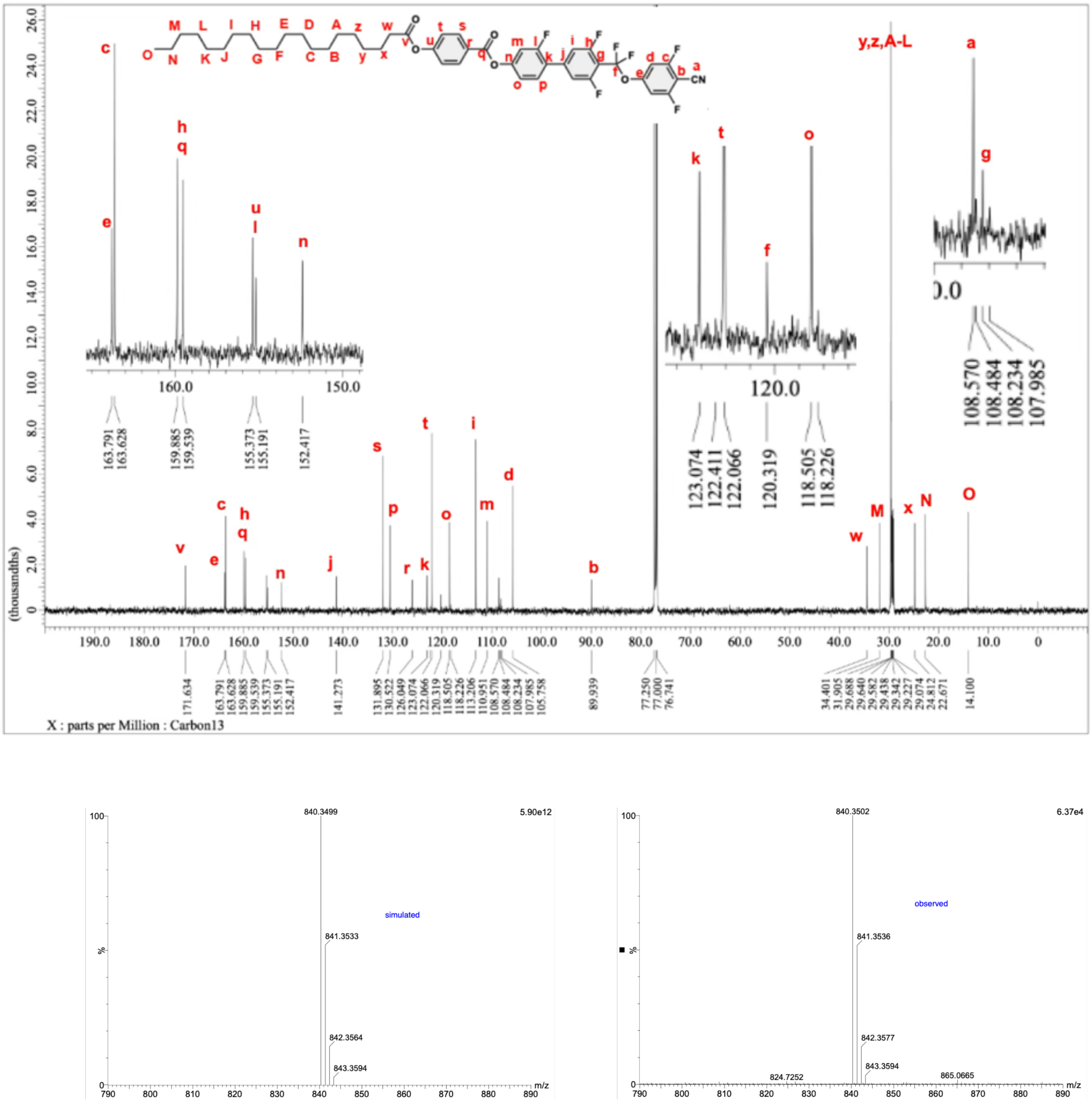

**Supplementary Fig. 103 |** $^{13}C\{1H,^{19}F\}$NMR (upper) and HR-ESI-TOF-MS spectra (bottom) for ***19*-012-CN**.

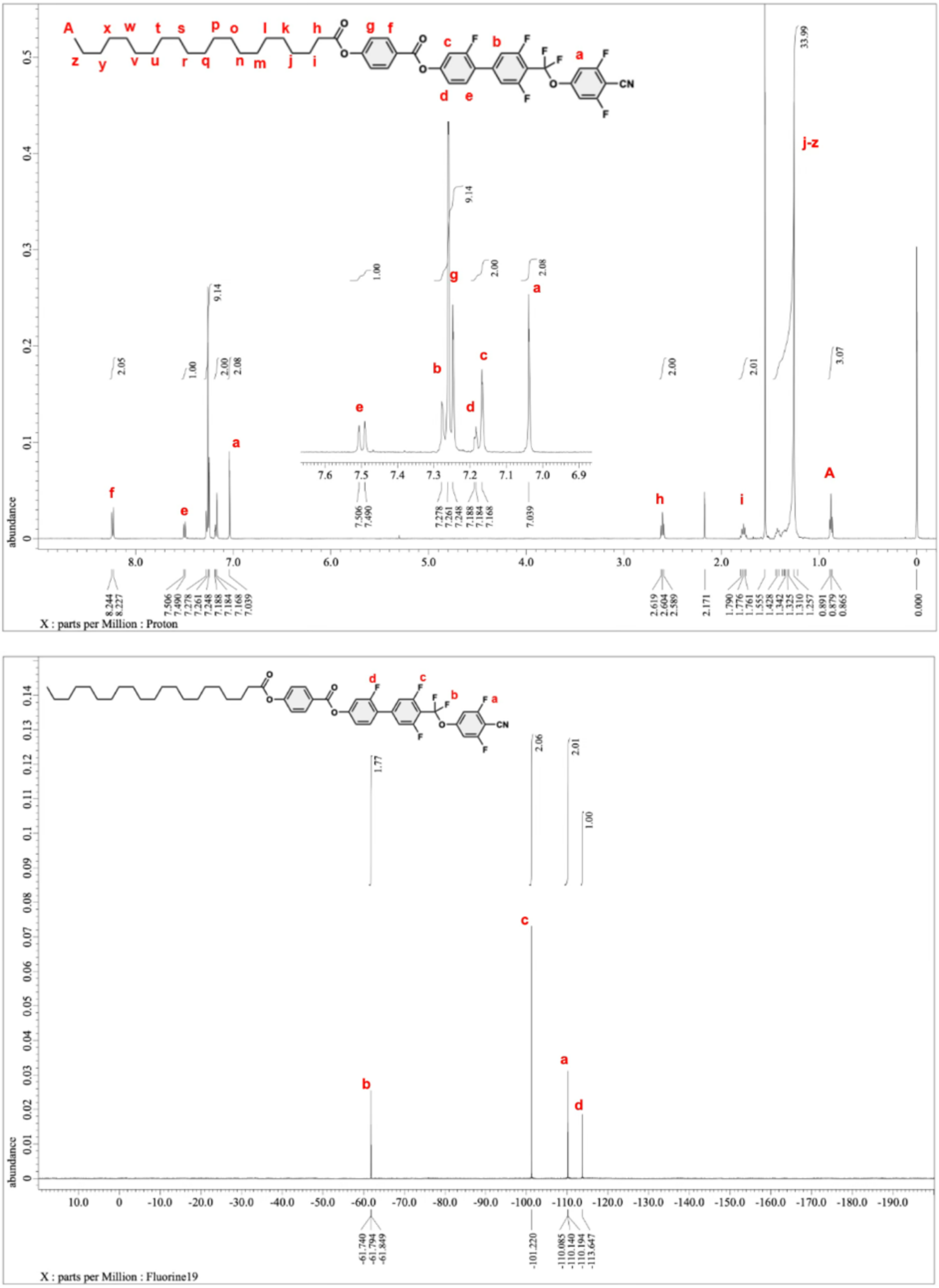

**Supplementary Fig. 104 |** $^{1}H\{^{19}F\}$ (upper) and $^{19}F\{^{1}H\}$ (bottom) NMR spectra for ***20*-012-CN**.

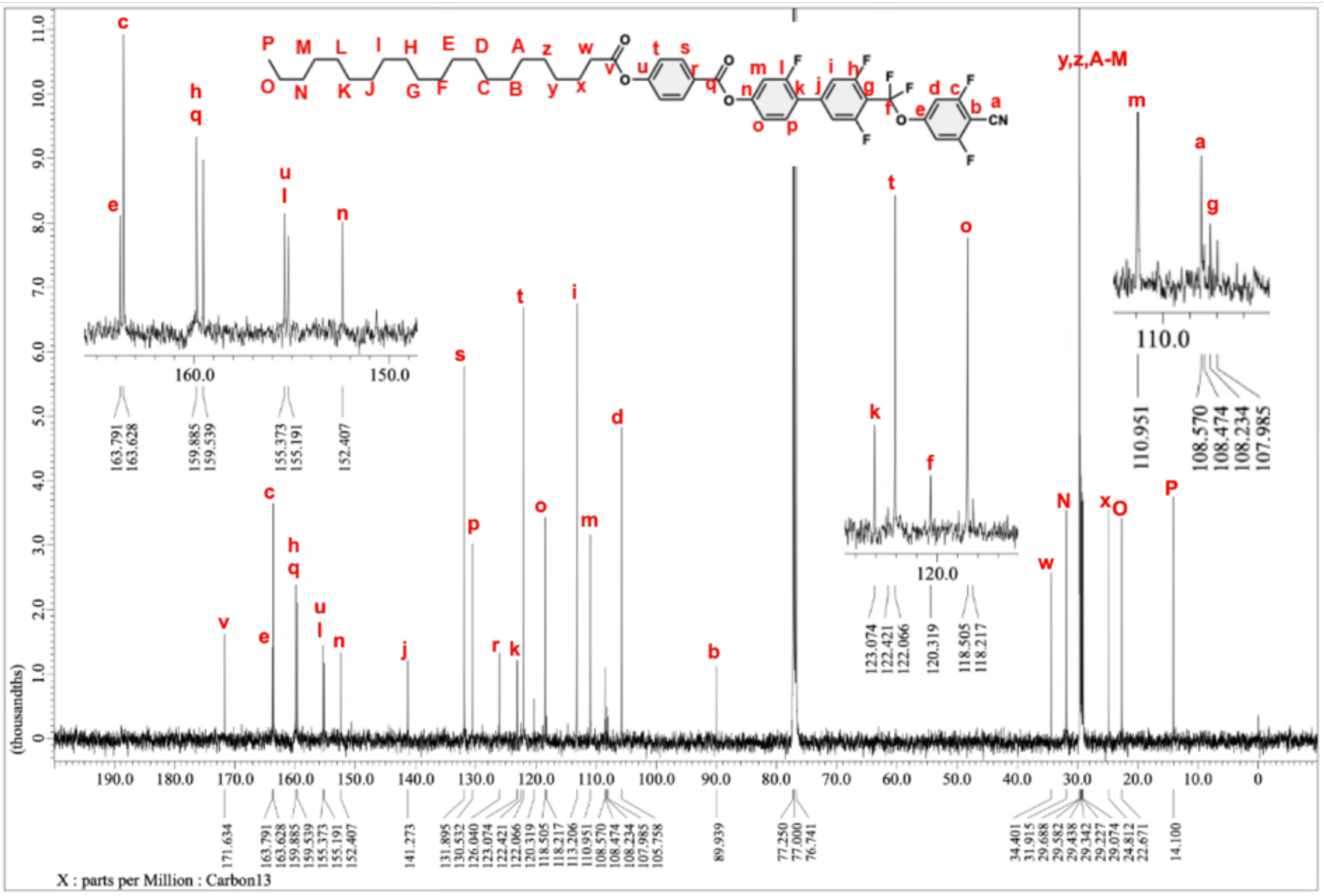

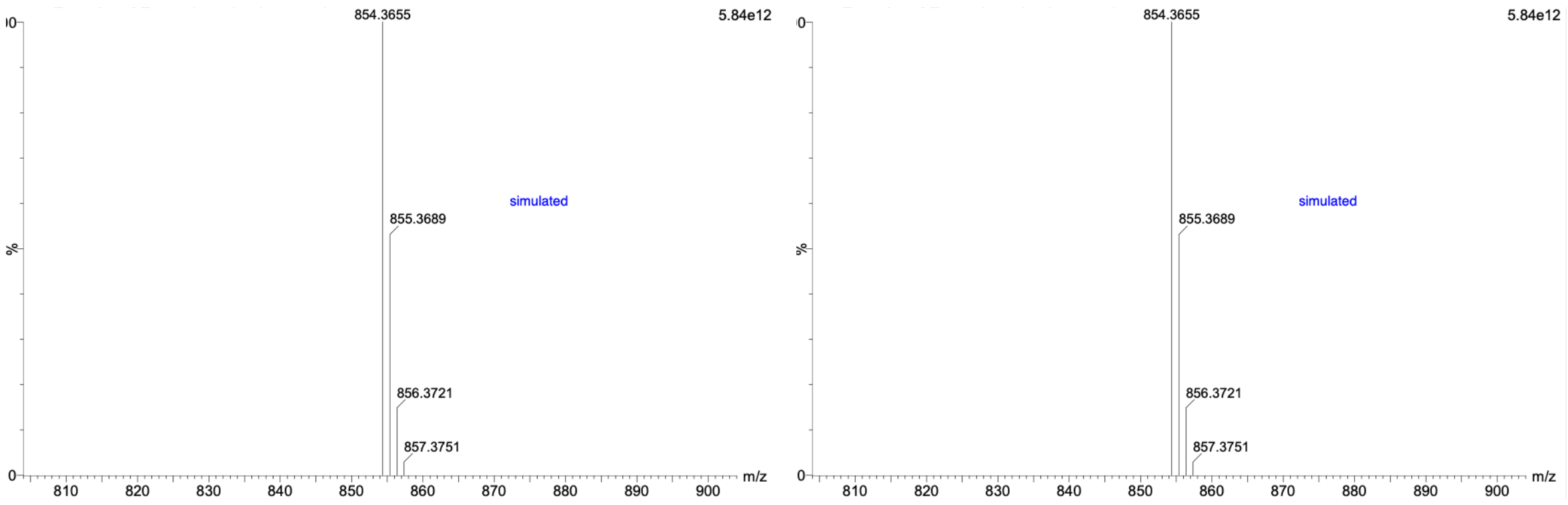

**Supplementary Fig. 105 |** $^{13}C\{1H,^{19}F\}$NMR (upper) and HR-ESI-TOF-MS spectra (bottom) for ***20*-012-CN**.

**Supplementary Tables (Table 1–11)**

**Supplementary Table 1** The molecular parameters of energy-minimized conformations calculated by DFT for ***n*-012-CN** (n = 1–20).

| Entry | Vector X | Vector Y | Vector Z | $\mu$ (D) | $\beta$ (deg)$^{\dagger}$ |
|---|---|---|---|---|---|
| ***1*-012-CN** | 12.90393 | 3.8181 | 0.1162 | 13.457 | 16.49 |
| ***2*-012-CN** | 13.113 | 3.7917 | 0.0079 | 13.650 | 16.13 |
| ***3*-012-CN** | 13.2316 | 3.7365 | 0.0127 | 13.749 | 15.77 |
| ***4*-012-CN** | 13.2999 | 3.8148 | 0.0547 | 13.836 | 16.01 |
| ***5*-012-CN** | 13.3289 | 3.7706 | 0.0738 | 13.852 | 15.80 |
| ***6*-012-CN** | 13.3591 | 3.8955 | 0.0491 | 13.916 | 16.26 |
| ***7*-012-CN** | 13.3624 | 3.8648 | 0.032 | 13.910 | 16.13 |
| ***8*-012-CN** | -13.3798 | -3.9846 | -0.1333 | 13.961 | 16.59 |
| ***9*-012-CN** | -13.3729 | -3.9353 | -0.1915 | 13.941 | 16.42 |
| ***10*-012-CN** | -13.3663 | -4.0544 | -0.5499 | 13.979 | 17.02 |
| ***11*-012-CN** | -13.3514 | -3.9696 | -0.6882 | 13.946 | 16.79 |
| ***12*-012-CN** | -13.3579 | -3.9565 | -1.4045 | 14.002 | 17.45 |
| ***13*-012-CN** | 13.3402 | -3.8523 | 1.5402 | 13.970 | 17.28 |
| ***14*-012-CN** | 13.3277 | -3.8703 | 1.9278 | 14.012 | 17.97 |
| ***15*-012-CN** | 13.31 | -3.6992 | 2.112 | 13.975 | 17.75 |
| ***16*-012-CN** | 13.3056 | -3.5398 | 2.5441 | 14.001 | 18.14 |
| ***17*-012-CN** | 13.2849 | -3.49 | 2.5629 | 13.973 | 18.05 |
| ***18*-012-CN** | 13.2983 | -3.6832 | 2.5188 | 14.027 | 18.55 |
| ***19*-012-CN** | 13.2819 | -3.6086 | 2.5629 | 14.000 | 18.43 |
| ***20*-012-CN** | -13.2614 | 3.7443 | 2.6528 | 14.033 | 19.09 |

† An angle between the permanent dipole moment (μ) and long molecular axis.

**Supplementary Table 2** Observed and calculated *d*-spacings and lattice parameters at different temperatures for ***9*-012-CN**.

| *T* (°C) | (*hk*) | $d_{obs.}$ (nm) | $d_{calc.}$ (nm) | *a* (nm) | *b* (nm) | *γ* (°) |
|---|---|---|---|---|---|---|
| 36 | (02) | 1.864 | 1.865 | 5.12 | 3.73 | 90 |
| | (12) | 1.751 | 1.752 | | | |
| | (11) | 3.013 | 3.015 | | | |
| | (21) | 2.109 | 2.111 | | | |
| | (20) | 2.560 | 2.560 | | | |

**Supplementary Table 3** Observed and calculated *d*-spacings and lattice parameters at different temperatures for ***10*-012-CN**.

| *T* (°C) | (*hk*) | $d_{obs.}$ (nm) | $d_{calc.}$ (nm) | *a* (nm) | *b* (nm) | *γ* (°) |
|---|---|---|---|---|---|---|
| 54 | (02) | 1.991 | 1.991 | 5.38 | 3.98 | 90 |
| | (12) | 1.864 | 1.867 | | | |
| | (11) | 3.200 | 3.200 | | | |
| | (21) | 2.225 | 2.228 | | | |
| | (20) | 2.688 | 2.688 | | | |

**Supplementary Table 4** Observed and calculated *d*-spacings and lattice parameters at different temperatures for ***11*-012-CN**.

| *T* (°C) | (*hk*) | $d_{obs.}$ (nm) | $d_{calc.}$ (nm) | *a* (nm) | *b* (nm) | *γ* (°) |
|---|---|---|---|---|---|---|
| 64 | (02) | 2.185 | 2.185 | 5.98 | 4.37 | 90 |
| | (12) | 2.052 | 2.052 | | | |
| | (11) | 3.526 | 3.527 | | | |
| | (21) | – | 2.466 | | | |
| | (20) | 2.988 | 2.988 | | | |
| 62 | (02) | 2.181 | 2.181 | 5.92 | 4.36 | 90 |
| | (12) | 2.046 | 2.047 | | | |
| | (11) | 3.509 | 3.511 | | | |
| | (21) | – | 2.448 | | | |
| | (20) | 2.958 | 2.958 | | | |
| 60 | (02) | 2.153 | 2.153 | 5.69 | 4.31 | 90 |
| | (12) | 2.013 | 2.014 | | | |
| | (11) | 3.429 | 3.434 | | | |
| | (21) | – | 2.374 | | | |
| | (20) | 2.845 | 2.845 | | | |
| 58 | (02) | 2.141 | 2.141 | 5.60 | 4.28 | 90 |
| | (12) | 2.000 | 2.000 | | | |
| | (11) | 3.397 | 3.402 | | | |
| | (21) | – | 2.345 | | | |
| | (20) | 2.802 | 2.802 | | | |
| 56 | (02) | 2.122 | 2.122 | 5.50 | 4.24 | 90 |
| | (12) | 1.978 | 1.980 | | | |
| | (11) | 3.359 | 3.360 | | | |
| | (21) | 2.307 | 2.308 | | | |
| | (20) | 2.750 | 2.750 | | | |

**Supplementary Table 4** Observed and calculated *d*-spacings and lattice parameters at different temperatures for ***11*-012-CN**. (*continued*)

| *T* (°C) | (*hk*) | $d_{obs.}$ (nm) | $d_{calc.}$ (nm) | *a* (nm) | *b* (nm) | *γ* (°) |
|---|---|---|---|---|---|---|
| 54 | (02) | 2.113 | 2.113 | 5.46 | 4.23 | 90 |
| | (12) | 1.970 | 1.971 | | | |
| | (11) | 3.336 | 3.342 | | | |
| | (21) | 2.292 | 2.293 | | | |
| | (20) | 2.730 | 2.730 | | | |
| 52 | (02) | 2.104 | 2.104 | 5.42 | 4.21 | 90 |
| | (12) | 1.963 | 1.962 | | | |
| | (11) | 3.321 | 3.324 | | | |
| | (21) | 2.275 | 2.279 | | | |
| | (20) | 2.710 | 2.710 | | | |
| 50 | (02) | 2.092 | 2.092 | 5.35 | 4.18 | 90 |
| | (12) | 1.947 | 1.949 | | | |
| | (11) | 3.292 | 3.297 | | | |
| | (21) | 2.254 | 2.255 | | | |
| | (20) | 2.676 | 2.676 | | | |
| 48 | (02) | 2.087 | 2.087 | 5.32 | 4.17 | 90 |
| | (12) | 1.942 | 1.943 | | | |
| | (11) | 3.285 | 3.284 | | | |
| | (21) | 2.244 | 2.244 | | | |
| | (20) | 2.662 | 2.662 | | | |
| 46 | (02) | 2.081 | 2.081 | 5.32 | 4.16 | 90 |
| | (12) | 1.937 | 1.938 | | | |
| | (11) | 3.270 | 3.279 | | | |
| | (21) | 2.234 | 2.242 | | | |
| | (20) | 2.662 | 2.662 | | | |
| 44 | (02) | 2.078 | 2.078 | 5.28 | 4.16 | 90 |
| | (12) | 1.932 | 1.933 | | | |
| | (11) | 3.263 | 3.265 | | | |
| | (21) | 2.224 | 2.227 | | | |
| | (20) | 2.638 | 2.638 | | | |
| 42 | (02) | 2.072 | 2.072 | 5.28 | 4.14 | 90 |
| | (12) | 1.925 | 1.929 | | | |
| | (11) | 3.249 | 3.259 | | | |
| | (21) | 2.214 | 2.225 | | | |
| | (20) | 2.638 | 2.638 | | | |
| 40 | (02) | 2.069 | 2.069 | 5.23 | 4.14 | 90 |
| | (12) | 1.922 | 1.924 | | | |
| | (11) | 3.242 | 3.245 | | | |
| | (21) | 2.207 | 2.211 | | | |
| | (20) | 2.615 | 2.615 | | | |
| 38 | (02) | 2.066 | 2.066 | 5.21 | 4.13 | 90 |
| | (12) | 1.920 | 1.921 | | | |
| | (11) | 3.235 | 3.238 | | | |
| | (21) | 2.204 | 2.204 | | | |
| | (20) | 2.606 | 2.606 | | | |

**Supplementary Table 5** Observed and calculated *d*-spacings and lattice parameters at different temperatures for ***12*-012-CN**.

| $T$ (°C) | ($hk$) | $d_{obs.}$ (nm) | $d_{calc.}$ (nm) | $a$ (nm) | $b$ (nm) | $\gamma$ (°) |
|---|---|---|---|---|---|---|
| 76 | (02) | 2.362 | 2.362 | 6.93 | 4.72 | 90 |
| | (12) | 2.235 | 2.236 | | | |
| | (11) | – | 3.903 | | | |
| | (21) | 2.847 | 2.795 | | | |
| | (20) | 3.466 | 3.466 | | | |
| 75 | (02) | 2.407 | 2.407 | 6.24 | 4.81 | 90 |
| | (12) | 2.189 | 2.246 | | | |
| | (11) | 3.753 | 3.812 | | | |
| | (21) | 2.595 | 2.618 | | | |
| | (20) | 3.120 | 3.120 | | | |
| 74 | (02) | 2.343 | 2.343 | 6.14 | 4.69 | 90 |
| | (12) | 2.189 | 2.189 | | | |
| | (11) | 3.725 | 3.725 | | | |
| | (21) | 2.568 | 2.567 | | | |
| | (20) | 3.069 | 3.069 | | | |
| 72 | (02) | 2.332 | 2.332 | 6.03 | 4.66 | 90 |
| | (12) | 2.173 | 2.175 | | | |
| | (11) | 3.679 | 3.689 | | | |
| | (21) | 2.529 | 2.531 | | | |
| | (20) | 3.013 | 3.013 | | | |
| 70 | (02) | 2.318 | 2.318 | 5.94 | 4.64 | 90 |
| | (12) | 2.161 | 2.159 | | | |
| | (11) | 3.652 | 3.655 | | | |
| | (21) | 2.500 | 2.501 | | | |
| | (20) | 2.971 | 2.971 | | | |
| 68 | (02) | 2.314 | 2.314 | 5.91 | 4.63 | 90 |
| | (12) | 2.154 | 2.155 | | | |
| | (11) | 3.635 | 3.643 | | | |
| | (21) | 2.487 | 2.490 | | | |
| | (20) | 2.953 | 2.953 | | | |
| 66 | (02) | 2.296 | 2.296 | 5.82 | 4.59 | 90 |
| | (12) | 2.136 | 2.136 | | | |
| | (11) | 3.600 | 3.604 | | | |
| | (21) | 2.455 | 2.457 | | | |
| | (20) | 2.908 | 2.908 | | | |
| 64 | (02) | 2.282 | 2.282 | 5.74 | 4.56 | 90 |
| | (12) | 2.121 | 2.121 | | | |
| | (11) | 3.565 | 3.572 | | | |
| | (21) | – | 2.429 | | | |
| | (20) | 2.869 | 2.869 | | | |
| 62 | (02) | 2.275 | 2.275 | 5.72 | 4.55 | 90 |
| | (12) | 2.112 | 2.114 | | | |
| | (11) | 3.557 | 3.560 | | | |
| | (21) | – | 2.420 | | | |
| | (20) | 2.858 | 2.858 | | | |

**Supplementary Table 5** Observed and calculated *d*-spacings and lattice parameters at different temperatures for ***12*-012-CN**. (*Continued*)

| $T$ (°C) | ($hk$) | $d_{obs.}$ (nm) | $d_{calc.}$ (nm) | $a$ (nm) | $b$ (nm) | $\gamma$ (°) |
|---|---|---|---|---|---|---|
| 60 | (02) | 2.258 | 2.258 | 5.65 | 4.52 | 90 |
| | (12) | 2.097 | 2.097 | | | |
| | (11) | 3.523 | 3.528 | | | |
| | (21) | – | 2.395 | | | |
| | (20) | 2.825 | 2.825 | | | |

**Supplementary Table 6** Observed and calculated *d*-spacings and lattice parameters at different temperatures for ***13*-012-CN**.

| *T* (°C) | (*hk*) | $d_{obs.}$ (nm) | $d_{calc.}$ (nm) | *a* (nm) | *b* (nm) | *γ* (°) |
|---|---|---|---|---|---|---|
| 74 | (02) | 3.256 | 3.256 | 4.98 | 6.51 | 90 |
| | (12) | – | 2.726 | | | |
| | (11) | 3.958 | 3.958 | | | |
| | (21) | – | 2.327 | | | |
| | (20) | – | 2.492 | | | |
| 72 | (02) | 3.193 | 3.193 | 4.98 | 6.39 | 90 |
| | (12) | – | 2.688 | | | |
| | (11) | 3.926 | 3.926 | | | |
| | (21) | – | 2.319 | | | |
| | (20) | – | 2.489 | | | |
| 70 | (02) | 3.081 | 3.081 | 4.94 | 6.16 | 90 |
| | (12) | 2.615 | 2.614 | | | |
| | (11) | 3.855 | 3.855 | | | |
| | (21) | 2.292 | 2.293 | | | |
| | (20) | – | 2.471 | | | |
| 68 | (02) | 3.062 | 3.062 | 4.92 | 6.12 | 90 |
| | (12) | 2.601 | 2.599 | | | |
| | (11) | 3.835 | 3.835 | | | |
| | (21) | 2.285 | 2.283 | | | |
| | (20) | – | 2.460 | | | |
| 66 | (02) | 3.012 | 3.012 | 4.89 | 6.02 | 90 |
| | (12) | 2.565 | 2.564 | | | |
| | (11) | 3.796 | 3.796 | | | |
| | (21) | 2.271 | 2.265 | | | |
| | (20) | – | 2.444 | | | |
| 64 | (02) | 2.988 | 2.988 | 4.87 | 5.98 | 90 |
| | (12) | 2.552 | 2.547 | | | |
| | (11) | 3.777 | 3.777 | | | |
| | (21) | 2.261 | 2.256 | | | |
| | (20) | – | 2.437 | | | |
| 62 | (02) | 2.970 | 2.970 | 4.87 | 5.94 | 90 |
| | (12) | 2.539 | 2.536 | | | |
| | (11) | 3.767 | 3.767 | | | |
| | (21) | 2.254 | 2.254 | | | |
| | (20) | – | 2.436 | | | |
| 60 | (02) | 2.946 | 2.946 | 4.84 | 5.89 | 90 |
| | (12) | 2.518 | 2.516 | | | |
| | (11) | 3.739 | 3.739 | | | |
| | (21) | 2.241 | 2.237 | | | |
| | (20) | 2.431 | 2.418 | | | |
| 58 | (02) | 2.923 | 2.923 | 4.83 | 5.85 | 90 |
| | (12) | – | 2.501 | | | |
| | (11) | 3.720 | 3.724 | | | |
| | (21) | 2.234 | 2.232 | | | |
| | (20) | 2.415 | 2.415 | | | |

**Supplementary Table 6** Observed and calculated *d*-spacings and lattice parameters at different temperatures for ***13*-012-CN** (*Continued*).

| $T$ (°C) | ($hk$) | $d_{obs.}$ (nm) | $d_{calc.}$ (nm) | $a$ (nm) | $b$ (nm) | $\gamma$ (°) |
|---|---|---|---|---|---|---|
| 56 | (02) | 2.912 | 2.912 | 4.82 | 5.82 | 90 |
| | (12) | – | 2.493 | | | |
| | (11) | 3.711 | 3.714 | | | |
| | (21) | 2.227 | 2.228 | | | |
| | (20) | 2.411 | 2.411 | | | |
| 54 | (02) | 2.889 | 2.889 | 4.80 | 5.78 | 90 |
| | (12) | – | 2.475 | | | |
| | (11) | 3.683 | 3.692 | | | |
| | (21) | 2.214 | 2.216 | | | |
| | (20) | 2.400 | 2.400 | | | |
| 52 | (02) | 2.867 | 2.867 | 4.77 | 5.73 | 90 |
| | (12) | – | 2.457 | | | |
| | (11) | 3.665 | 3.666 | | | |
| | (21) | 2.201 | 2.202 | | | |
| | (20) | 2.384 | 2.384 | | | |
| 50 | (02) | 2.889 | 2.889 | 4.75 | 5.78 | 90 |
| | (12) | – | 2.469 | | | |
| | (11) | 3.647 | 3.671 | | | |
| | (21) | 2.191 | 2.198 | | | |
| | (20) | 2.377 | 2.377 | | | |
| 48 | (02) | 2.840 | 2.840 | 4.70 | 5.68 | 90 |
| | (12) | – | 2.431 | | | |
| | (11) | 3.629 | 3.621 | | | |
| | (21) | 2.181 | 2.172 | | | |
| | (20) | 2.350 | 2.350 | | | |

**Supplementary Table 7** Observed and calculated *d*-spacings and lattice parameters at different temperatures for ***14*-012-CN**.

| *T* (°C) | (*hk*) | $d_{obs.}$ (nm) | $d_{calc.}$ (nm) | *a* (nm) | *b* (nm) | *γ* (°) |
|---|---|---|---|---|---|---|
| 66 | (02) | 2.638 | 2.638 | 6.52 | 5.28 | 90 |
| | (12) | 2.442 | 2.446 | | | |
| | (11) | 4.101 | 4.102 | | | |
| | (21) | 2.769 | 2.773 | | | |
| | (20) | 3.260 | 3.260 | | | |
| 64 | (02) | 2.625 | 2.625 | 6.35 | 5.25 | 90 |
| | (12) | 2.426 | 2.426 | | | |
| | (11) | 4.046 | 4.047 | | | |
| | (21) | 2.719 | 2.718 | | | |
| | (20) | 3.177 | 3.177 | | | |
| 62 | (02) | 2.620 | 2.620 | 6.35 | 5.24 | 90 |
| | (12) | 2.418 | 2.422 | | | |
| | (11) | 4.036 | 4.043 | | | |
| | (21) | 2.714 | 2.717 | | | |
| | (20) | 3.177 | 3.177 | | | |
| 60 | (02) | 2.611 | 2.611 | 6.26 | 5.22 | 90 |
| | (12) | 2.407 | 2.410 | | | |
| | (11) | 4.014 | 4.011 | | | |
| | (21) | 2.685 | 2.685 | | | |
| | (20) | 3.131 | 3.131 | | | |
| 58 | (02) | 2.602 | 2.602 | 6.17 | 5.20 | 90 |
| | (12) | 2.395 | 2.398 | | | |
| | (11) | 3.972 | 3.979 | | | |
| | (21) | 2.652 | 2.655 | | | |
| | (20) | 3.086 | 3.086 | | | |
| 56 | (02) | 2.598 | 2.598 | 6.16 | 5.20 | 90 |
| | (12) | 2.392 | 2.393 | | | |
| | (11) | 3.972 | 3.971 | | | |
| | (21) | 2.648 | 2.649 | | | |
| | (20) | 3.080 | 3.080 | | | |
| 54 | (02) | 2.589 | 2.589 | 6.16 | 5.18 | 90 |
| | (12) | 2.380 | 2.386 | | | |
| | (11) | 3.941 | 3.963 | | | |
| | (21) | 2.625 | 2.647 | | | |
| | (20) | 3.080 | 3.080 | | | |
| 52 | (02) | 2.580 | 2.580 | 6.07 | 5.16 | 90 |
| | (12) | 2.373 | 2.374 | | | |
| | (11) | 3.931 | 3.932 | | | |
| | (21) | 2.611 | 2.617 | | | |
| | (20) | 3.037 | 3.037 | | | |

**Supplementary Table 8** Observed and calculated *d*-spacings and lattice parameters at different temperatures for ***15*-012-CN**.

| *T* (°C) | (*hk*) | $d_{obs.}$ (nm) | $d_{calc.}$ (nm) | *a* (nm) | *b* (nm) | *γ* (°) |
|---|---|---|---|---|---|---|
| 63 | (02) | 2.774 | 2.774 | 6.52 | 5.55 | 90 |
| | (12) | 2.554 | 2.553 | | | |
| | (11) | 4.228 | 4.225 | | | |
| | (21) | 2.810 | 2.811 | | | |
| | (20) | 3.260 | 3.260 | | | |
| 61 | (02) | 2.764 | 2.764 | 6.42 | 5.53 | 90 |
| | (12) | 2.541 | 2.539 | | | |
| | (11) | 4.193 | 4.190 | | | |
| | (21) | 2.774 | 2.777 | | | |
| | (20) | 3.211 | 3.211 | | | |
| 59 | (02) | 2.759 | 2.759 | 6.33 | 5.52 | 90 |
| | (12) | 2.528 | 2.529 | | | |
| | (11) | 4.158 | 4.159 | | | |
| | (21) | 2.744 | 2.745 | | | |
| | (20) | 3.164 | 3.164 | | | |
| 57 | (02) | 2.754 | 2.754 | 6.30 | 5.51 | 90 |
| | (12) | 2.524 | 2.524 | | | |
| | (11) | 4.147 | 4.147 | | | |
| | (21) | 2.734 | 2.735 | | | |
| | (20) | 3.151 | 3.151 | | | |
| 55 | (02) | 2.749 | 2.749 | 6.25 | 5.50 | 90 |
| | (12) | 2.515 | 2.516 | | | |
| | (11) | 4.124 | 4.128 | | | |
| | (21) | 2.714 | 2.717 | | | |
| | (20) | 3.125 | 3.125 | | | |
| 53 | (02) | 2.744 | 2.744 | 6.21 | 5.49 | 90 |
| | (12) | 2.511 | 2.510 | | | |
| | (11) | 4.113 | 4.113 | | | |
| | (21) | 2.705 | 2.703 | | | |
| | (20) | 3.105 | 3.105 | | | |
| 51 | (02) | 2.739 | 2.739 | 6.19 | 5.48 | 90 |
| | (12) | 2.503 | 2.504 | | | |
| | (11) | 4.101 | 4.101 | | | |
| | (21) | 2.690 | 2.693 | | | |
| | (20) | 3.093 | 3.093 | | | |
| 49 | (02) | 2.734 | 2.734 | 6.15 | 5.47 | 90 |
| | (12) | 2.499 | 2.498 | | | |
| | (11) | 4.079 | 4.086 | | | |
| | (21) | 2.681 | 2.679 | | | |
| | (20) | 3.074 | 3.074 | | | |
| 47 | (02) | 2.729 | 2.729 | 6.14 | 5.46 | 90 |
| | (12) | 2.494 | 2.494 | | | |
| | (11) | 4.079 | 4.078 | | | |
| | (21) | 2.671 | 2.674 | | | |
| | (20) | 3.068 | 3.068 | | | |

**Supplementary Table 8** Observed and calculated *d*-spacings and lattice parameters at different temperatures for ***15*-012-CN** (*Continued*).

| *T* (°C) | (*hk*) | $d_{obs.}$ (nm) | $d_{calc.}$ (nm) | *a* (nm) | *b* (nm) | *γ* (°) |
|---|---|---|---|---|---|---|
| 45 | (02) | 2.724 | 2.724 | 6.11 | 5.45 | 90 |
| | (12) | 2.486 | 2.488 | | | |
| | (11) | 4.057 | 4.067 | | | |
| | (21) | 2.662 | 2.665 | | | |
| | (20) | 3.055 | 3.055 | | | |

**Supplementary Table 9** Observed and calculated *d*-spacings and lattice parameters at different temperatures for ***16*-012-CN**.

| $T$ (°C) | ($hk$) | $d_{obs.}$ (nm) | $d_{calc.}$ (nm) | $a$ (nm) | $b$ (nm) | $\gamma$ (°) |
|---|---|---|---|---|---|---|
| 66 | (02) | 2.650 | 2.650 | 14.21 | 5.78 | 113.56 |
| | (01) | 5.311 | 5.300 | | | |
| | (11) | 4.341 | 4.341 | | | |
| | (31) | 2.896 | 2.847 | | | |
| | (30) | 4.341 | 4.341 | | | |
| | (40) | 3.277 | 3.256 | | | |
| 64 | (02) | 2.641 | 2.641 | 14.17 | 5.78 | 113.96 |
| | (11) | 4.316 | 4.316 | | | |
| | (31) | 2.891 | 2.826 | | | |
| | (30) | 4.316 | 4.316 | | | |
| | (40) | 3.241 | 3.237 | | | |
| 62 | (02) | 2.636 | 2.636 | 14.09 | 5.77 | 114.02 |
| | (11) | 4.303 | 4.303 | | | |
| | (31) | 2.885 | 2.814 | | | |
| | (30) | 4.291 | 4.291 | | | |
| | (40) | 3.213 | 3.218 | | | |
| 60 | (02) | 2.632 | 2.632 | 14.15 | 5.80 | 114.88 |
| | (11) | 4.279 | 4.279 | | | |
| | (31) | 2.885 | 2.794 | | | |
| | (30) | 4.279 | 4.279 | | | |
| | (40) | 3.193 | 3.209 | | | |
| 58 | (02) | 2.627 | 2.627 | 14.13 | 5.80 | 115.08 |
| | (11) | 4.266 | 4.266 | | | |
| | (31) | 2.885 | 2.784 | | | |
| | (30) | 4.266 | 4.266 | | | |
| | (40) | 3.179 | 3.200 | | | |

**Supplementary Table 10** Observed and calculated *d*-spacings and lattice parameters at different temperatures for ***17*-012-CN**.

| *T* (°C) | (*hk*) | $d_{obs.}$ (nm) | $d_{calc.}$ (nm) | *a* (nm) | *b* (nm) | *γ* (°) |
|---|---|---|---|---|---|---|
| 60 | (02) | 2.790 | 2.790 | 11.36 | 5.77 | 104.70 |
| | (11) | 4.533 | 4.533 | | | |
| | $(2\bar{1})$ | 4.533 | 4.533 | | | |
| | $(3\bar{1})$ | 3.354 | 3.497 | | | |
| 59 | (02) | 2.790 | 2.790 | 11.27 | 5.77 | 104.83 |
| | (11) | 4.519 | 4.519 | | | |
| | $(2\bar{1})$ | 4.506 | 4.519 | | | |
| | $(3\bar{1})$ | 3.332 | 3.478 | | | |
| 58 | (02) | 2.785 | 2.785 | 11.22 | 5.76 | 104.88 |
| | (11) | 4.506 | 4.506 | | | |
| | $(2\bar{1})$ | 4.506 | 4.506 | | | |
| | $(3\bar{1})$ | 3.310 | 3.465 | | | |
| 57 | (02) | 2.785 | 2.785 | 11.22 | 5.76 | 104.88 |
| | (11) | 4.506 | 4.506 | | | |
| | $(2\bar{1})$ | 4.506 | 4.506 | | | |
| | $(3\bar{1})$ | 3.303 | 3.465 | | | |
| 56 | (02) | 2.790 | 2.790 | 11.18 | 5.78 | 104.96 |
| | (11) | 4.506 | 4.506 | | | |
| | $(2\bar{1})$ | 4.492 | 4.506 | | | |
| | $(3\bar{1})$ | 3.295 | 3.460 | | | |
| 55 | (02) | 2.790 | 2.790 | 11.18 | 5.78 | 104.96 |
| | (11) | 4.506 | 4.506 | | | |
| | $(2\bar{1})$ | 4.492 | 4.506 | | | |
| | $(3\bar{1})$ | 3.288 | 3.460 | | | |

**Supplementary Table 11** The total number of molecules in the MD simulation box and the initial sizes of the MD simulation box for ***12*-012-CN** and ***16*-012-CN**.

| Entry | Number of molecules | Size of MD simulation box | | | | | |
|---|---|---|---|---|---|---|---|
| | | *a* (nm) | *b* (nm) | *c* (nm) | *α* (deg) | *β* (deg) | *γ* (deg) |
| ***12*-012-CN** | 560 | 11.82 | 4.40 | 9.26 | 90 | 90 | 90 |
| ***16*-012-CN** | 1280 | 28.20 | 5.00 | 11.54 | 90 | 66 | 90 |

**Supplementary References**